\documentclass{article}

\usepackage{PRIMEarxiv}

\usepackage[utf8]{inputenc}
\usepackage[T1]{fontenc}    
\usepackage{hyperref}       
\usepackage{url}           
\usepackage{booktabs}      
\usepackage{amsfonts}       
\usepackage{nicefrac}      
\usepackage{microtype}     
\usepackage{lipsum}
\usepackage{algorithm}
\usepackage{algorithmic}
\usepackage{fancyhdr}       
\usepackage{graphicx}       
\graphicspath{{media/}}     
\usepackage{enumitem}
\usepackage{multirow} 
\usepackage{caption}  
\usepackage{booktabs}
\usepackage{amsmath}
\usepackage{xcolor}        

\usepackage[numbers,sort&compress]{natbib}
\newcommand{\ourmethod}{AURAD }
\title{\ourmethod: Anatomy–Pathology Unified Radiology Synthesis with Progressive Representations}

\author{%
  \textbf{Shuhan Ding}\textsuperscript{1}\thanks{Equal contribution.} \hspace{1mm}\thanks{Work done during an internship at Microsoft Research Asia.} \quad
  \textbf{Jingjing Fu}\textsuperscript{2}\footnotemark[1] \quad
  \textbf{Yu Gu}\textsuperscript{2}\footnotemark[1] \quad
  \textbf{Naiteek Sangani}\textsuperscript{2} \quad
  \textbf{Mu Wei}\textsuperscript{2} \quad
  \textbf{Paul Vozila}\textsuperscript{2} \\
  \textbf{Nan Liu}\textsuperscript{1}\thanks{Corresponding author.} \quad
  \textbf{Jiang Bian}\textsuperscript{2}\footnotemark[3] \quad
  \textbf{Hoifung Poon}\textsuperscript{2}\footnotemark[3] \\
  \textsuperscript{1}Duke-NUS Medical School \quad
  \textsuperscript{2}Microsoft
}

\begin{document}
\maketitle

\begin{abstract}
Medical image synthesis has become an essential strategy for augmenting datasets and improving model generalization in data-scarce clinical settings. However, fine-grained and controllable synthesis remains difficult due to limited high-quality annotations and domain shifts across datasets. Existing methods, often designed for natural images or well-defined tumors, struggle to generalize to chest radiographs, where disease patterns are morphologically diverse and tightly intertwined with anatomical structures. To address these challenges, we propose \textbf{\ourmethod}, a controllable radiology synthesis framework that jointly generates high-fidelity chest X-rays and pseudo semantic masks. Unlike prior approaches that rely on randomly sampled masks—limiting diversity, controllability, and clinical relevance—our method learns to generate masks that capture \textbf{multi-pathology coexistence and anatomical–pathological consistency}. It follows a progressive pipeline: pseudo masks are first generated from clinical prompts conditioned on anatomical structures, and then used to guide image synthesis.  We also leverage pretrained expert medical models to filter outputs and ensure clinical plausibility. Beyond visual realism, the synthesized masks also serve as labels for downstream tasks such as detection and segmentation, bridging the gap between generative modeling and real-world clinical applications. Extensive experiments and blinded radiologist evaluations demonstrate the effectiveness and generalizability of our method across tasks and datasets. In particular, 78\% of our synthesized images are classified as authentic by board-certified radiologists, and over 40\% of predicted segmentation overlays are rated as clinically useful. All code, pre-trained models, and the synthesized dataset will be released upon publication.
\end{abstract}

\section{Introduction}\label{sec:introduction}

\begin{figure}[t]
\centering
\includegraphics[width=0.9\linewidth]{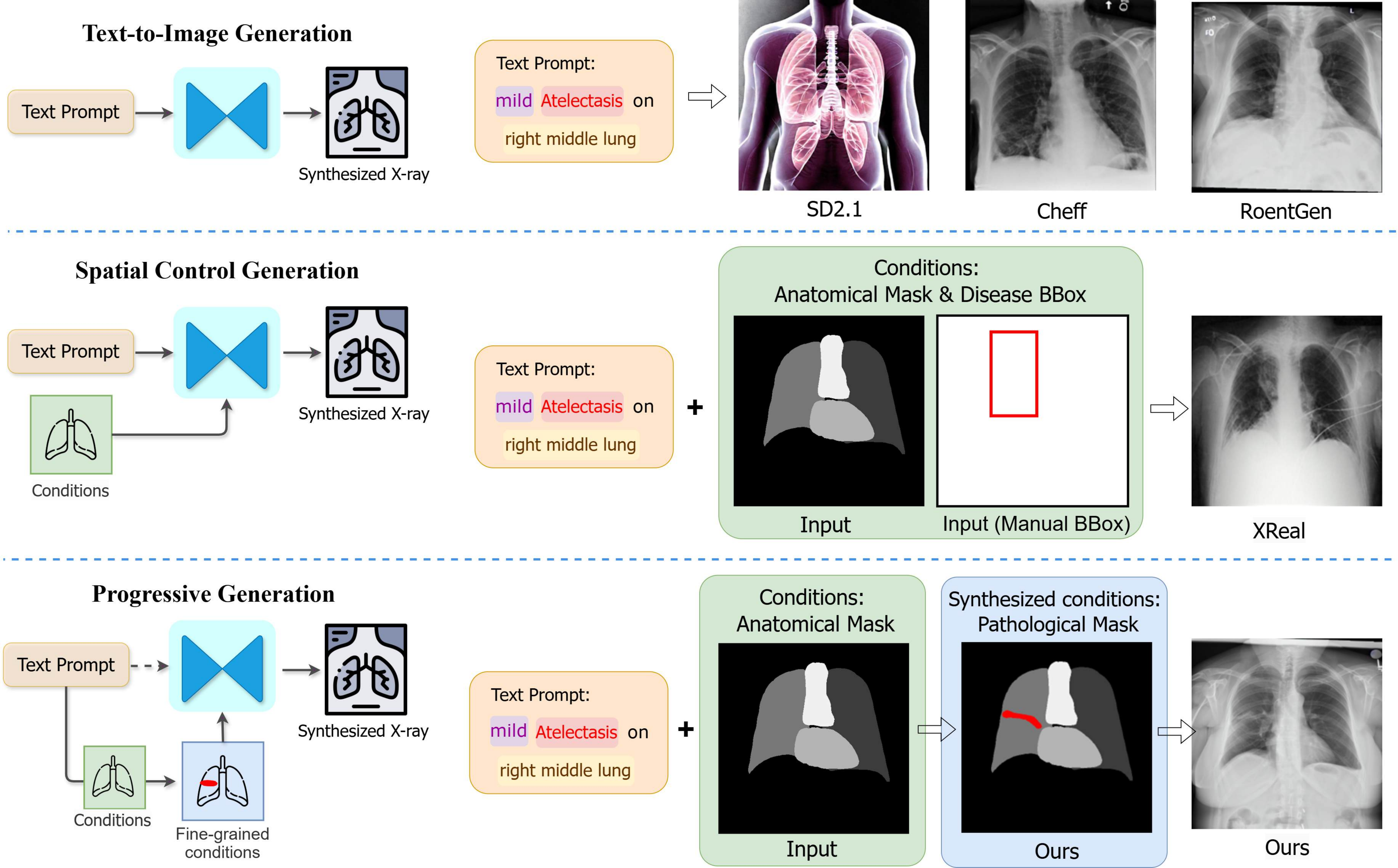}
    \caption{Overview of CXR synthesis strategies. Text-driven methods lack spatial control and medical priors; conditional methods depend on manual inputs. Our framework infers pathology–anatomy relations from text and organ masks, enabling clinically consistent, mask-guided generation.}
\label{fig:teaser}
\end{figure}

Medical image analysis faces challenges due to complex data acquisition, privacy constraints, and the high cost of expert annotations~\cite{prevedello2019challenges}. To mitigate these challenges, synthesized data generation has emerged as a promising strategy to expand datasets, enhance diversity, and reduce reliance on large-scale manual annotation~\cite{kokosi2022synthetic, giuffre2023harnessing}.

Chest X-ray (CXR), a widely used imaging modality, has been a focal point for medical image synthesis~\cite{bluethgen2024vision, huang2024chest, gu2023biomedjourney}. While GAN-based approaches~\cite{madani2018semi, yang2020xraygan} have made progress, they often struggle with fidelity, diversity, and spatial controllability~\cite{albahli2020efficient}. Recently, diffusion models~\cite{rombach2022high, dhariwal2021diffusion} have demonstrated superior performance, with models like Cheff~\cite{weber2023cascaded} and RoentGen~\cite{bluethgen2024vision} producing more realistic, free-form text-guided CXR images.
However, text prompts alone are insufficient to capture the complex spatial relationships between anatomical structures and disease patterns. Recent methods incorporating spatial conditioning—such as manual sketches or bounding boxes~\cite{perez2024radedit, hashmi2024XReal}—remain coarse, anatomically unspecific, and clinically limited.

Inspired by the success of mask-based generation in tumor synthesis~\cite{hu2023label, chen2024towards} and spatial controllability achieved in text-to-layout pipelines for natural images~\cite{zheng2023layoutdiffusion, qu2023layoutllm}, we attempt to obtain disease masks through a progressive generation process, which then serve as structural priors to guide image synthesis. However, existing assumptions are often based on relatively clear boundaries and spatially independent regions. In CXR, pathological patterns exhibit ambiguous edges, coexisting diseases, and strong anatomical dependencies. This raises the key challenge: how can we generate clinically meaningful masks that preserve the intricate pathology–anatomy consistency of real CXRs, and use them to control and enhance image synthesis?

Furthermore, the utility of synthesized datasets remains a significant bottleneck. While classification has advanced~\cite{rajpurkar2017chexnet, irvin2019chexpert}, tasks like detection and segmentation still lack sufficient annotated data~\cite{lian2021structure, islam2017abnormality, hao2024yolo}. Prior studies~\cite{prakash2024evaluating} explored downstream use but were restricted to single-disease, well-annotated cases.

To overcome these limitations, we propose \textbf{A}natomy–Pathology \textbf{U}nified \textbf{RAD}iology Synthesis, referred to as \textbf{\ourmethod}, a novel generative framework designed to synthesize high-fidelity CXR images with explicit spatial and semantic control. Our approach utilizes a progressive pipeline. By incorporating intermediate mask representations, \ourmethod facilitates accurate disease localization and enhances model interpretability. Furthermore, it can be effectively trained with limited annotated data and generate diverse, reusable labels, making it well-suited for low-resource clinical environments. We conduct extensive experiments and expert radiologist evaluations to evaluate the quality of our synthesized data both quantitatively and qualitatively.

The main contributions of our work are summarized as follows:

\begin{itemize}[leftmargin=1cm]

\item \textbf{Anatomy–pathology integration:} \ourmethod is the first framework to explicitly model spatial relationships between pulmonary diseases and anatomical structures, generating clinically interpretable dense multi-disease masks. These masks preserve \textbf{pathology–anatomy consistency}, support coexisting conditions, and serve as spatial priors to guide controllable and anatomically grounded CXR synthesis.

\item \textbf{Paired data synthesis:} It jointly generates paired CXR images and mask-level pathology annotations, reducing annotation burden and enabling supervised downstream training.

\item \textbf{Bias-aware augmentation:} By generating diverse samples across disease types and anatomical variants, \ourmethod  enhances model robustness across heterogeneous clinical populations.

\item \textbf{Medical knowledge guidance:} Leveraging domain-specific priors and pretrained models mitigates error accumulation across the multi-stage pipeline and boosts both visual fidelity and task-specific effectiveness.

\end{itemize}

\section{Related Work}\label{sec:related_work}

\subsection{Medical Image Synthesis Models}

Generative models have played a pivotal role in medical image synthesis, spanning modalities such as CT, MRI, and X-ray. Early efforts leveraged GANs for tasks such as modality translation~\cite{yang2020mri, zhang2018translating}, lesion simulation~\cite{qin2020gan, bissoto2021gan}, and data augmentation~\cite{frid2018gan, chlap2021review}. 

More recently, diffusion models have emerged as a more stable and controllable alternative across diverse modalities and pathologies. Subsequent advances have integrated text conditioning~\cite{bluethgen2024vision, weber2023cascaded} and spatial priors~\cite{han2023medgen3d,du2024boosting, jin2024maskmedpaint}, enabling synthesis guided by radiology reports or masks. These guidance strategies have also been explored in dermatology~\cite{du2024boosting} and colonoscopy~\cite{jin2024maskmedpaint}, where fine-grained lesion editing. These advances underscore a shift from purely realistic generation toward controllable and semantically aligned synthesis.

\subsection{Synthesized Data for Image Analysis}

Beyond generation itself, synthesized medical images have been increasingly used to enhance downstream image analysis, particularly in scenarios of data scarcity or imbalance. In the industrial domain, synthesized images are widely used for defect detection and visual quality control, where collecting diverse examples of rare defects is often infeasible. Prior work~\cite{lu2023removing, haselmann2019pixel} has shown that generative models can convert unsupervised anomaly detection into a supervised task, enabling finer-grained scoring of visual anomalies. Similarly, in medical imaging, synthetic data has been leveraged to enhance classification, detection, and segmentation performance~\cite{qin2022learning, hamghalam2024medical, wolleb2022diffusion, zhang2024diffboost}.

While most existing CXR studies focus on classification~\cite{rajpurkar2017chexnet, irvin2019chexpert}, more complex tasks such as abnormality detection and segmentation remain underexplored due to the lack of fine-grained annotations~\cite{lian2021structure, hao2024yolo}. Our intermediate representation not only enables anatomically grounded image generation, but also serves as explicit supervision, effectively bridging the gap in segmentation and detection research.

\subsection{Modeling Anatomy–Pathology Relationships}

For clinical utility, a significant challenge is the realistic simulation of diseases. Prior mask-guided methods such as MedGen3D~\cite{han2023medgen3d} focus solely on organ masks without modeling diseased regions, limiting their diagnostic value. Disease simulation has been more actively explored in oncology, yet existing tumor synthesis approaches remain rule-based, generating tumor masks through deformation or sampling, and they fail to ensure alignment with anatomical structures. COMG~\cite{gu2024complex} attempts to bridge this gap by linking disease names to organ masks via a manually designed knowledge graph. However, this name-to-organ mapping is overly coarse and unable to capture distinctive morphopathological characteristics.

A side-by-side comparison of these representative approaches with ours is summarized in Table~\ref{tab:related_work_comparison}, underscoring that, unlike organ-only or rule-based designs, AURAD explicitly models diseases in addition to anatomy, enforces pathology–anatomy consistency, and enables fine-grained spatial control—three aspects that remain insufficiently addressed in prior work.

\begin{table}[t]
\centering
\small
\caption{
Comparison of representative methods in terms of disease modeling, pathology–anatomy alignment, and fine-grained controllability.
}
\vspace{2pt}
\resizebox{0.9\linewidth}{!}{
\setlength{\tabcolsep}{3pt}
\renewcommand{\arraystretch}{1.15}
\begin{tabular}{lcccc}
\toprule
\textbf{Aspect} & \textbf{MedGen3D~\cite{han2023medgen3d}} & \textbf{DiffTumor~\cite{chen2024towards}} & \textbf{COMG~\cite{gu2024complex}} & \textbf{Ours (AURAD)} \\
\midrule
Task & Organ mask + CT gen. & Rule-based tumor mask + CT gen. & Report generation & Disease–organ mask + CXR gen. \\
Mask Conditions & Organ only & Tumor only & Organ only & Syn. disease + organ mask \\
Disease–Organ Relationship & \texttimes & \texttimes & Disease-to-organ lookup & Dense spatial mask (learned) \\
Fine-grained Control & \texttimes & \texttimes\ Early-stage tumor only & \texttimes\ Disease type only & \checkmark\ 14-disease, 11-zone, 3-severity \\
\bottomrule
\end{tabular}}
\label{tab:related_work_comparison}
\end{table}

\section{Method}\label{sec:method}

\subsection{Limitations of Existing Medical Image Generation Approaches} 

\begin{figure}[t]
  \centering
  \includegraphics[width=0.8\textwidth]{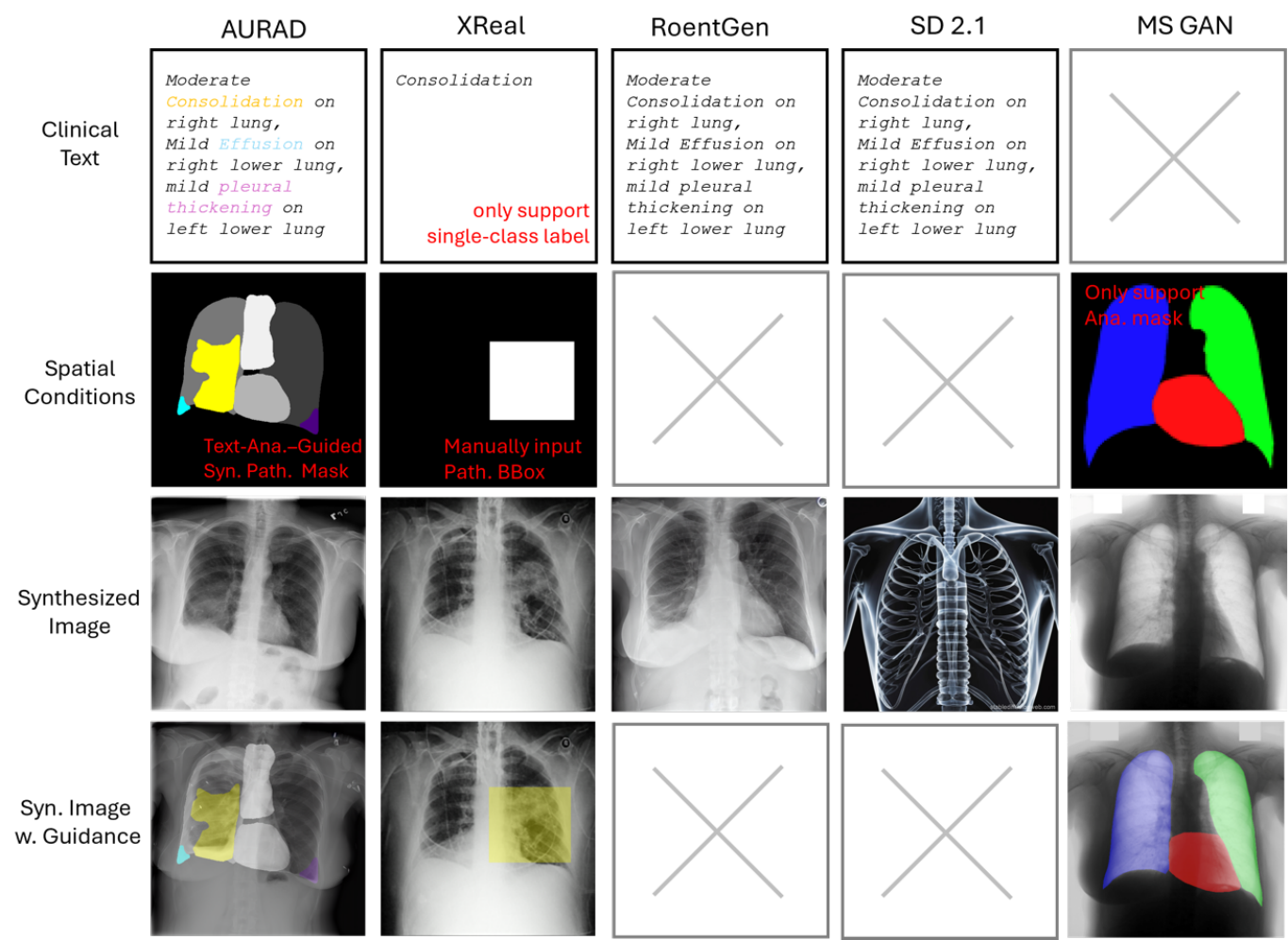}
  \caption{Comparison of synthesis capabilities and limitations across existing CXR generation methods. \textit{Path.: pathological; Ana.: anatomical; MS-GAN: from~\cite{ciano2021multi}}}
  \label{fig:method_diff}
\end{figure}

Recent advances in generative modeling—from GANs to text-conditioned diffusion models—have enabled realistic medical image synthesis, primarily to address data scarcity and enhance training workflows. However, current approaches often fall short in several critical aspects required for clinical relevance and robustness. For reference, Figure~\ref{fig:method_diff} demonstrates a comparison of synthesis results from recent works.
\paragraph{Limitation 1: Lack of Semantic Control in Image-Only Generation.}
Early medical image generation methods, particularly those based on GANs~\cite{ciano2021multi}, focus on visual fidelity without semantic conditioning. These models often generate plausible images but lack alignment with diagnostic concepts such as pathology type, anatomical location, or severity. In clinical settings, where interpretability is paramount, such unstructured synthesis holds limited practical value.

\paragraph{Limitation 2: Domain Gap in General-Purpose Text-to-Image Models.}
Diffusion models pretrained on general-domain datasets (e.g., DALL·E~\cite{radford2021learning}, Stable Diffusion~\cite{podell2023sdxl}) exhibit limited transfer to medical imaging. These models struggle with grayscale textures, modality-specific patterns, and medical terminology, often hallucinating features or misrepresenting pathology~\cite{ali2022spot, chambon2022adapting}. Domain-adapted variants like RoentGen~\cite{bluethgen2024vision} and BiomedJourney~\cite{gu2023biomedjourney} improve relevance via continued pretraining, but still lack mechanisms to accurately encode spatial relations, leading to anatomically inconsistent outputs.

\paragraph{Limitation 3: Inadequacy of Textual Prompts for Spatial Precision.}
Clinical image interpretation hinges on precise anatomical localization—subtle differences in lesion placement can imply drastically different diagnoses~\cite{lee2022localization}. However, free-text prompts are inherently under-specified for capturing such fine-grained spatial detail. Most text-to-image pipelines treat language as flat input, without enforcing alignment between phrases and spatial layouts, resulting in outputs that often misplace or blur critical findings.

\paragraph{Limitation 4: Bounding-Box Guidance is Coarse and Non-Scalable.}
To introduce spatial control, several methods (e.g., X-ReaL~\cite{hashmi2024XReal}, RadEdit~\cite{perez2024radedit}) use bounding boxes as generation constraints. While this improves anatomical targeting, the guidance is coarse, typically limited to region- or organ-level annotations. Fine-grained pathologies such as spiculated nodules or infiltrative patterns fall outside this control. Moreover, bounding boxes are usually manually annotated or heuristically derived, limiting scalability and full automation.

\paragraph{Limitation 5: Segmentation-Based Guidance Adds Accuracy—at a Cost.}
Segmentation masks provide pixel-level spatial structure, enabling high-fidelity control over anatomical and pathological regions. Methods like MedSegDiff~\cite{wu2024medsegdiff} leverage such masks for precise and interpretable synthesis. However, these approaches require accurate, often disease-specific segmentation models—introducing annotation burden, propagation of upstream errors, and greater system complexity. This reduces robustness and scalability, particularly in settings with limited ground-truth labels or unseen pathologies.

\subsection{Method Overview}
\begin{figure}[t]
\centering
\includegraphics[width=0.9\linewidth]{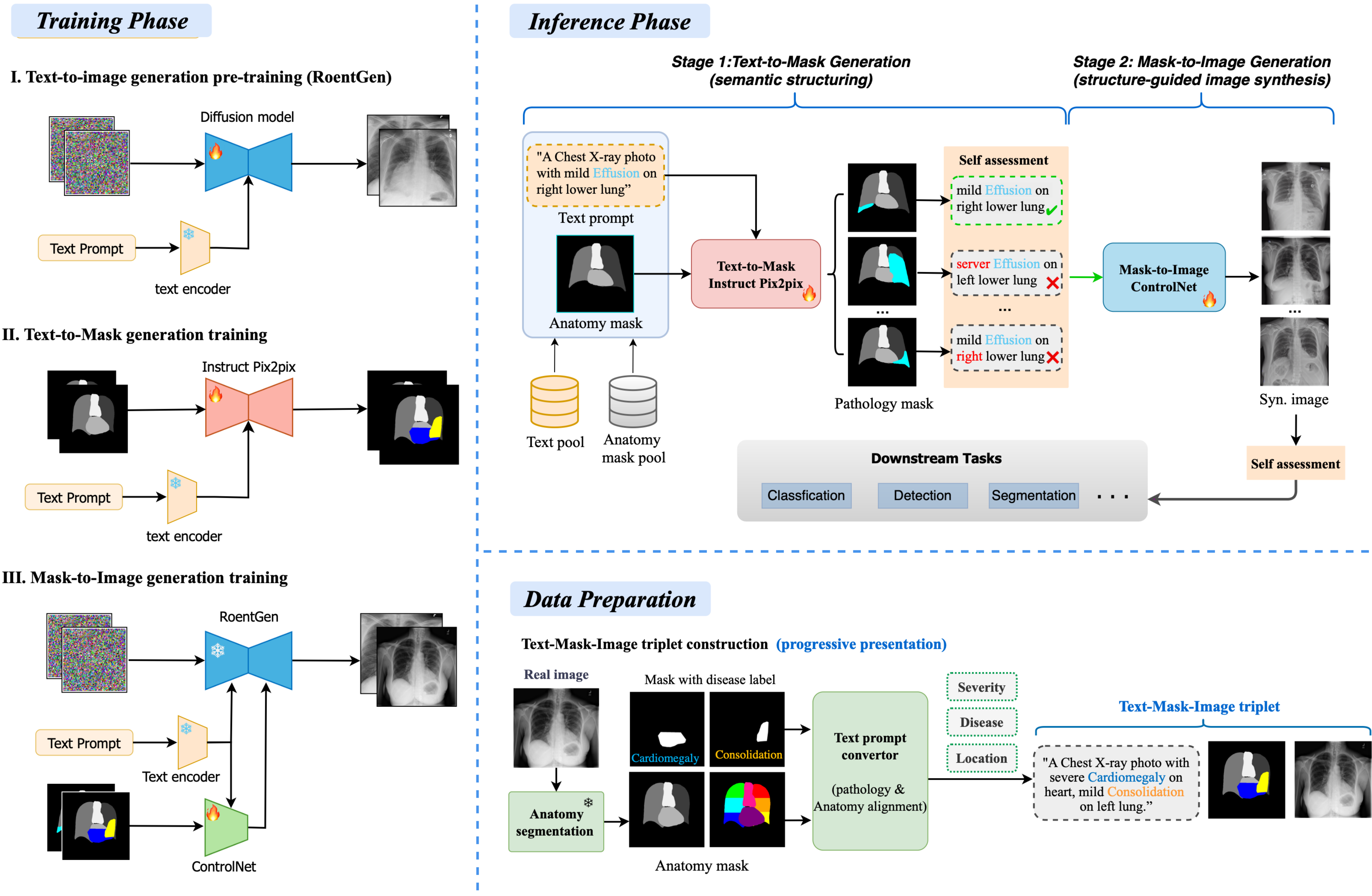}
\caption{The \ourmethod pipeline consists of three phases: data preparation, training and inference. Data preparation forms the foundation of the two-stage generation process, creating data triplets that map from text prompts to masks and from masks to images. Additionally, a self-assessment mechanism is integrated at each generation stage to ensure the quality of synthesized outputs. (Phase~I illustrates the RoentGen training process, but we adopt the pre-trained model without fine-tuning.)}
\label{fig:pipeline}
\end{figure}

To address the challenges outlined in Section~3.1—specifically the lack of semantic alignment (\textit{Limit. 1}), poor domain adaptation in general text-to-image models (\textit{Limit. 2}), and limited spatial control (\textit{Limit. 3–5})—we propose \textbf{\ourmethod} (Anatomy–Pathology Unified Radiology Synthesis), a fully automatic pipeline for medical image synthesis, shown in Figure~\ref{fig:pipeline}. \ourmethod decomposes the generation task into two stages: (1) \textit{semantic structuring}, which captures clinical semantics and spatial layout via segmentation masks, and (2) \textit{structure-guided image synthesis}, which renders high-fidelity images from those masks. This explicit intermediate representation improves interpretability and control, allowing \ourmethod to align image content with both textual and anatomical intent.

\subsubsection{Semantic Structuring: Text-to-Mask Generation}

To overcome \textbf{Limit. 1–3}, we introduce a text-to-mask generation stage that captures both the presence (\textit{what}) and location (\textit{where}) of pathologies. This structured intermediate allows the model to reason spatially and semantically before rendering pixel-level images.
While text-to-layout models for natural scenes~\cite{zheng2023layoutdiffusion, chai2023layoutdm} provide inspiration, they fall short in medical domains, where pathology distribution is heavily governed by anatomical priors—e.g., pleural effusion typically occurs in lung base zones, and cardiomegaly is diagnosed by heart-to-thorax width ratios.

We adapt InstructPix2Pix~\cite{brooks2023instructpix2pix}, a diffusion-based instruction-tuned model, to transform organ-only anatomical masks into pathology-augmented segmentation maps. Disease descriptions are treated as natural language instructions, guiding the placement of pathology relative to anatomy. This results in masks that are semantically rich and spatially plausible.
To ensure quality and alignment, we introduce a self-assessment mechanism. The generated masks are converted into captions using our custom text prompt converter, which leverages the overlay of pathological and organ masks. These predicted captions are then compared to the original input prompts. Any inconsistencies identified in the generation are discarded, and the process is repeated until alignment is achieved. This process yields masks that are interpretable, accurate, and tightly aligned with clinical prompts.

\subsubsection{Structure-Guided Image Synthesis: Conditional Image Generation}

To address \textbf{Limit. 4 and 5}, we convert semantic masks into high-resolution radiological images using a spatially controllable generative model. Unlike methods that depend on manually provided bounding boxes~\cite{hashmi2024XReal} or segmentation maps~\cite{prakash2024evaluating}, \ourmethod learns spatial guidance automatically and applies it consistently.
We extend RoentGen~\cite{bluethgen2024vision},  a domain-specific text-to-image model, with ControlNet~\cite{zhang2023adding}—a plug-in module that conditions generation on segmentation masks. The mask is encoded via a dedicated convolutional branch, and control signals are injected into RoentGen's frozen U-Net through zero convolution layers.

This architecture preserves RoentGen’s high-quality image generation while enforcing pixel-level spatial control. It ensures that pathology is rendered in the correct location and with correct morphology—essential in clinical use cases where anatomical accuracy is non-negotiable.
Furthermore, the model supports \textit{interactive generation}: users can provide custom masks to generate specific findings or simulate counterfactual scenarios. This flexibility makes \ourmethod suitable for data augmentation, rare disease modeling, and hypothesis-driven image synthesis.

\section{Data}\label{sec:data}

\subsection{Dataset Construction: Structured Prompts and Segmentation Masks}

To support spatially and semantically grounded image synthesis, we construct a multi-modal dataset comprising CXR images, anatomical segmentation masks, pathology masks, and structured textual prompts. Unlike prior mask-based datasets~\cite{lian2021structure, feng2021curation}, which lack aligned clinical reports, we design a simple rule-based conversion tool (Appendix \ref{appendix:prompt_tool}) that calculates the overlay area between pathology and organ masks to derive three key attributes: \textbf{Class} (e.g., effusion, cardiomegaly), \textbf{Location} (e.g., left upper lobe, right middle lung), and \textbf{Severity} (mild, moderate, severe). This process produces fine-grained, clinically meaningful structured prompts.

In the constructed dataset, pathology masks are sourced from ChestX-Det~\cite{liu2020chestxdet10} and CANDID-PTX~\cite{feng2021curation}, while the corresponding anatomical masks are extracted using a pre-trained segmentation model~\cite{cohen2022torchxrayvision}. Disease severity is calculated using geometric heuristics; for example, cardiomegaly is defined as:
\textbf{Mild:} cardiothoracic ratio $<$ 0.50;
\textbf{Moderate:} 0.50–0.55;
\textbf{Severe:} $>$ 0.55. 
For more implementation details on attribute extraction, please refer to the Appendix.

The ChestX-Det and CANDID-PTX datasets provide pathology masks but lack corresponding clinical reports. To address this, we generate text prompts in a structured, report-style format using the following fixed template: \textit{``A photo of a Chest X-ray with [SEVERITY1] [CLASS1] on [LOC1], [SEVERITY2] [CLASS2] on [LOC2], ...''}. This structured format eliminates linguistic ambiguity and enhances learnability, particularly in low-data regimes. Finally, we construct high-quality (prompt, mask, image) triplets. Additionally, to improve anatomical diversity during inference, we sample supplementary anatomical masks from the MIMIC-CXR dataset~\cite{johnson2019mimic}.

\subsection{Quality Filtering and Augmentation Pipeline}
Given the subtle nature of radiographic features, stringent quality filtering is crucial for maintaining anatomical and semantic consistency in synthesized samples. To achieve this, we introduce a self-assessment mechanism to ensure high-quality outputs at each stage of the generation process. This mechanism leverages available medical tools and domain knowledge to evaluate the synthesized images across multiple dimensions. Consequently, our filtering pipeline consists of the following components:

\begin{itemize}[leftmargin=0.8cm]
    \item \textbf{Prompt-Match Validation}: Each generated mask is re-captioned using the prompt tool and compared against the original input. Mismatched samples are discarded and regenerated.

    \noindent\item \textbf{Realism Assessment}: A lightweight binary discriminator filters out low-fidelity images, reducing artifacts and unrealistic text markers (e.g., “L” or “R” labels) introduced during synthesis.
    
    \item \textbf{Diagnostic Reliability}: A pre-trained multi-label CXR classifier~\cite{cohen2022torchxrayvision} verifies the presence and anatomical consistency of pathologies in the generated images.
    
\end{itemize}

We avoid general foundation model filters (e.g., BioMedCLIP~\cite{zhang2023biomedclip}), which have been shown to reduce sample diversity and downstream performance~\cite{prakash2024evaluating}. Instead, we prioritizes realism and pathology-specific filtering to retain clinically meaningful variation. This is exemplified in Figure~\ref{fig:pipeline}, where synthesized pathology masks undergo re-captioning to ensure alignment validation.

\vspace{-2mm}

\section{Experiments}\label{sec:exp}

\subsection{Experimental Setup}

\paragraph{Datasets.} 
We utilize two CXR datasets, namely ChestX-Det~\cite{liu2020chestxdet10} and CANDID-PTX~\cite{feng2021curation}, for training, and ChestX-Det, Chest X-ray14~\cite{wang2017chestx}, VinDr-CXR~\cite{nguyen2020vindr}, and MIMIC-CXR~\cite{johnson2019mimic} for testing. The datasets used in our experiments are comprehensively detailed in Appendix \ref{appendix:dataset}. For the training phase, we integrated the ChestX-Det training set with 1000 randomly selected abnormal pairs from the CANDID-PTX dataset, culminating in a total of 4001 images. The ChestX-Det test set, comprising 542 images, is reserved for evaluating detection and segmentation tasks. To further validate the model's performance on downstream tasks, we employ the Chest X-ray14 and VinDr-CXR test sets for multi-disease detection. In addition, we assess classification performance using the MIMIC-CXR test set.

\paragraph{Metrics.}
The quality and clinical relevance of generated chest X-rays are evaluated using four complementary metrics. Image fidelity is measured by Fréchet Inception Distance (FID)\cite{heusel2018ganstrainedtimescaleupdate}, computed with an XRV encoder\cite{Cohen2022xrv} to better capture domain-specific features, instead of the standard ImageNet-pretrained InceptionV3. \textbf{Structural consistency} is assessed via MS-SSIM, reflecting anatomical preservation. Semantic alignment is quantified using CLIP~\cite{radford2021clip} scores (ViT-B/32), based on image–text embedding similarity. Diagnostic accuracy is evaluated through multi-label disease classification using a pretrained model from TorchXRayVision~\cite{Cohen2022xrv}, indicating whether generated images exhibit meaningful pathological features. These metrics jointly capture fidelity, structure, semantics, and task-specific relevance.

\subsection{Radiologist Study: Realism and Diagnostic Value}\label{sec:radio_study}

\paragraph{Radiologist Cohort.} We recruited 3 radiologists with 5-15 years of post-certification experience in thoracic imaging. All were active practitioners uninvolved in model development.

\paragraph{Evaluation Data.} Two evaluation sets were used. The first included 20 patients to evaluate visual realism. The second study included 50 patients, where radiologists evaluated disease masks predicted by a downstream Mask R-CNN trained with our synthetic data augmentation. Importantly, they assessed these model-predicted masks rather than the raw synthetic masks generated by our framework. Both sets are randomly sampled and preserved the disease distribution of the original set, including normal, single-disease, and multi-disease cases. 

\paragraph{Study Protocol.} Each radiologist reviewed a randomized mix of real and \ourmethod synthesized images, with sources blinded. More details of the implementation and the interface are shown in Appendix \ref{appendix:radiologist}. For each image, they completed tasks:

\begin{itemize}[leftmargin=0.3cm]
    \item \textbf{Realism Rating:} Rate whether the image appears real or synthesized on a 5-point scale: (1) Definitely synthesized, (2) Probably synthesized, (3) Not sure, (4) Probably real, (5) Definitely real.
    \item \textbf{Segmentation Helpfulness:} Rate how helpful the overlaid disease-area segmentation is for diagnosis (0 = Not helpful, 1 = Helpful).
\end{itemize}

\begin{figure*}[!t]
    \scriptsize
    \centering
    \begin{minipage}[t]{0.99\linewidth}  
    
        \begin{minipage}[t]{0.139\linewidth}
            \centerline{\includegraphics[width=1\linewidth]{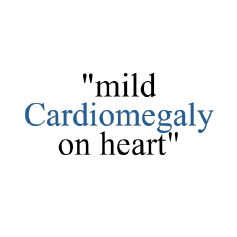}}
        \end{minipage}
        \begin{minipage}[t]{0.139\linewidth}
            \centering
            \centerline{\includegraphics[width=1\linewidth]{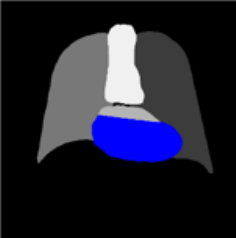}}
        \end{minipage}
        \begin{minipage}[t]{0.139\linewidth}
            \centering
            \centerline{\includegraphics[width=1\linewidth]{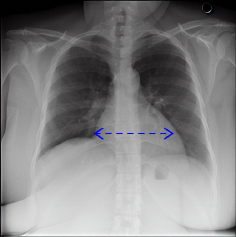}}
        \end{minipage}
        \begin{minipage}[t]{0.139\linewidth}
            \centering
            \centerline{\includegraphics[width=1\linewidth]{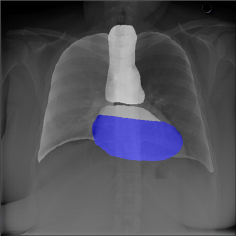}}
        \end{minipage}
        \begin{minipage}[t]{0.139\linewidth}
            \centering
            \centerline{\includegraphics[width=1\linewidth]{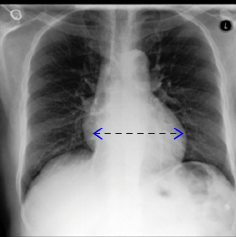}}
        \end{minipage}
        \begin{minipage}[t]{0.139\linewidth}
            \centering
            \centerline{\includegraphics[width=1\linewidth]{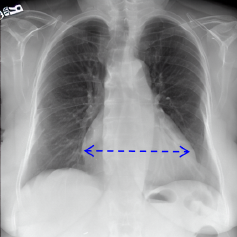}}
        \end{minipage}
        \begin{minipage}[t]{0.139\linewidth}
            \centering
            \centerline{\includegraphics[width=1\linewidth]{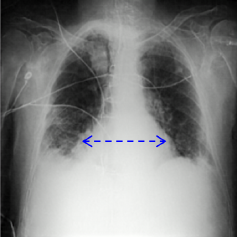}}
        \end{minipage}

        \begin{minipage}[t]{0.139\linewidth}
            \centerline{\includegraphics[width=1\linewidth]{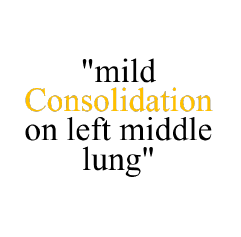}}
        \end{minipage}
        \begin{minipage}[t]{0.139\linewidth}
            \centering
            \centerline{\includegraphics[width=1\linewidth]{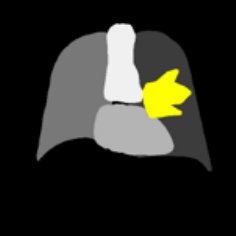}}
        \end{minipage}
        \begin{minipage}[t]{0.139\linewidth}
            \centering
            \centerline{\includegraphics[width=1\linewidth]{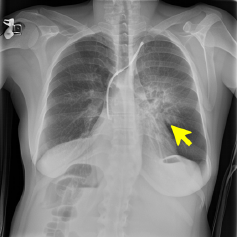}}
        \end{minipage}
        \begin{minipage}[t]{0.139\linewidth}
            \centering
            \centerline{\includegraphics[width=1\linewidth]{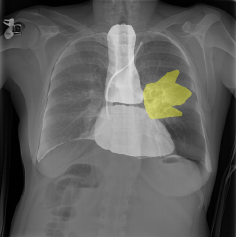}}
        \end{minipage}
        \begin{minipage}[t]{0.139\linewidth}
            \centering
            \centerline{\includegraphics[width=1\linewidth]{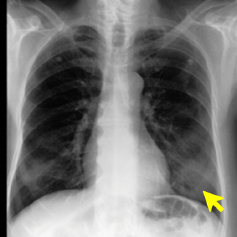}}
        \end{minipage}
        \begin{minipage}[t]{0.139\linewidth}
            \centering
            \centerline{\includegraphics[width=1\linewidth]{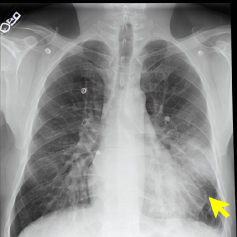}}
        \end{minipage}
        \begin{minipage}[t]{0.139\linewidth}
            \centering
            \centerline{\includegraphics[width=1\linewidth]{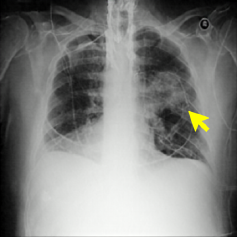}}
        \end{minipage}

        \begin{minipage}[t]{0.139\linewidth}
            \centerline{\includegraphics[width=1\linewidth]{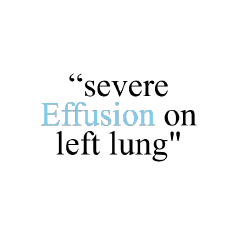}}
        \end{minipage}
        \begin{minipage}[t]{0.139\linewidth}
            \centering
            \centerline{\includegraphics[width=1\linewidth]{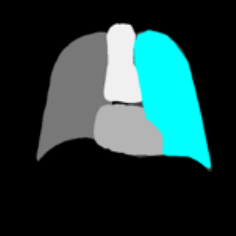}}
        \end{minipage}
        \begin{minipage}[t]{0.139\linewidth}
            \centering
            \centerline{\includegraphics[width=1\linewidth]{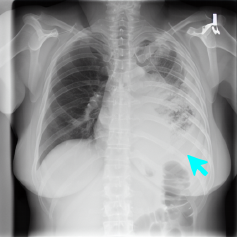}}
        \end{minipage}
        \begin{minipage}[t]{0.139\linewidth}
            \centering
            \centerline{\includegraphics[width=1\linewidth]{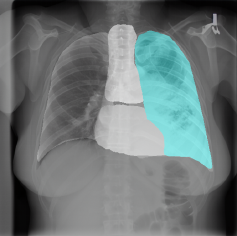}}
        \end{minipage}
        \begin{minipage}[t]{0.139\linewidth}
            \centering
            \centerline{\includegraphics[width=1\linewidth]{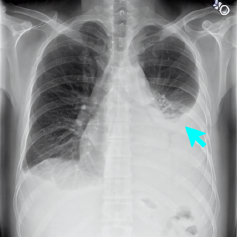}}
        \end{minipage}
        \begin{minipage}[t]{0.139\linewidth}
            \centering
            \centerline{\includegraphics[width=1\linewidth]{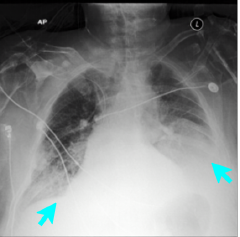}}
        \end{minipage}
        \begin{minipage}[t]{0.139\linewidth}
            \centering
            \centerline{\includegraphics[width=1\linewidth]{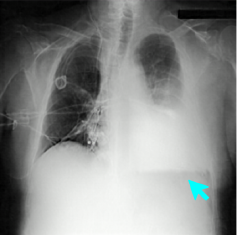}}
        \end{minipage}

        \begin{minipage}[t]{0.139\linewidth}
            \centering
            \centerline{\includegraphics[width=1\linewidth]{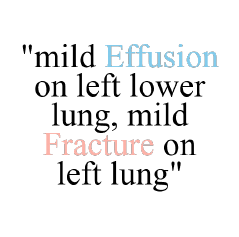}}
            \centerline{\parbox{1\linewidth}{\centering }}
        \end{minipage}
        \begin{minipage}[t]{0.139\linewidth}
            \centering
            \centerline{\includegraphics[width=1\linewidth]{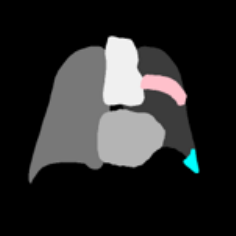}}
            \centerline{\parbox{1\linewidth}{\centering(a) Our Syn. Mask}}
        \end{minipage}
        \begin{minipage}[t]{0.139\linewidth}
            \centering
            \centerline{\includegraphics[width=1\linewidth]{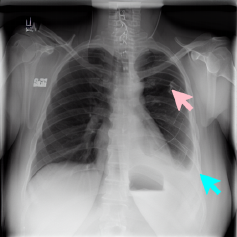}}
            \centerline{\parbox{1\linewidth}{\centering(b) Our Syn. Image}}
        \end{minipage}
        \begin{minipage}[t]{0.139\linewidth}
            \centering
            \centerline{\includegraphics[width=1\linewidth]{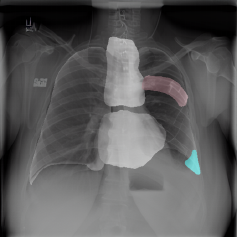}}
            \centerline{\parbox{1\linewidth}{\centering(c) Our Overlay}}
        \end{minipage}
        \begin{minipage}[t]{0.139\linewidth}
            \centering
            \centerline{\includegraphics[width=1\linewidth]{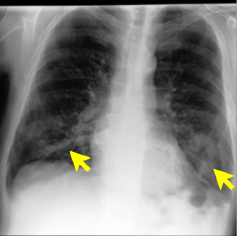}}
            \centerline{\parbox{1\linewidth}{\centering(d) Cheff}}
        \end{minipage}
        \begin{minipage}[t]{0.139\linewidth}
            \centering
            \centerline{\includegraphics[width=1\linewidth]{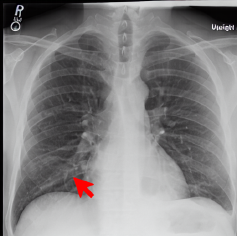}}
            \centerline{\parbox{1\linewidth}{\centering(e) RoentGen}}
        \end{minipage}
        \begin{minipage}[t]{0.139\linewidth}
            \centering
            \centerline{\includegraphics[width=1\linewidth]{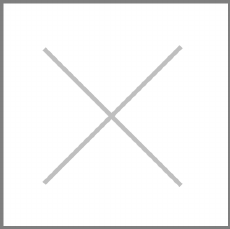}}
            \centerline{\parbox{1\linewidth}{\centering(f) XReal}}
        \end{minipage}
    \end{minipage}

    \caption{Qualitative comparison of CXR synthesis methods. Colors denote different disease types; arrows of the same color indicate corresponding regions.}
    \vspace{-2mm}
    \label{fig:vis_main}
\end{figure*}

\begin{table}[t]
    \centering
    \caption{Quantitative evaluation of generated CXR images. Classification AUC is averaged over five diseases. Best in \textbf{bold}.}
    \resizebox{0.5\textwidth}{!}{
    \begin{tabular}{lccccc}
        \toprule
        Method & FID↓ & CLIP Score↑ & MS-SSIM↑ & Path. AUC↑ \\
        \midrule
        Real & - & 29.59 & - & 57.64 \\
        Cheff & 59.37 & \textbf{30.64} & 0.4290 & 52.34 (↓5.30)\\
        RoentGen & 56.39 & 29.65 & 0.4031 & 56.51 (↓1.13)  \\
        Ours & \textbf{32.83} & 30.14 & \textbf{0.5968} & \textbf{59.31} (↑1.67)\\
        \bottomrule
    \end{tabular}}
    \label{tab:quantitative_evaluation}
    \vspace{-6pt}
\end{table}

\subsection{Main Results}
We compare \ourmethod with recent CXR synthesis approaches, including Cheff~\cite{weber2023cascaded}, RoentGen~\cite{bluethgen2024vision}, and XReal~\cite{hashmi2024XReal}, using both quantitative metrics and qualitative assessments.

\paragraph{Quantitative Results.} 
As shown in Table~\ref{tab:quantitative_evaluation}, \ourmethod achieves the lowest FID of 32.83 using the in-domain XRV encoder, outperforming both RoentGen and Cheff in image fidelity. While CLIP similarity scores are comparable across methods, \ourmethod significantly surpasses others in MS-SSIM, indicating stronger preservation of anatomical structures—an advantage attributable to our anatomy-aware conditioning. In addition, our approach yields the highest classification accuracy of 59.31\%, slightly exceeding the real-data baseline. Importantly, all results are obtained using only 4k training samples, underscoring the efficiency and effectiveness of \ourmethod in generating high-fidelity, anatomically coherent, and diagnostically meaningful CXRs.

\paragraph{Qualitative Evaluation.}
In medical imaging, realism cannot be fully assessed through quantitative metrics alone. Visual inspection remains essential for evaluating anatomical plausibility and pathological fidelity, as illustrated in Figure~\ref{fig:vis_main}. For fair comparison, we use identical text prompts for Cheff and RoentGen, and provide XReal with ground-truth anatomical masks and randomly sampled disease bounding boxes. The results highlight the advantages of \ourmethod. First, the synthesized masks exhibit accurate anatomical alignment, with pathologies localized to clinically plausible regions. The generated disease patterns closely resemble real-world cases in spatial distribution. Unlike XReal, which supports only single-disease synthesis, \ourmethod can generate multiple coexisting diseases with distinct spatial locations, effectively modeling the complexity of real-world clinical comorbidities.

\paragraph{Radiologist Assessment.}
\ourmethod-generated images were perceived as real in approximately 78\% of cases, demonstrating high perceptual realism. Moreover, 41\% of the corresponding segmentation masks were rated as providing meaningful diagnostic guidance, under a strict evaluation setting where correctness required accurate segmentation of all coexisting diseases. These findings suggest that \ourmethod not only produces visually convincing radiographs, but also generates spatial annotations that enhance clinical interpretability. By integrating semantic prompts with anatomically structured guidance, \ourmethod effectively bridges the gap between generative modeling and diagnostic decision support.

\begin{table}[!t]
\centering
\caption{Detection performance (mAP@50) on three test sets. Best in \textbf{bold}.}
\resizebox{0.98\textwidth}{!}{%
\begin{tabular}{l l c c c c c c}
\toprule
Test Set & Train Set & Atelectasis & Cardiomegaly & Consolidation & Effusion & Pneumothorax & Overall \\
\midrule
\multirow{4}{*}{ChestX-Det (542)} 
& Real & 43.60 & \textbf{99.43} & 22.74 & 23.66 & 28.82 & 43.65\\
& Real+1xCheff & 37.96 & 99.02 & \textbf{22.77} & 23.13 & 33.52 & 43.28 (↓0.37) \\
& Real+1xRoentGen & 44.58 & 97.24 & \textbf{22.77} & \textbf{25.02} & 19.02 & 41.73 (↓1.92) \\
& Real+1xOurs & \textbf{51.91} & 99.18 & \textbf{22.77} & 23.32 & \textbf{37.49} & \textbf{46.94} (↑3.28) \\
\midrule
\multirow{4}{*}{VinDr-CXR (2938)} 
& Real & 8.57 & 32.67 & 16.19 & \textbf{30.13} & 14.38 & 20.39 \\
& Real+1xCheff & 3.94 & 32.67 & 23.18 & 18.44 & 15.77 & 18.80 (↓1.59) \\
& Real+1xRoentGen & \textbf{9.68} & 32.67 & 12.13 & 16.80 & 12.02 & 16.66 (↓3.73) \\
& Real+1xOurs & 7.17 & 32.67 & \textbf{26.13} & 21.73 & \textbf{21.24} & \textbf{21.79} (↑1.40) \\
\midrule
\multirow{4}{*}{ChestX-ray14 (741)} 
& Real & 17.08 & \textbf{56.67} & - & \textbf{11.58} & 22.40 & 26.93 \\
& Real+1xCheff & 17.88 & 53.76 & - & 8.26 & 33.82 & 28.43 (↑1.50) \\
& Real+1xRoentGen & 18.83 & 53.32 & - & 11.26 & 29.81 & 28.31 (↑1.37) \\
& Real+1xOurs & \textbf{21.39} & 55.93 & - & 11.02 & \textbf{39.47} & \textbf{31.96} (↑5.02) \\
\bottomrule
\end{tabular}
}
\label{tab:det_result}
\vspace{-12pt}
\end{table}

\begin{table}[!t]
\centering
\caption{Segmentation performance (Dice score) on ChestX-Det test set. Best in \textbf{bold}.}
\resizebox{0.98\textwidth}{!}{%
\begin{tabular}{l l c c c c c c}
\toprule
Test Set & Train Set & Atelectasis & Cardiomegaly & Consolidation & Effusion & Pneumothorax & Overall \\
\midrule
\multirow{4}{*}{ChestX-Det (542)} 
& Real & 22.15 & 70.01 & \textbf{45.90} & \textbf{42.36} & 8.75 & 37.83 \\
& Real+1xCheff & 25.68 & 57.49 & 43.03 & 39.63 & 9.12 & 34.99 (↓2.84) \\
& Real+1xRoentGen & 25.56 & 65.81 & 38.97 & 33.85 & 5.92 & 34.02 (↓3.81) \\
& Real+1xOurs & \textbf{25.70} & \textbf{72.62} & 45.82 & 42.09 & \textbf{9.57} & \textbf{39.16} (↑1.33) \\     
\bottomrule
\end{tabular}
}
\label{tab:seg_result}
\end{table}

\begin{table}[!t]
\centering
\caption{Classification performance (AUC) on MIMIC-CXR test set. Best in \textbf{bold}.}
\resizebox{0.98\textwidth}{!}{%
\begin{tabular}{l l c c c c c c}
\toprule
Test Set & Train Set & Atelectasis & Cardiomegaly & Consolidation & Effusion & Pneumothorax & Overall \\
\midrule
\multirow{4}{*}{MIMIC-CXR (2675)} 
& Real  & 58.84 & 68.38 & 66.38 & \textbf{77.07} & 59.26 & 65.99 \\
& Real+1xCheff & 50.29 & 69.61 & 58.19 & 55.46 & \textbf{68.83} & 60.47 (↓5.52) \\
& Real+1xRoentGen  & 50.68 & 61.92 & 62.86 & 55.80 & 65.67 & 59.39 (↓6.60) \\
& Real+1xOurs  & \textbf{59.86} & \textbf{79.23} & \textbf{67.85} & 75.62 & 66.27 & \textbf{69.77} (↑3.78) \\
\bottomrule
\end{tabular}
}
\label{tab:cls_result}
\vspace{-12pt}
\end{table}

\subsection{Evaluating Downstream Generalization}

To assess the utility of our synthesized data for medical image analysis, we conduct a comprehensive set of experiments covering detection, segmentation, and classification tasks under both in-domain and out-of-domain settings. The effectiveness of \ourmethod is further evaluated under data-scarce conditions and long-tail disease distributions.

\paragraph{In-Domain Detection and Segmentation.}
Experiments are conducted on the ChestX-Det dataset to evaluate detection and segmentation performance under standard supervised settings. For detection, a Mask R-CNN with Feature Pyramid Networks (FPN)~\cite{He_2017_ICCV, Lin_2017_FPN} is trained from scratch using combinations of real and synthesized data produced by different CXR synthesis methods.

As shown in Tables~\ref{tab:det_result} and \ref{tab:seg_result}, models trained with \ourmethod-augmented data outperform all baselines. Specifically, detection performance (mAP@50) improves by +3.28\% and segmentation accuracy (Dice) increases by +1.33\% over the real-only baseline, and outperforms both RoentGen and Cheff. These results demonstrate that the synthesized data from \ourmethod not only maintains anatomical and pathological consistency, but also provides effective supervision for downstream tasks.

\paragraph{Out-of-Domain Generalization.}
Cross-domain performance is evaluated on three external datasets: ChestX-ray14 and VinDr-CXR for lesion detection, and MIMIC-CXR for disease classification (Tables~\ref{tab:det_result} and~\ref{tab:cls_result}). Augmenting training with \ourmethod-generated samples improves performance, with gains of +5.02\% mAP on ChestX-ray14, +1.40\% on VinDr-CXR, and +3.78\% classification accuracy on MIMIC-CXR. These results highlight the robustness of our approach and its ability to mitigate dataset-specific overfitting.

\begin{figure}[t]
  \centering
  \begin{minipage}[t]{0.52\textwidth}
    \centering
    \includegraphics[width=\linewidth]{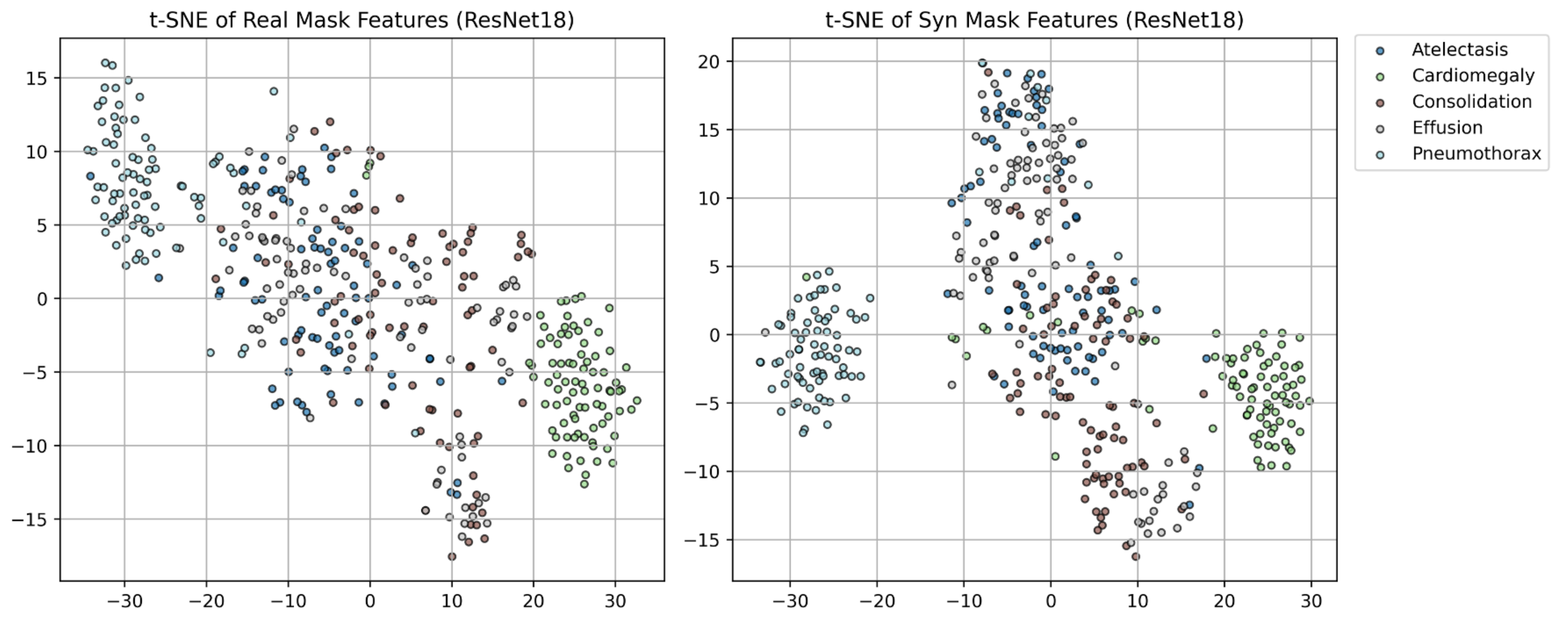}
    \captionof{figure}{Pathology-driven t-SNE visualization of ResNet-18 features from real and synthesis masks.}
    \label{fig:tsne}
  \end{minipage}
  \hfill
  \begin{minipage}[t]{0.45\textwidth}
    \centering
    \includegraphics[width=\linewidth]{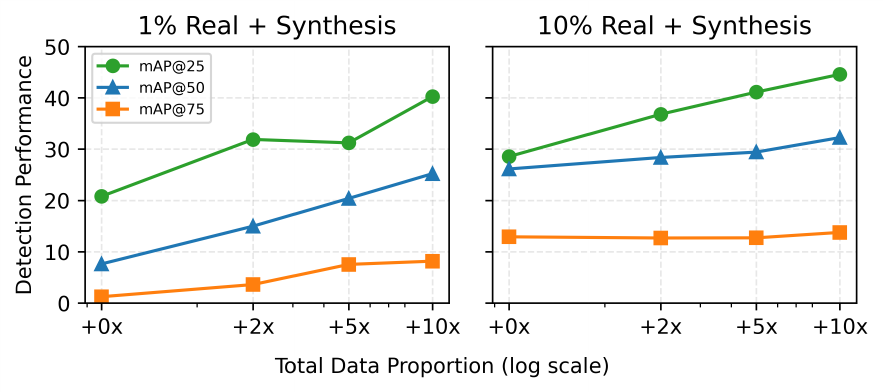}
    \captionof{figure}{Detection performance with varying ratios of real and synthesis.}
    \label{fig:1_10}
  \end{minipage}
\end{figure}

\paragraph{Data Efficiency Under Scarcity.}
We examine how synthesized data complements limited real annotations by evaluating models trained on 1\%, and 10\% of the real data, augmented with 2$\times$, 5$\times$, and 10$\times$ synthesized samples. As shown in Figure~\ref{fig:1_10}, \ourmethod provides substantial gains in low-resource settings. In the extreme 1\% case, adding 5$\times$ synthesized data increases mAP@50 from 7.61\% to 25.21\%—a 3$\times$ improvement. Gains remain consistent even in moderate and full-data regimes, highlighting \ourmethod’s scalability.

\begin{figure}[t]
  \centering
  \begin{minipage}[t]{0.52\textwidth}
    \captionof{table}{Ablation on key modules. Best in bold.}
    \label{tab:ablation}
    \resizebox{\linewidth}{!}{%
        \begin{tabular}{@{}cccccccc@{}}
            \toprule
            Real Mask & Syn. Mask & Expert filter &  FID↓ & MS-SSIM↑ & Path. AUC↑ & mAP@50↑ \\
            \midrule
            \checkmark &  &   & 35.78 & 0.56 & 56.72 & 42.54 \\
            & \checkmark &   & 34.00 & 0.57 & 57.37 & 44.60 \\
            & \checkmark & \checkmark & \textbf{32.83} & \textbf{0.60} & \textbf{59.31} & \textbf{46.94} \\
            \bottomrule
          \end{tabular}
    }
  \end{minipage}
  \hfill
  \begin{minipage}[t]{0.45\textwidth}
    \captionof{table}{Ablation on mask conditions. Best in bold.}
    \label{tab:mask_ablation}
    \resizebox{\linewidth}{!}{%
        \begin{tabular}{lcccc}
            \toprule
            Mask Condition & FID ↓ & CLIP ↑ & MS-SSIM ↑ & Path. AUC ↑ \\
            \midrule
            no organ mask & 56.39 & 29.65 & 0.4031 & 56.51 \\
            only organ mask & 33.45 & 28.93 & 0.5709 & 55.96 \\
            disease + organ mask (Ours) & \textbf{32.83} & \textbf{30.14} & \textbf{0.5968} & \textbf{59.31} \\
            \bottomrule
            \end{tabular}
        }
  \end{minipage}
\end{figure}

\paragraph{Long-Tail Class Augmentation.}
ChestX-Det exhibits long-tail class imbalance, particularly in underrepresented conditions like Atelectasis. We generate additional \ourmethod samples targeting this class, improving mAP@50 from 43.60\% to 48.08\%, and raising overall test performance by +1.63\%. These results suggest \ourmethod can serve as a targeted augmentation strategy to address class imbalance in clinical datasets.

\subsection{Ablation Studies}

To assess the contributions of the three core components—Text-to-Mask, Mask-to-Image, and the Expert Filter—an ablation was conducted. With Text-to-Mask enabled, masks are synthesized from clinical prompts; otherwise, ground-truth masks are used. As shown in Table~\ref{tab:ablation}, synthetic masks yield better fidelity (lower FID), structural consistency (higher MS-SSIM), and pathology alignment (higher AUC), leading to a +2.06\% mAP@50 gain. These results indicate that generative masks enhance generalization beyond radiologist annotations, further strengthened by progressive modeling of anatomy–pathology dependencies. A pathology-driven feature analysis further supports this: ResNet-18 features visualized with t-SNE (Figure~\ref{fig:tsne}) show that synthesized masks form disease-specific clusters comparable to those of real masks—and in some cases more distinct. The Expert Filter further improves all metrics, underscoring its role in enforcing clinically meaningful generation. The complete pipeline achieves the best overall performance, validating the necessity of each module.

An additional ablation unraveled the anatomical and pathological conditioning (Table~\ref{tab:mask_ablation}). Compared to a text-only baseline, organ-only masks improved realism and structure but offered no pathology gains. Adding disease masks restored and surpassed semantic alignment while further boosting realism.

Further analyses are provided in Appendix~\ref{appendix:free_text}, free-form prompts show that AURAD remains competitive with less structured input: the mask-to-image stage generalizes well, whereas text-to-mask is more sensitive to linguistic variance. Comparisons with alternative mask-to-image baselines (Appendix~\ref{appendix:instructpix2pix}) demonstrate that InstructPix2Pix is the most reliable choice for text-to-mask generation in CXR, achieving higher fidelity, better anatomy–pathology consistency, and more efficient inference. Together, these ablations underscore both the robustness and the unique contributions of our framework.

\section{Conclusion}\label{sec:conclusion}
This work addresses the challenge of controllable, annotation-aware CXR synthesis with \ourmethod, a two-stage framework that integrates structural and semantic guidance into the generation process. It enables fine-grained control over disease appearance and location. Quantitative results and expert evaluations confirm the realism and prompt alignment of generated images, while downstream detection and segmentation tasks benefit from the synthesized supervision. By unifying generation and annotation, \ourmethod provides a scalable alternative to manual labeling and supports robust medical vision development under limited data. Several limitations remain. The method assumes standardized labels across datasets and struggles with subtle or fine-grained pathological features, such as nodules and fractures. It also does not yet model complex disease interactions. Future work may address these challenges through label harmonization, higher-resolution modeling, and the incorporation of relational disease representations. 

\newpage
\bibliography{references}  

\newpage

\appendix
\section*{Appendix}\label{sec:appendix}

\section{Datasets: Statistics and Annotation Examples}\label{appendix:dataset}

\begin{figure}[h]
    \centering
        \begin{minipage}[b]{0.136\linewidth}
            \centering
            \includegraphics[width=\linewidth]{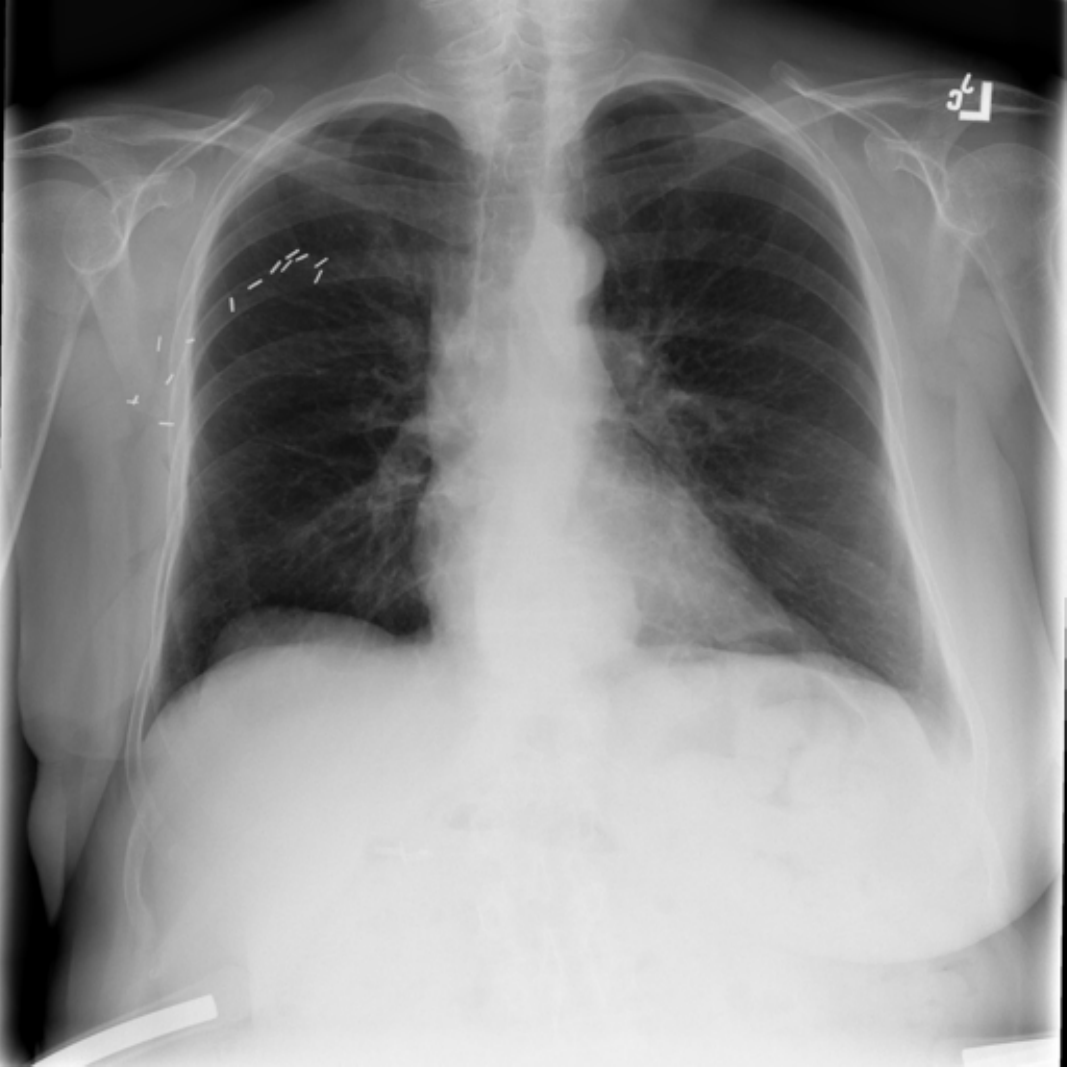}
            \centerline{Normal}
        \end{minipage}
        \begin{minipage}[b]{0.136\linewidth}
            \centering
            \includegraphics[width=\linewidth]{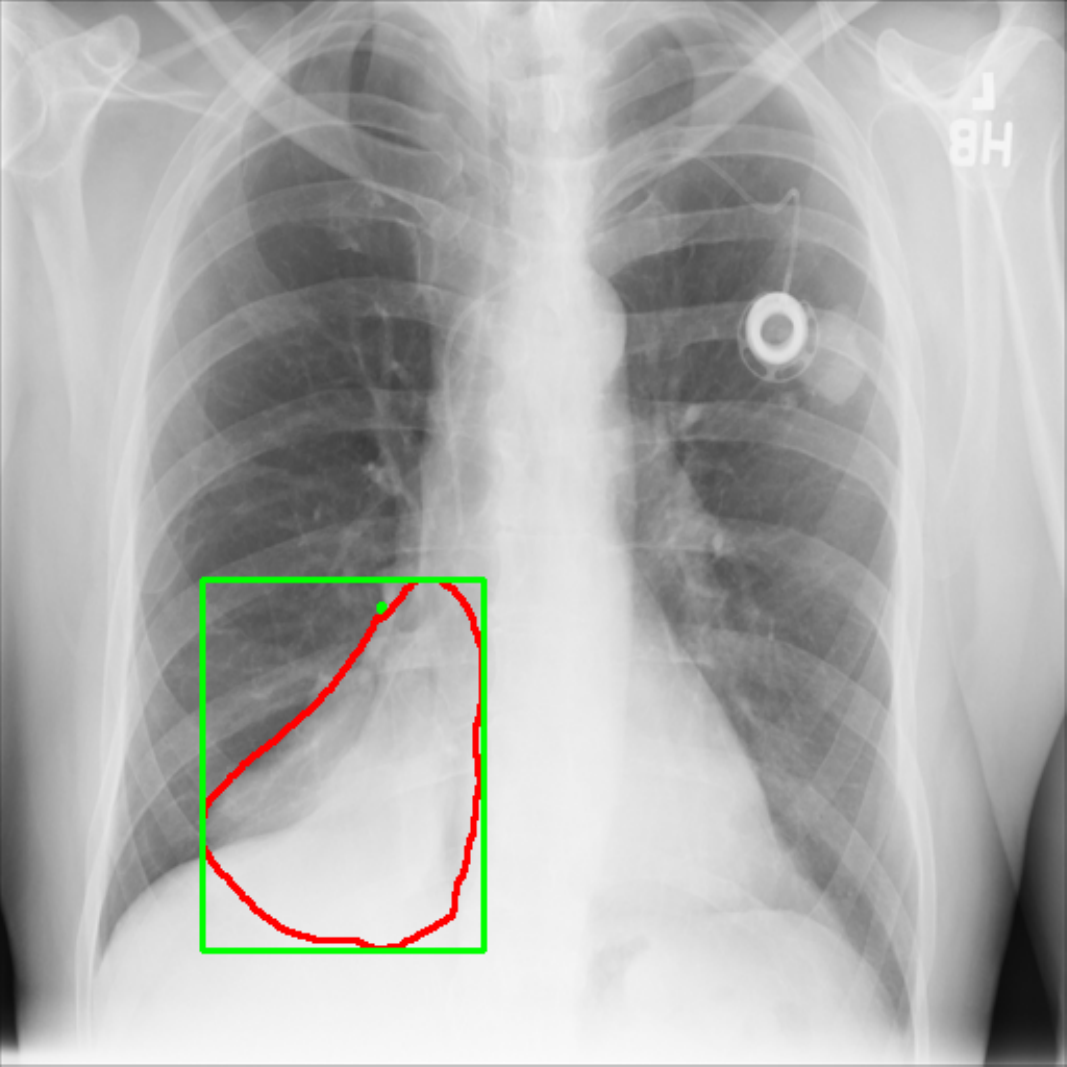}
            \centerline{Atelectasis}
        \end{minipage}
        \begin{minipage}[b]{0.136\linewidth}
            \centering
            \includegraphics[width=\linewidth]{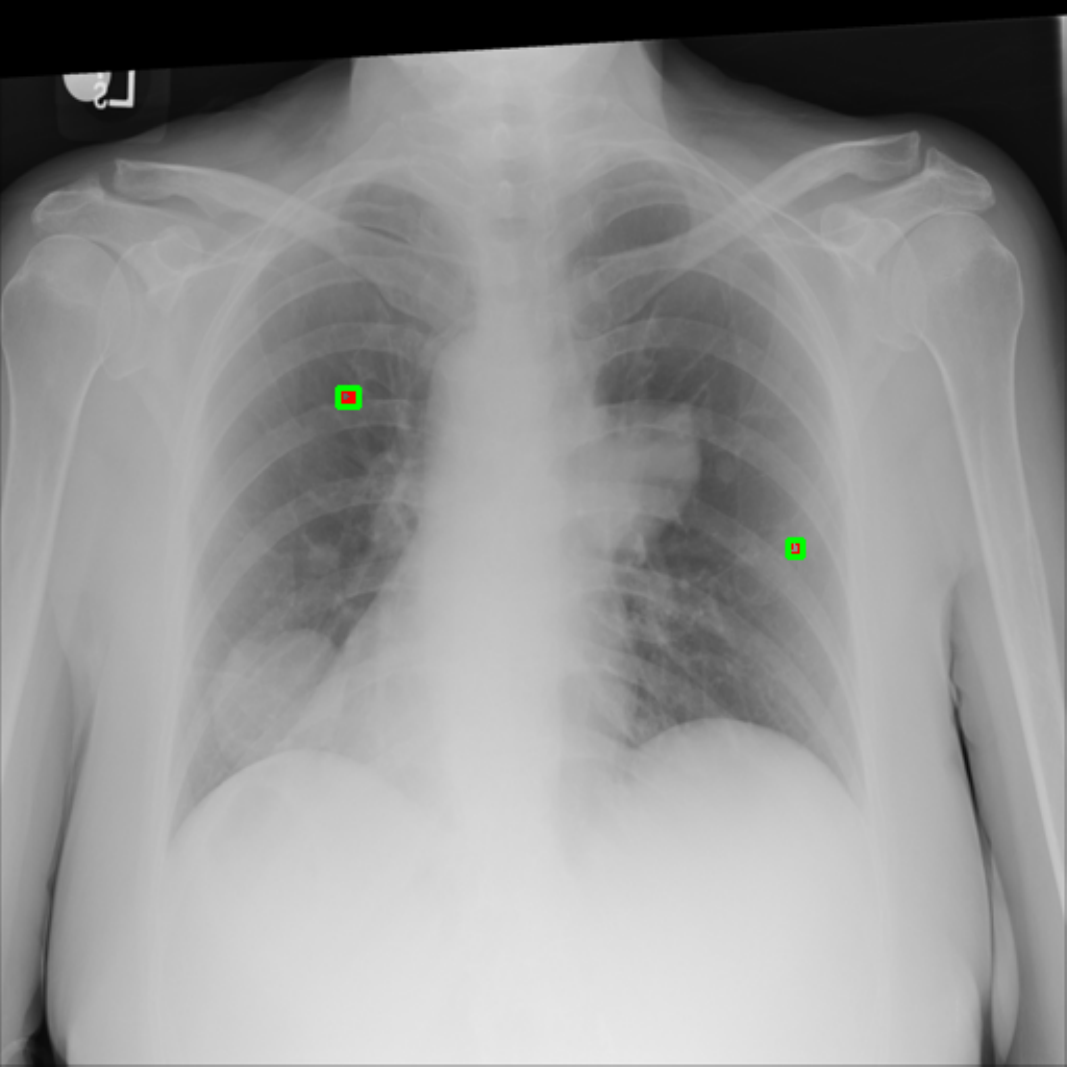}
            \centerline{Calcification}
        \end{minipage}
        \begin{minipage}[b]{0.136\linewidth}
            \centering
            \includegraphics[width=\linewidth]{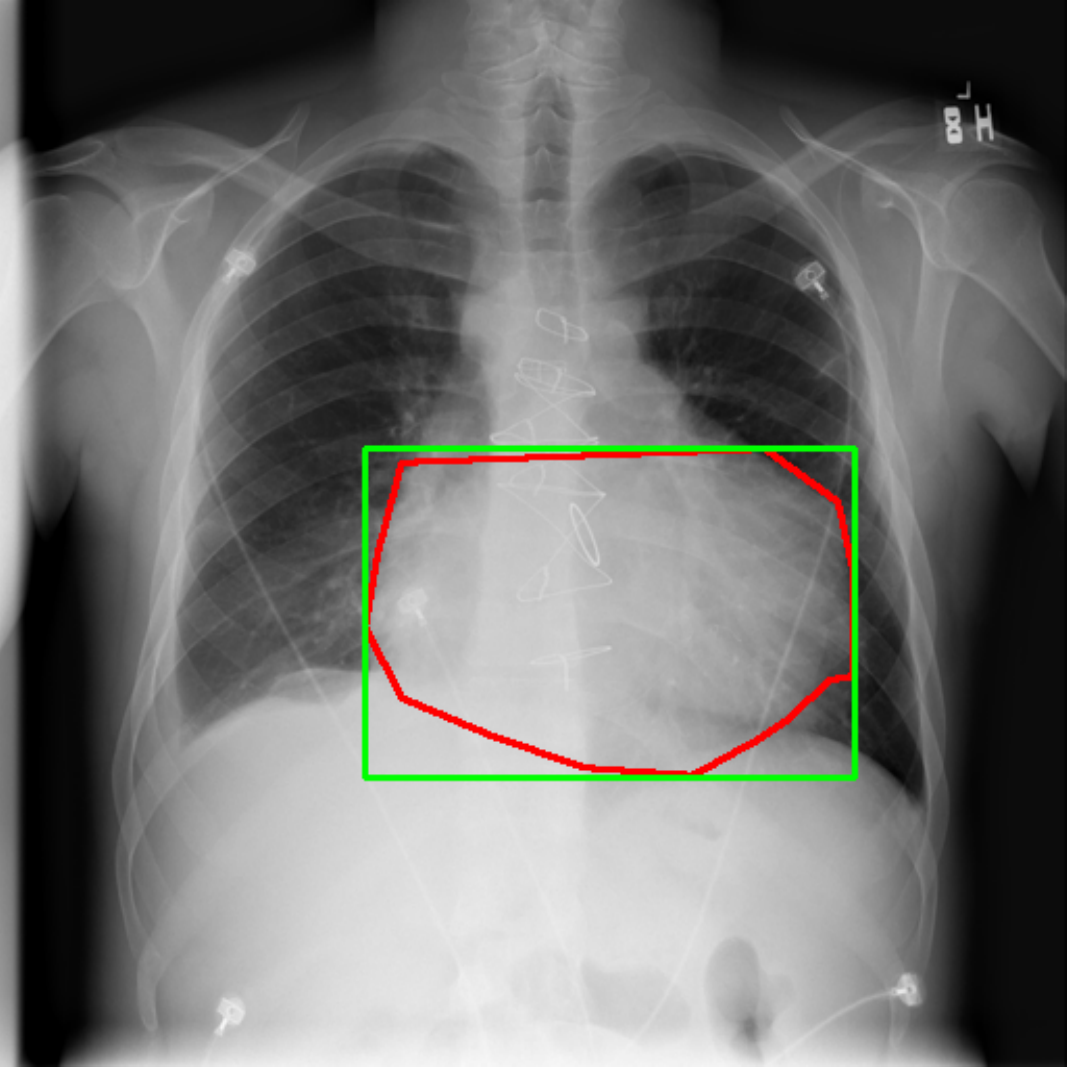}
            \centerline{Cardiomegaly}
        \end{minipage}
        \begin{minipage}[b]{0.136\linewidth}
            \centering
            \includegraphics[width=\linewidth]{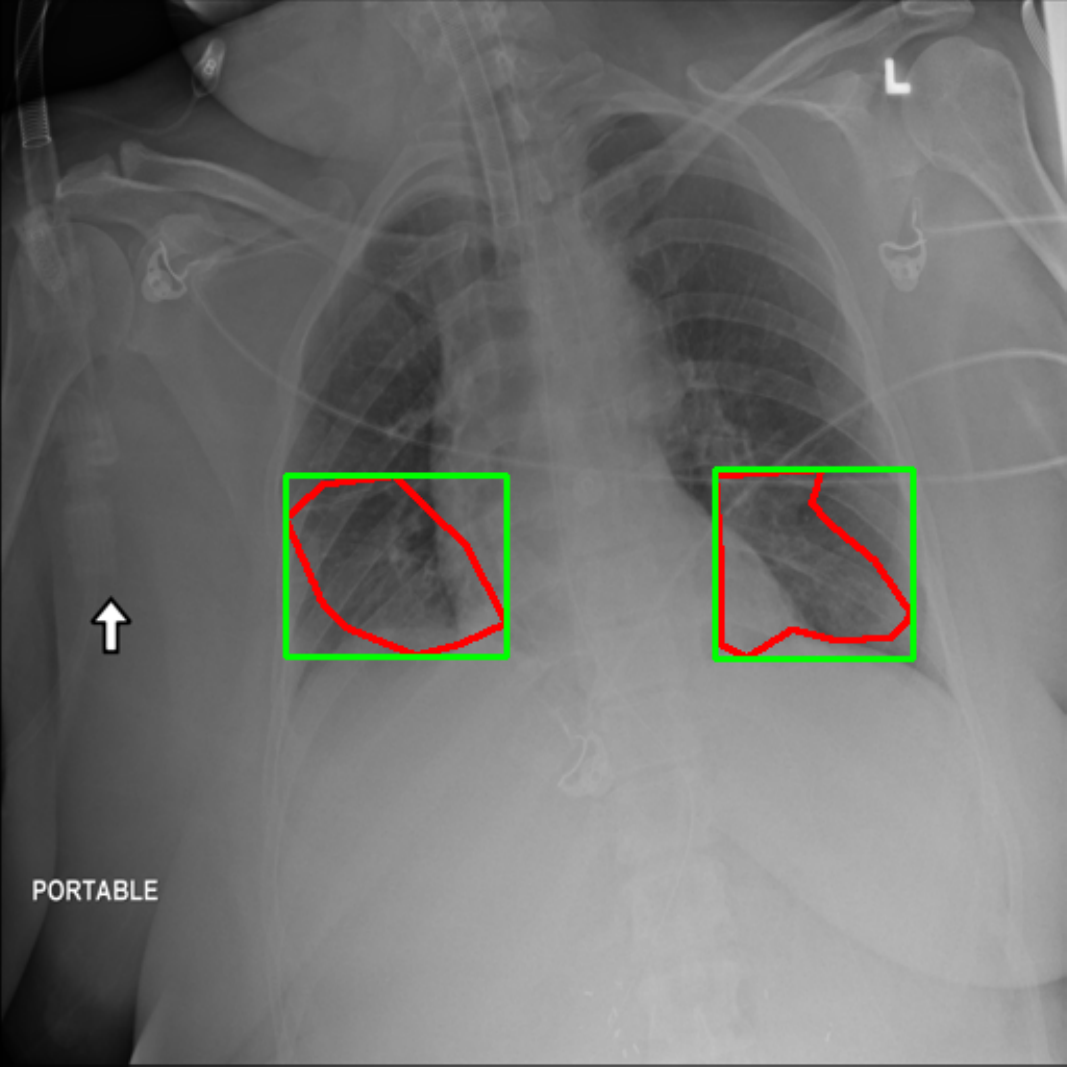}
            \centerline{Consolidation}
        \end{minipage}
        \begin{minipage}[b]{0.136\linewidth}
            \centering
            \includegraphics[width=\linewidth]{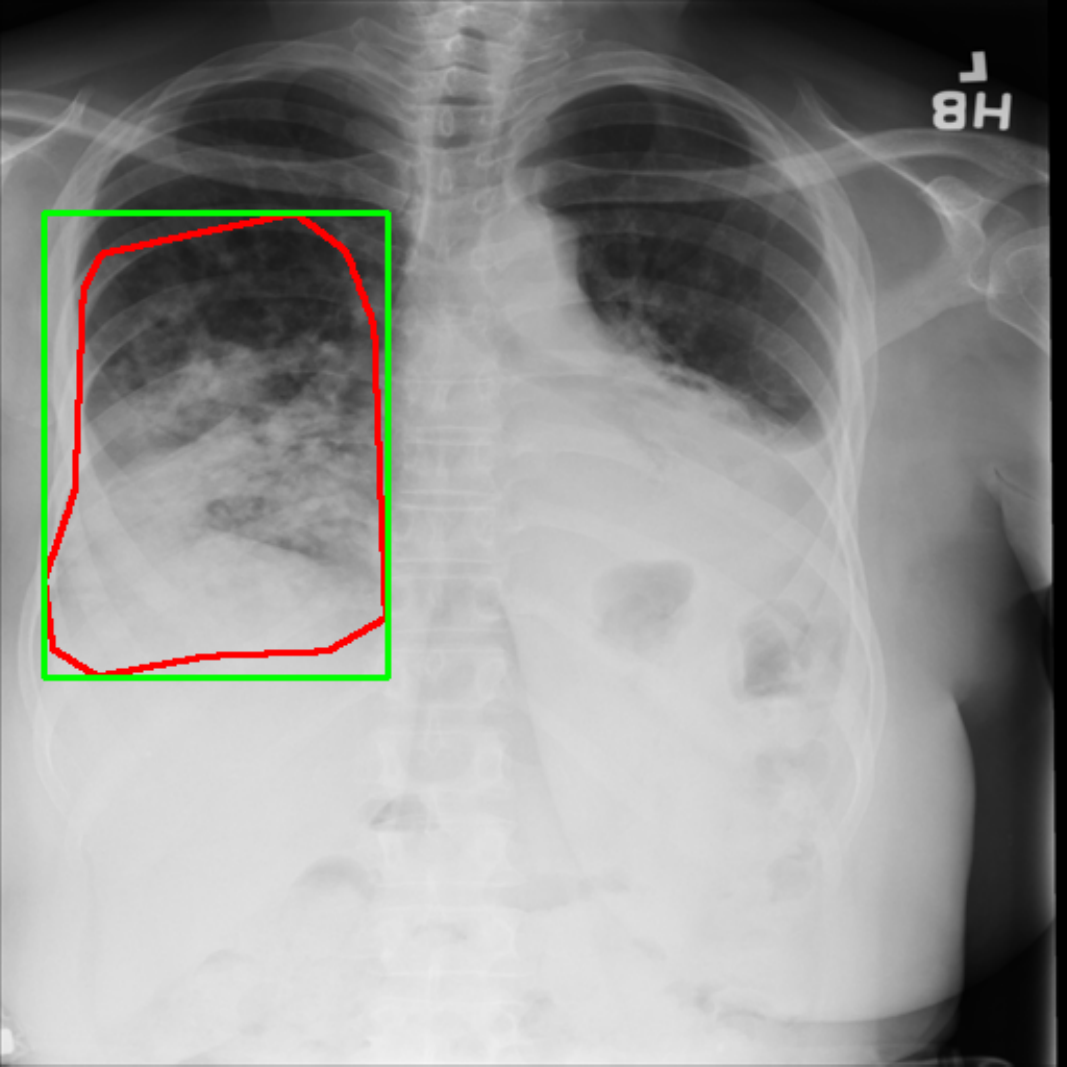}
            \centerline{Diffuse Nodule}
        \end{minipage}
        \begin{minipage}[b]{0.136\linewidth}
            \centering
            \includegraphics[width=\linewidth]{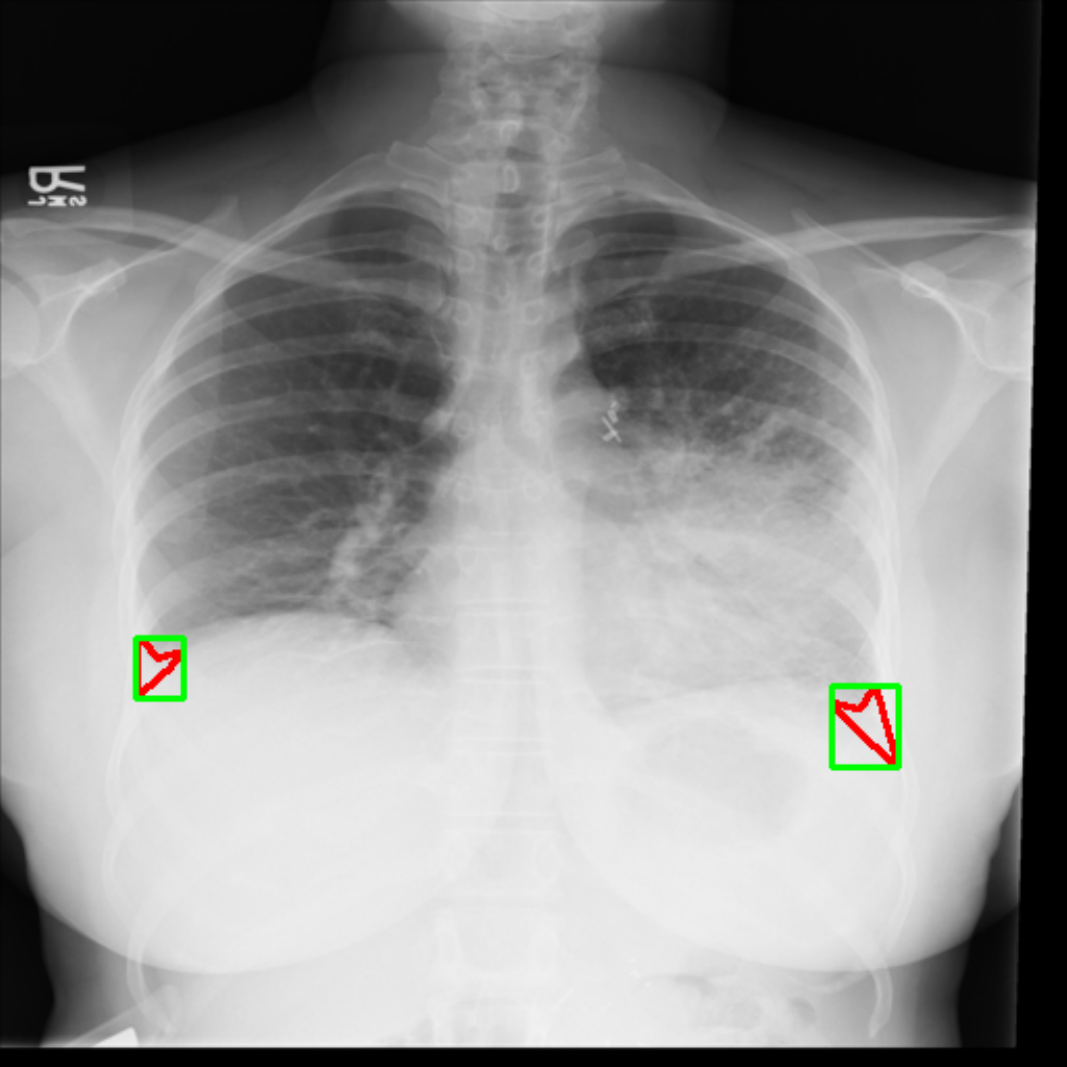}
            \centerline{Effusion}
        \end{minipage} 
        
        \begin{minipage}[b]{0.136\linewidth}
            \centering
            \includegraphics[width=\linewidth]{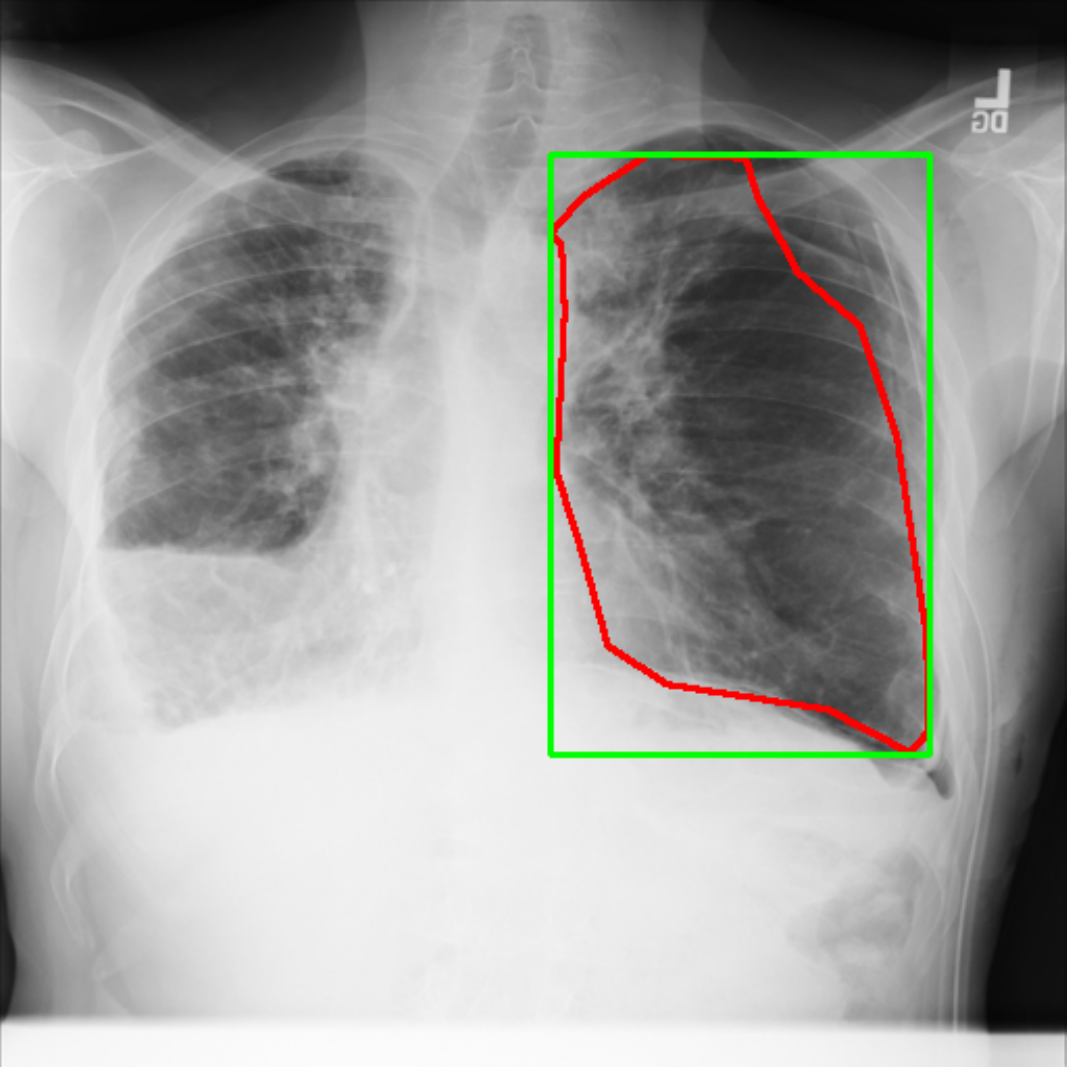}
            \centerline{Emphysema}
        \end{minipage} 
        \begin{minipage}[b]{0.136\linewidth}
            \centering
            \includegraphics[width=\linewidth]{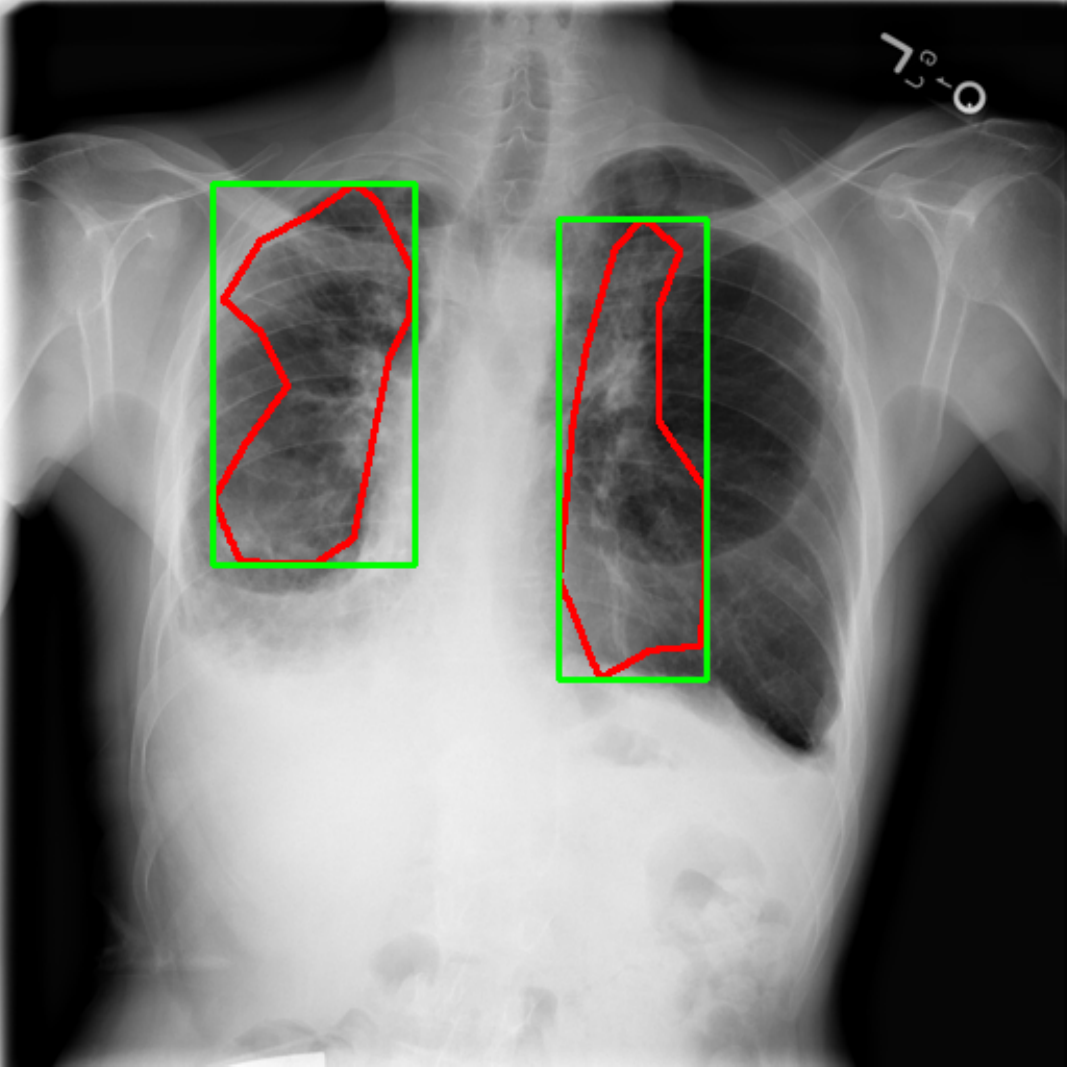}
            \centerline{Fibrosis}
        \end{minipage}
        \begin{minipage}[b]{0.136\linewidth}
            \centering
            \includegraphics[width=\linewidth]{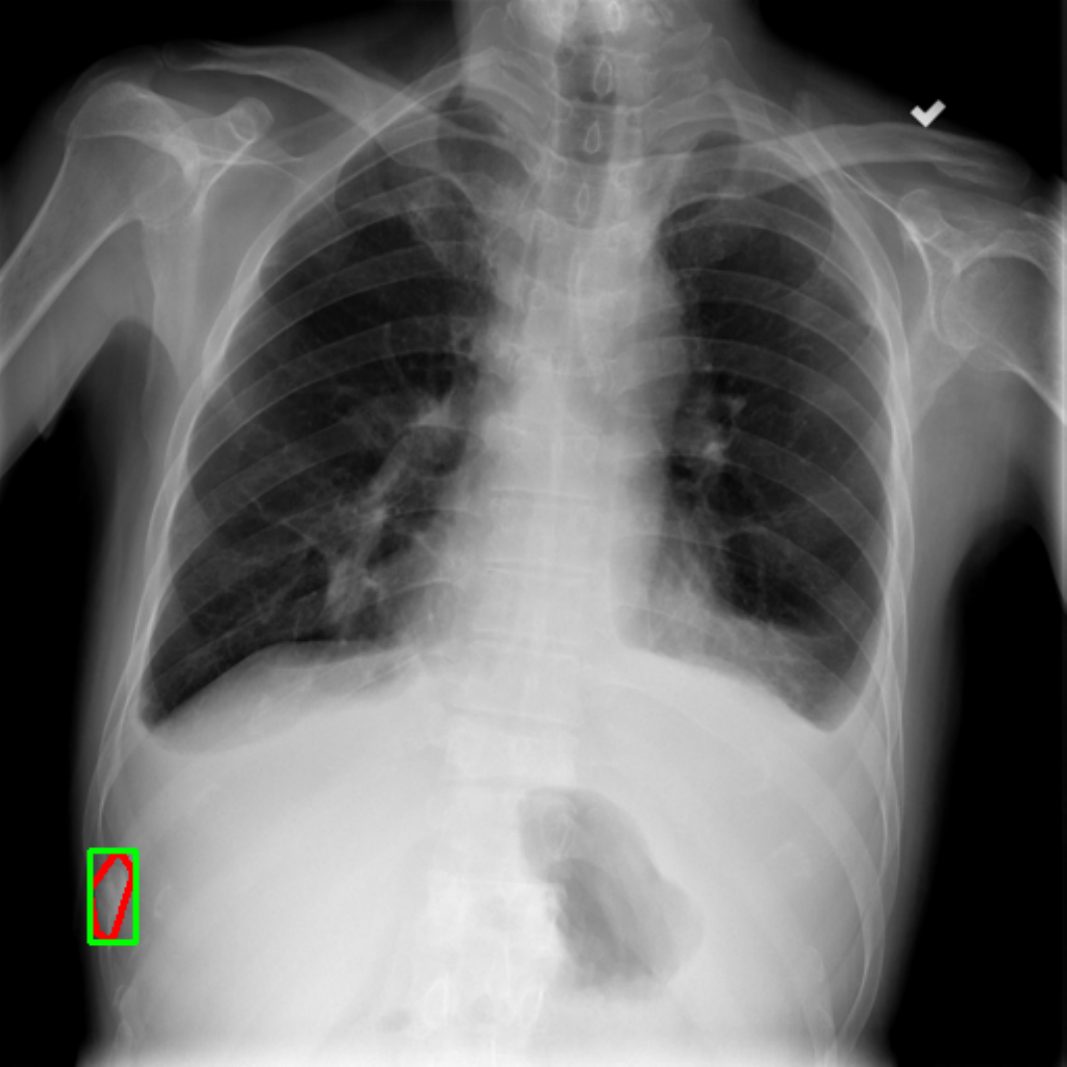}
            \centerline{Fracture}
        \end{minipage}
        \begin{minipage}[b]{0.136\linewidth}
            \centering
            \includegraphics[width=\linewidth]{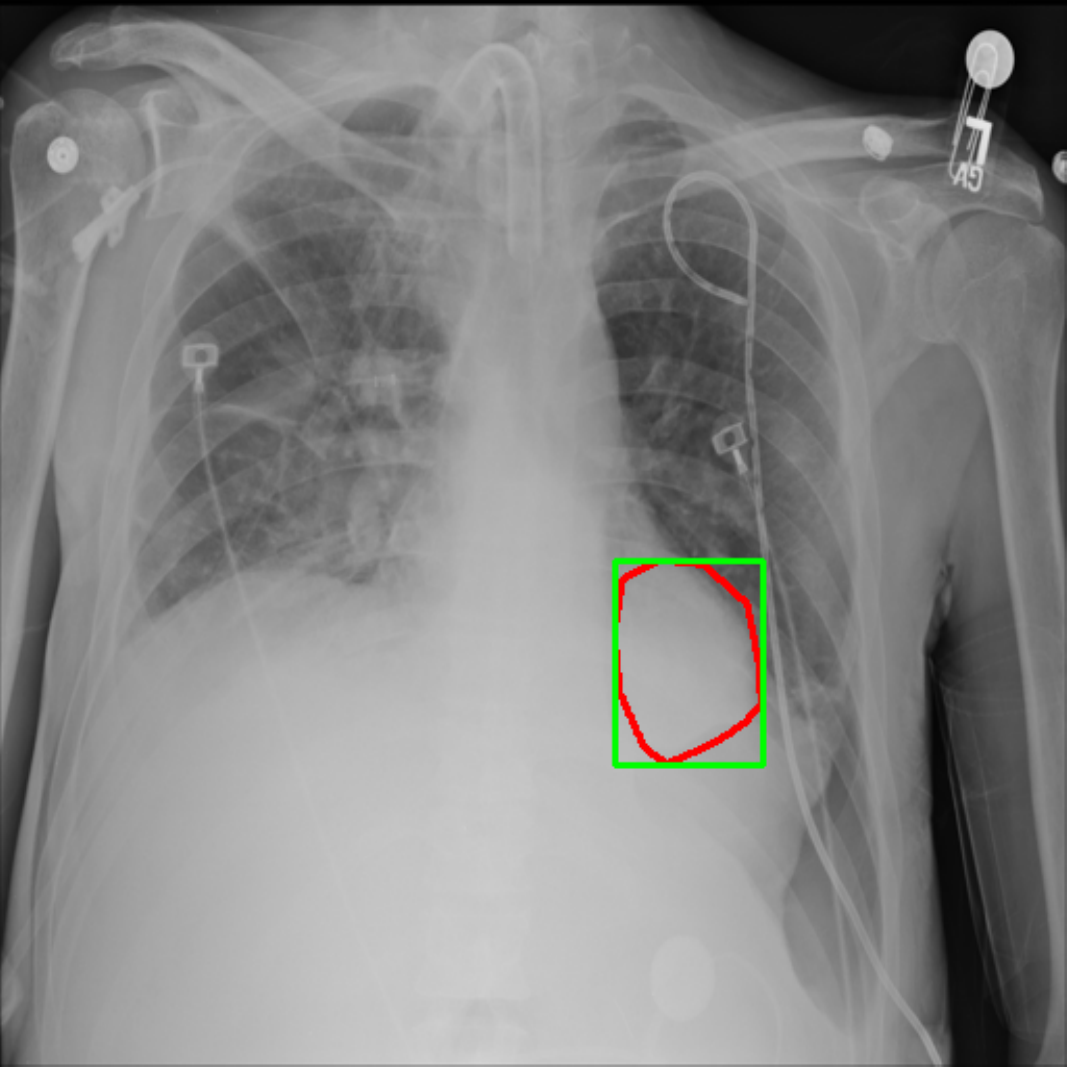}
            \centerline{Mass}
        \end{minipage}
        \begin{minipage}[b]{0.136\linewidth}
            \centering
            \includegraphics[width=\linewidth]{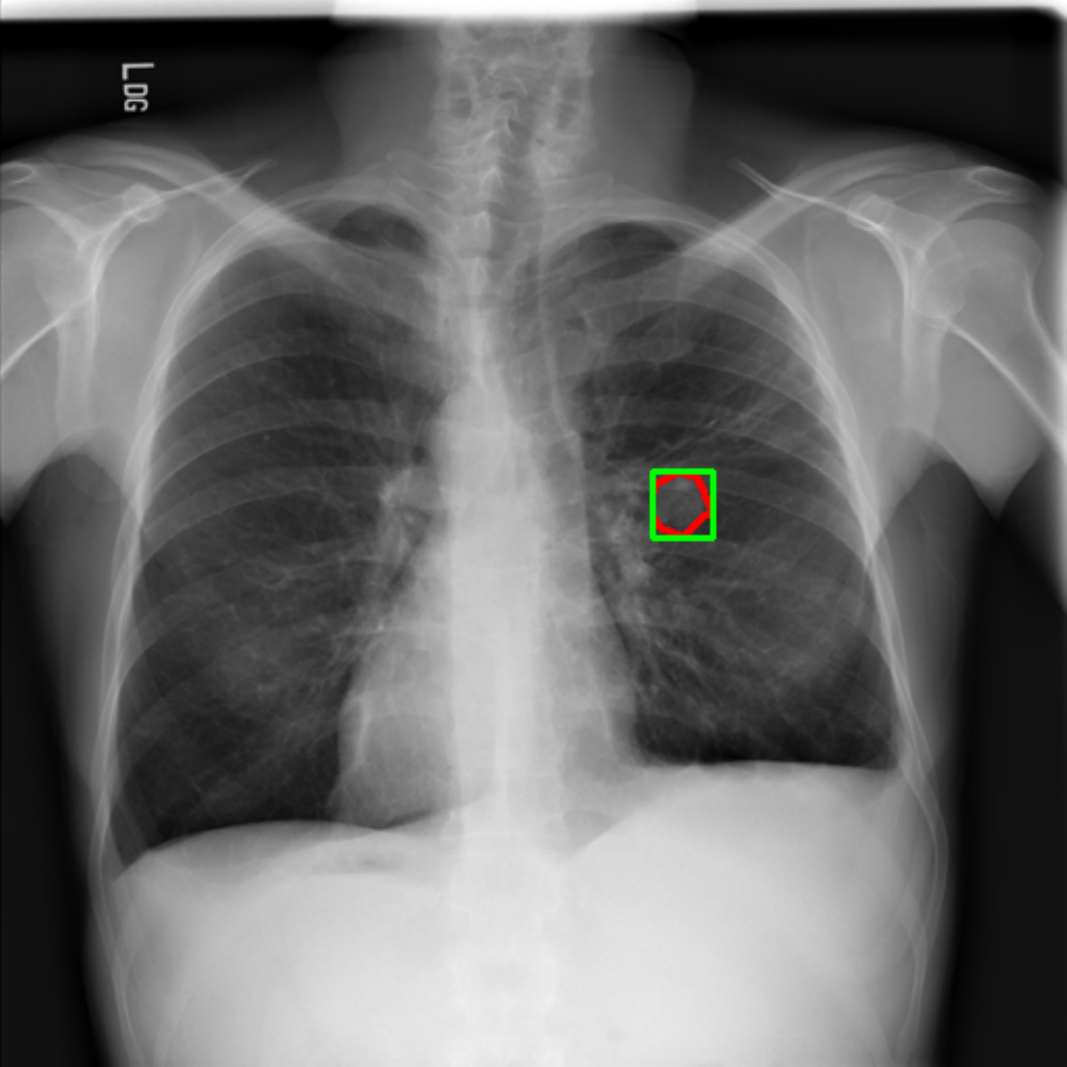}
            \centerline{Nodule}
        \end{minipage}
        \begin{minipage}[b]{0.136\linewidth}
            \centering
            \includegraphics[width=\linewidth]{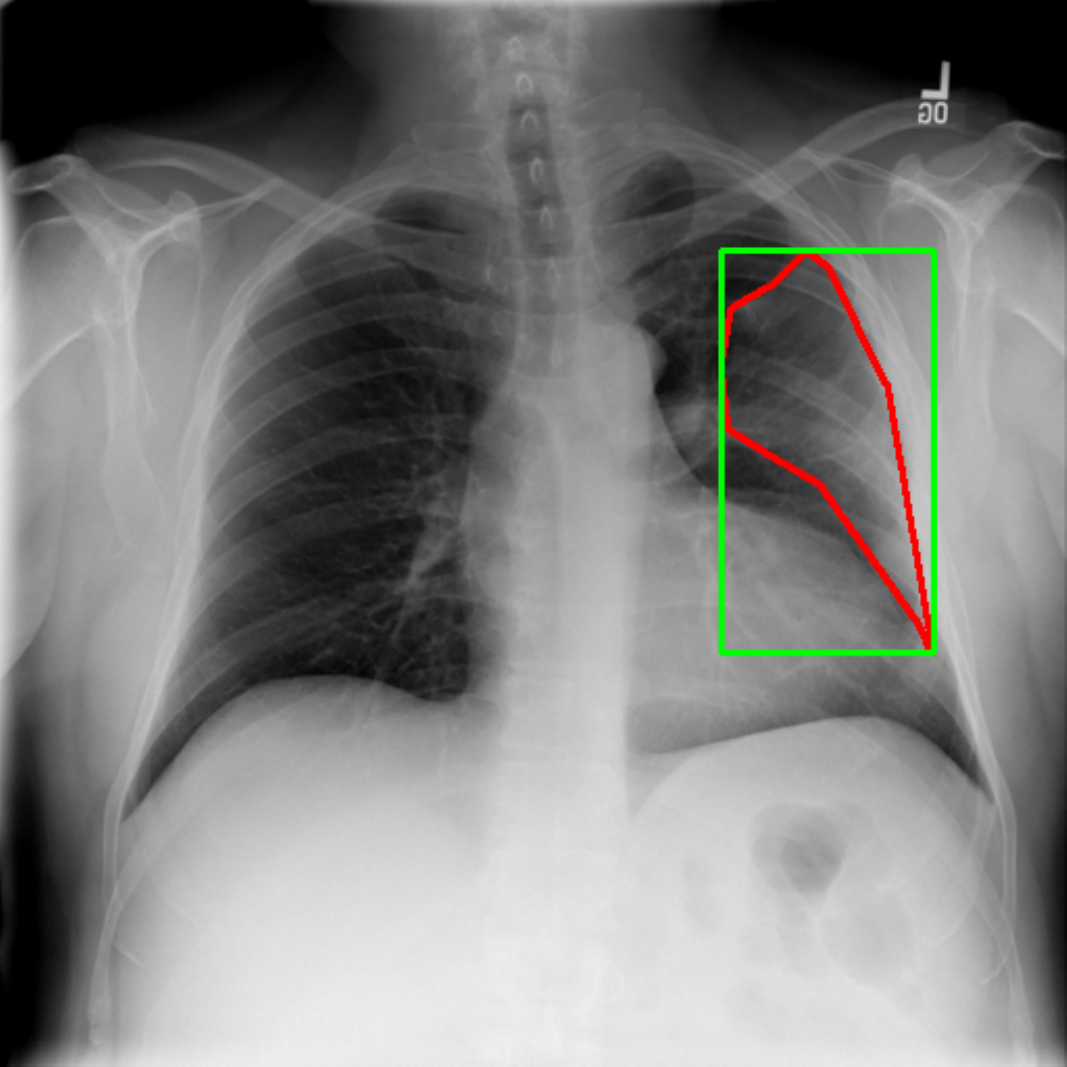}
            \centerline{Pleural.}
        \end{minipage}
        \begin{minipage}[b]{0.136\linewidth}
            \centering
            \includegraphics[width=\linewidth]{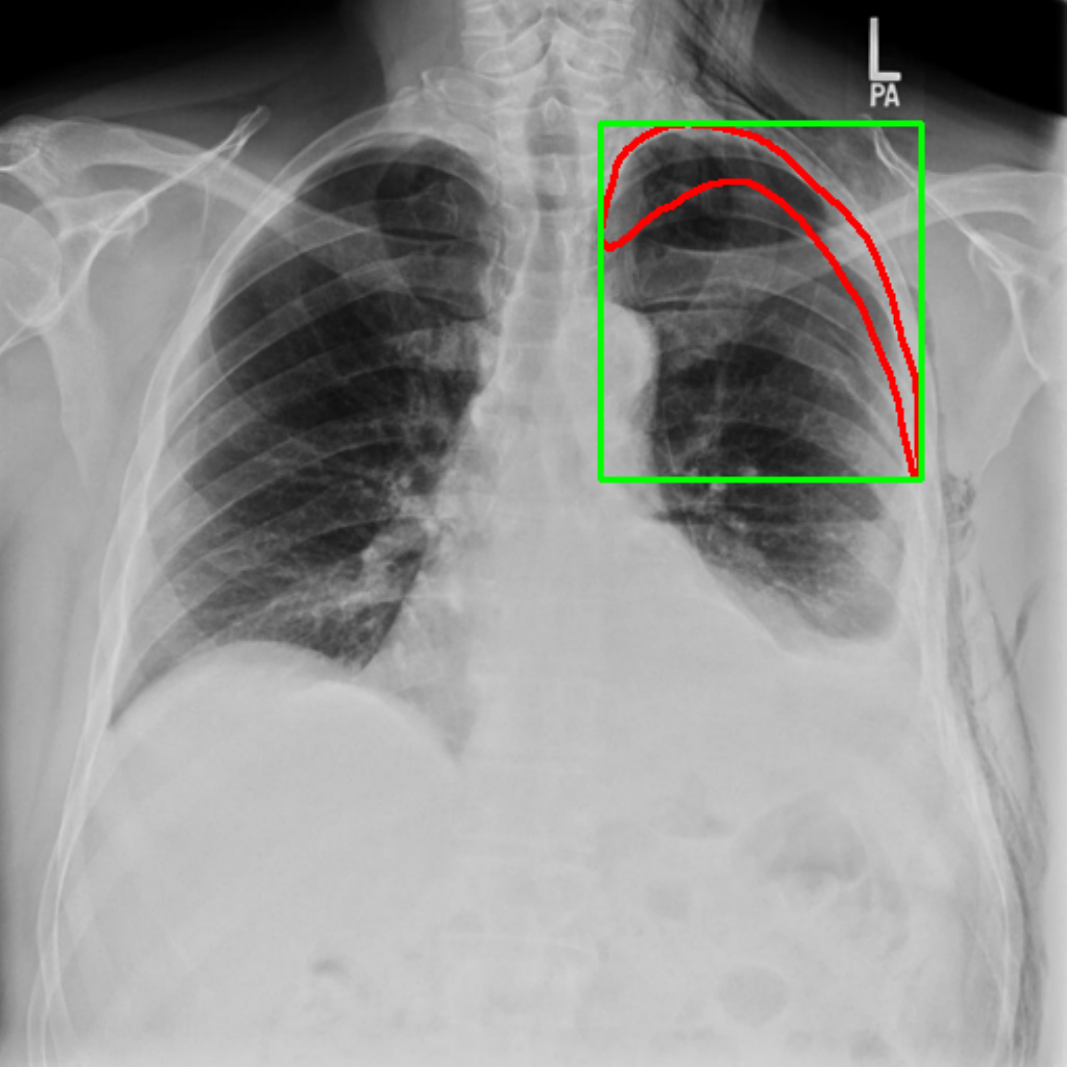}
            \centerline{Pneumothorax}
        \end{minipage} 
        \vspace{2mm}
        \centerline{(a) ChestX-Det}

        \begin{minipage}[b]{0.136\linewidth}
            \centering
            \includegraphics[width=\linewidth]{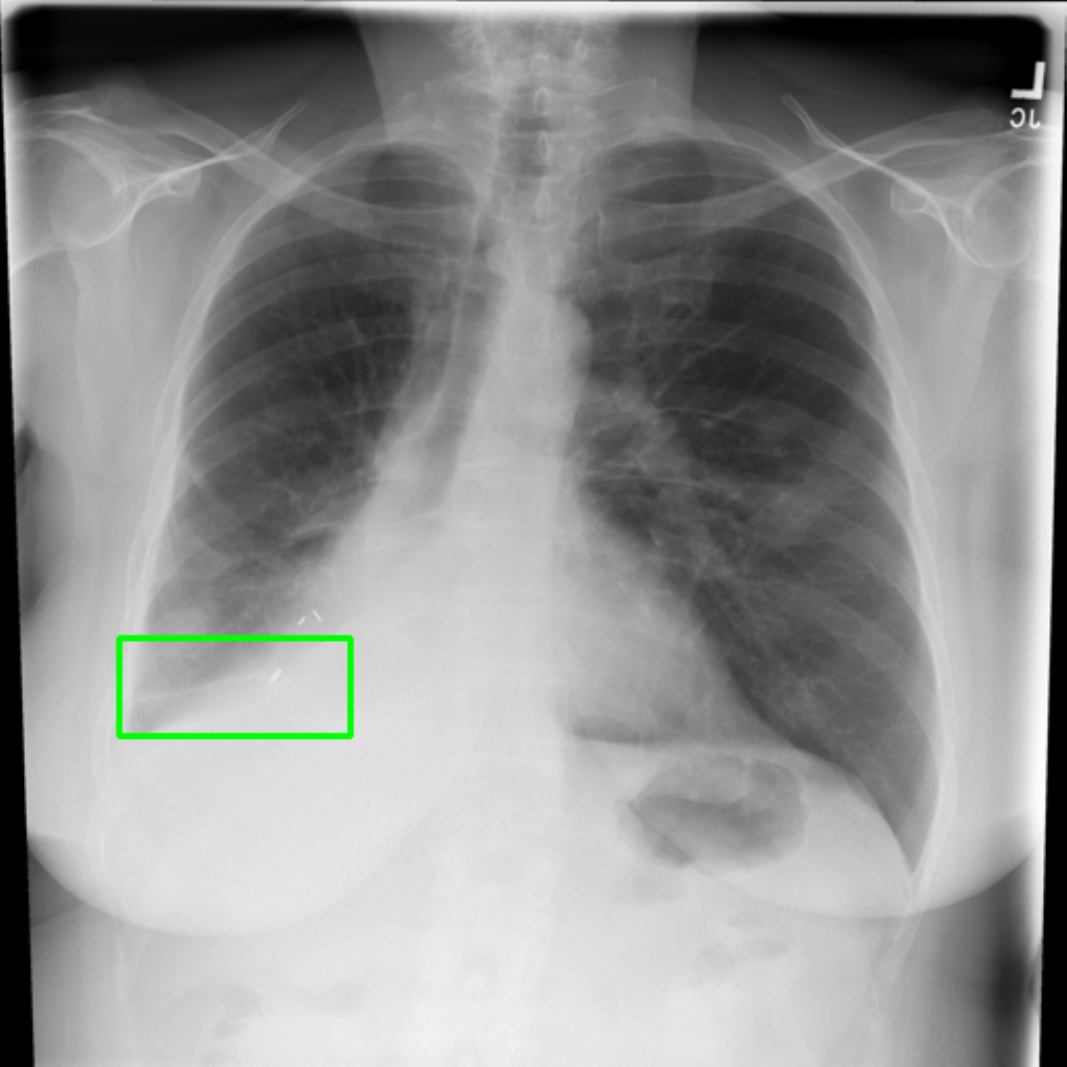}
            \centerline{Atelectasis}
        \end{minipage}
        \begin{minipage}[b]{0.136\linewidth}
            \centering
            \includegraphics[width=\linewidth]{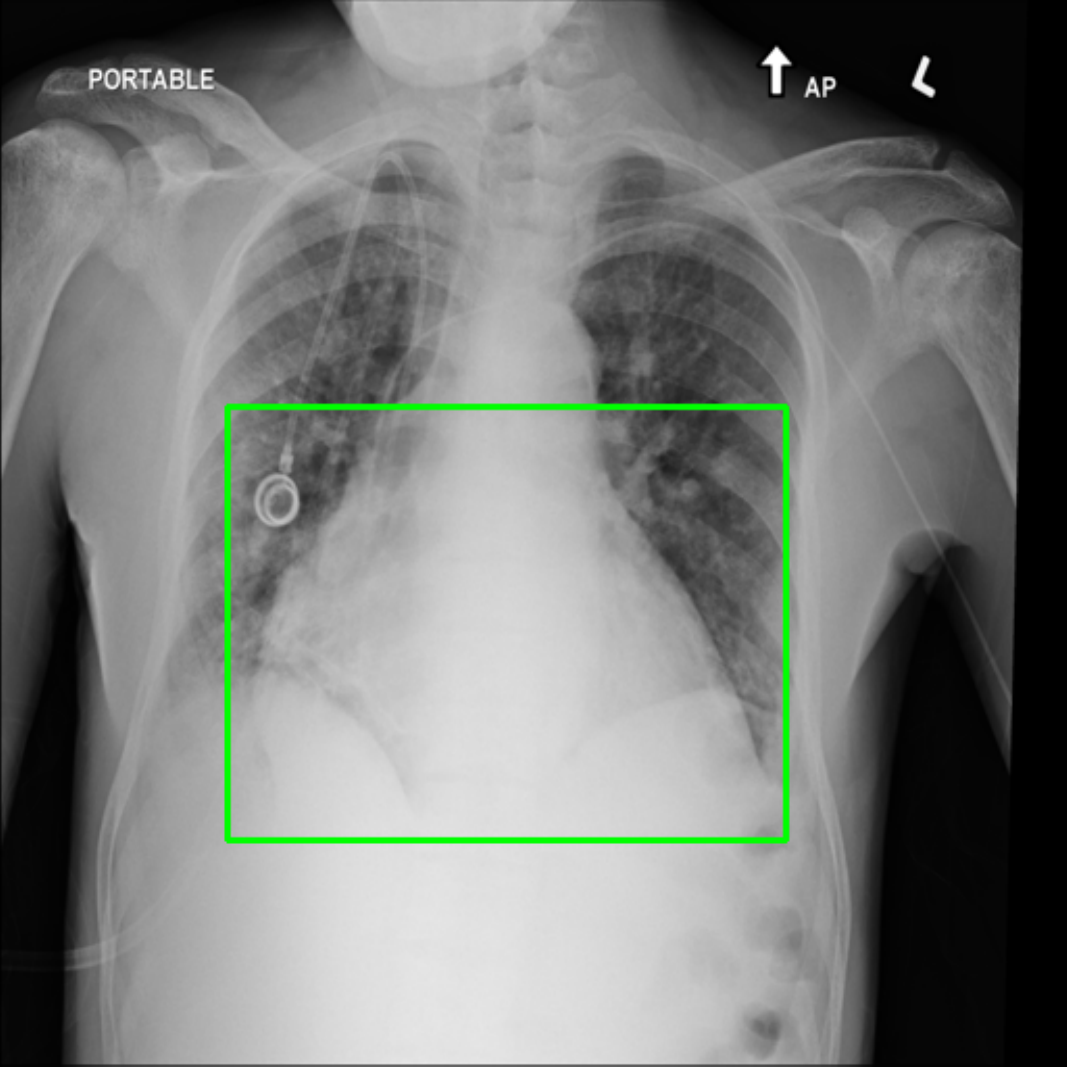}
            \centerline{Cardiomegaly}
        \end{minipage}
        \begin{minipage}[b]{0.136\linewidth}
            \centering
            \includegraphics[width=\linewidth]{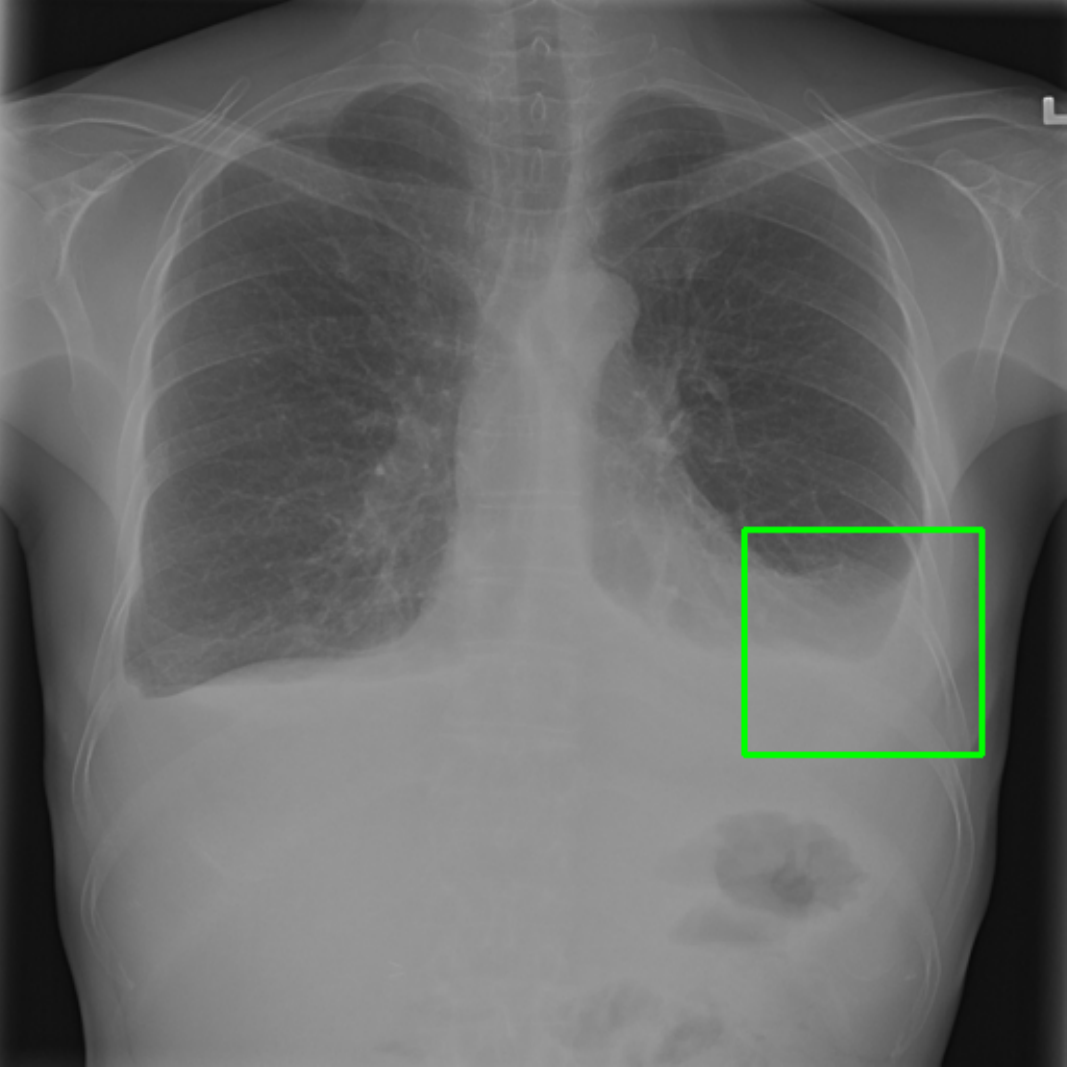}
            \centerline{Effusion}
        \end{minipage} 
        \begin{minipage}[b]{0.136\linewidth}
            \centering
            \includegraphics[width=\linewidth]{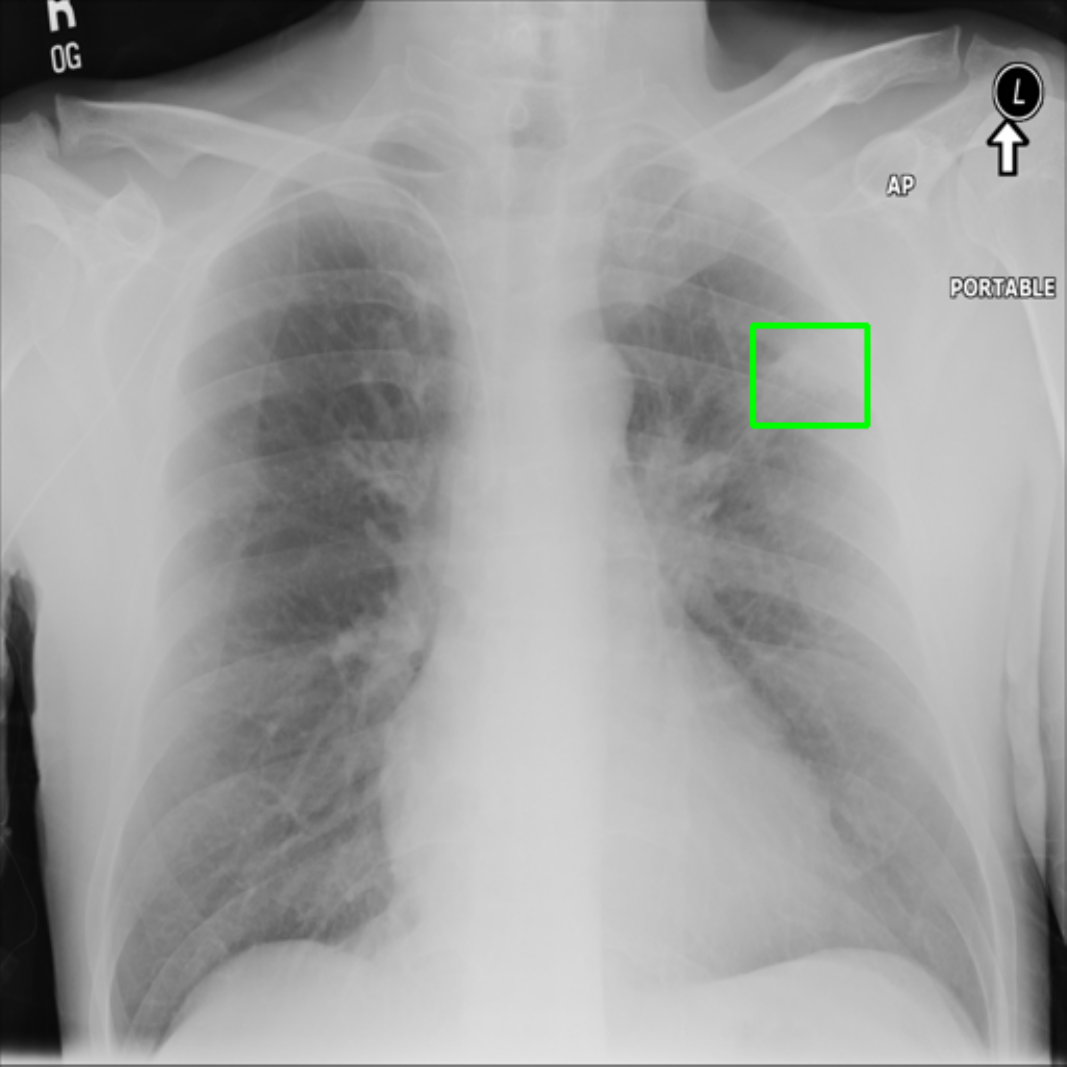}
            \centerline{Mass}
        \end{minipage}
        \begin{minipage}[b]{0.136\linewidth}
            \centering
            \includegraphics[width=\linewidth]{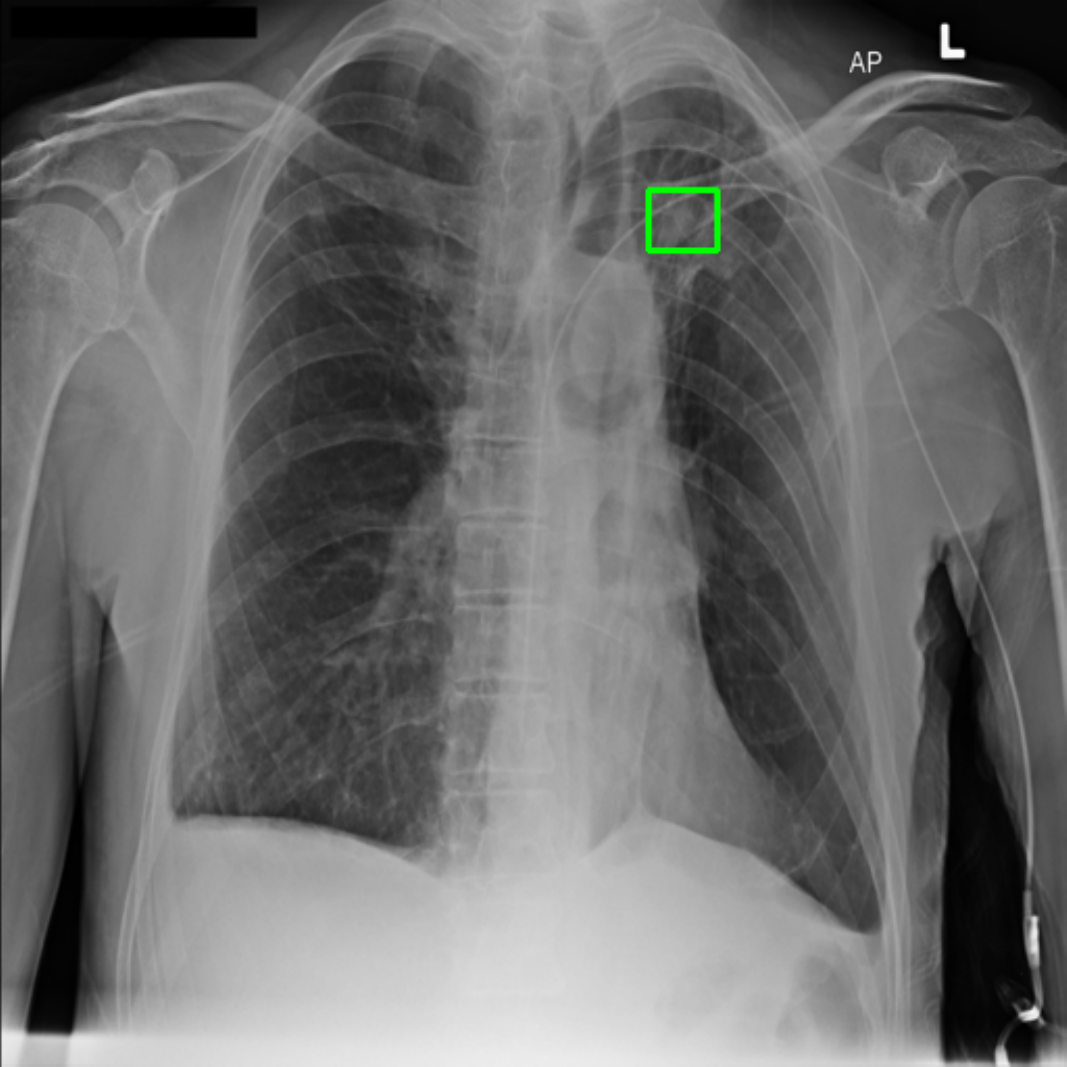}
            \centerline{Nodule}
        \end{minipage}
        \begin{minipage}[b]{0.136\linewidth}
            \centering
            \includegraphics[width=\linewidth]{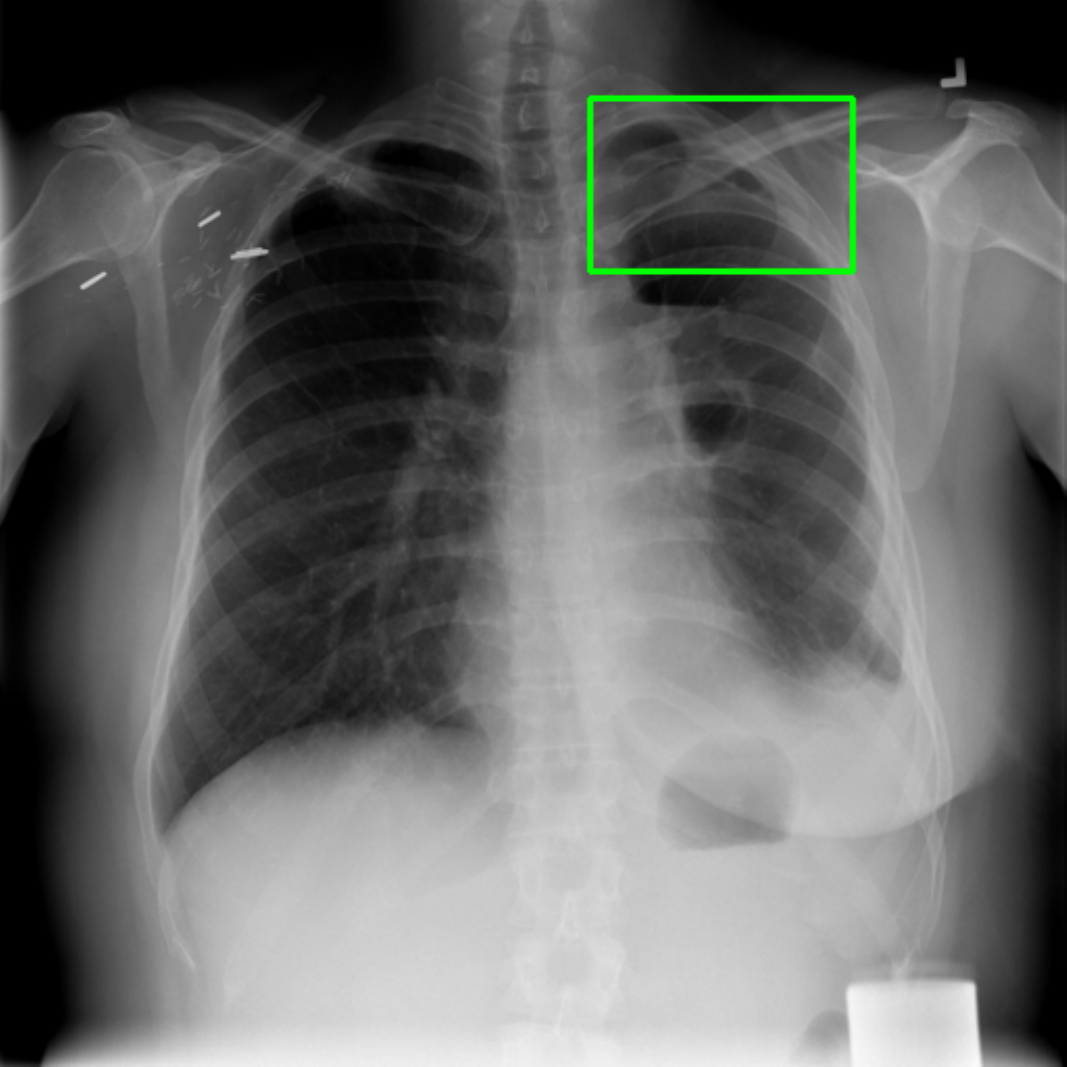}
            \centerline{Pneumothorax}
        \end{minipage} 
        \vspace{2mm}
        \centerline{(b) ChestX-ray14}

        \begin{minipage}[b]{0.136\linewidth}
            \centering
            \includegraphics[width=\linewidth]{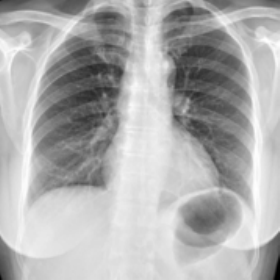}
            \centerline{Normal}
        \end{minipage}
        \begin{minipage}[b]{0.136\linewidth}
            \centering
            \includegraphics[width=\linewidth]{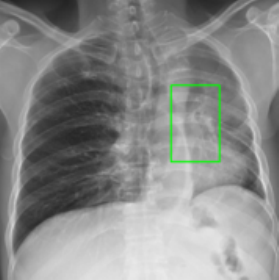}
            \centerline{Atelectasis}
        \end{minipage}
        \begin{minipage}[b]{0.136\linewidth}
            \centering
            \includegraphics[width=\linewidth]{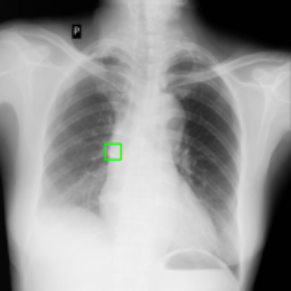}
            \centerline{Calcification}
        \end{minipage}
        \begin{minipage}[b]{0.136\linewidth}
            \centering
            \includegraphics[width=\linewidth]{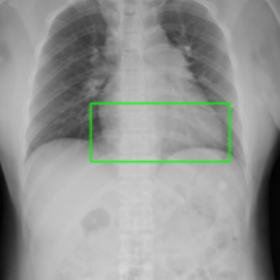}
            \centerline{Cardiomegaly}
        \end{minipage}
        \begin{minipage}[b]{0.136\linewidth}
            \centering
            \includegraphics[width=\linewidth]{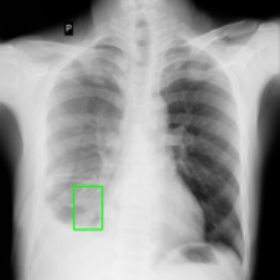}
            \centerline{Consolidation}
        \end{minipage}
        \begin{minipage}[b]{0.136\linewidth}
            \centering
            \includegraphics[width=\linewidth]{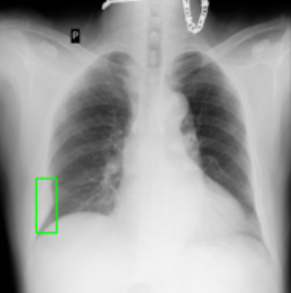}
            \centerline{Diffuse Nodule}
        \end{minipage}
        \begin{minipage}[b]{0.136\linewidth}
            \centering
            \includegraphics[width=\linewidth]{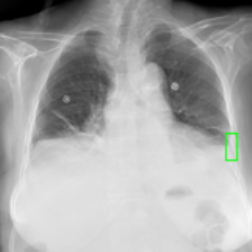}
            \centerline{Effusion}
        \end{minipage} 
        \begin{minipage}[b]{0.136\linewidth}
            \centering
            \includegraphics[width=\linewidth]{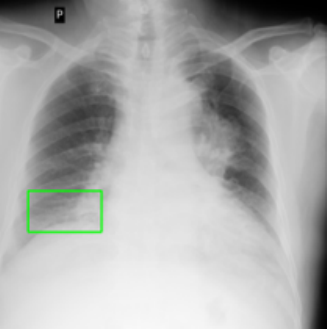}
            \centerline{Fibrosis}
        \end{minipage}
        \begin{minipage}[b]{0.136\linewidth}
            \centering
            \includegraphics[width=\linewidth]{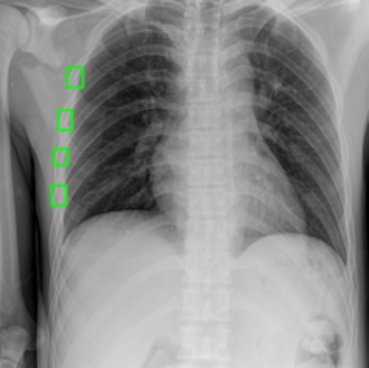}
            \centerline{Fracture}
        \end{minipage}
        \begin{minipage}[b]{0.136\linewidth}
            \centering
            \includegraphics[width=\linewidth]{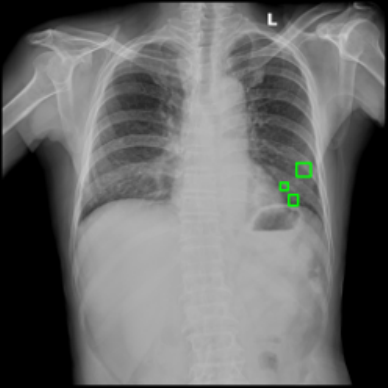}
            \centerline{Mass}
        \end{minipage}
        \begin{minipage}[b]{0.136\linewidth}
            \centering
            \includegraphics[width=\linewidth]{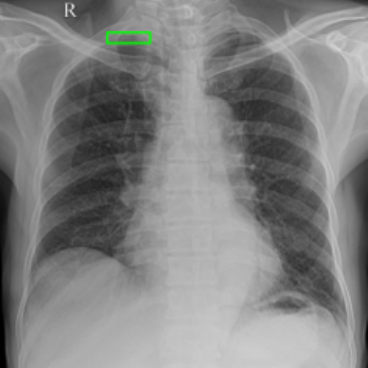}
            \centerline{Pleural.}
        \end{minipage}
        \begin{minipage}[b]{0.136\linewidth}
            \centering
            \includegraphics[width=\linewidth]{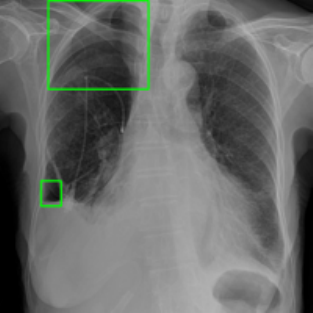}
            \centerline{Pneumothorax}
        \end{minipage} 
        \vspace{2mm}
        \centerline{(c) VinDr-CXR}
        \vspace{-4mm}
        \caption{Examples of annotated diseases in Chest X-ray datasets.}
        \vspace{-4mm}
    \label{fig:dataset}
\end{figure}

We utilize five publicly available CXR datasets in our experiments, including ChestX-Det~\cite{liu2020chestxdet10}, CANDID-PTX~\cite{feng2021curation}, Chest X-ray14~\cite{wang2017chestx}, VinDr-CXR~\cite{nguyen2020vindr}, and MIMIC-CXR~\cite{johnson2019mimic}. All images were selected from frontal PA/AP views, with the shortest edge rescaled to 512 pixels, followed by a center crop to a uniform size of 512 × 512 pixels. The distribution of disease categories is summarized in Table~\ref{tab:dataset_statistics}, and representative annotation examples are shown in Figure~\ref{fig:dataset}.

\begin{table}[h]
    \centering
    \caption{Overview of the datasets utilized for model training and evaluation.} 
    \resizebox{0.8\linewidth}{!}{
    \begin{tabular}{lcccccc}
        \toprule
        \multirow{2}{*}{\textbf{Dataset}} & ChestX-Det & CANDID-PTX & ChestX-Det & ChestX-ray14 & VinDr-CXR & MIMIC-CXR \\
        \cmidrule(lr){2-3} \cmidrule(lr){4-7}
        & \multicolumn{2}{c}{Train} & \multicolumn{4}{c}{Test} \\
        \midrule
        Normal & 547 & - & 64 & - & 2051 & 609 \\
        Atelectasis & 249 & - & 48 & 180 & 20 & 789 \\
        Calcification & 163 & - & 38 & - & 122 & - \\
        Cardiomegaly & 223 & - & 68 & 146 & 244 & 640 \\
        Consolidation & 1340 & - & 289 & - & 41 & 146 \\
        Diffuse Nodule & 109 & - & 25 & - & 52 & - \\
        Effusion & 1177 & - & 252 & 153 & 62 & 394 \\
        Emphysema & 137 & - & 39 & - & - & - \\
        Fibrosis & 417 & - & 82 & - & 170 & - \\
        Fracture & 313 & - & 76 & - & 6 & 41 \\
        Mass & 112 & - & 30 & 85 & 96 & - \\
        Nodule & 307 & - & 77 & 79 & - & - \\
        Pleural Thickening & 412 & - & 85 & - & 65 & - \\
        Pneumothorax & 154 & 1000 & 35 & 98 & 9 & 56 \\
        \midrule
        \textbf{Total} & 3001 & 1000 & 542 & 741 & 2938 & 2675 \\
        \bottomrule
    \end{tabular}
    }
    \label{tab:dataset_statistics}
\end{table}

We keep emphasizing anatomical and pathological alignment in this paper, because it ensures that generated abnormalities appear in medically plausible locations. As effusion typically appears at the lung base, cardiomegaly depends on the heart, and pneumothorax manifests along the pleural edge. Leveraging these anatomical constraints allows the model to focus generation on relevant regions rather than unrelated areas like the spine or clavicle. This alignment helps the model learn meaningful spatial priors, ensuring that synthesis remains clinically relevant rather than randomly dispersed. Consequently, our approach succeeds even with a small dataset by effectively capturing and utilizing these essential medical priors.

\section{Implementation Details}\label{appendix:implementation}

All experiments were conducted on four NVIDIA 80G A100 GPUs and implemented using PyTorch, along with the TorchXRayVision~\cite{cohen2022torchxrayvision} and Diffusers.

We utilized several pre-trained models in this paper. The anatomy segmentation and classification models were obtained from the TorchXRayVision library~\cite{cohen2022torchxrayvision}, using PSPNet for segmentation and ``densenet121-res224-all'' weights for classification. The RoentGen model~\cite{bluethgen2024vision} was used with default settings, with inference performed using the PNDM noise scheduler for 75 steps and a guidance scale of 4.

Our framework consists of two stages: Text-to-Mask and Mask-to-Image. Both stages are initialized from InstructPix2Pix~\cite{brooks2023instructpix2pix} and ControlNet~\cite{zhang2023adding}, respectively. Training configuration and wall-clock time are summarized in Table~\ref{tab:train_config}. 

\begin{table}[h]
\centering
\small
\caption{Training configuration (4 $\times$ 80 GB NVIDIA A100).}
\label{tab:train_config}
\resizebox{0.8\linewidth}{!}{
\begin{tabular}{lcccccc}
\toprule
\textbf{Stage} & \textbf{Base method} & \textbf{Steps} & \textbf{LR} & \textbf{Batch (total/per-GPU)} & \textbf{GPUs} & \textbf{Wall-clock} \\
\midrule
Stage 1 (Text-to-Mask) & InstructPix2Pix & 6,000 & $1\times10^{-5}$ & 128 / 32 & 4 & $\sim$7 h \\
Stage 2 (Mask-to-Image) & ControlNet & 6,000 & $1\times10^{-5}$ & 128 / 32 & 4 & $\sim$7 h \\
\bottomrule
\end{tabular}
}
\end{table}

\begin{table}[h]
\centering
\small
\caption{Inference configuration (single NVIDIA A100, peak GPU memory $\sim$23 GB).}
\label{tab:inference_config}
\resizebox{0.55\linewidth}{!}{
\begin{tabular}{lccc}
\toprule
\textbf{Stage} & \textbf{Scheduler} & \textbf{Steps} & \textbf{Time per attempt} \\
\midrule
Stage 1 (Text-to-Mask) & UniPCMultistep & 20 & $\sim$5 s \\
Stage 2 (Mask-to-Image) & UniPCMultistep & 70 & $\sim$11 s \\
\bottomrule
\end{tabular}
}
\end{table}

For inference, we adopted UniPCMultistepScheduler with 20 steps for text-to-mask and 70 steps for mask-to-image generation (Table~\ref{tab:inference_config}). We further employ a self-assessment loop with early stopping, where the number of attempts is configurable by \texttt{max\_attempts}. Each attempt takes $\sim$16 s end-to-end; thus, in the worst case, the latency scales linearly with \texttt{max\_attempts} (e.g., 1 attempt $\sim$16 s, 5 attempts $\sim$80 s). In practice, early stopping reduces the average to $\sim$60 s per sample ($\sim$3–4 attempts).

For downstream tasks, we adopted the baseline approach from ChestX-Det~\cite{liu2020chestxdet10}, utilizing a Mask R-CNN + FPN architecture with an ImageNet-pretrained backbone. 

\section{Prompt Tool} \label{appendix:prompt_tool}
The pseudocode for the text prompt generator is provided in Algorithm \ref{alg:prompt_generator}. We utilize pixel-level abnormal annotations from Chest-Det and CANDID-PTX dataset, anatomical masks obtained by a pre-trained organ segmentation model~\cite{cohen2022torchxrayvision}. The lungs are segmented into left and right lungs, with each further divided into upper, middle, and lower regions. Ultimately partitioning the CXR into 6 sub-regions and 5 major regions (left lung, right lung, bilateral lung, mediastinum, and heart). The reason for using major regions is that severe diseases often occupy one full side of the lung or both lungs, thus requiring major lung regions for better description.

For severity assessment, we classify conditions into mild, moderate, and severe based on the overlap area between the lesion and organ masks. For cardiomegaly, a different criterion was used, calculating the ratio between the heart-to-chest ratio, according to the medical definition. We admit the limitations of this rule-based approach—it relies solely on spatial overlap and does not tell multiple disease relative location. However, it is a effective way to provide a structured and standardized text prompt for pathology mask generation. Our final prompt format follows the structure: \textit{``A photo of a chest X-ray with {[SEVERITY1] [CLASS1] on [LOC1]}, {[SEVERITY2] [CLASS2] on [LOC2]}, ..."}, Where: 
\begin{itemize}
    \item SEVERITY tokens include: mild, moderate, severe
\item CLASS tokens include: pneumonia, atelectasis, cardiomegaly, effusion, infiltration, mass, nodule, pneumothorax, consolidation, edema, emphysema, fibrosis, pleural thickening
\item LOC tokens include: left upper lung, left middle lung, left lower lung, right upper lung, right middle lung, right lower lung, mediastinum, heart, left lung, right lung, bilateral lung
\end{itemize}

\begin{figure}[t]
\centering
\resizebox{0.8\linewidth}{!}{%
\begin{minipage}{\linewidth}
\begin{algorithm}[H]
\caption{Prompt Tool}
\label{alg:prompt_generator}
\begin{algorithmic}[1]
\REQUIRE Disease $C$, organ mask $O$, path mask $P$
\ENSURE Standardized text prompt $T$ (severity $S$, disease $C$, location $L$)
\STATE $O_i \gets \text{DefineOrganParts}(O)$
\STATE Initialize empty prompt $T$
\FOR{each pathology mask $P_j$ in $P$}
    \STATE Identify overlapping regions $R_i = P_j \cap O_i$
    \STATE Calculate overlap area $A_i = \text{Area}(R_i)$
    \IF{$A_i \neq \emptyset$}
        \IF{$C = Cardiomegaly$}
            \STATE $L \gets heart$
            \STATE $S$ based on heart-to-thorax ratio
        \ELSE
            \STATE main part $\gets \arg\max(A_i)$
            \IF{both lungs overlap}
                \STATE $L \gets \text{bilateral lung}$
            \ELSE
                \STATE $L \gets$ main part
                \STATE $S$ based on $A_i$ according to $C$ threshold
            \ENDIF
        \ENDIF
    \ELSE
        \STATE $C \gets Normal$
    \ENDIF
    \STATE Append $S C$ on $L$ to $T$
\ENDFOR
\RETURN "A Chest x-ray photo with " + $T$
\end{algorithmic}
\end{algorithm}
\end{minipage}%
}
\vspace{-2mm}
\end{figure}

\section{Free-Text Prompt Compatibility}\label{appendix:free_text}

The prompts used in our main experiments are derived from structured templates and geometric heuristics. However, a natural question is whether AURAD can also operate on less structured, report-style text, as found in radiology reports.

Publicly available datasets that provide disease masks (e.g., ChestDet, CANDID-PTX) do not release the corresponding reports, limiting access to clinician-authored free-text. To approximate this setting, we used GPT to rewrite our structured prompts into free-form, MIMIC-style ``Impression'' statements (Table \ref{tab:free_form_examples}). These free-form rewrites serve as proxies for radiology reports and allow us to stress-test the model under both zero-shot and from-scratch settings. No architectural changes were required; the same AURAD pipeline was used.

\begin{table}[h]
\centering
\small
\caption{Examples of structured prompts rewritten into free-form prompt.}
\label{tab:free_form_examples}
\resizebox{0.95\linewidth}{!}{
\begin{tabular}{p{0.48\linewidth} p{0.50\linewidth}}
\toprule
\textbf{Structured Prompt} & \textbf{Free-form Prompt} \\
\midrule
mild Pneumothorax on left lung & Mild left-sided pneumothorax is present. \\
\midrule
moderate Cardiomegaly on heart, mild Fibrosis on right middle lung & The heart is moderately enlarged. Mild fibrosis is present in the right middle lung. \\
\midrule
severe Consolidation on bilateral lung, mild Effusion on right lower lung, mild Nodule on right lower lung, mild Pneumothorax on right lower lung & There is severe consolidation present in both lungs. Additionally, there is a pleural effusion and pneumothorax noted in the right lower lung, as well as a small nodule in the same region. \\
\bottomrule
\end{tabular}
}
\end{table}

As shown in Table~\ref{tab:free_form_results}, end-to-end free-form training (Exp C) is competitive with the structured baseline (AUC 57.64 vs. 59.31), indicating robustness of the CLIP text encoder to natural phrasing. Stage II (mask-to-image) generalizes well to free-form text, reflecting its RoentGen backbone trained on report-style inputs; zero-shot (Exp A) performs reasonably, and retraining (Exp B) nearly closes the gap to the baseline. Stage I (text-to-mask) is more sensitive: free-form training slightly reduces performance, as dense text-to-mask generation is domain-specific and benefits from reduced linguistic variance under limited data. Overall, the structured template baseline remains optimal, but AURAD is fully compatible with free-text supervision and would directly benefit from datasets that provide aligned report–mask–image pairs.

\begin{table}[h]
\centering
\small
\caption{Performance under free-form prompts vs. structured baseline.``Free-form*'' indicates the stage was trained from scratch with free-form prompts; otherwise the model was frozen.}
\label{tab:free_form_results}
\resizebox{0.8\linewidth}{!}{
\begin{tabular}{lcccccc}
\toprule
\textbf{ID} & \textbf{Stage I (Train/Infer)} & \textbf{Stage II (Train/Infer)} & \textbf{FID ↓} & \textbf{CLIP ↑} & \textbf{MS-SSIM ↑} & \textbf{Path. AUC ↑} \\
\midrule
A & Structured / Structured & Structured / Free-form & 40.42 & 29.74 & 0.5663 & 56.69 \\
B & Structured / Structured & Free-form* / Free-form & 37.63 & 28.97 & 0.5594 & 58.11 \\
C & Free-form* / Free-form & Free-form* / Free-form & 37.40 & 29.07 & 0.5629 & 57.64 \\
D (Ours) & Structured / Structured & Structured / Structured & \textbf{32.83} & \textbf{30.14} & \textbf{0.5968} & \textbf{59.31} \\
\bottomrule
\end{tabular}
}

\end{table}

\section{Choice of Text-to-Mask Backbone}\label{appendix:instructpix2pix}

A key design choice in AURAD is the use of InstructPix2Pix as the backbone for text-to-mask generation. Alternative approaches such as text-to-layout are not suitable in this context: layout models typically predict bounding boxes or keypoints rather than dense masks, and box-level layout is insufficient for capturing organ–lesion geometry in CXR (e.g., differentiating pleural effusion in the pleural space with a meniscus versus intraparenchymal consolidation with distinct shapes and locations). 

To validate this choice, we compared InstructPix2Pix against state-of-the-art instruction-guided editing baselines, including DiffEdit~\cite{couairon2022diffedit} and Add-It~\cite{tewel2024add}. Results are summarized in Table~\ref{tab:pix2pix_comparison}.

\begin{table}[h]
\centering
\small
\caption{Comparison of instruction-guided editing methods for CXR mask generation.}
\label{tab:pix2pix_comparison}
\resizebox{0.95\linewidth}{!}{
\begin{tabular}{lcccccc}
\toprule
\textbf{Method} & \textbf{IoU \(\uparrow\)} & \textbf{Dice \(\uparrow\)} & \textbf{FID \(\downarrow\)} & \textbf{MS-SSIM \(\uparrow\)} & \textbf{Qualitative (Anatomy--Pathology Consistency)} & \textbf{Inference Time (A100)} \\
\midrule
InstructPix2Pix & \textbf{0.6897} & \textbf{0.7772} & \textbf{15.19} & \textbf{0.9286} & \(\checkmark\)\,consistent & \(\sim\)5\,s \\
DiffEdit        & 0.4596 & 0.5935 & 277.53 & 0.7691 & spurious border edits & \(\sim\)10\,s \\
Add-It          & 0.3901 & 0.5074 & 297.79 & 0.7387 & mislocalized attention & \(\sim\)3\,min \\
\bottomrule
\end{tabular}
}
\end{table}

Despite strong results on natural images, DiffEdit and Add-It underperform on CXR masks, yielding lower overlap (IoU/Dice), degraded fidelity (FID), and spurious or mislocalized edits, while also being more computationally expensive. By contrast, InstructPix2Pix proves most stable and efficient for disease-aware mask generation. It is well-established, integrated in the Diffusers library, easily transferable to medical settings, and data-/compute-efficient, making it the most appropriate choice for our framework.

\section{Detection and Segmentation: all diseases}

Although our main analysis focuses on five major disease categories, we conducted data augmentation experiments across all available classes (Table~\ref{tab:full_results}). The performance trends are consistent with those observed in the main categories, demonstrating that our synthesized data effectively improves detection performance.

\begin{table*}[ht]
\centering
\vspace{-2mm}
\caption{13-Class Performance comparison of ChestX-Det dataset on detection (mAP@50), segmentation (Dice), and segmentation (IoU).}
\resizebox{\textwidth}{!}{%
    \begin{tabular}{lcccccccccccccc}
        \toprule
        & Atelectasis & Calcific & Cardio. & Consolid. & Diffuse & Effusion & Emphysem. & Fibrosis & Fracture & Mass & Nodule & Pleural & Pneumo. & Overall \\
        \midrule
        \multicolumn{15}{c}{\textbf{Detection - mAP@50}} \\
        \midrule
        Real & 43.60 & 18.37 & \textbf{99.43} & 22.74 & 90.96 & 23.66 & 92.43 & 44.46 & \textbf{31.15} & 13.55 & 17.67 & \textbf{28.78} & 28.82 & 42.74  \\
        Real+1xCheff & 37.96 & 15.79 & 99.02 & \textbf{22.77} & 80.34 & 23.13 & 91.02 & \textbf{45.26} & 22.43 & 10.31 & 13.15 & 24.57 & \textbf{33.52} & 39.94 \\
        Real+1xRoentGen & 44.58 & \textbf{20.36} & 97.24 & \textbf{22.77} & 86.20 & \textbf{25.02} & 92.28 & 44.87 & 20.02 & 12.73 & \textbf{19.42} & 21.15 & 19.02 & 40.43  \\
        Real+1xOurs & \textbf{51.91} & 20.24 & 99.17 & \textbf{22.77} & \textbf{93.87} & 23.32 & \textbf{93.52} & 43.71 & 22.76 & \textbf{18.17} & 14.49 & 30.91 & 37.49 & \textbf{44.02}  \\
        \midrule
        \multicolumn{15}{c}{\textbf{Segmentation - Dice}} \\
        \midrule
        Real & 22.15 & 16.05 & 70.01 & \textbf{45.90} & \textbf{50.14} & \textbf{42.36} & \textbf{55.95} & \textbf{31.68} & \textbf{22.39} & 8.86 & 24.49 & \textbf{24.41} & 8.75 & 32.55 \\
        Real+1xCheff & 25.68 & \textbf{16.80} & 57.49 & 43.03 & 36.42 & 39.63 & 50.98 & 27.30 & 19.68 & \textbf{15.83} & 25.50 & 19.61 & 9.12 & 29.78 \\
        Real+1xRoentGen & 25.56 & 15.24 & 65.81 & 38.97 & 45.96 & 33.85 & 51.43 & 23.24 & 17.10 & 14.60 & 28.00 & 18.16 & 5.92 & 29.52 \\
        Real+1xOurs & \textbf{25.70} & 13.12 & \textbf{72.62} & 45.82 & 47.50 & 42.09 & 54.53 & 27.99 & 19.04 & 15.29 & \textbf{28.63} & 21.44 & \textbf{9.57} & \textbf{32.56} \\
        \midrule
        \multicolumn{15}{c}{\textbf{Segmentation - IoU}} \\
        \midrule
        Real & 15.48 & 10.94 & 59.58 & 34.32 & \textbf{39.27} & \textbf{30.29} & \textbf{43.63} & \textbf{23.93} & \textbf{16.52} & 6.44 & 18.54 & \textbf{15.98} & 5.40 & \textbf{24.64} \\
        Real+1xCheff & \textbf{18.32} & \textbf{11.48} & 48.70 & 32.27 & 27.86 & 28.44 & 40.06 & 20.19 & 14.56 & \textbf{11.81} & 19.11 & 12.69 & 5.55 & 22.39 \\
        Real+1xRoentGen & 17.77 & 10.82 & 55.97 & 29.08 & 34.36 & 23.97 & 39.60 & 16.66 & 12.74 & 11.13 & \textbf{21.44} & 11.85 & 3.42 & 22.22 \\
        Real+1xOurs & 18.28 & 9.36 & \textbf{62.08} & \textbf{34.52} & 36.10 & 30.14 & 42.57 & 20.17 & 13.76 & 11.53 & 21.56 & 13.91 & \textbf{5.84} & 24.60 \\
        \bottomrule
    \end{tabular}}
    \vspace{-2mm}
    \label{tab:full_results}
\end{table*}

However, segmentation tasks—which demand precise, pixel-level annotations—are more sensitive to label quality. In some cases, incorporating synthesized masks with imperfect alignment or semantic noise can lead to performance degradation. Models trained solely on real data tend to better match the test distribution, while synthesized samples may introduce distributional gaps. Interestingly, although our synthesized masks are sometimes subjectively more plausible, discrepancies with the ground truth labels reduce their measured accuracy, underscoring the broader challenges of annotation quality and the need for standardized labeling in medical imaging.

\section{Augmented with imbalanced Data}

Table \ref{tab:imbalance} presents the results from our previous experiments on long-tail data, where we focused on addressing the challenges posed by imbalanced data distribution. The results demonstrate a enhancement in detection tasks, with the synthesized data contributing to a more balanced performance across all categories, mitigating the impact of long-tail data.

\begin{table}[h]
  \centering
  \vspace{-4mm}
  \caption{Synthesis data augmentation for long-tail problems.}
  \resizebox{0.75\columnwidth}{!}{
  \begin{tabular}{@{}llccccl@{}}
    \toprule
    Train Set & Atelectasis & Cardio. & Consolid. & Effusion & Pneumo. & Overall \\
    \midrule
    Real & 43.60 & 99.43 & 22.74 & 23.66 & 28.82 & 43.65 \\
    Real+Syn. Atelectasis & 48.08(↑3.28) & 96.36 & 22.77 & 23.85 & 35.35 & 45.28(↑1.63) \\
    \bottomrule
  \end{tabular}}
  \vspace{-4mm}
  \label{tab:imbalance}
\end{table}

\section{Radiologist Evaluation}\label{appendix:radiologist}

\paragraph{Results.}
As shown in Table \ref{tab:eval_summary}), for Task 1 (Realism Rating), each image was rated on a 1–5 Likert scale by three board-certified radiologists, where 5 indicates "Definitely real". To compute an overall realism score, we first binarized each rating using a threshold of $\geq 4$ (i.e., probably realistic), and then averaged the binary ratings across the three annotators. This resulted in a continuous score between 0 and 1, reflecting the proportion of raters who found the image realistic. Using this criterion, \textbf{78\%} of the synthesized images were judged as realistic, indicating high visual fidelity.

For Task 2 (Segmentation Usefulness), radiologists provided binary labels (1 = helpful, 0 = not helpful) indicating whether the predicted segmentation masks would assist clinical interpretation. An image was considered \textit{useful} only if \textbf{all disease masks were correct} (i.e., a strict macro-level criterion across multi-label cases). The binary ratings were averaged across annotators and the resulting score was used to determine usefulness. Based on this, \textbf{41\%} of the segmentations were considered clinically useful. This reflects a promising level of practical utility, given the strict evaluation criteria and the complexity of overlapping pathologies.

\paragraph{Inter-Annotator Agreement (IAA).}
To assess the reliability of expert ratings, we computed inter-annotator agreement for both tasks. For Task 1 (realism rating, 1–5 Likert scale), we used the \textbf{Intra-class Correlation Coefficient (ICC)}, which measures absolute agreement among raters. It is defined as:
\begin{equation}
\text{ICC}(2,1) = \frac{MS_R - MS_E}{MS_R + (k - 1) \cdot MS_E}
\end{equation}
where $MS_R$ and $MS_E$ are the between-target and residual mean squares estimated from a two-way random-effects ANOVA, and $k$ is the number of raters.

For Task 2 (binary segmentation usefulness), we used \textbf{Fleiss' Kappa}, which generalizes Cohen's Kappa to multiple raters:
\begin{equation}
\kappa = \frac{\bar{P} - \bar{P}_e}{1 - \bar{P}_e}
\end{equation}
where $\bar{P}$ is the observed agreement and $\bar{P}_e$ is the agreement expected by chance.

The resulting agreement scores were \textbf{0.40} for ICC and \textbf{0.53} for Fleiss’ Kappa, indicating moderate consistency among raters across both tasks (Table \ref{tab:eval_summary}).

Despite the modest sample size, inter-rater reliability showed statistically robust agreement: ICC = \textbf{0.40} ($p=3\times10^{-12}$) and Fleiss’ $\kappa=\textbf{0.53}$ ($p=1\times10^{-9}$), both highly significant ($p \ll 0.05$) and within the moderate agreement range (0.40–0.60) across both tasks (Table~\ref{tab:eval_summary}). While the limited scale constrains formal statistical claims, the evaluations provide strong directional evidence: radiologist feedback consistently reflected realism and utility trends aligned with downstream task metrics.

\begin{figure}[h]
    \centering
    \begin{minipage}[c]{0.38\textwidth}
        \centering
        \includegraphics[height=6cm]{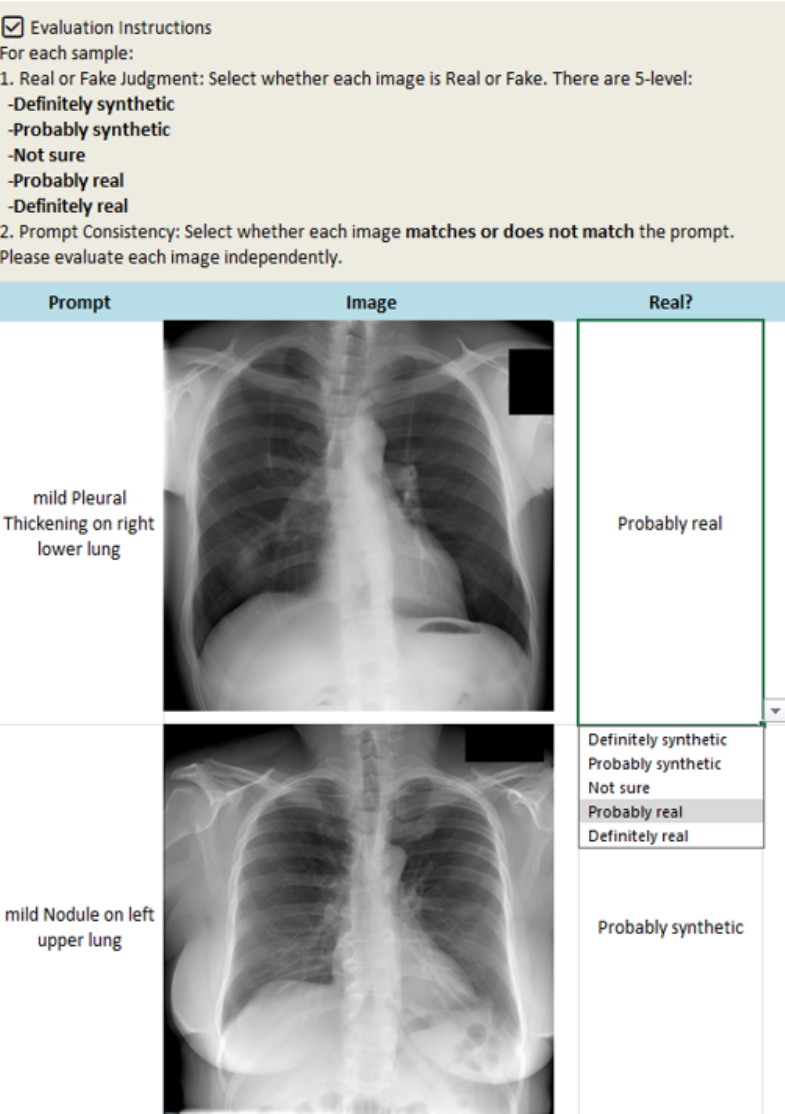}
        \caption*{(a)}
    \end{minipage}
    \hfill
    \begin{minipage}[c]{0.48\textwidth}
        \centering
        \includegraphics[height=6cm]{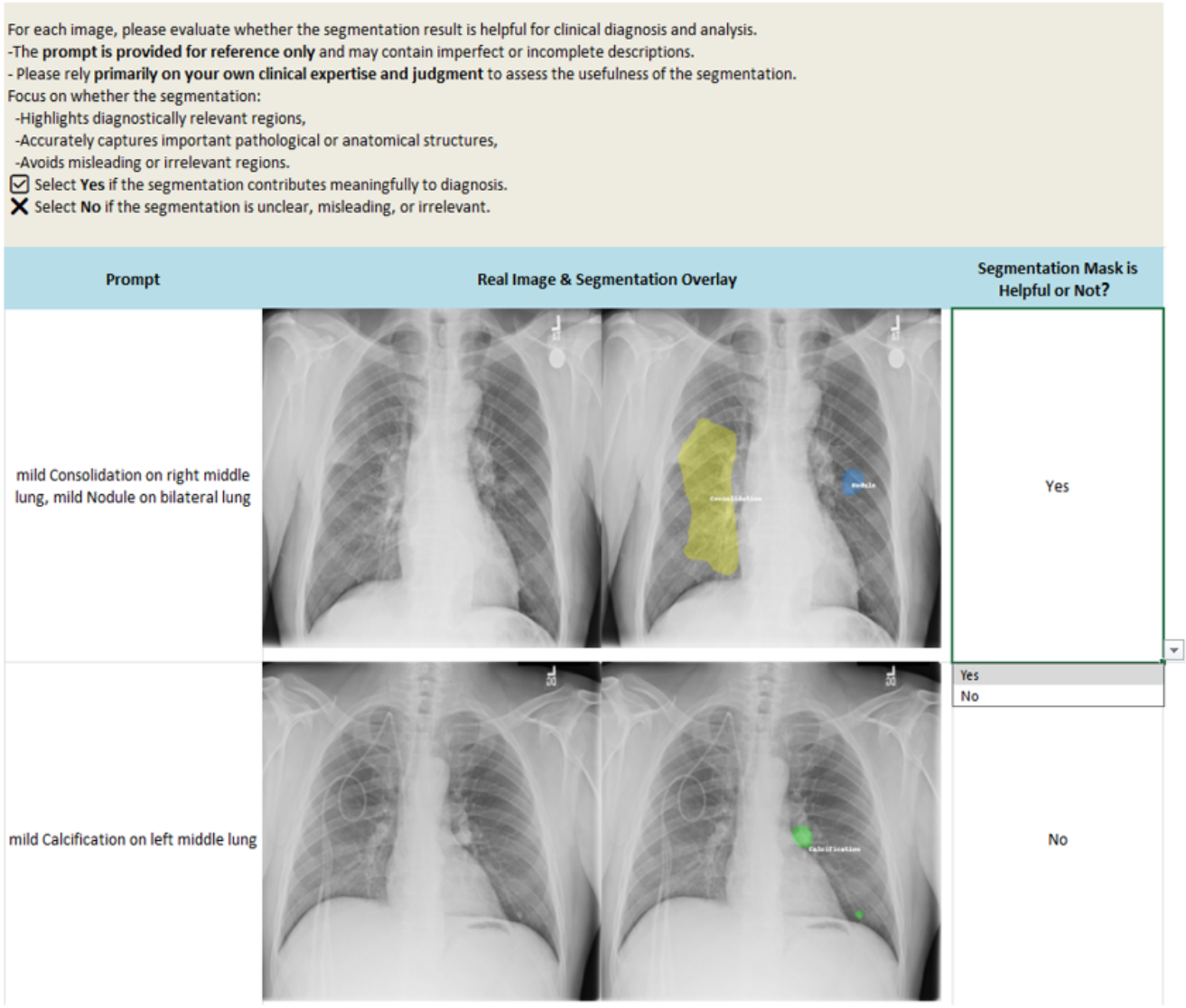}
        \caption*{(b)}
    \end{minipage}
    \vspace{-2mm}
    \caption{Screenshot of the evaluation interface.(a) Realism options for real and synthesized images. (b) Segmentation evaluation using models trained with our synthesized data.}
    \label{fig:radio_eval}
\end{figure}
\vspace{-5mm}

\begin{table}[ht]
\centering
\caption{Radiologist evaluation results.}
\label{tab:eval_summary}
\resizebox{0.55\linewidth}{!}{
\begin{tabular}{lcc}
\toprule
\textbf{Radiologist ID} & \textbf{Task I} & \textbf{Task II} \\
\midrule
A & 0.82 & 0.34 \\
B & 0.71 & 0.56 \\
C & 0.71 & 0.34 \\
\midrule
Avg. & 0.78 & 0.41 \\
Consensus rate & 0.60 & 0.42 \\
\midrule
IAA & ICC = 0.40 ($p=3\times10^{-12}$) & $\kappa=0.50$ ($p=1\times10^{-9}$) \\
\bottomrule
\end{tabular}
}
\end{table}
 
\paragraph{Evaluation Interface.}
We designed a table-based interface to standardize the assessment process. The display order was randomized to ensure blinding. To eliminate potential bias from synthesized artifacts, all textual overlays or image-embedded markings (e.g., patient identifiers, orientation labels) were masked. A screenshot of the evaluation interface is shown in Figure~\ref{fig:radio_eval}. Responses were collected in CSV format and included radiologist IDs, timestamps, and selected options for each task.

\section{Supplementary Visualizations: Representative Successes and Failures}

We present additional qualitative results, with each color denoting a specific disease: 
\textcolor[rgb]{1,0,0}{Atelectasis}, 
\textcolor[rgb]{0,1,0}{Calcification}, 
\textcolor[rgb]{0,0,1}{Cardiomegaly}, 
\textcolor[rgb]{1,1,0}{Consolidation}, 
\textcolor[rgb]{1,0.65,0}{Diffuse Nodule}, 
\textcolor[rgb]{0,1,1}{Effusion}, 
\textcolor[rgb]{1,0,1}{Emphysema}, 
\textcolor[rgb]{0.5,0,0.5}{Fibrosis}, 
\textcolor[rgb]{1,0.75,0.8}{Fracture}, 
\textcolor[rgb]{0.68,1,0.18}{Mass}, 
\textcolor[rgb]{0,0.5,1}{Nodule}, 
\textcolor[rgb]{0.29,0,0.51}{Pleural Thickening}, 
\textcolor[rgb]{1,0.41,0.71}{Pneumothorax}.

\begin{figure*}[t]
    \scriptsize
    \centering
    
    \begin{minipage}[t]{0.16\linewidth}
        \centering
        \includegraphics[width=\linewidth]{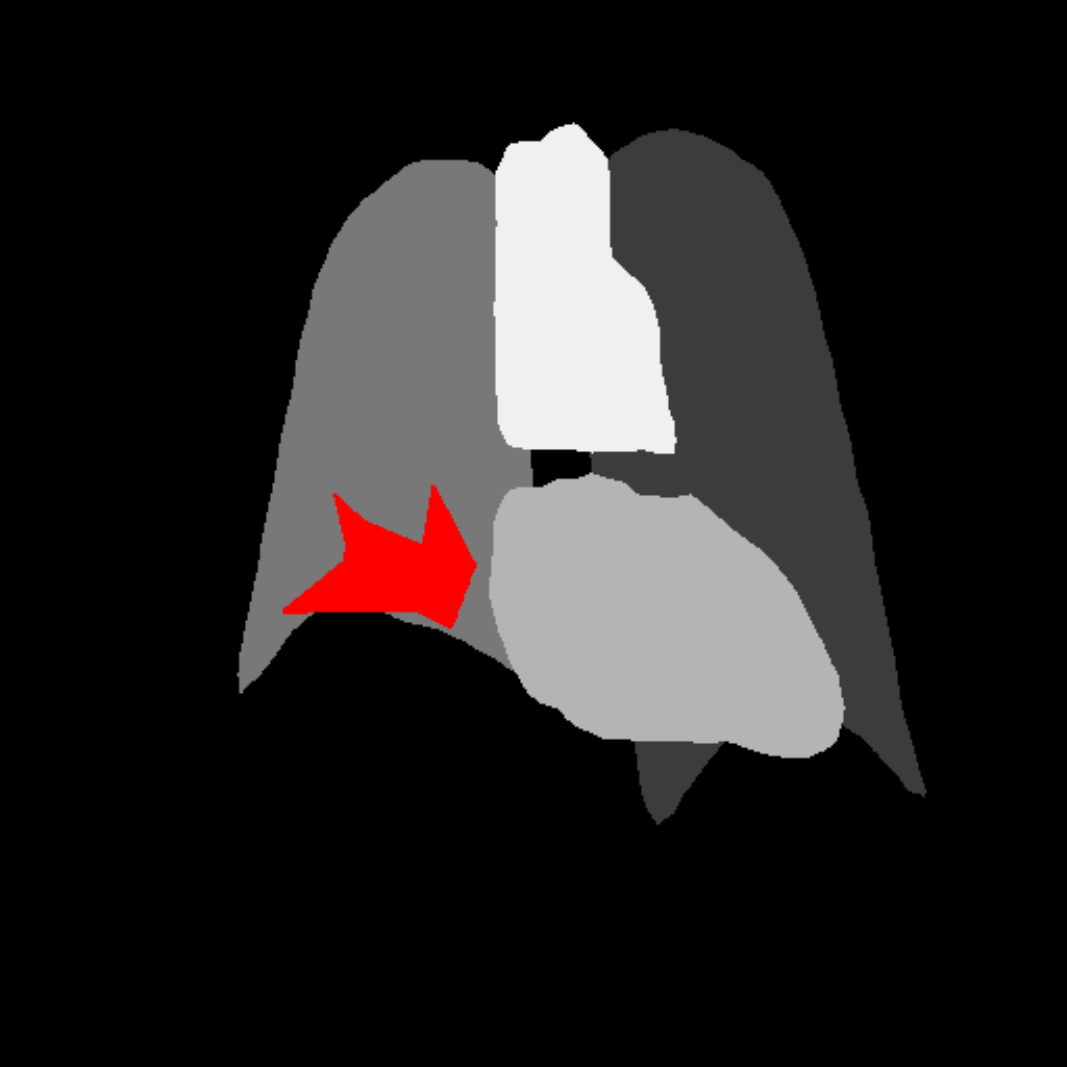}
    \end{minipage}
    \begin{minipage}[t]{0.16\linewidth}
        \centering
        \includegraphics[width=\linewidth]{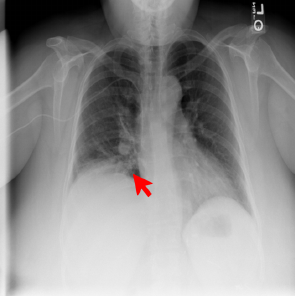}
    \end{minipage}
    \begin{minipage}[t]{0.16\linewidth}
        \centering
        \includegraphics[width=\linewidth]{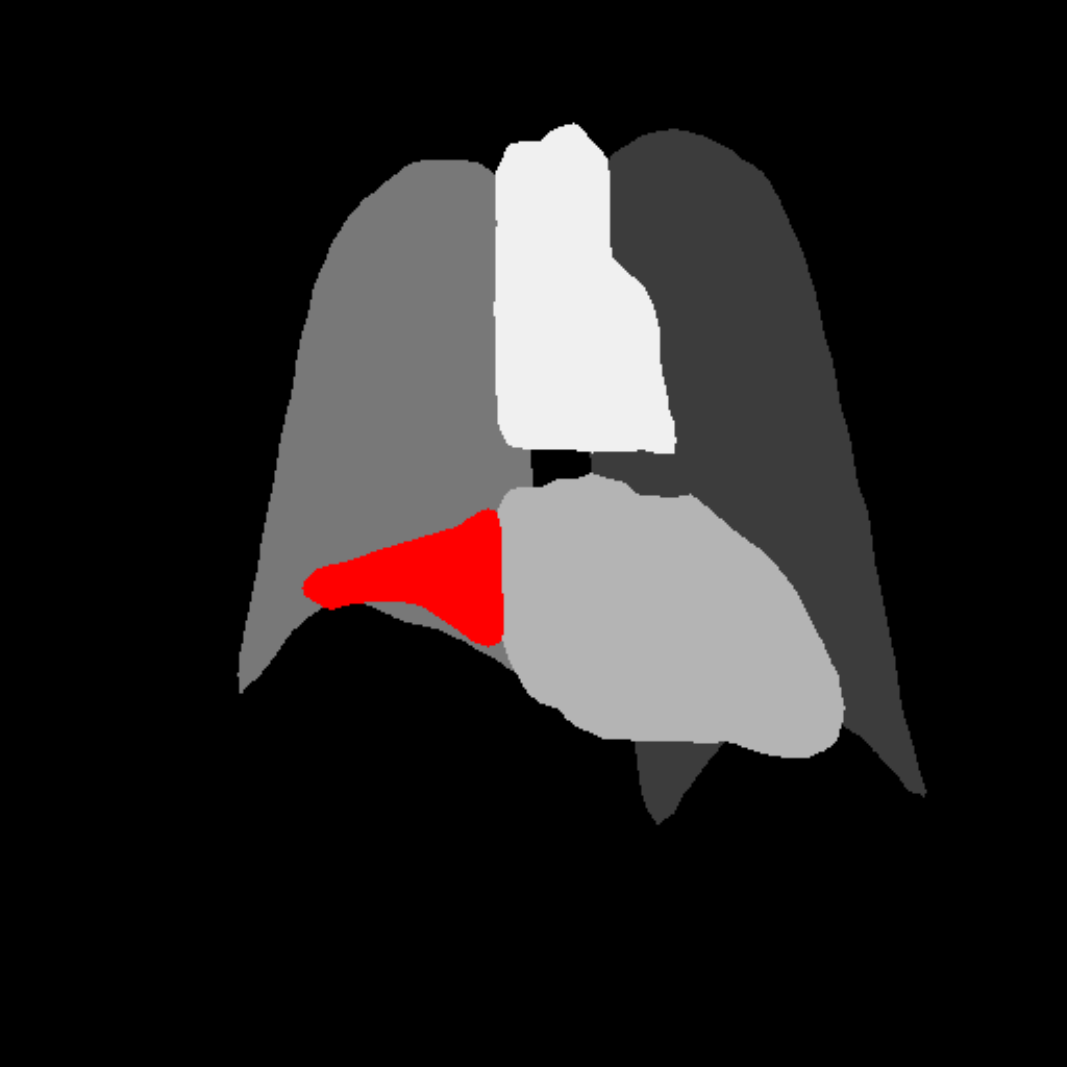}
    \end{minipage}
    \begin{minipage}[t]{0.16\linewidth}
        \centering
        \includegraphics[width=\linewidth]{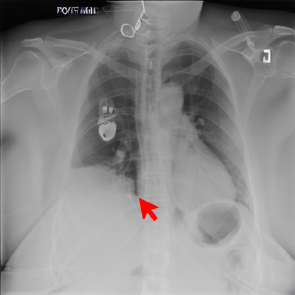}
    \end{minipage}
    \begin{minipage}[t]{0.16\linewidth}
        \centering
        \includegraphics[width=\linewidth]{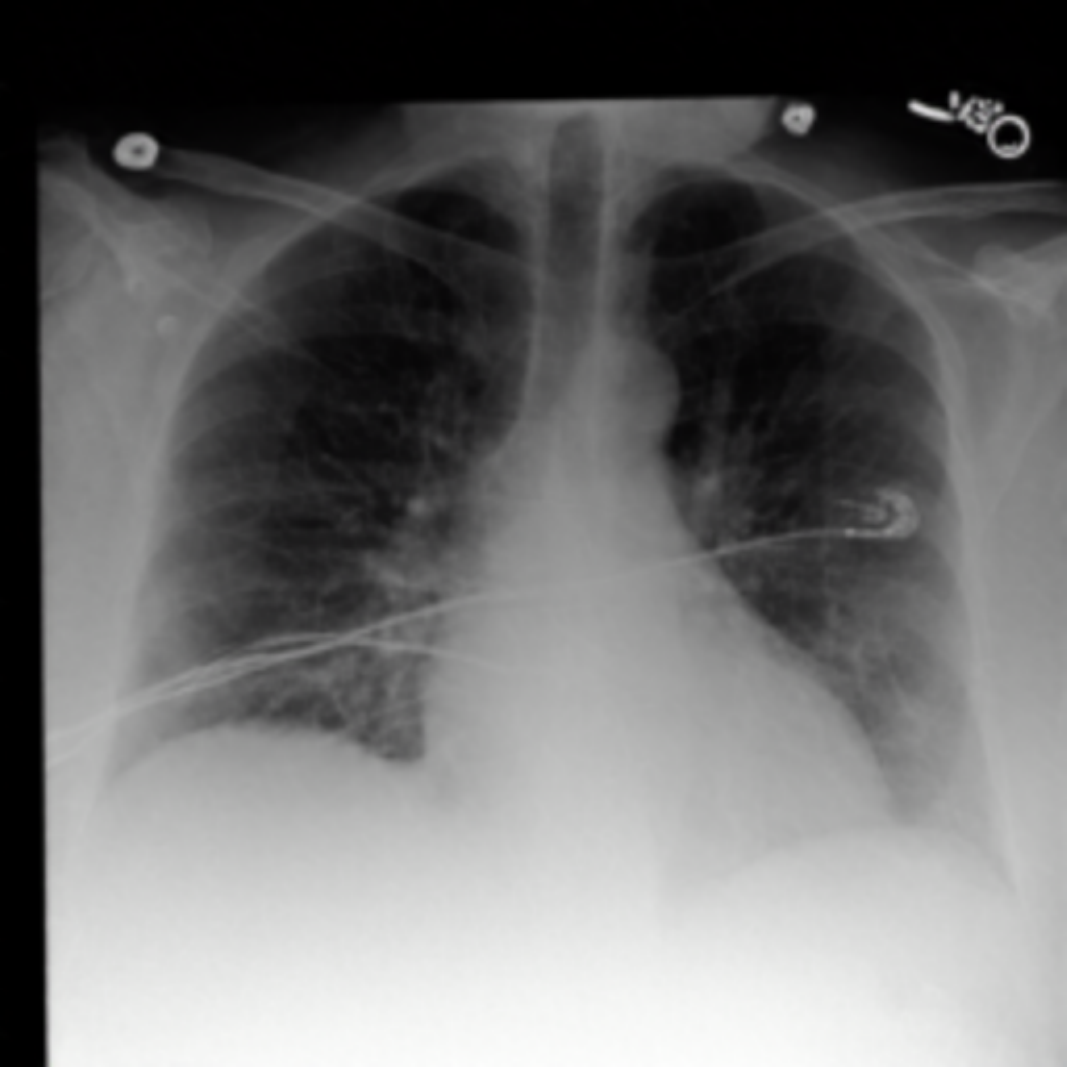}
    \end{minipage}
    \begin{minipage}[t]{0.16\linewidth}
        \centering
        \includegraphics[width=\linewidth]{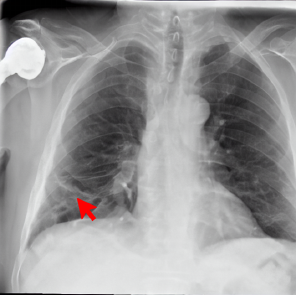}
    \end{minipage}

    \begin{minipage}[t]{0.16\linewidth}
        \centering
        \includegraphics[width=\linewidth]{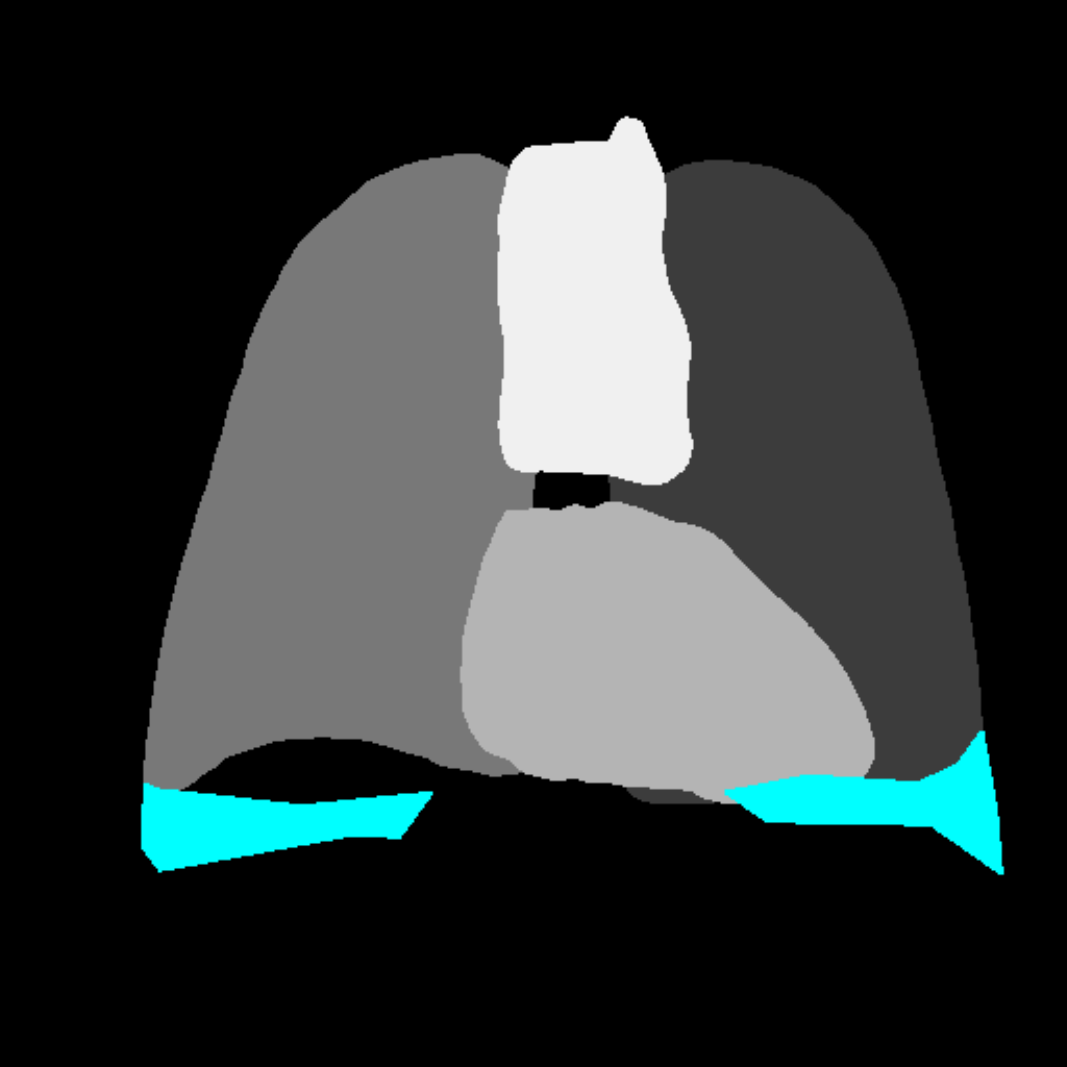}
    \end{minipage}
    \begin{minipage}[t]{0.16\linewidth}
        \centering
        \includegraphics[width=\linewidth]{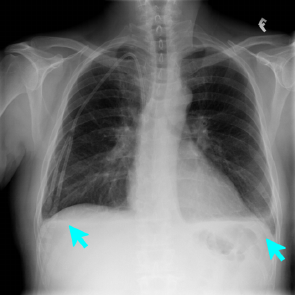}
    \end{minipage}
    \begin{minipage}[t]{0.16\linewidth}
        \centering
        \includegraphics[width=\linewidth]{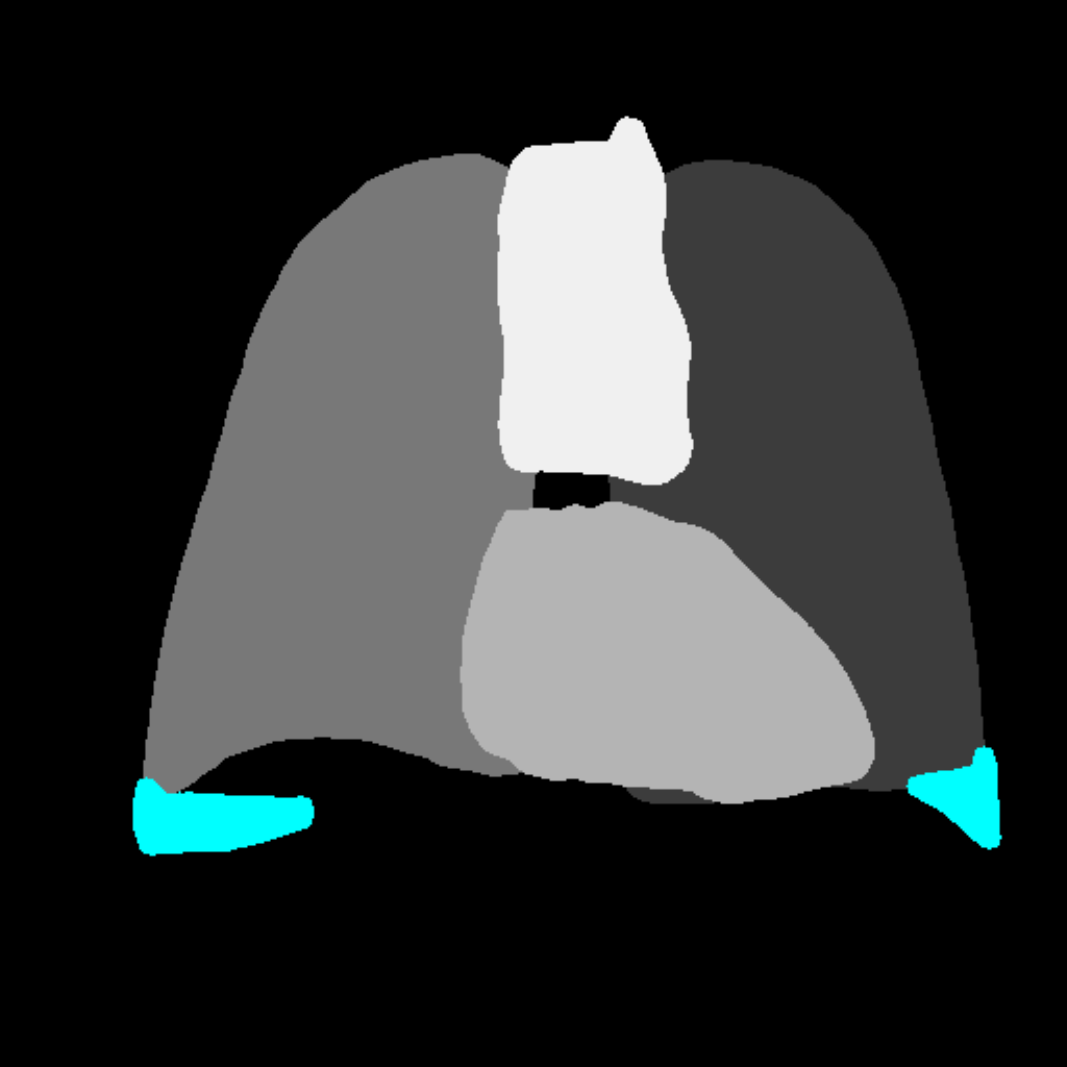}
    \end{minipage}
    \begin{minipage}[t]{0.16\linewidth}
        \centering
        \includegraphics[width=\linewidth]{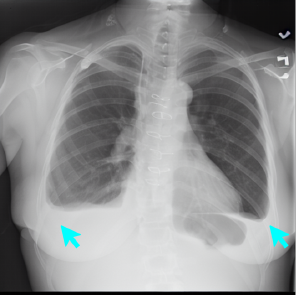}
    \end{minipage}
    \begin{minipage}[t]{0.16\linewidth}
        \centering
        \includegraphics[width=\linewidth]{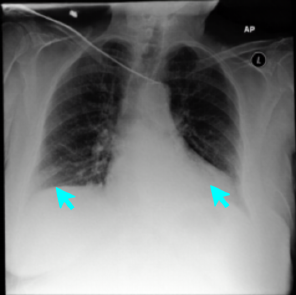}
    \end{minipage}
    \begin{minipage}[t]{0.16\linewidth}
        \centering
        \includegraphics[width=\linewidth]{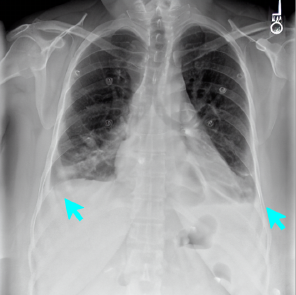}
    \end{minipage}

    \begin{minipage}[t]{0.16\linewidth}
        \centering
        \includegraphics[width=\linewidth]{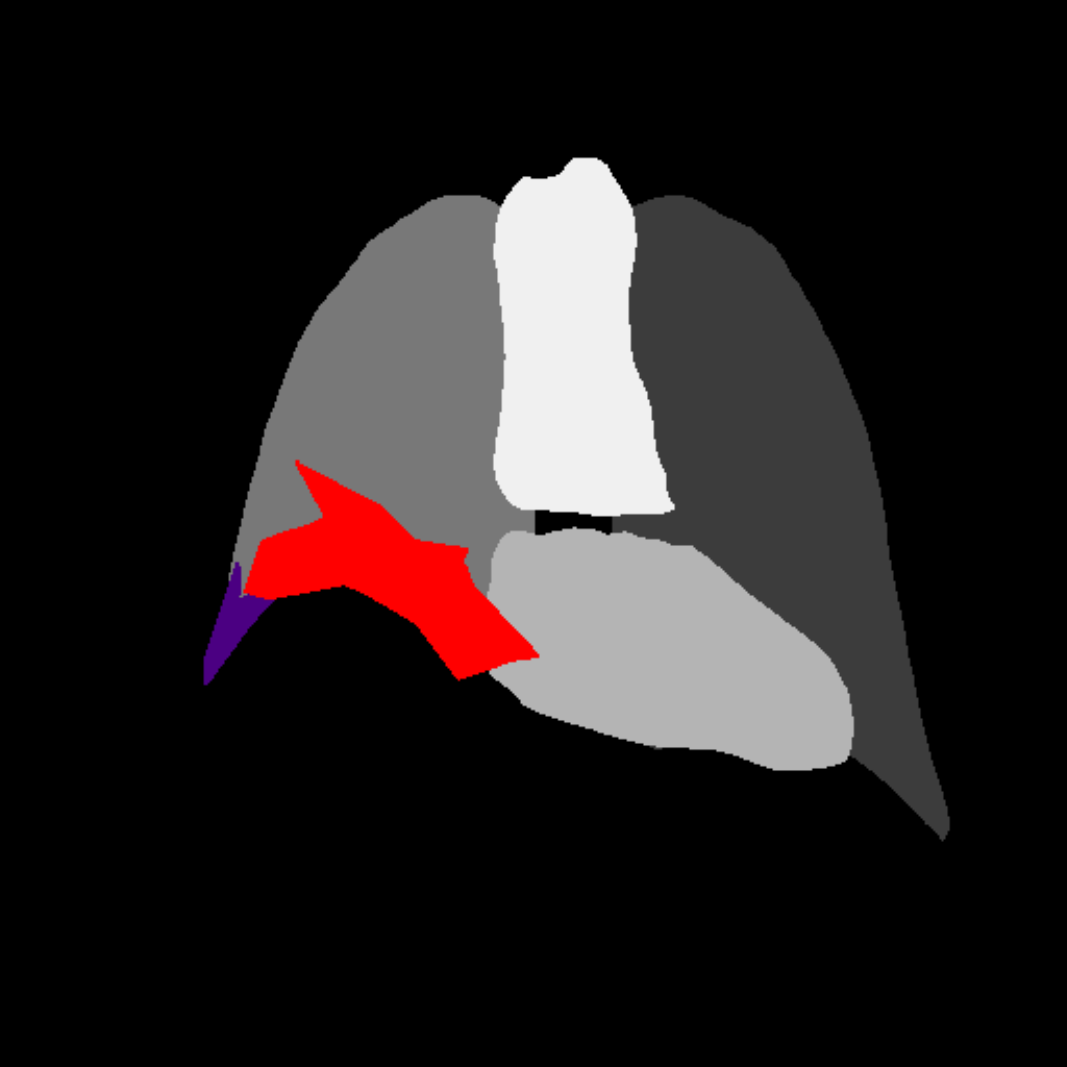}
    \end{minipage}
    \begin{minipage}[t]{0.16\linewidth}
        \centering
        \includegraphics[width=\linewidth]{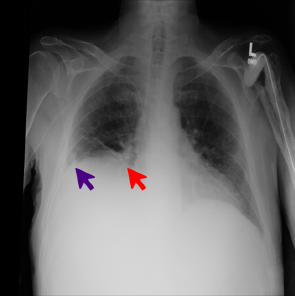}
    \end{minipage}
    \begin{minipage}[t]{0.16\linewidth}
        \centering
        \includegraphics[width=\linewidth]{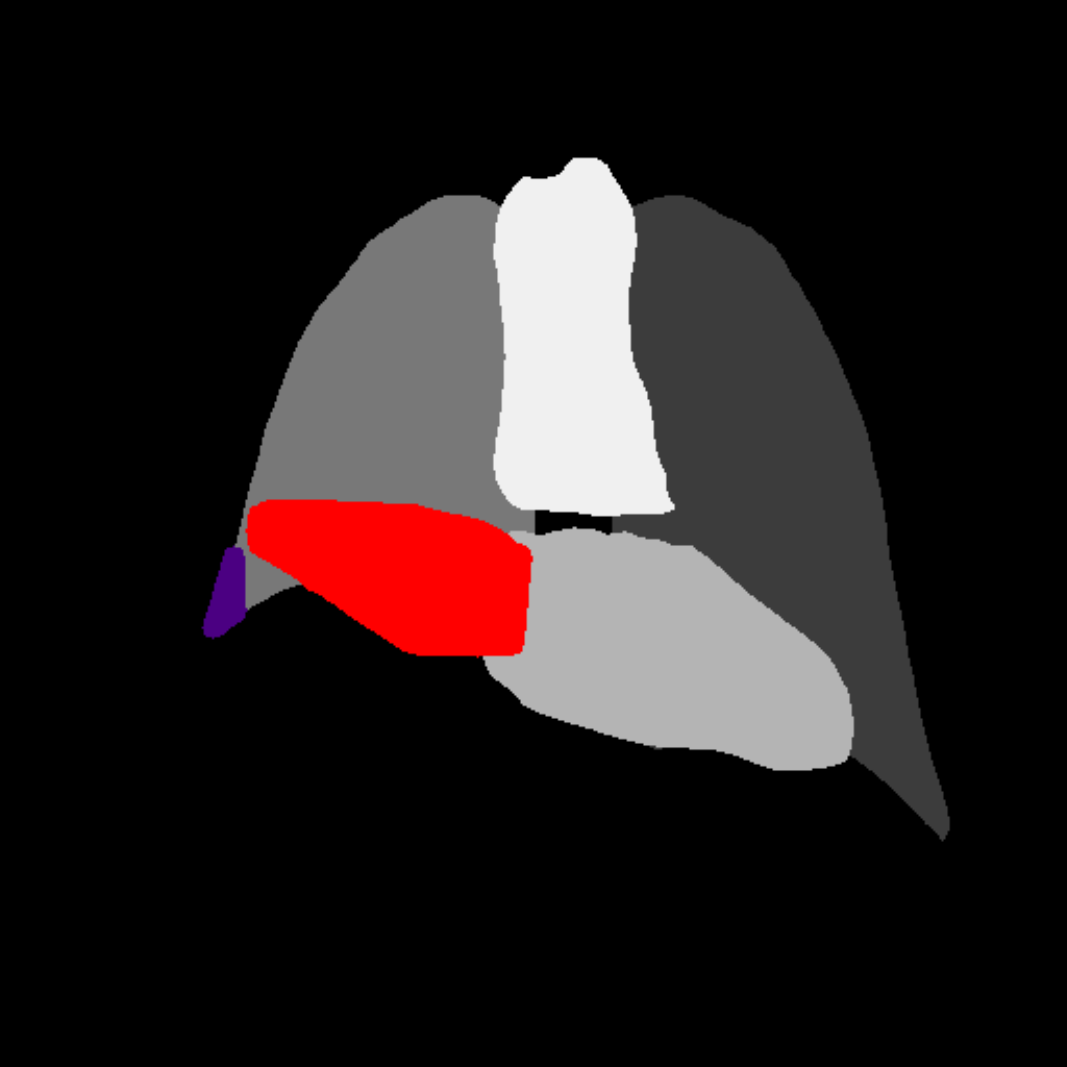}
    \end{minipage}
    \begin{minipage}[t]{0.16\linewidth}
        \centering
        \includegraphics[width=\linewidth]{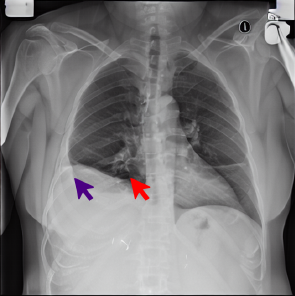}
    \end{minipage}
    \begin{minipage}[t]{0.16\linewidth}
        \centering
        \includegraphics[width=\linewidth]{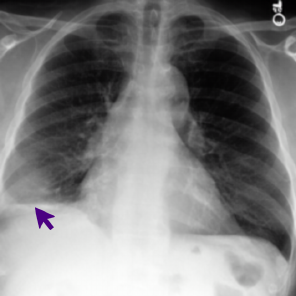}
    \end{minipage}
    \begin{minipage}[t]{0.16\linewidth}
        \centering
        \includegraphics[width=\linewidth]{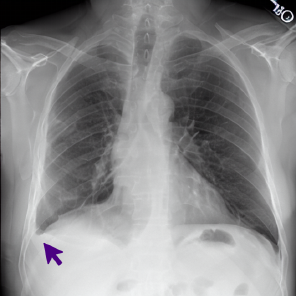}
    \end{minipage}

    \begin{minipage}[t]{0.16\linewidth}
        \centering
        \includegraphics[width=\linewidth]{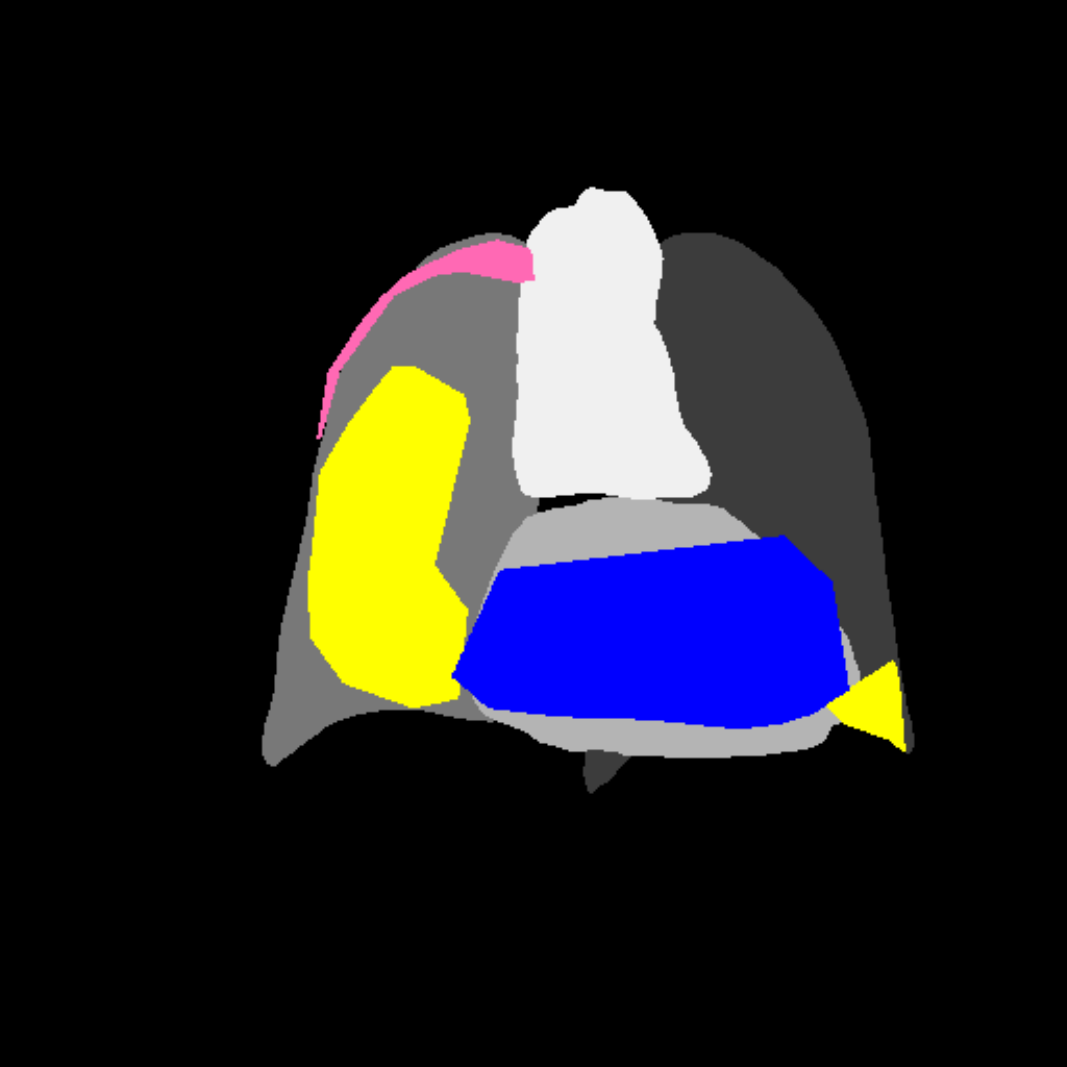}
    \end{minipage}
    \begin{minipage}[t]{0.16\linewidth}
        \centering
        \includegraphics[width=\linewidth]{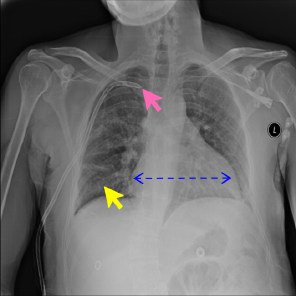}
    \end{minipage}
    \begin{minipage}[t]{0.16\linewidth}
        \centering
        \includegraphics[width=\linewidth]{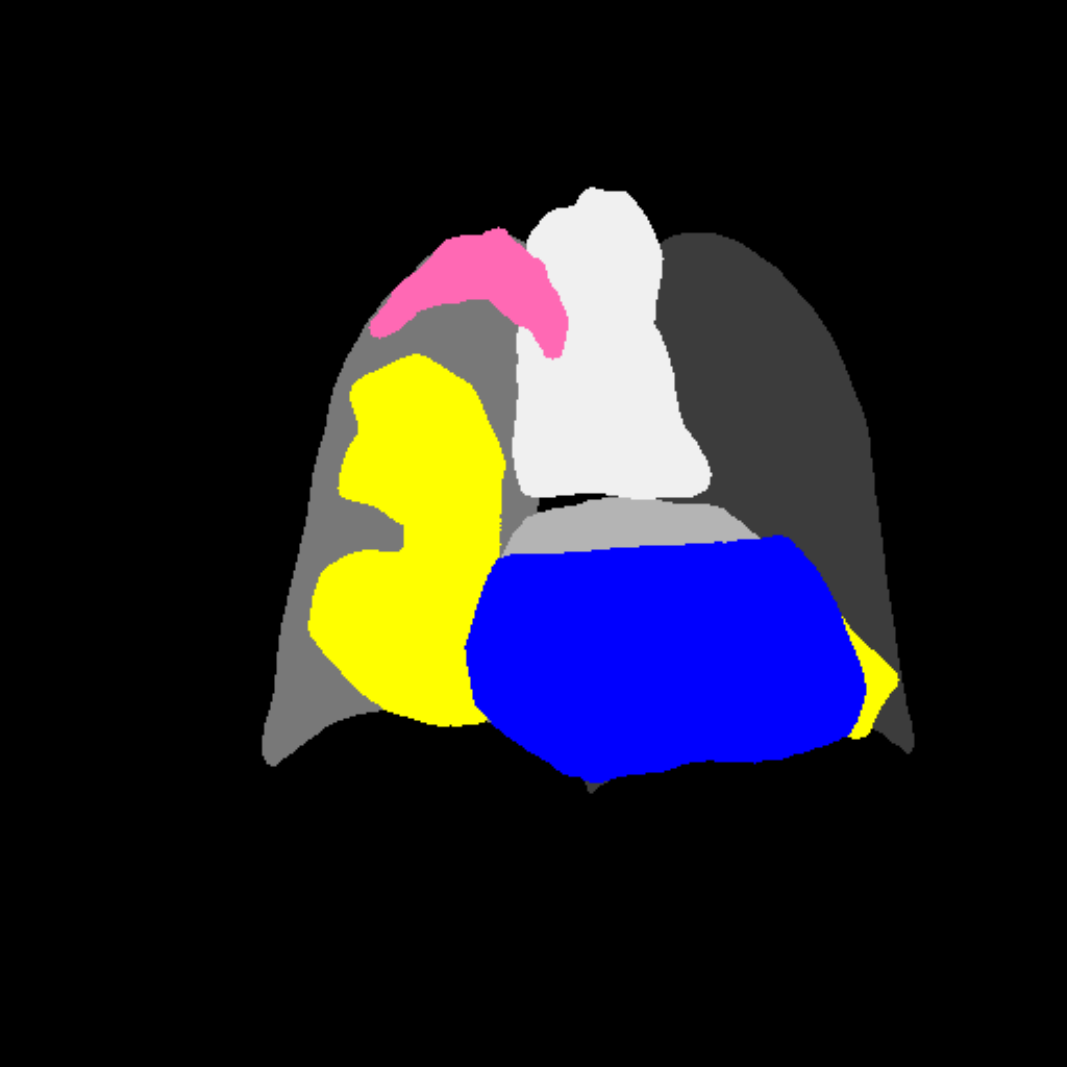}
    \end{minipage}
    \begin{minipage}[t]{0.16\linewidth}
        \centering
        \includegraphics[width=\linewidth]{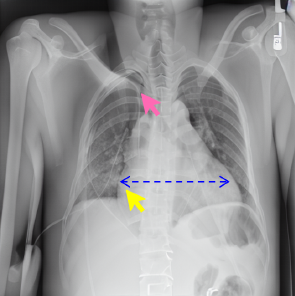}
    \end{minipage}
    \begin{minipage}[t]{0.16\linewidth}
        \centering
        \includegraphics[width=\linewidth]{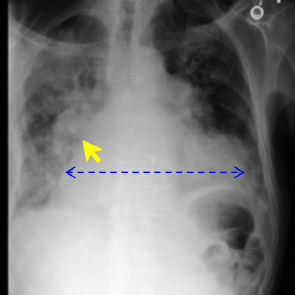}
    \end{minipage}
    \begin{minipage}[t]{0.16\linewidth}
        \centering
        \includegraphics[width=\linewidth]{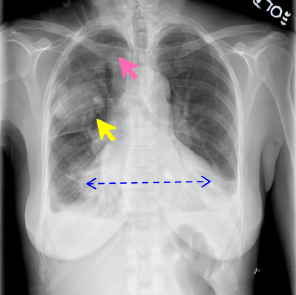}
    \end{minipage}
    
    \begin{minipage}[t]{0.16\linewidth}
        \centering
        \centerline{\includegraphics[width=1\linewidth]{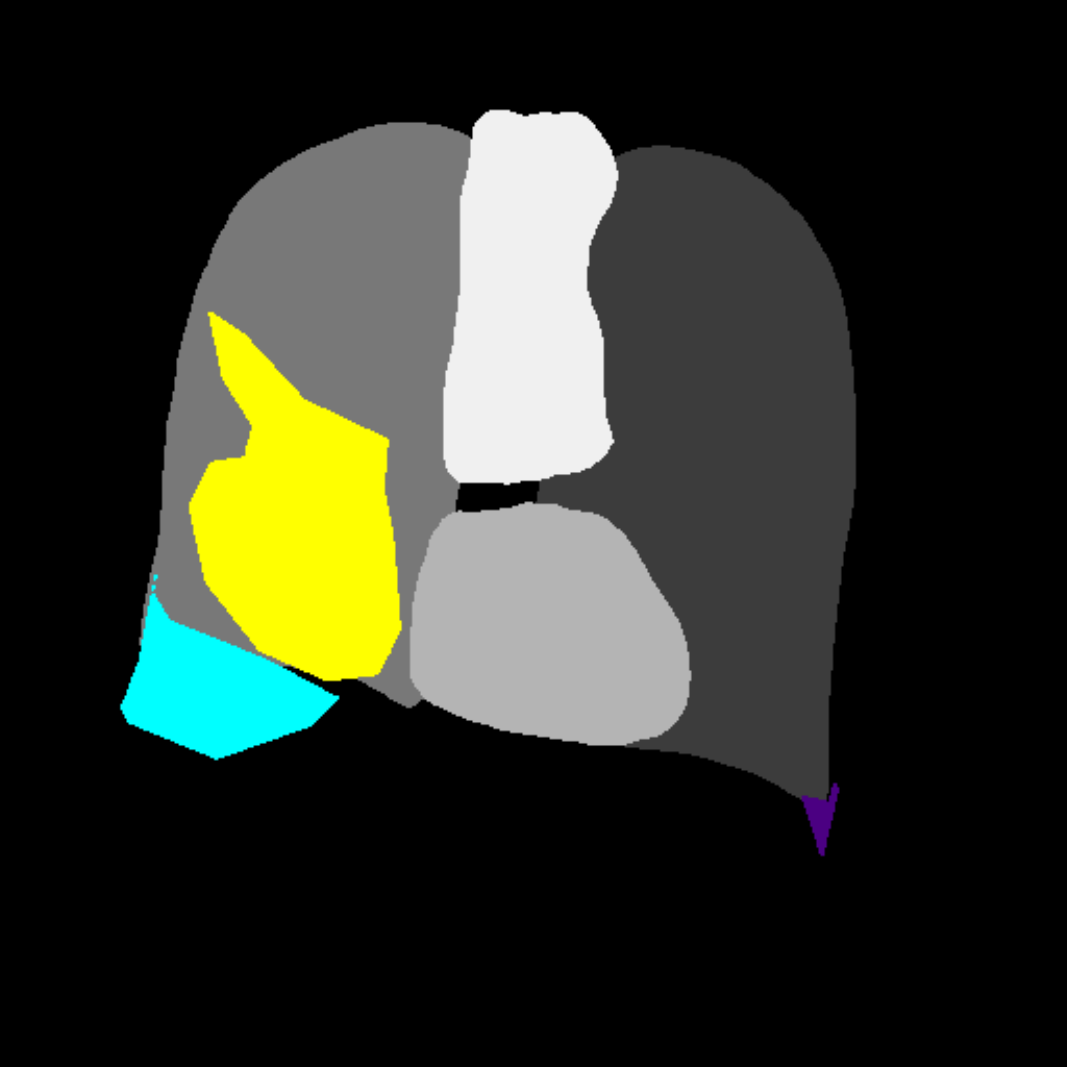}}
        \centerline{\parbox{1\linewidth}{\centering(a) Real Mask}}
    \end{minipage}
    \begin{minipage}[t]{0.16\linewidth}
        \centering
        \centerline{\includegraphics[width=1\linewidth]{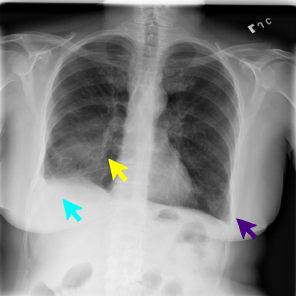}}
        \centerline{\parbox{1\linewidth}{\centering(b) Real Image}}
    \end{minipage}
    \begin{minipage}[t]{0.16\linewidth}
        \centering
        \centerline{\includegraphics[width=1\linewidth]{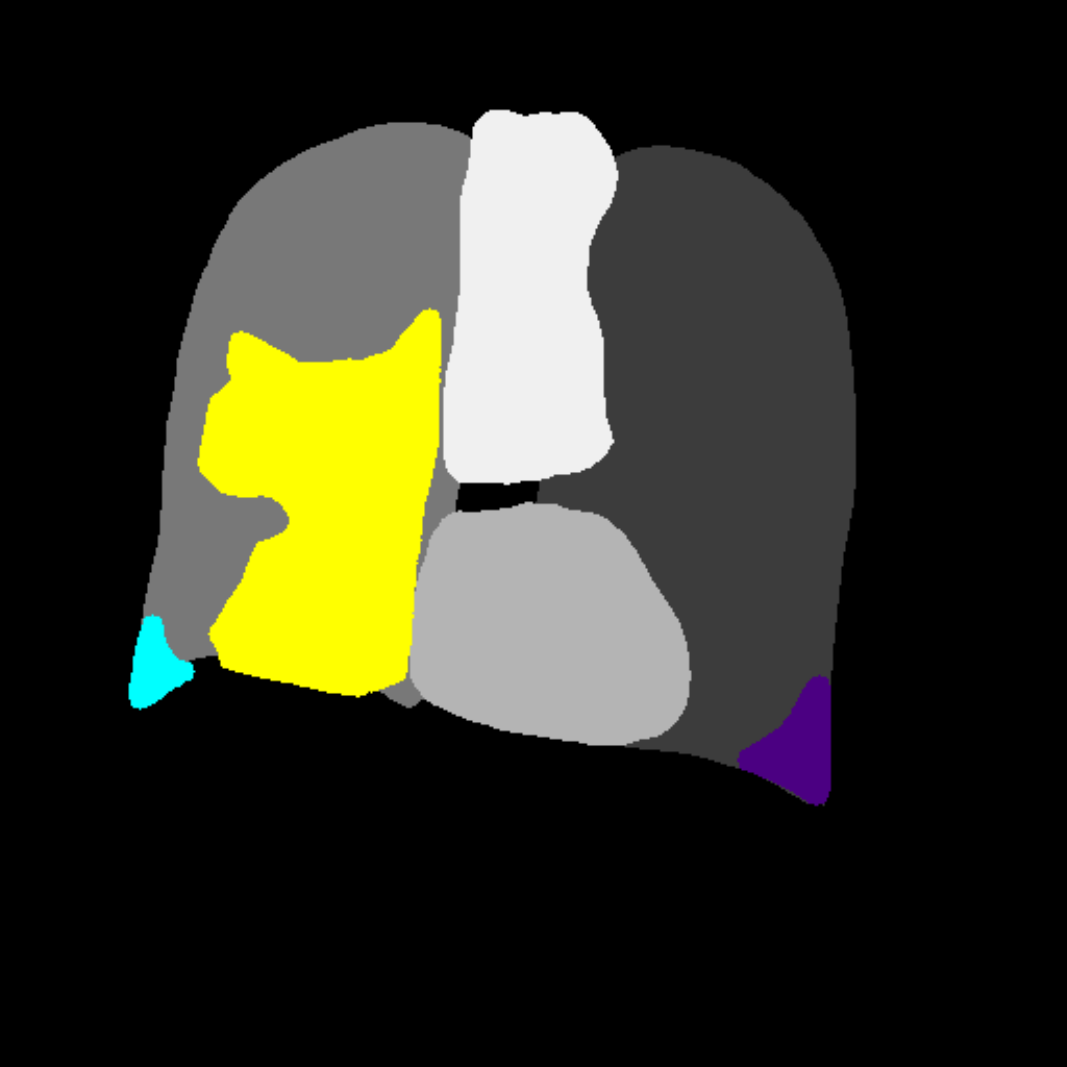}}
        \centerline{\parbox{1\linewidth}{\centering(c) Syn. Mask (Ours)}}
    \end{minipage}
    \begin{minipage}[t]{0.16\linewidth}
        \centering
        \centerline{\includegraphics[width=1\linewidth]{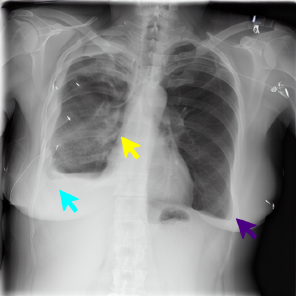}}
        \centerline{\parbox{1\linewidth}{\centering(d) Syn. Image (Ours)}}
    \end{minipage}
    \begin{minipage}[t]{0.16\linewidth}
        \centering
        \centerline{\includegraphics[width=1\linewidth]{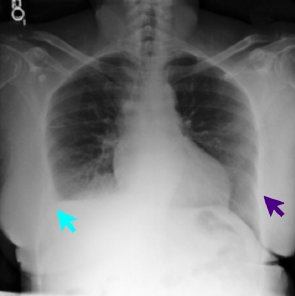}}
        \centerline{\parbox{1\linewidth}{\centering(e) Cheff}}
    \end{minipage}
    \begin{minipage}[t]{0.16\linewidth}
        \centering
        \centerline{\includegraphics[width=1\linewidth]{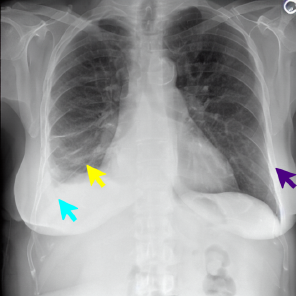}}
        \centerline{\parbox{1\linewidth}{\centering(f) RoentGen}}
    \end{minipage}
    \caption{Examples of our synthesized data comparing with real data, Cheff and RoentGen.}
    \label{fig:vis_app2}

    \vspace{5mm}

    \begin{minipage}[t]{0.16\linewidth}
        \centering
        \includegraphics[width=\linewidth]{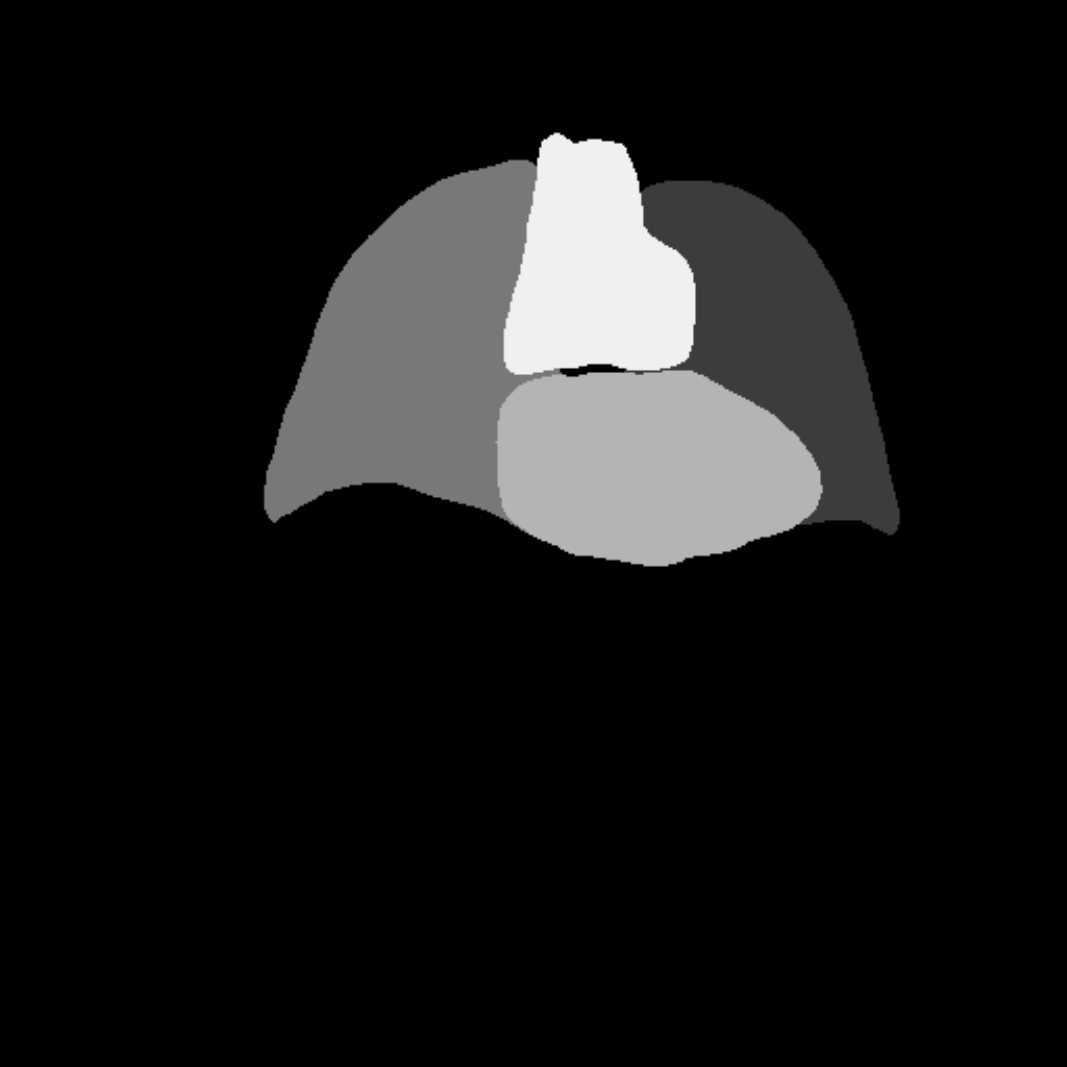}
    \end{minipage}
    \begin{minipage}[t]{0.16\linewidth}
        \centering
        \includegraphics[width=\linewidth]{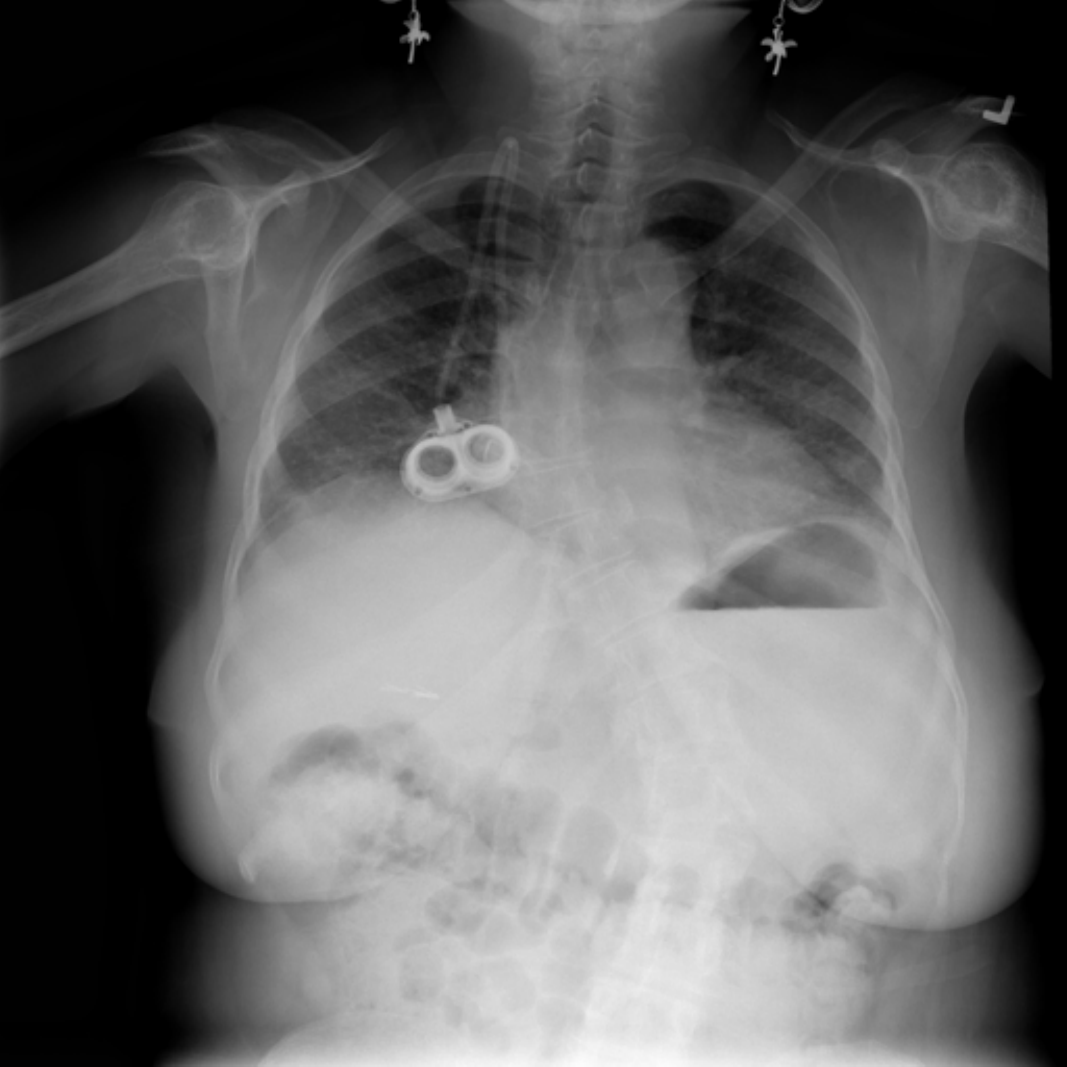}
    \end{minipage}
    \begin{minipage}[t]{0.16\linewidth}
        \centering
        \includegraphics[width=\linewidth]{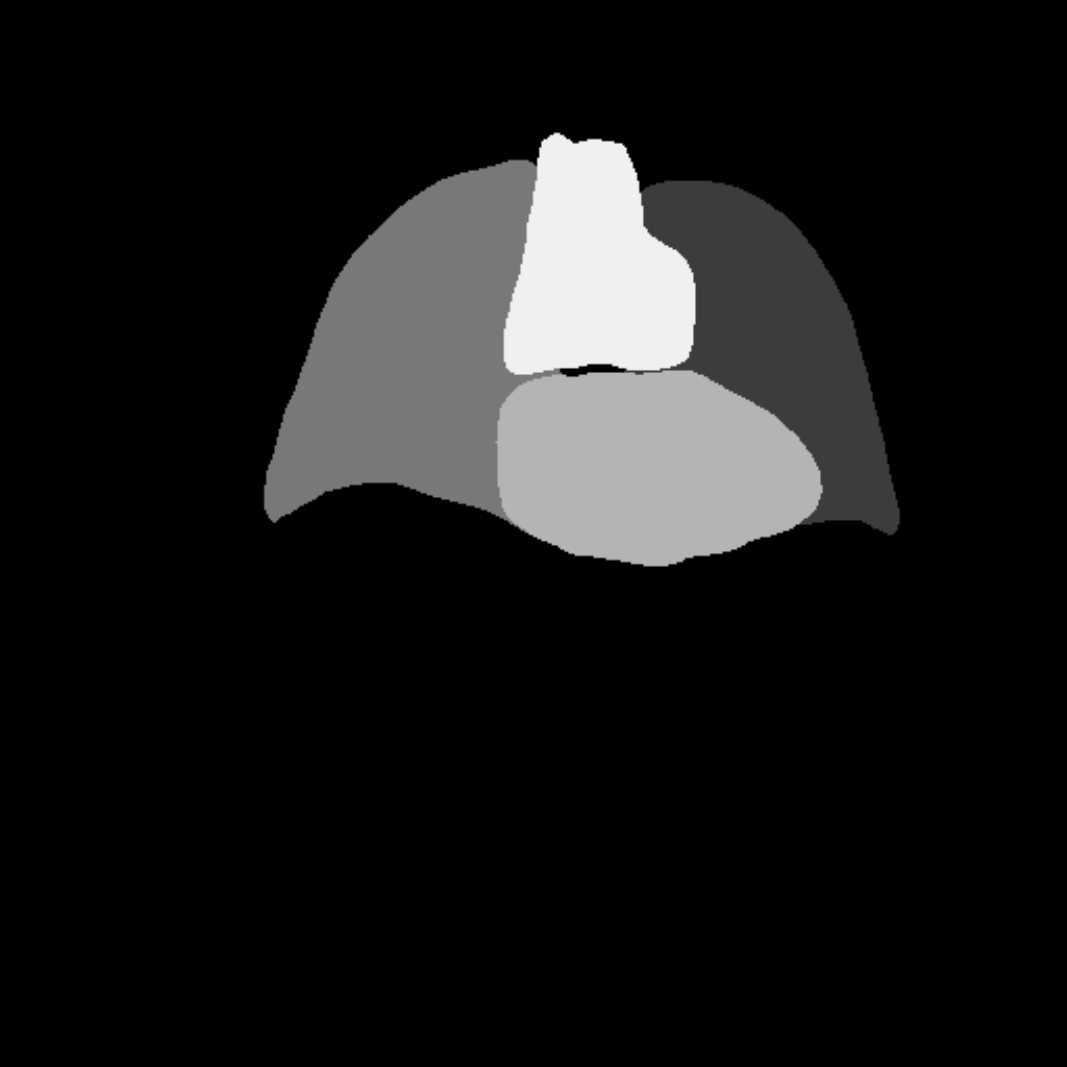}
    \end{minipage}
    \begin{minipage}[t]{0.16\linewidth}
        \centering
        \includegraphics[width=\linewidth]{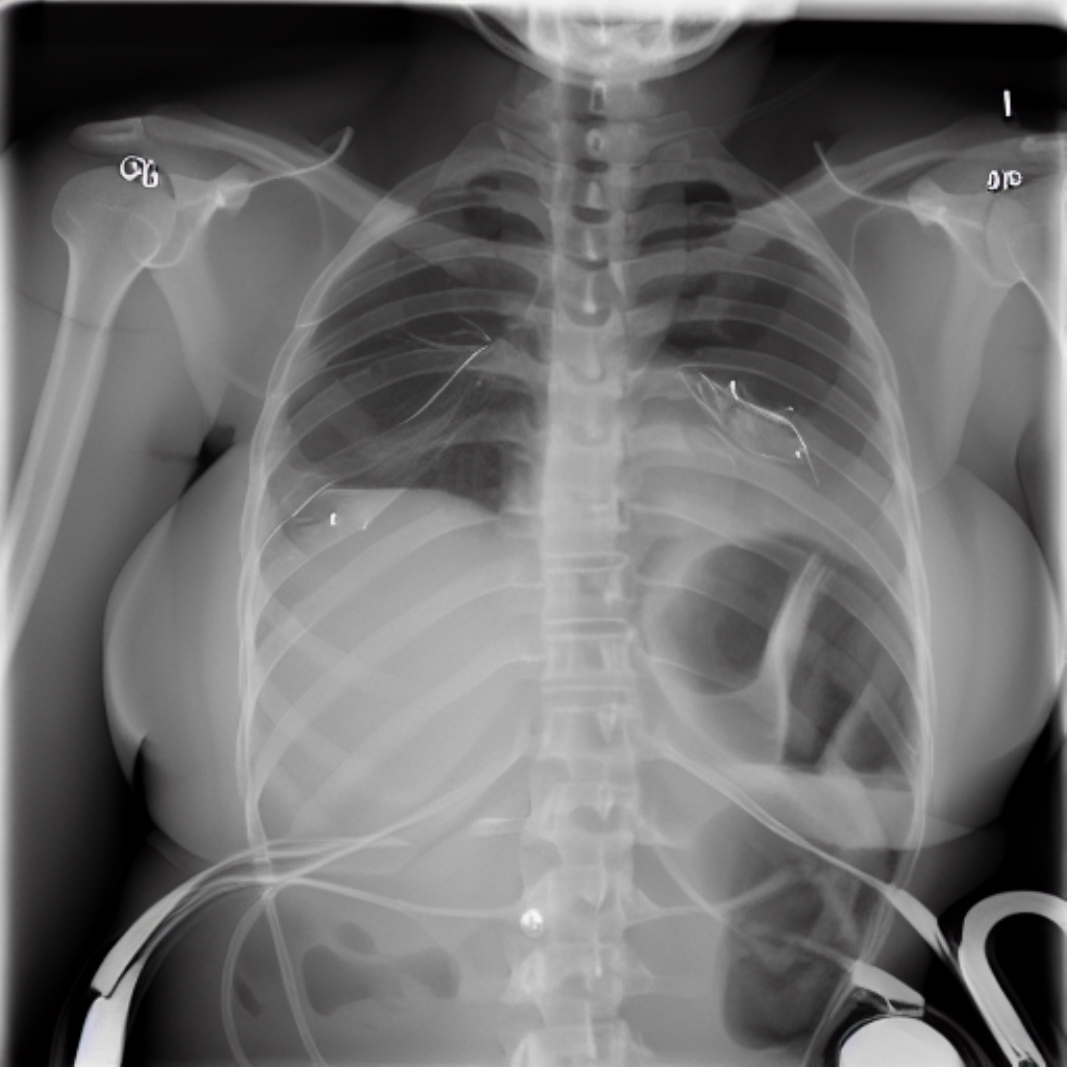}
    \end{minipage}
    \begin{minipage}[t]{0.16\linewidth}
        \centering
        \includegraphics[width=\linewidth]{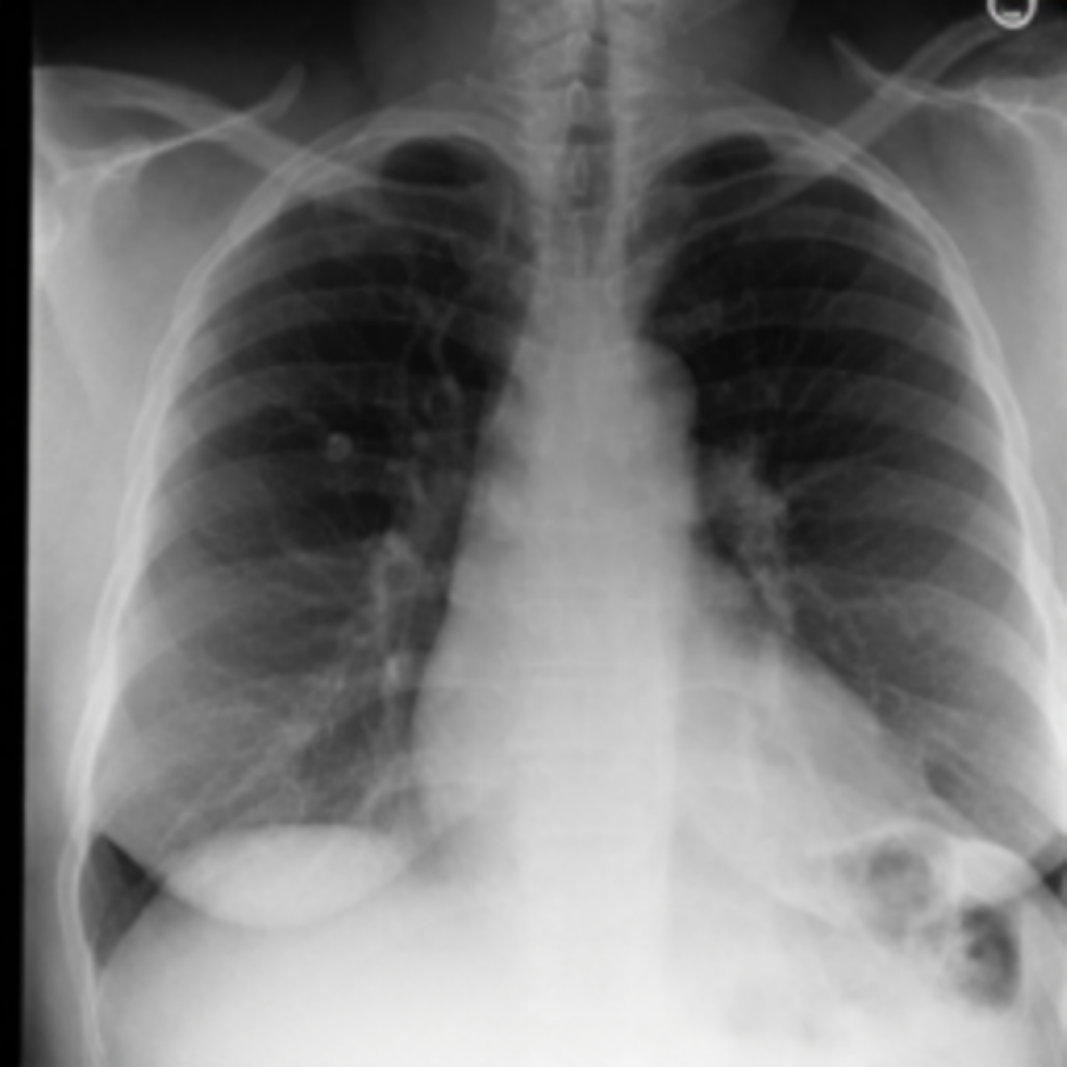}
    \end{minipage}
    \begin{minipage}[t]{0.16\linewidth}
        \centering
        \includegraphics[width=\linewidth]{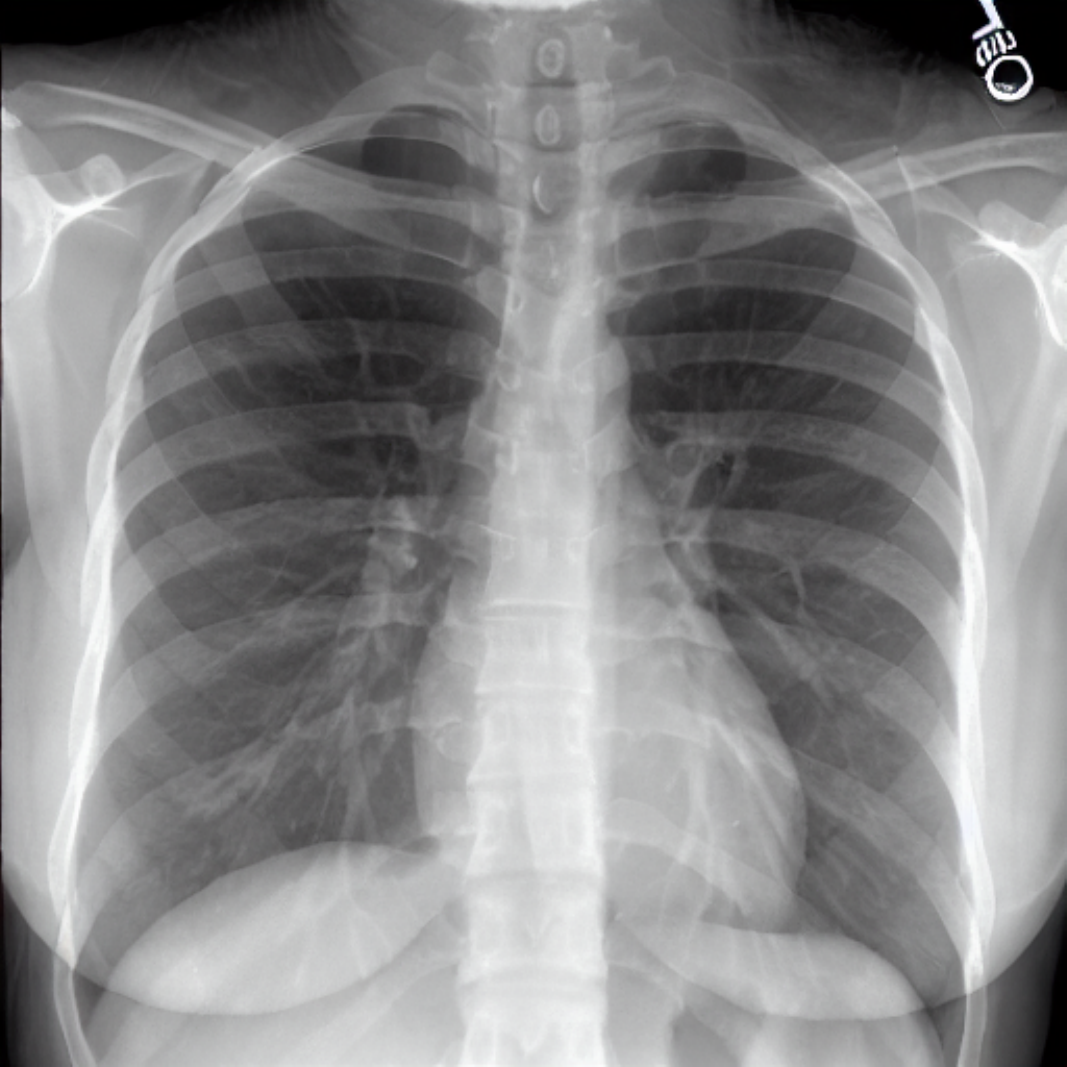}
    \end{minipage}

    \begin{minipage}[t]{0.16\linewidth}
        \centering
        \includegraphics[width=\linewidth]{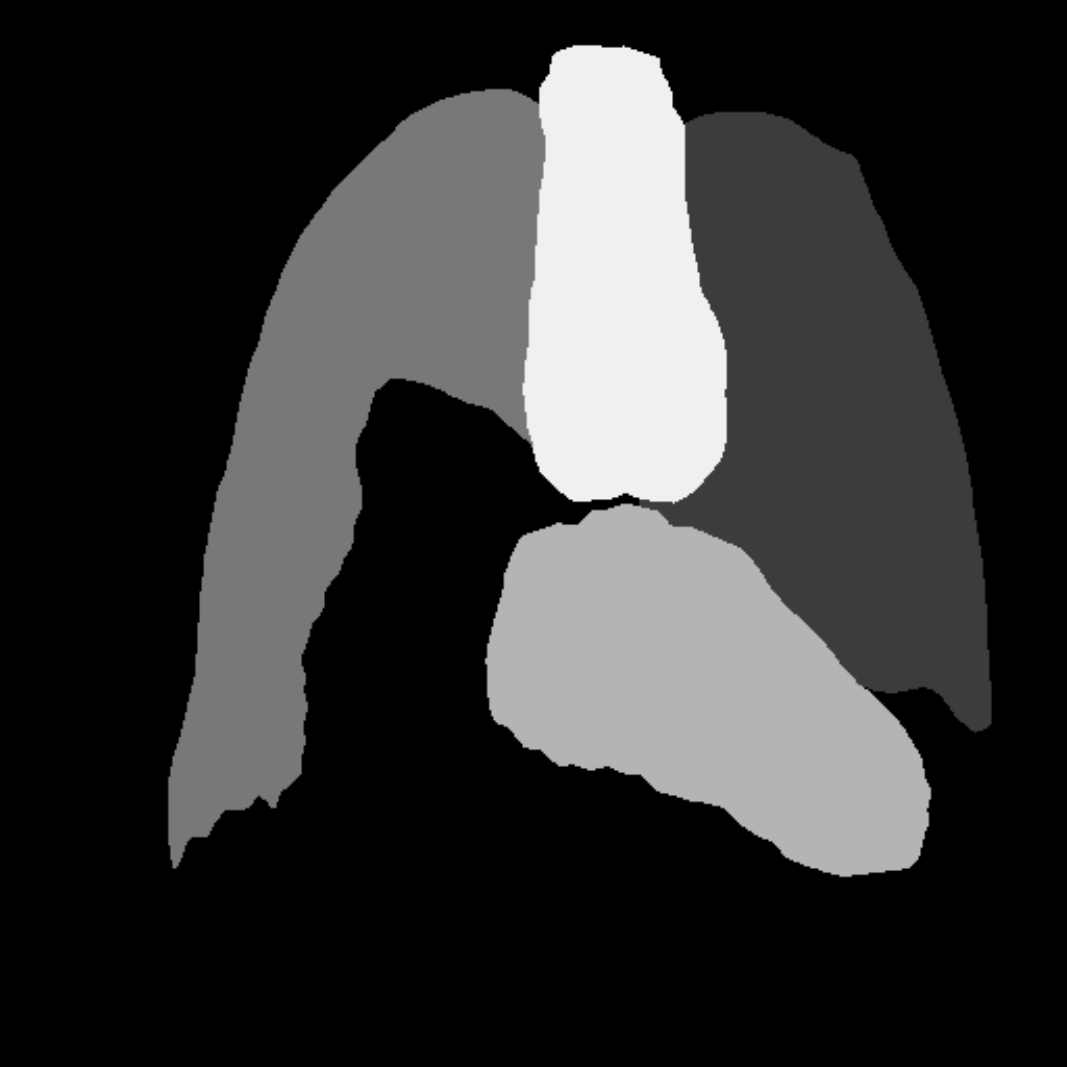}
    \end{minipage}
    \begin{minipage}[t]{0.16\linewidth}
        \centering
        \includegraphics[width=\linewidth]{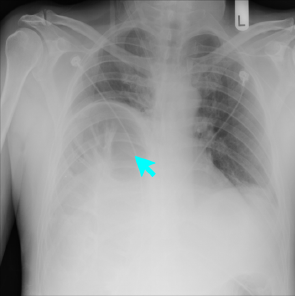}
    \end{minipage}
    \begin{minipage}[t]{0.16\linewidth}
        \centering
        \includegraphics[width=\linewidth]{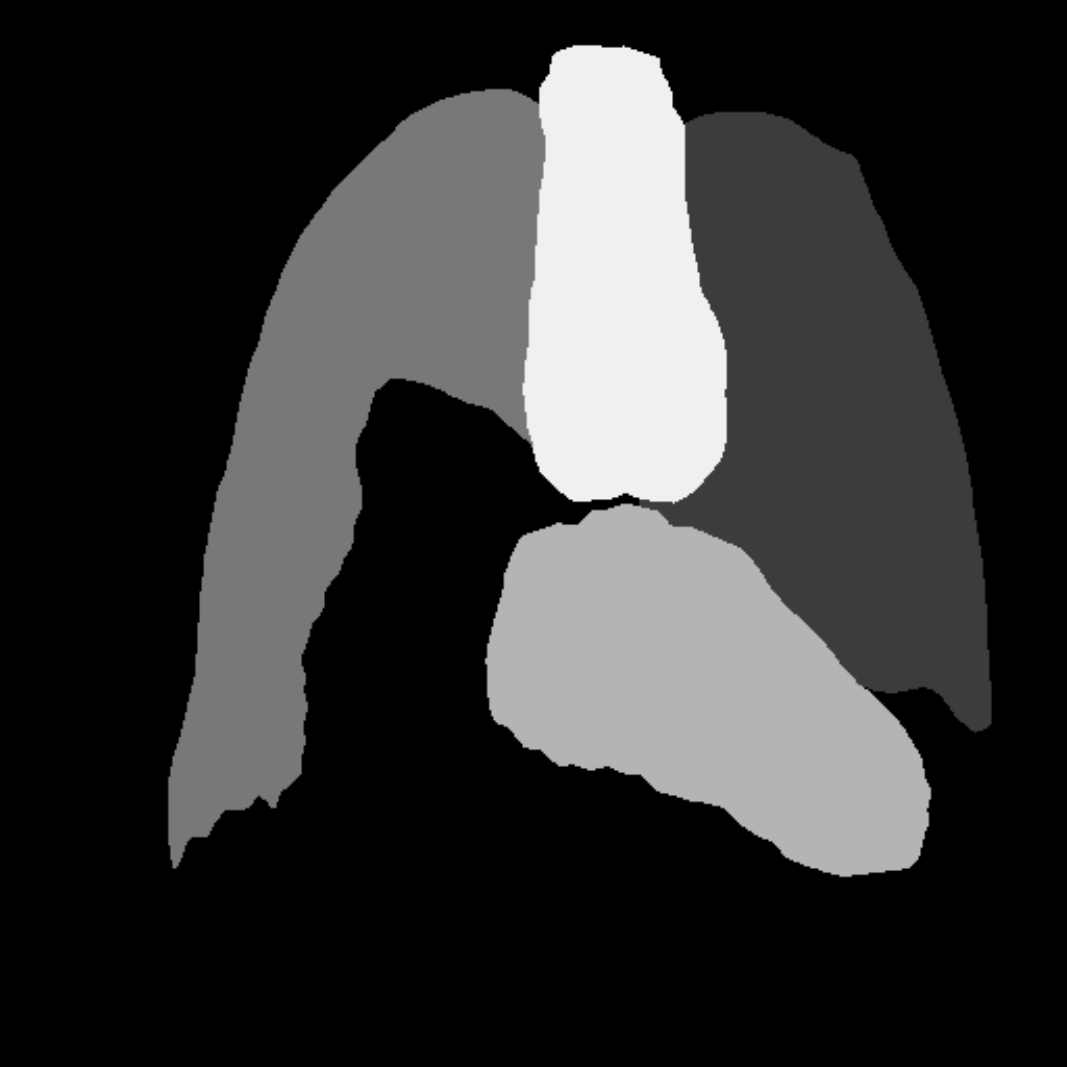}
    \end{minipage}
    \begin{minipage}[t]{0.16\linewidth}
        \centering
        \includegraphics[width=\linewidth]{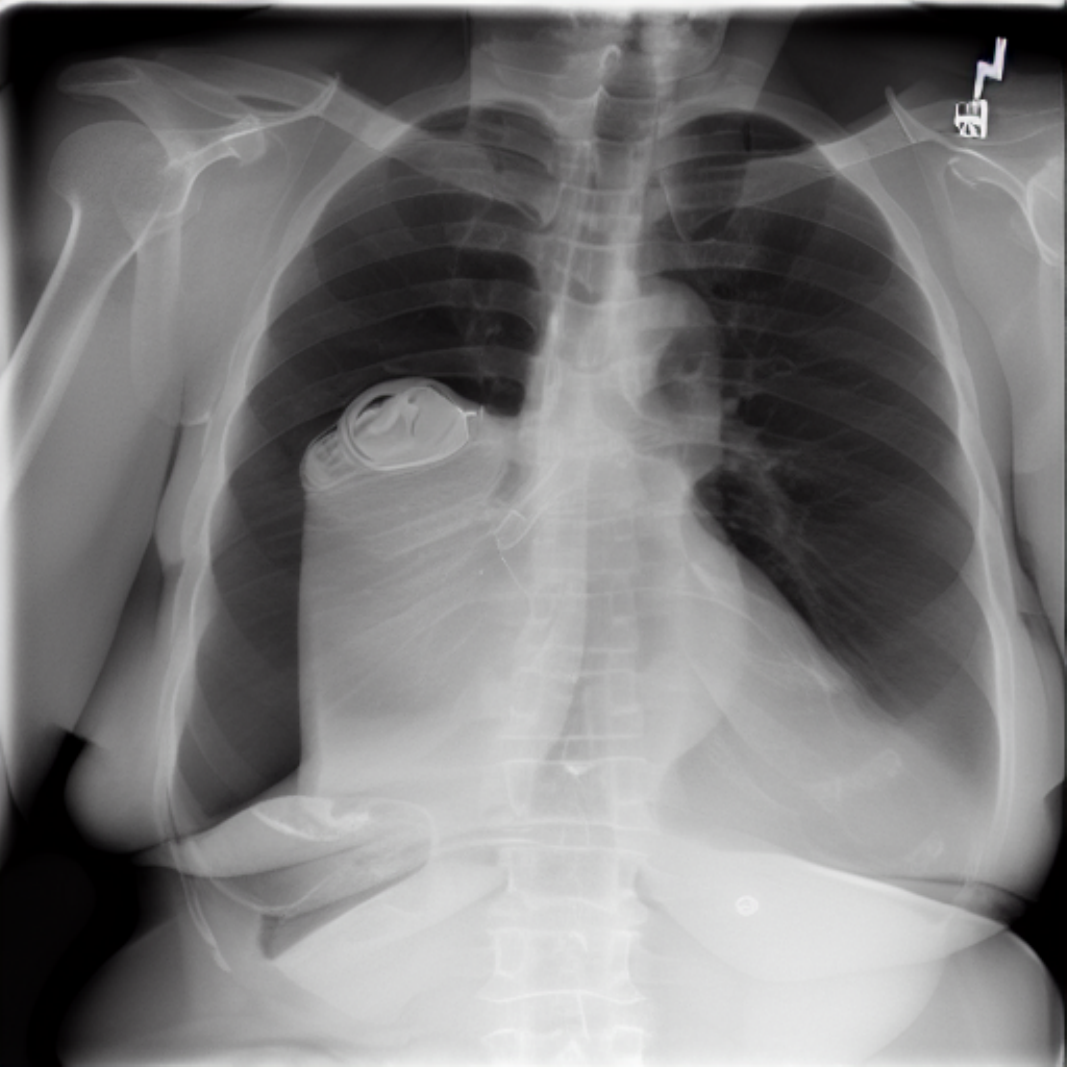}
    \end{minipage}
    \begin{minipage}[t]{0.16\linewidth}
        \centering
        \includegraphics[width=\linewidth]{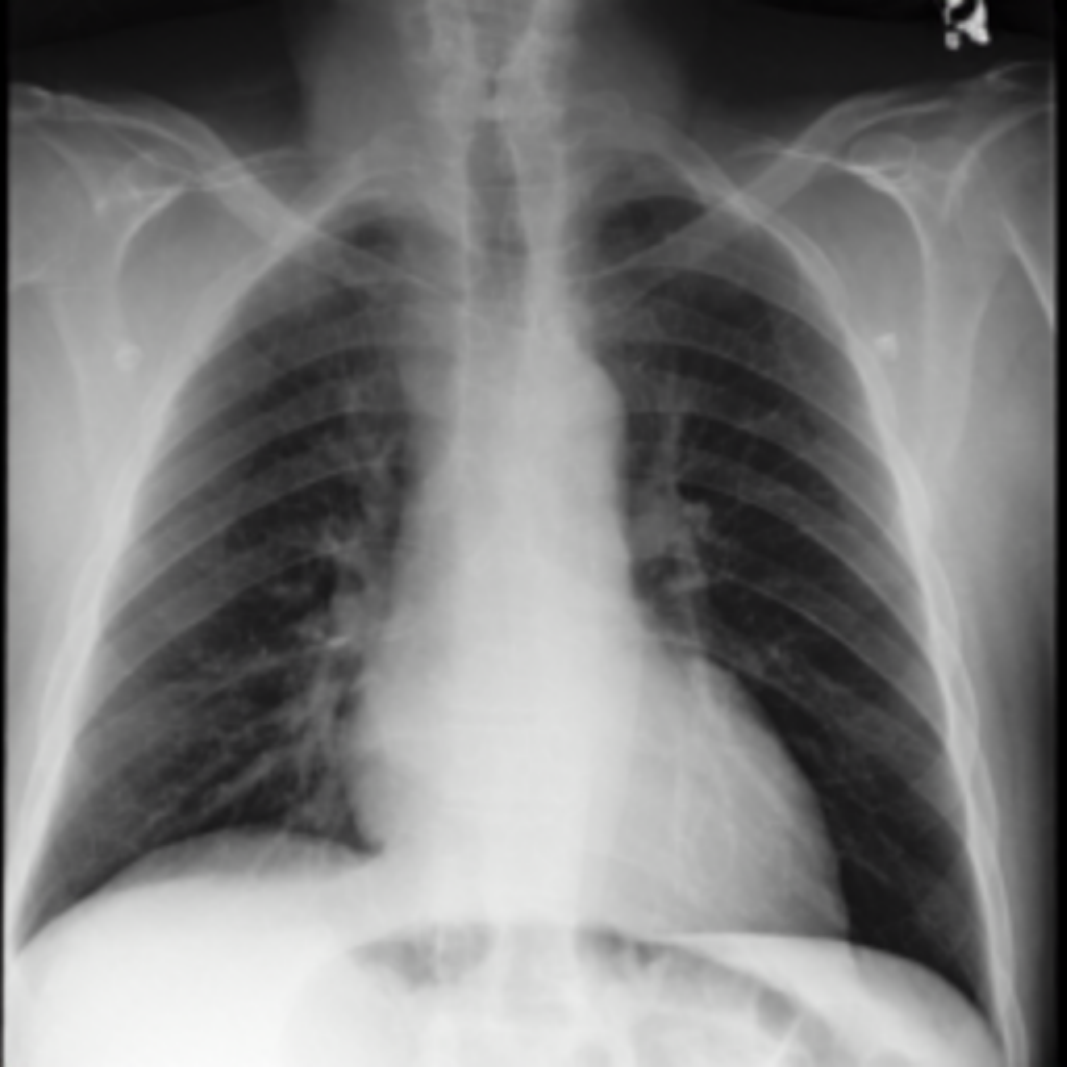}
    \end{minipage}
    \begin{minipage}[t]{0.16\linewidth}
        \centering
        \includegraphics[width=\linewidth]{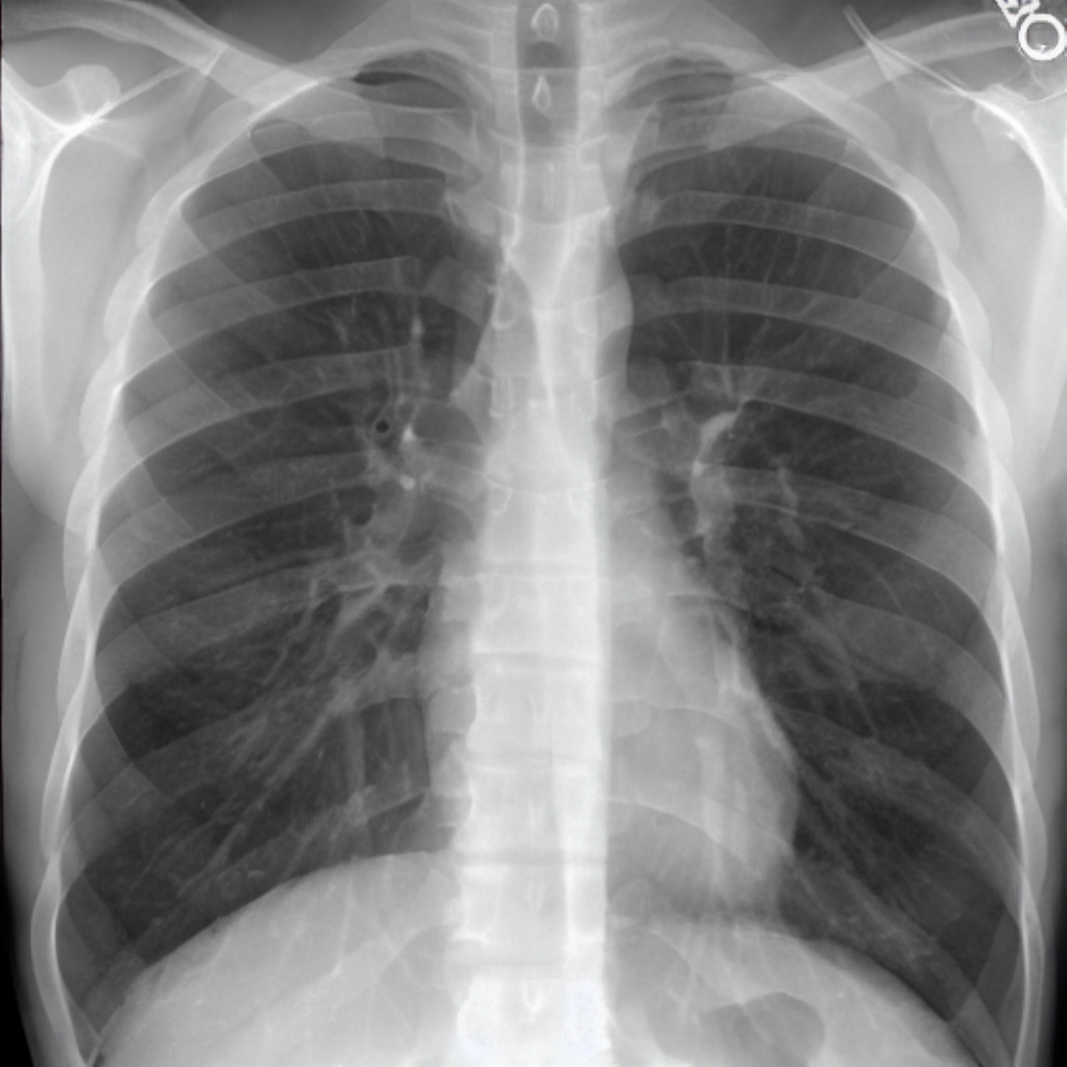}
    \end{minipage}

\end{figure*}

As shown in Figures \ref{fig:vis_app2}, our method produces results that exhibit strong anatomical consistency with real CXRs, along with diverse and clinically meaningful pathological patterns. The model is capable of generating fine-grained details such as subtle fractures and small nodules, which are often missed by baseline methods. In particular, under complex comorbidity conditions, our synthesized results more faithfully retain multiple disease features compared to Cheff and RoentGen, which tend to lose or blur certain pathological cues.

\begin{figure*}[t]
    \scriptsize
    \centering

    \begin{minipage}[t]{0.16\linewidth}
        \centering
        \includegraphics[width=\linewidth]{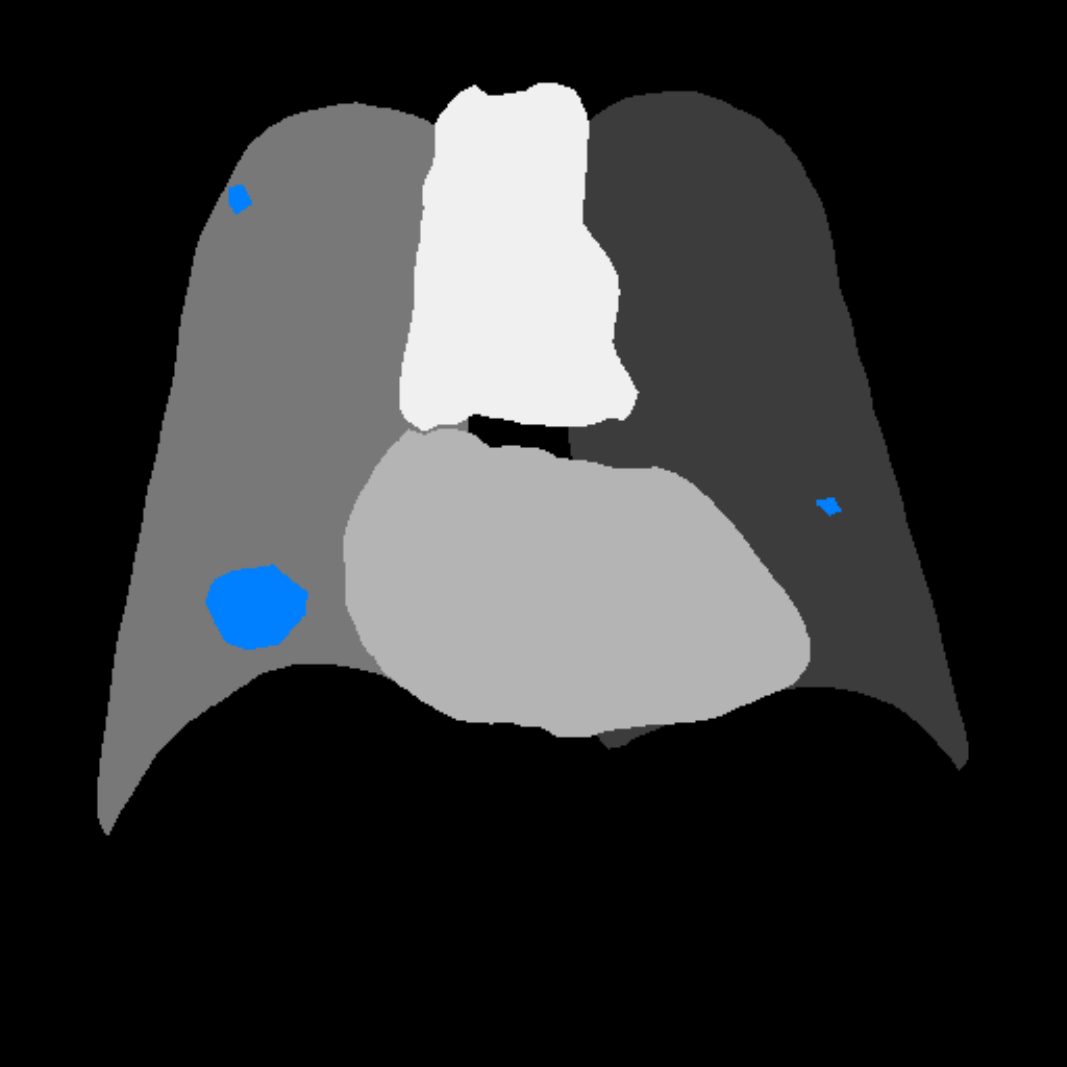}
    \end{minipage}
    \begin{minipage}[t]{0.16\linewidth}
        \centering
        \includegraphics[width=\linewidth]{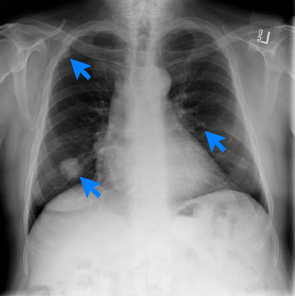}
    \end{minipage}
    \begin{minipage}[t]{0.16\linewidth}
        \centering
        \includegraphics[width=\linewidth]{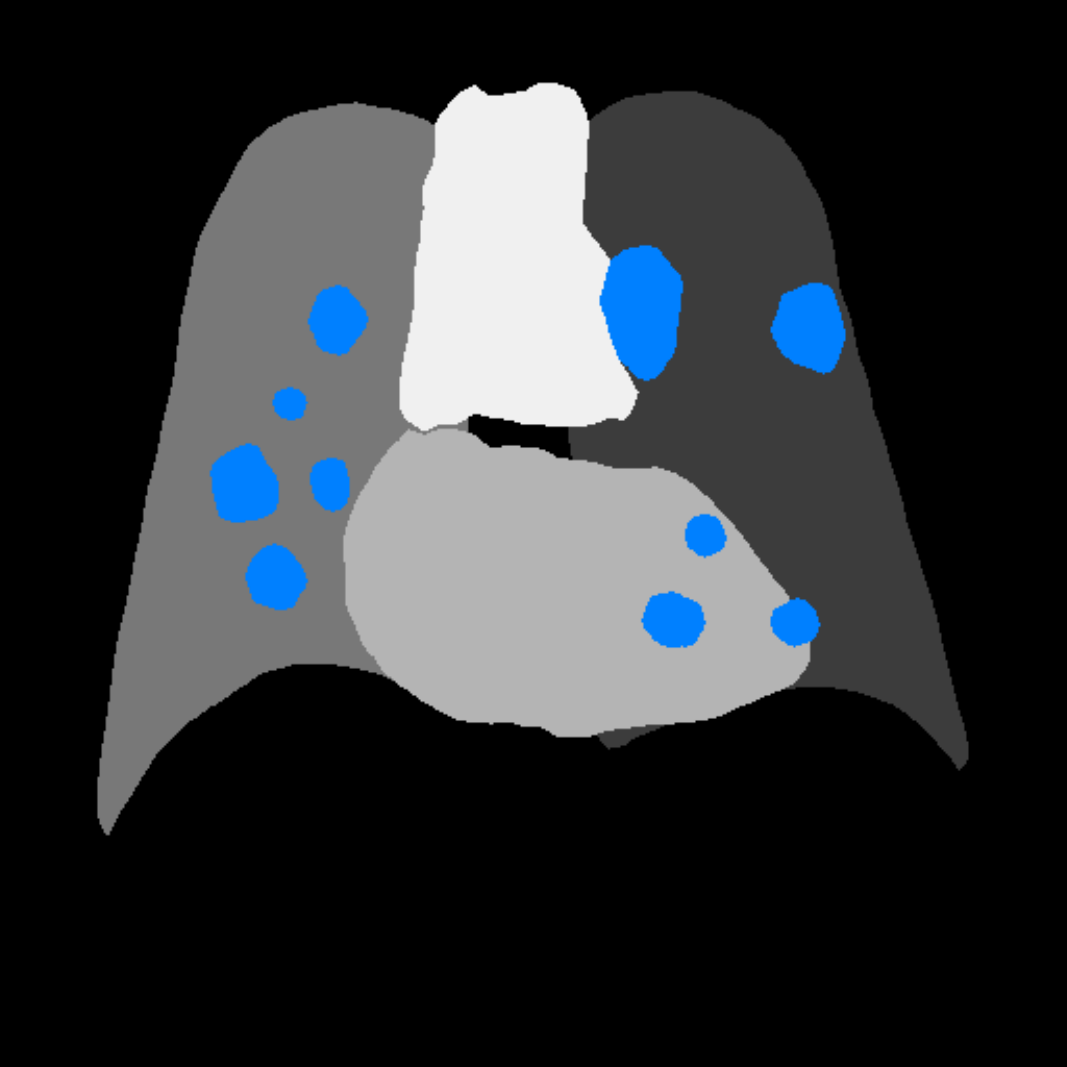}
    \end{minipage}
    \begin{minipage}[t]{0.16\linewidth}
        \centering
        \includegraphics[width=\linewidth]{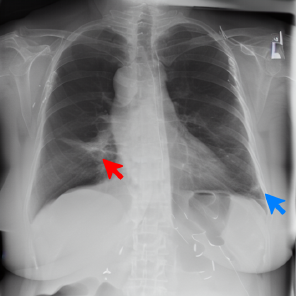}
    \end{minipage}
    \begin{minipage}[t]{0.16\linewidth}
        \centering
        \includegraphics[width=\linewidth]{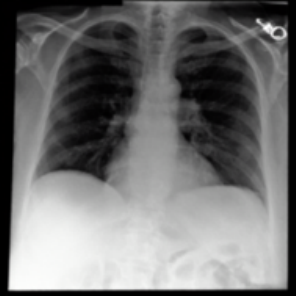}
    \end{minipage}
    \begin{minipage}[t]{0.16\linewidth}
        \centering
        \includegraphics[width=\linewidth]{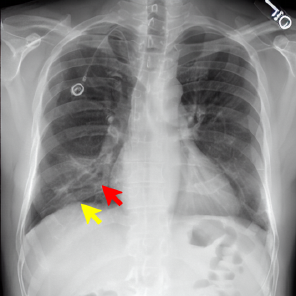}
    \end{minipage}

    \begin{minipage}[t]{0.16\linewidth}
        \centering
        \includegraphics[width=\linewidth]{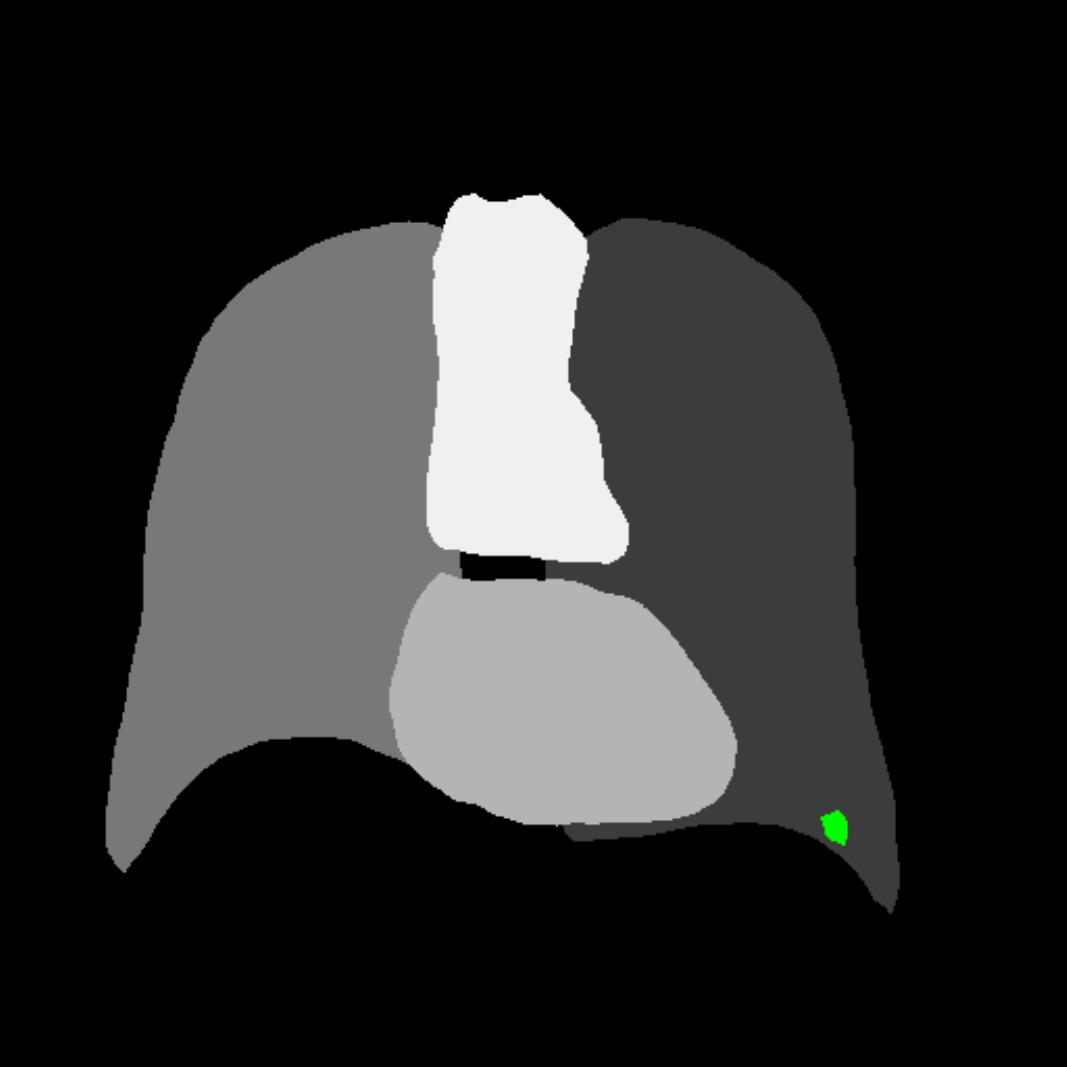}
    \end{minipage}
    \begin{minipage}[t]{0.16\linewidth}
        \centering
        \includegraphics[width=\linewidth]{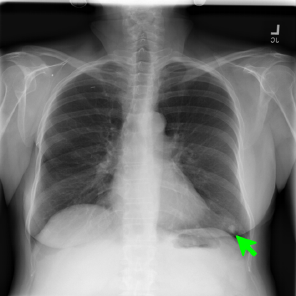}
    \end{minipage}
    \begin{minipage}[t]{0.16\linewidth}
        \centering
        \includegraphics[width=\linewidth]{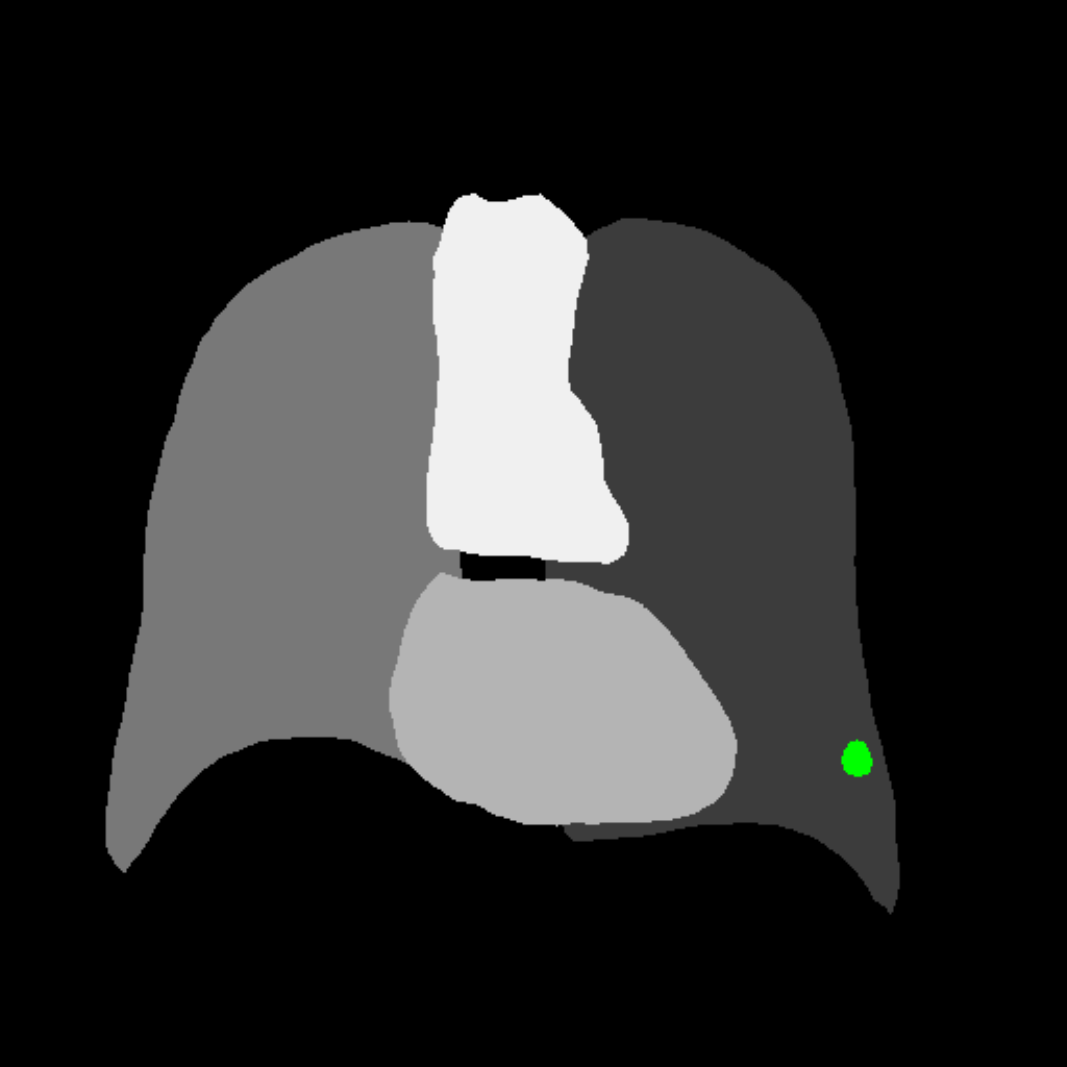}
    \end{minipage}
    \begin{minipage}[t]{0.16\linewidth}
        \centering
        \includegraphics[width=\linewidth]{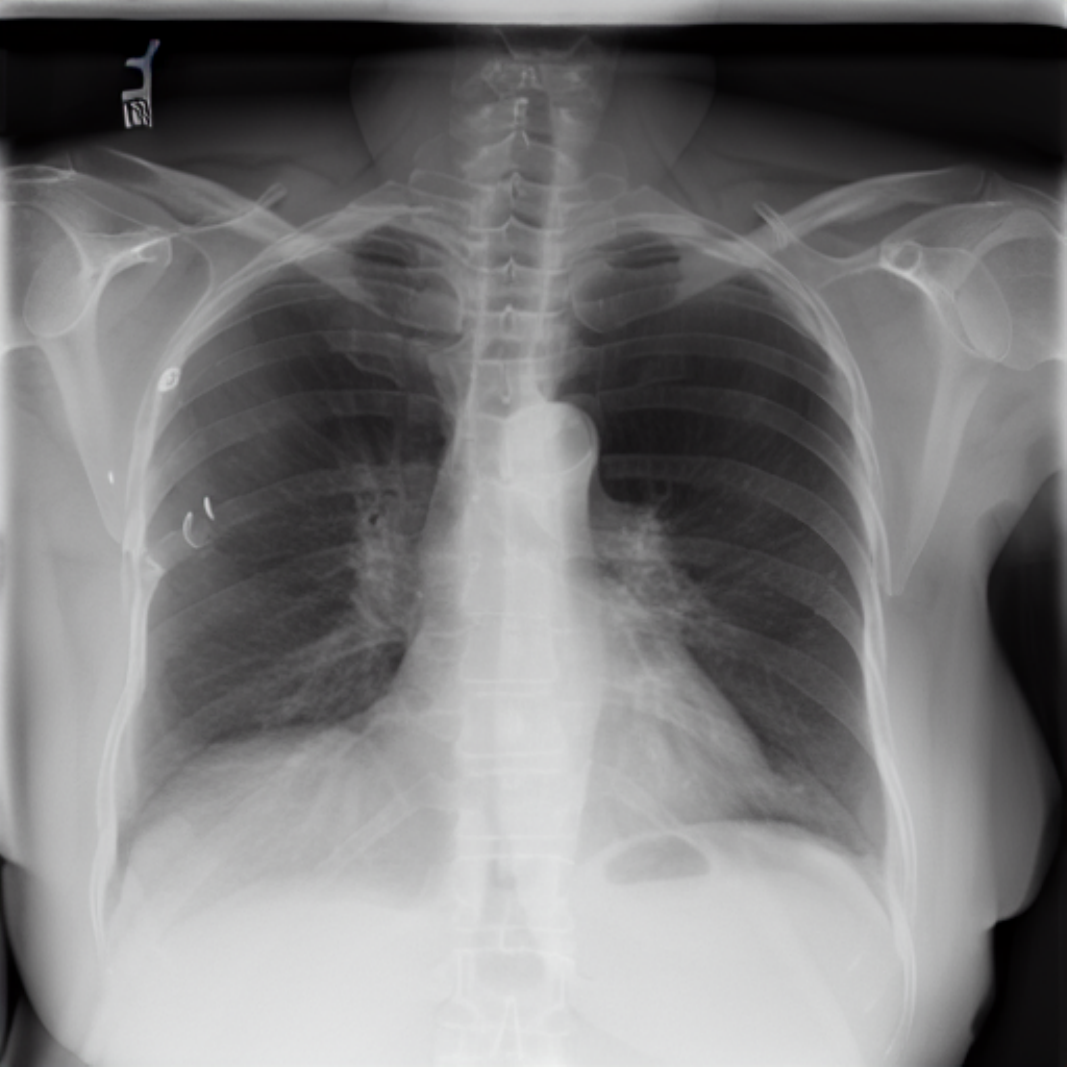}
    \end{minipage}
    \begin{minipage}[t]{0.16\linewidth}
        \centering
        \includegraphics[width=\linewidth]{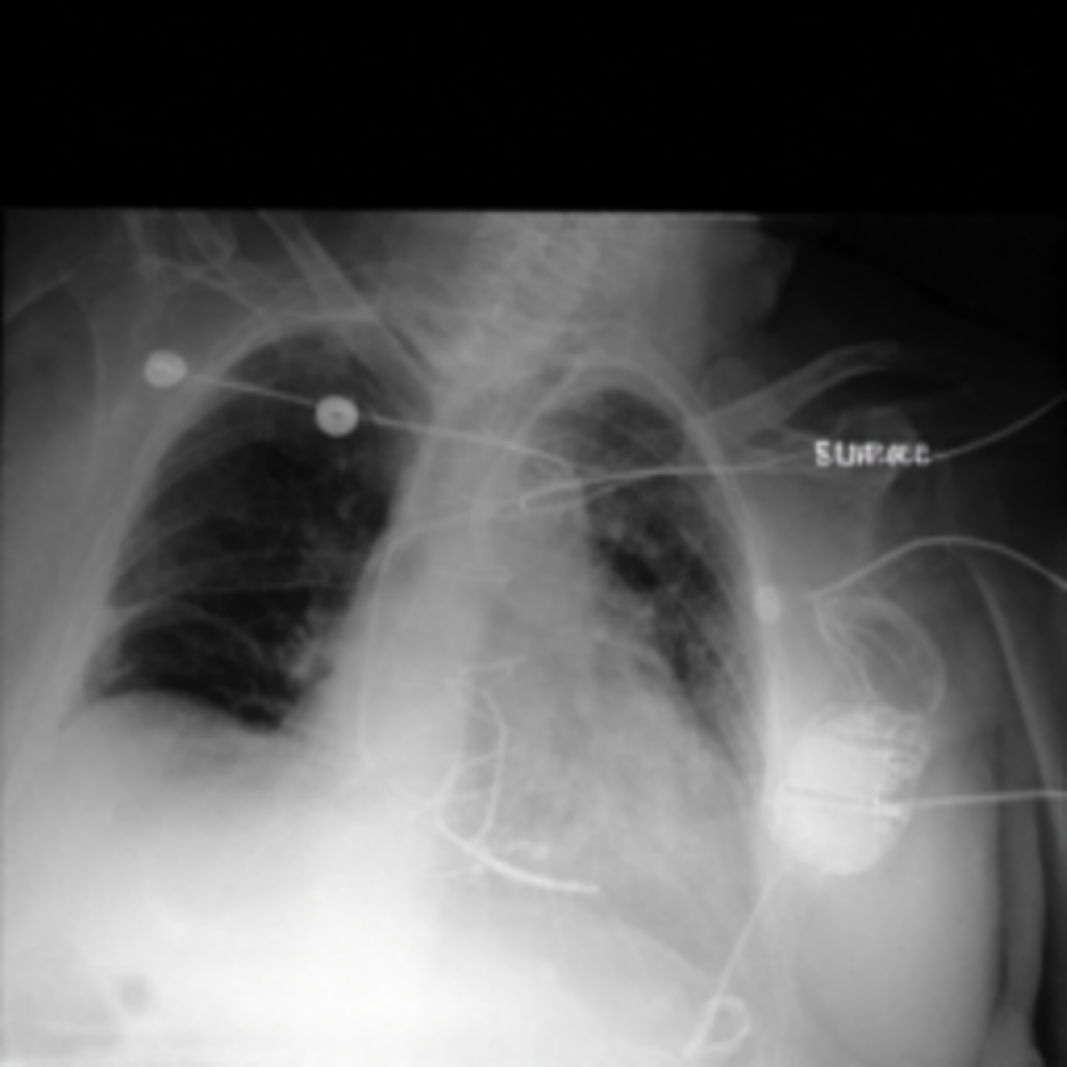}
    \end{minipage}
        \begin{minipage}[t]{0.16\linewidth}
        \centering
        \includegraphics[width=\linewidth]{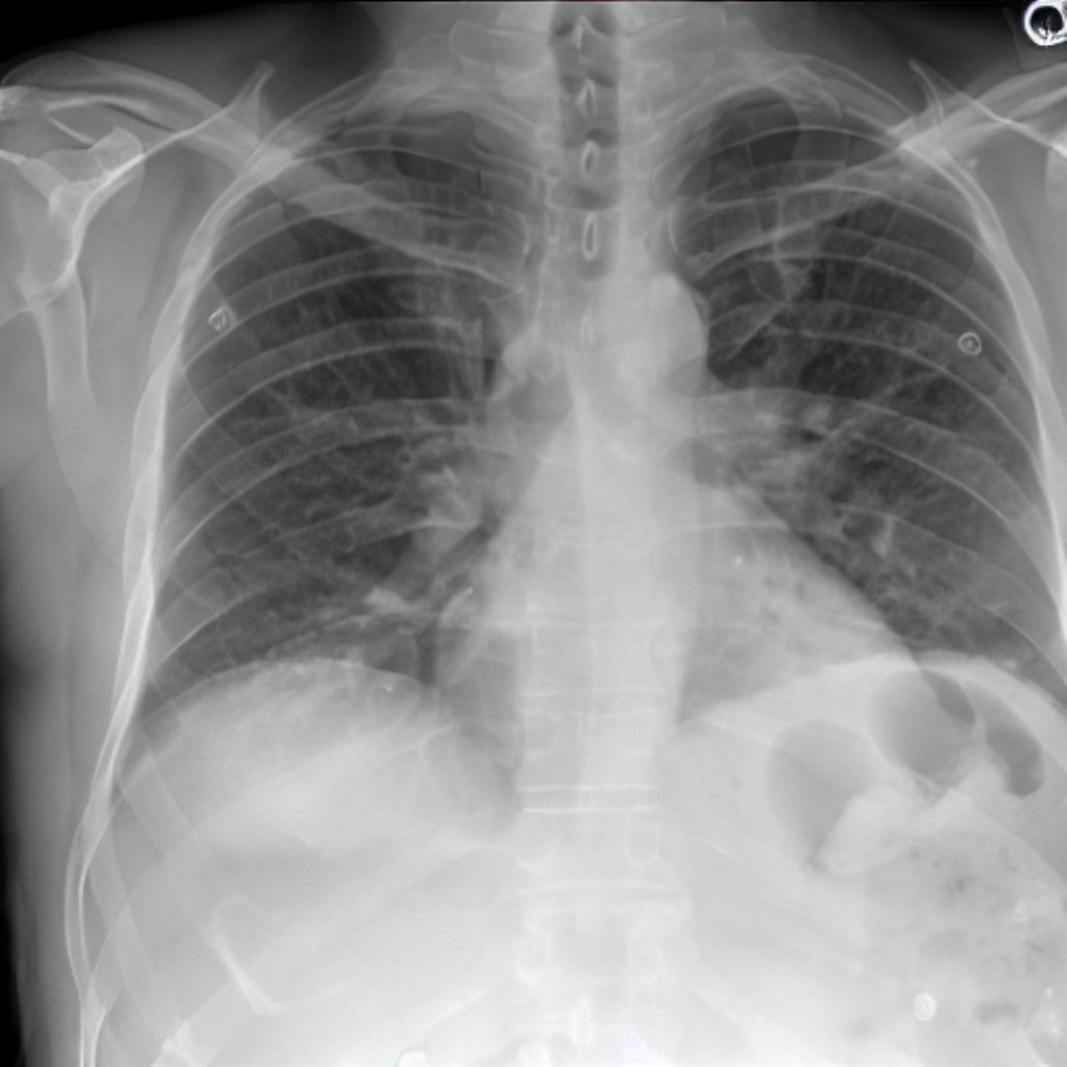}
    \end{minipage}

    \begin{minipage}[t]{0.16\linewidth}
        \centering
        \includegraphics[width=\linewidth]{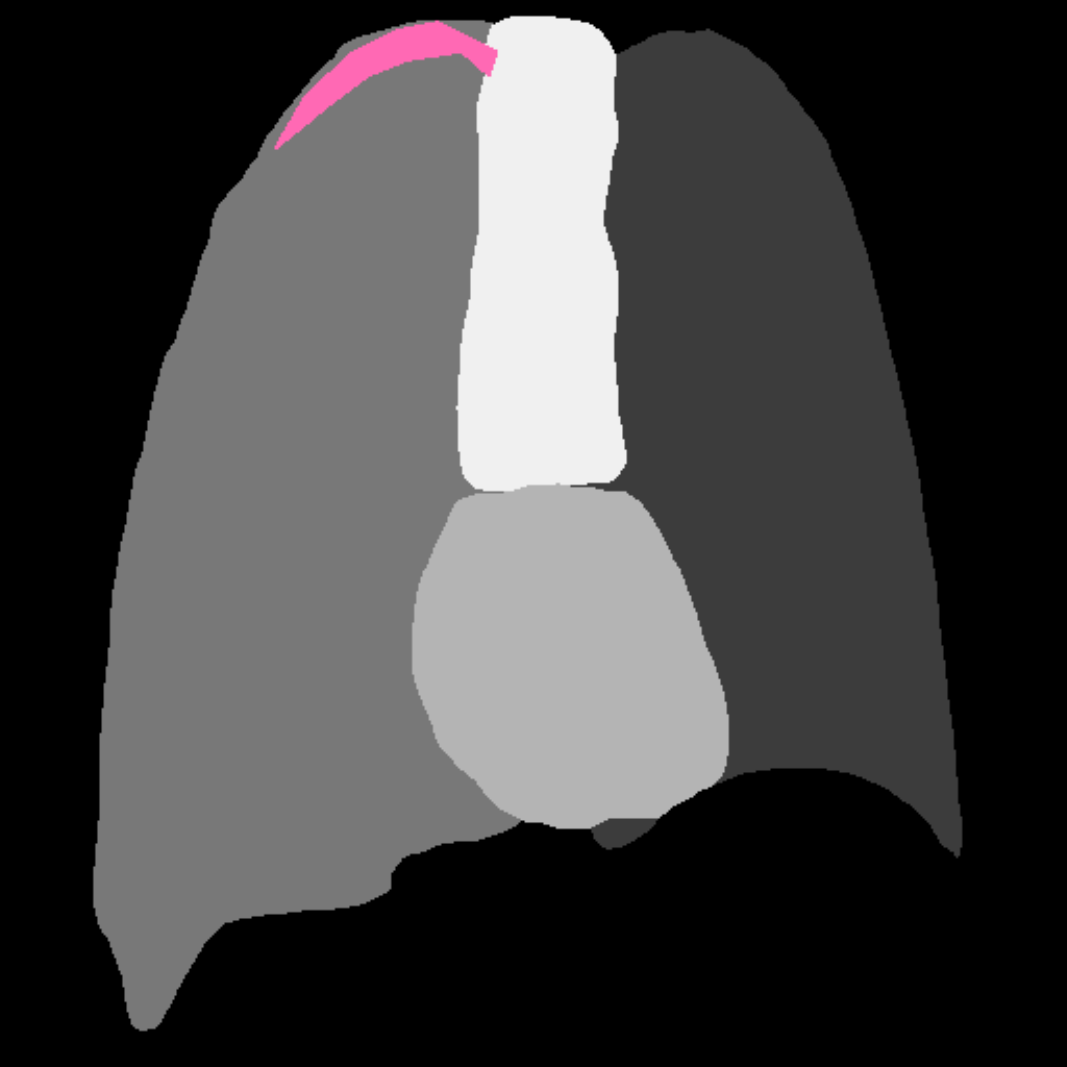}
    \end{minipage}
    \begin{minipage}[t]{0.16\linewidth}
        \centering
        \includegraphics[width=\linewidth]{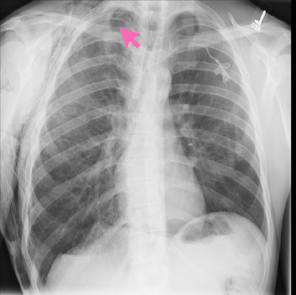}
    \end{minipage}
    \begin{minipage}[t]{0.16\linewidth}
        \centering
        \includegraphics[width=\linewidth]{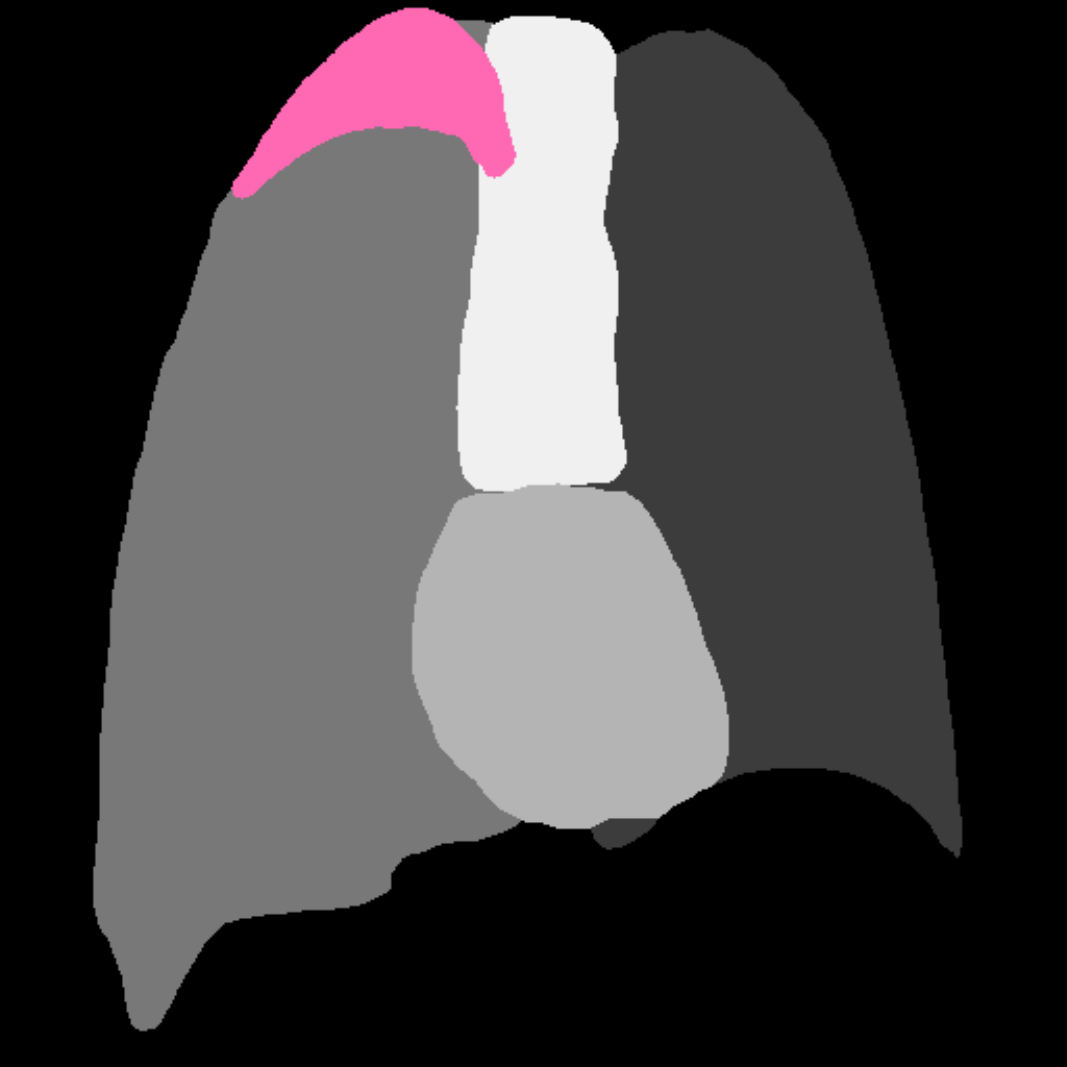}
    \end{minipage}
    \begin{minipage}[t]{0.16\linewidth}
        \centering
        \includegraphics[width=\linewidth]{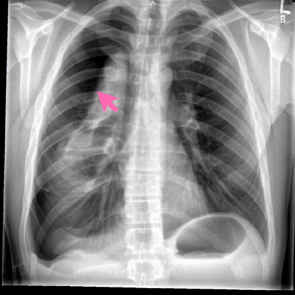}
    \end{minipage}
    \begin{minipage}[t]{0.16\linewidth}
        \centering
        \includegraphics[width=\linewidth]{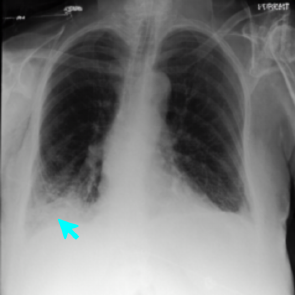}
    \end{minipage}
    \begin{minipage}[t]{0.16\linewidth}
        \centering
        \includegraphics[width=\linewidth]{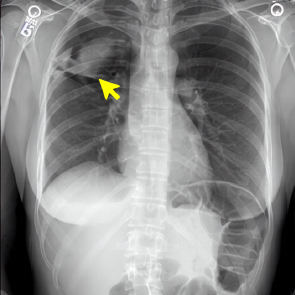}
    \end{minipage}

    \begin{minipage}[t]{0.16\linewidth}
        \centering
        \includegraphics[width=\linewidth]{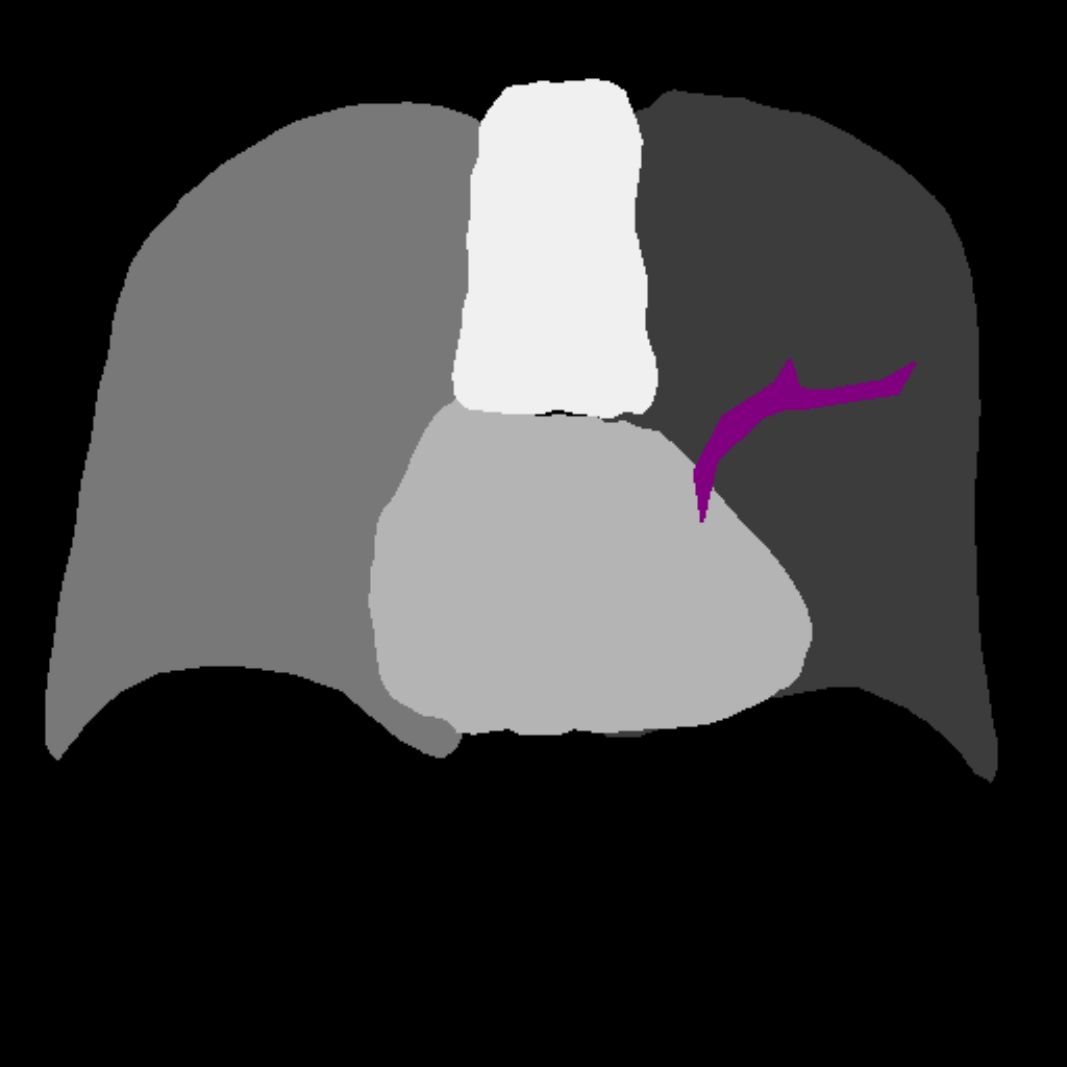}
    \end{minipage}
    \begin{minipage}[t]{0.16\linewidth}
        \centering
        \includegraphics[width=\linewidth]{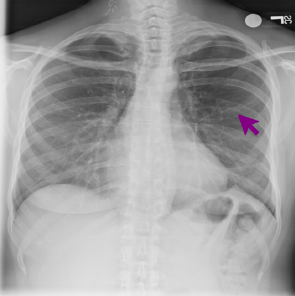}
    \end{minipage}
    \begin{minipage}[t]{0.16\linewidth}
        \centering
        \includegraphics[width=\linewidth]{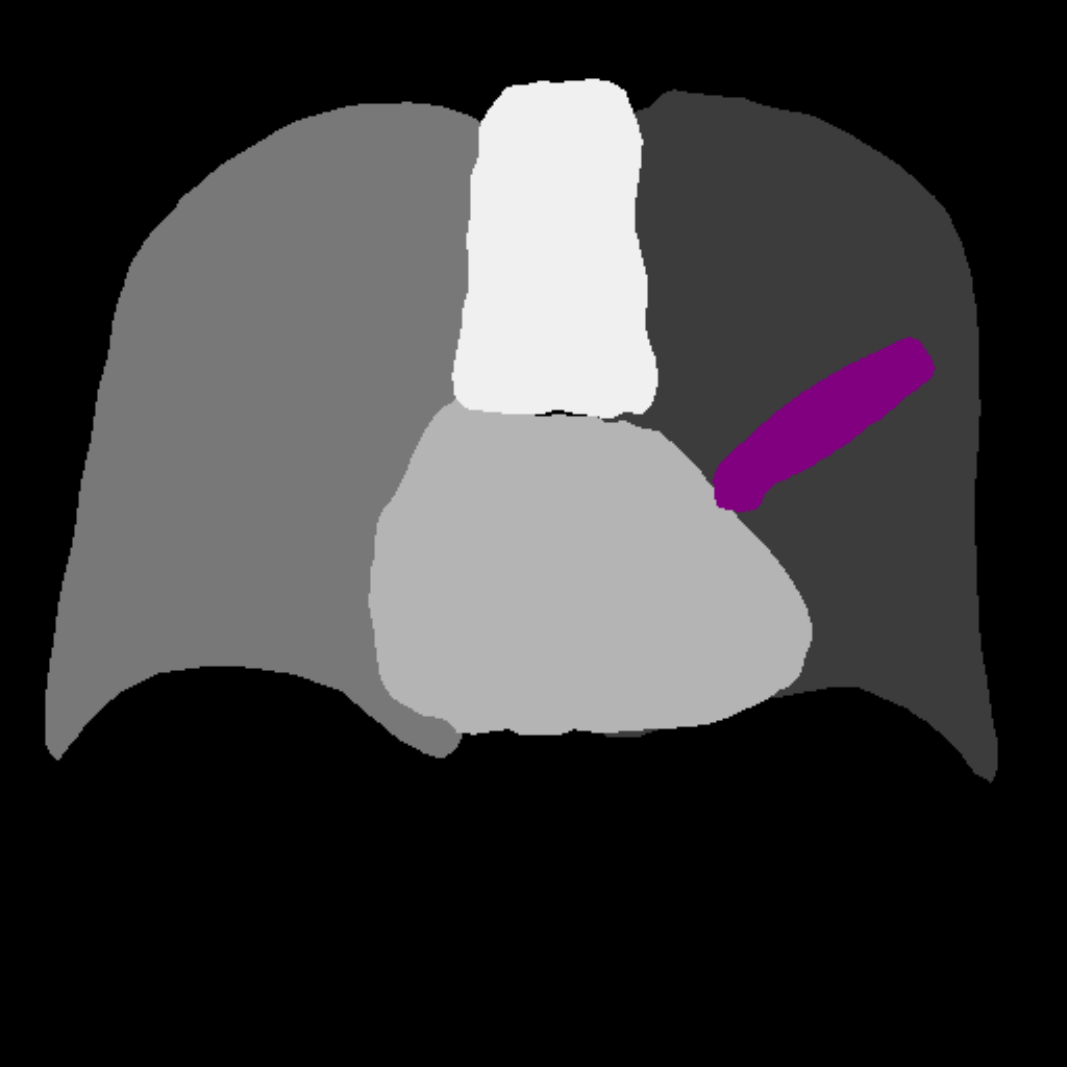}
    \end{minipage}
    \begin{minipage}[t]{0.16\linewidth}
        \centering
        \includegraphics[width=\linewidth]{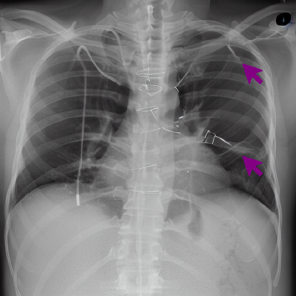}
    \end{minipage}
    \begin{minipage}[t]{0.16\linewidth}
        \centering
        \includegraphics[width=\linewidth]{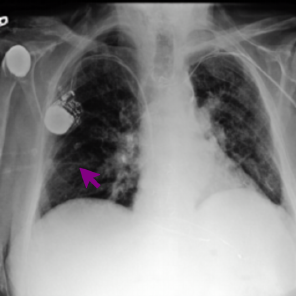}
    \end{minipage}
    \begin{minipage}[t]{0.16\linewidth}
        \centering
        \includegraphics[width=\linewidth]{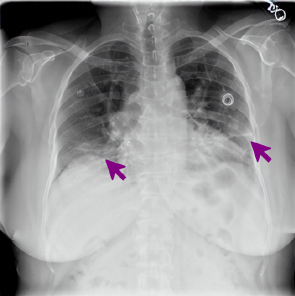}
    \end{minipage}

    \begin{minipage}[t]{0.16\linewidth}
        \centering
        \centerline{\includegraphics[width=1\linewidth]{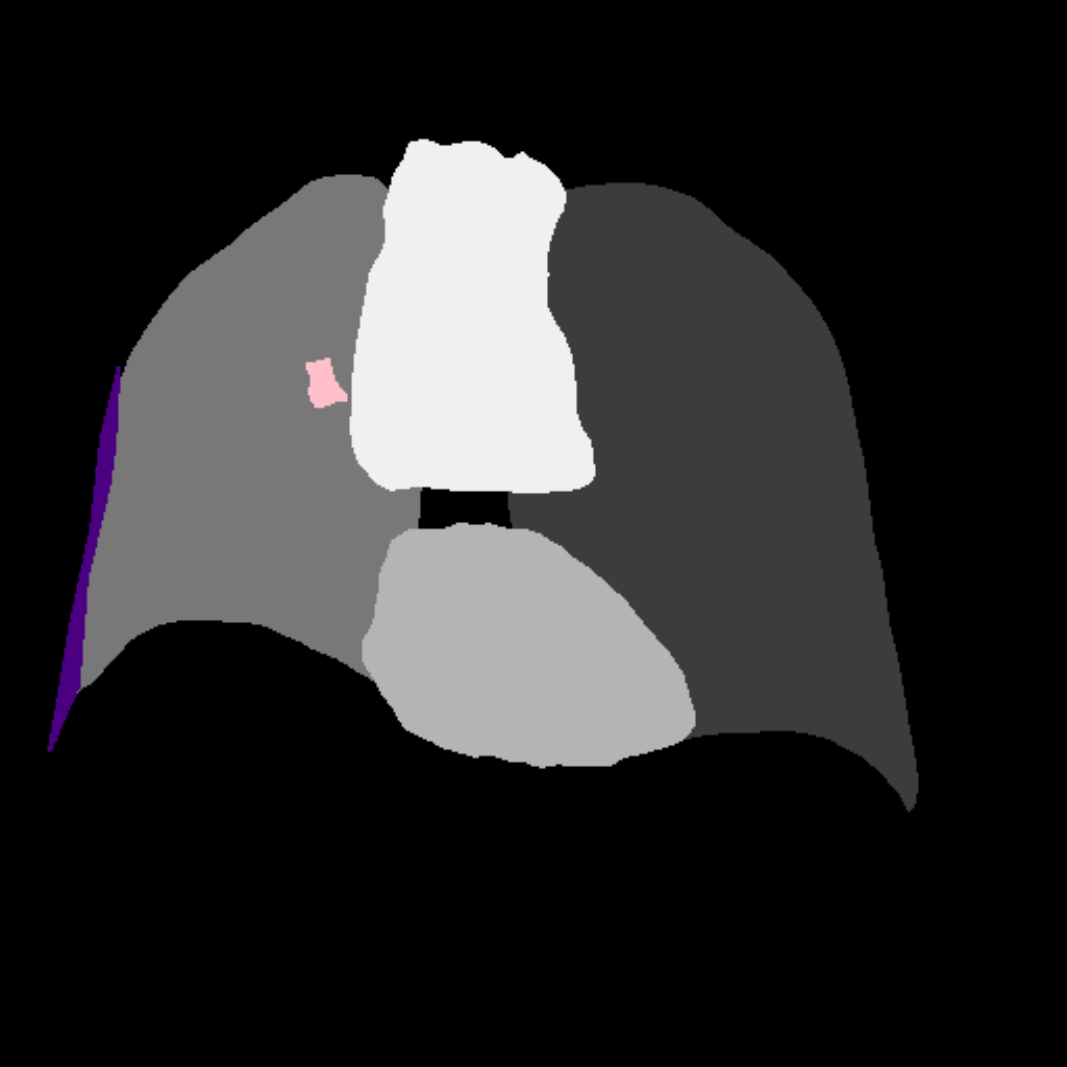}}
        \centerline{\parbox{1\linewidth}{\centering(a) Real Mask}}
    \end{minipage}
    \begin{minipage}[t]{0.16\linewidth}
        \centering
        \centerline{\includegraphics[width=1\linewidth]{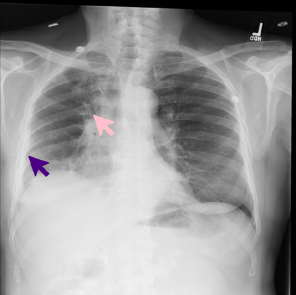}}
        \centerline{\parbox{1\linewidth}{\centering(b) Real Image}}
    \end{minipage}
    \begin{minipage}[t]{0.16\linewidth}
        \centering
        \centerline{\includegraphics[width=1\linewidth]{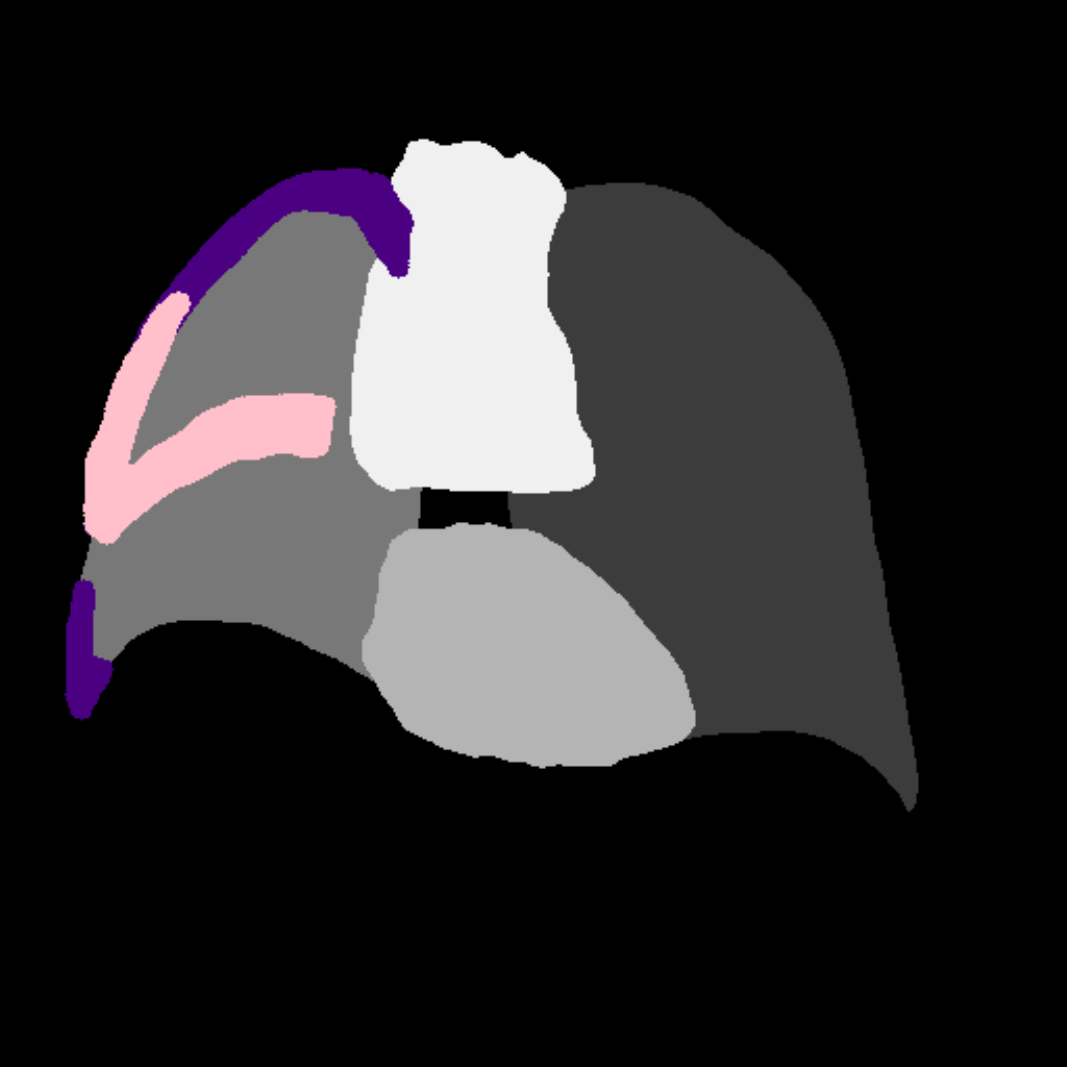}}
        \centerline{\parbox{1\linewidth}{\centering(c) Syn. Mask (Ours)}}
    \end{minipage}
    \begin{minipage}[t]{0.16\linewidth}
        \centering
        \centerline{\includegraphics[width=1\linewidth]{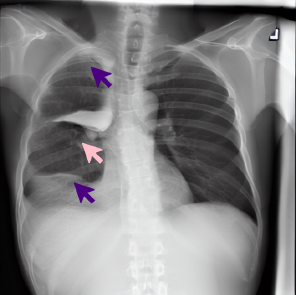}}
        \centerline{\parbox{1\linewidth}{\centering(d) Syn. Image (Ours)}}
    \end{minipage}
    \begin{minipage}[t]{0.16\linewidth}
        \centering
        \centerline{\includegraphics[width=1\linewidth]{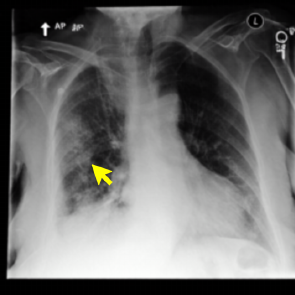}}
        \centerline{\parbox{1\linewidth}{\centering(e) Cheff}}
    \end{minipage}
    \begin{minipage}[t]{0.16\linewidth}
        \centering
        \centerline{\includegraphics[width=1\linewidth]{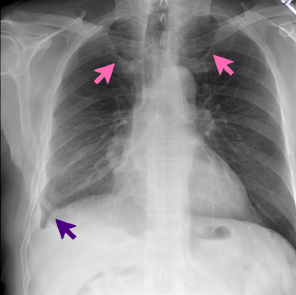}}
        \centerline{\parbox{1\linewidth}{\centering(f) RoentGen}}
        \end{minipage}
    \caption{Examples of failure cases in our generated results.}
    \label{fig:vis_bad}
\end{figure*}

Figure \ref{fig:vis_bad} highlights several failure cases. These errors typically arise from inaccurate organ segmentation, which leads to imprecise anatomical boundaries and affects the overall spatial correctness of the generated image. In some instances, the spatial layout of the generated pathology does not align well with the underlying anatomical structures, thereby reducing clinical reliability. Furthermore, certain generated masks fail to reflect realistic pathological appearances, resulting in medically implausible features. These limitations underscore the difficulty of simultaneously achieving spatial fidelity and clinical validity in medical image synthesis.

\begin{figure*}[t]
    \scriptsize
    \centering
    
     \begin{minipage}[t]{0.16\linewidth}
        \centering
        \includegraphics[width=\linewidth]{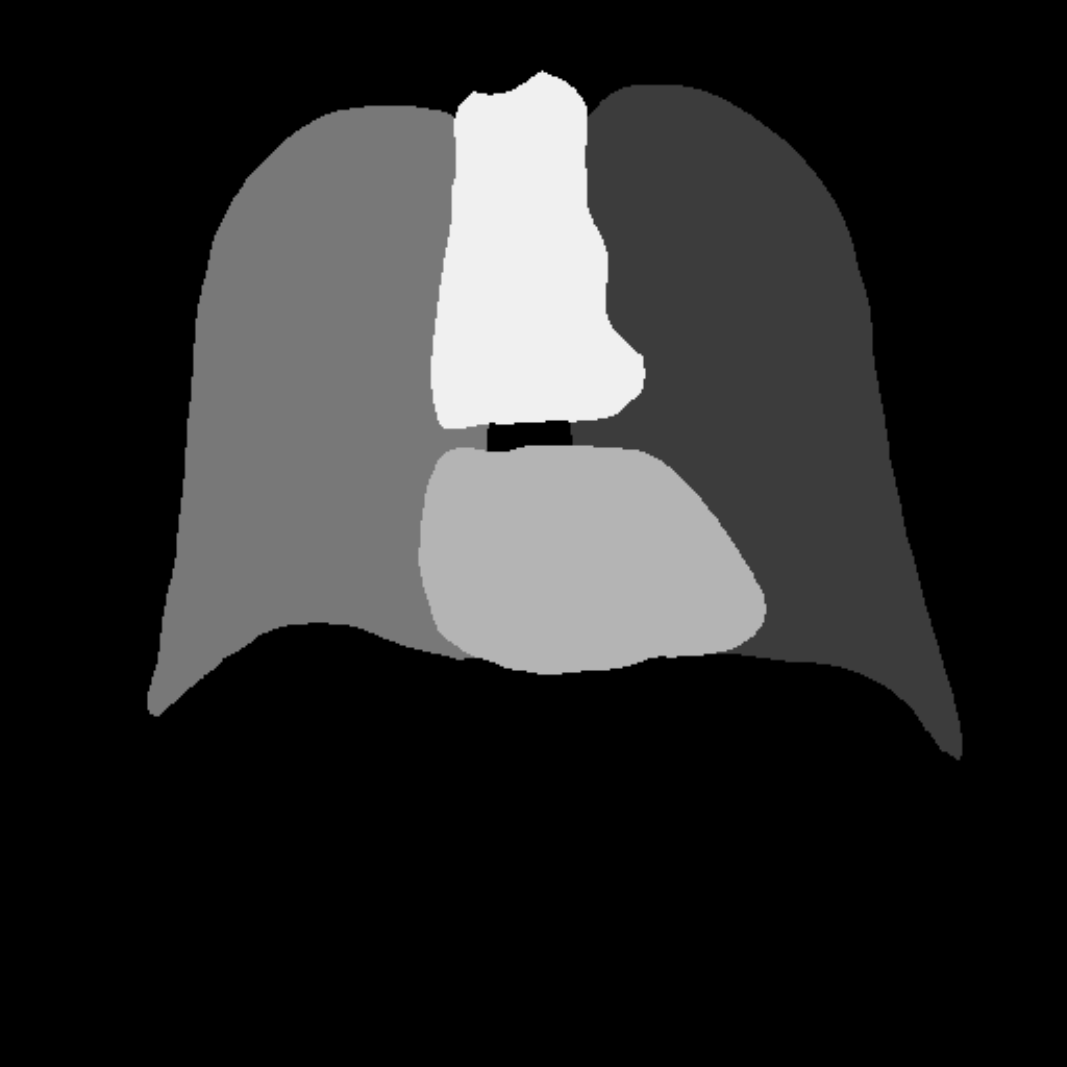}
    \end{minipage}
    \begin{minipage}[t]{0.16\linewidth}
        \centering
        \includegraphics[width=\linewidth]{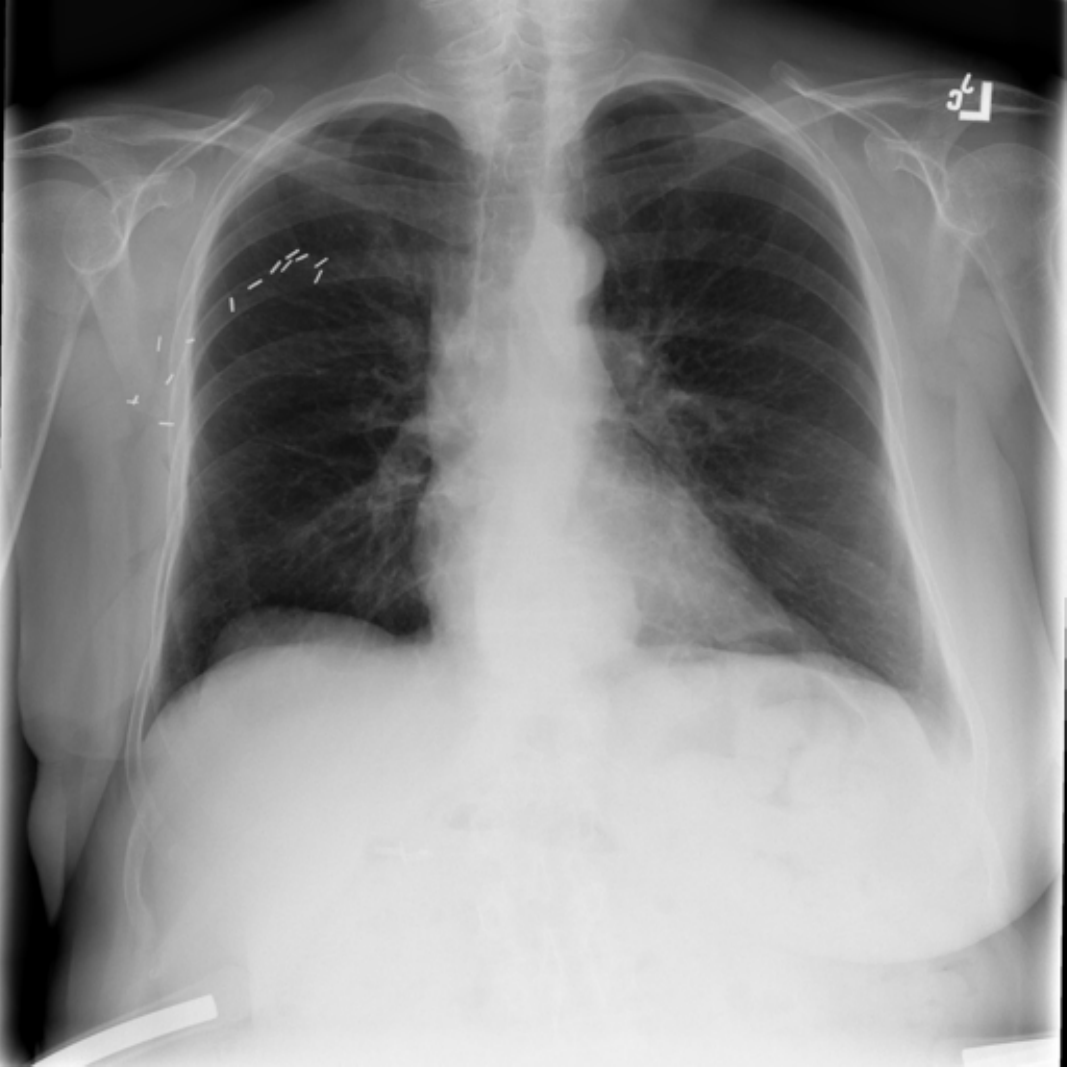}
    \end{minipage}
    \begin{minipage}[t]{0.16\linewidth}
        \centering
        \includegraphics[width=\linewidth]{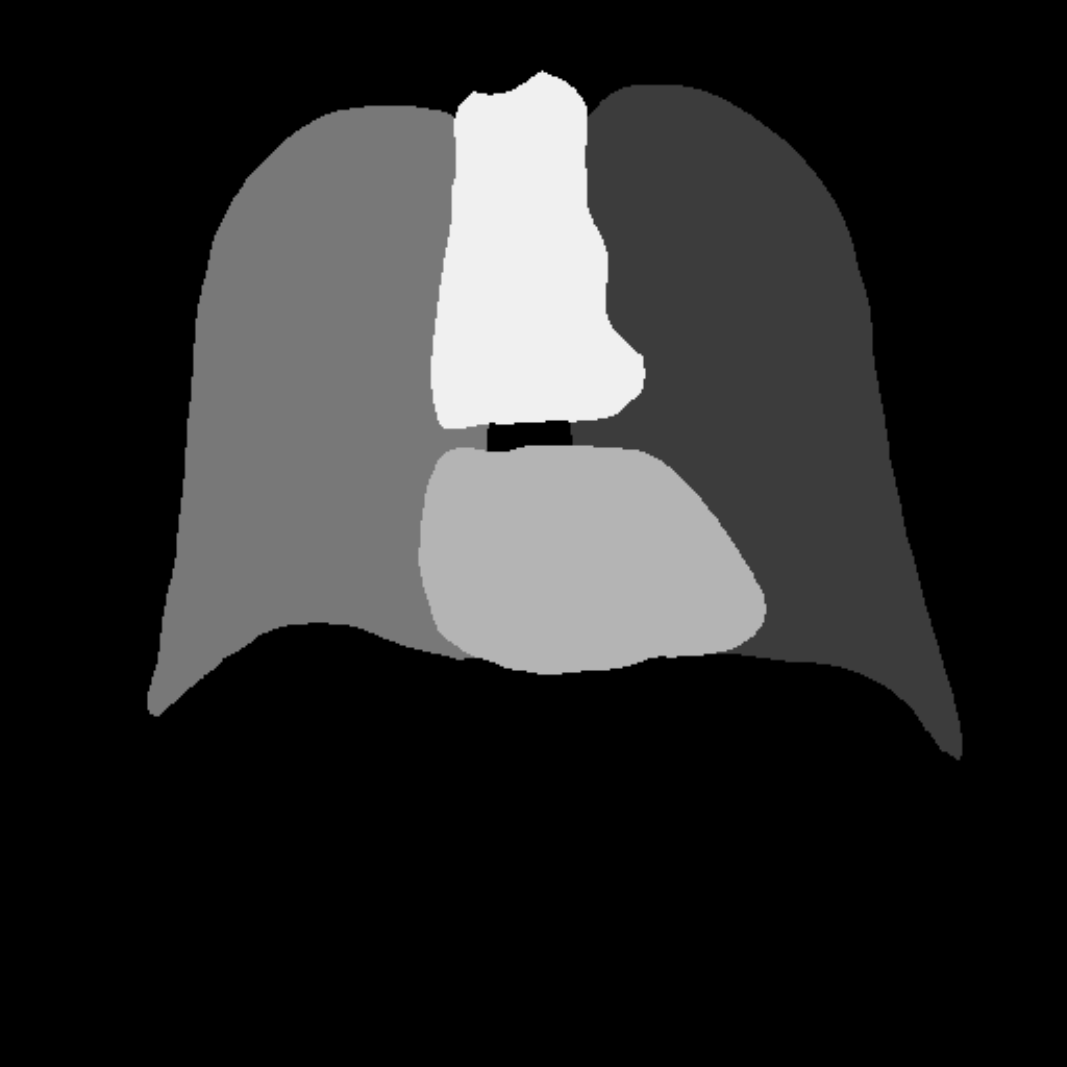}
    \end{minipage}
    \begin{minipage}[t]{0.16\linewidth}
        \centering
        \includegraphics[width=\linewidth]{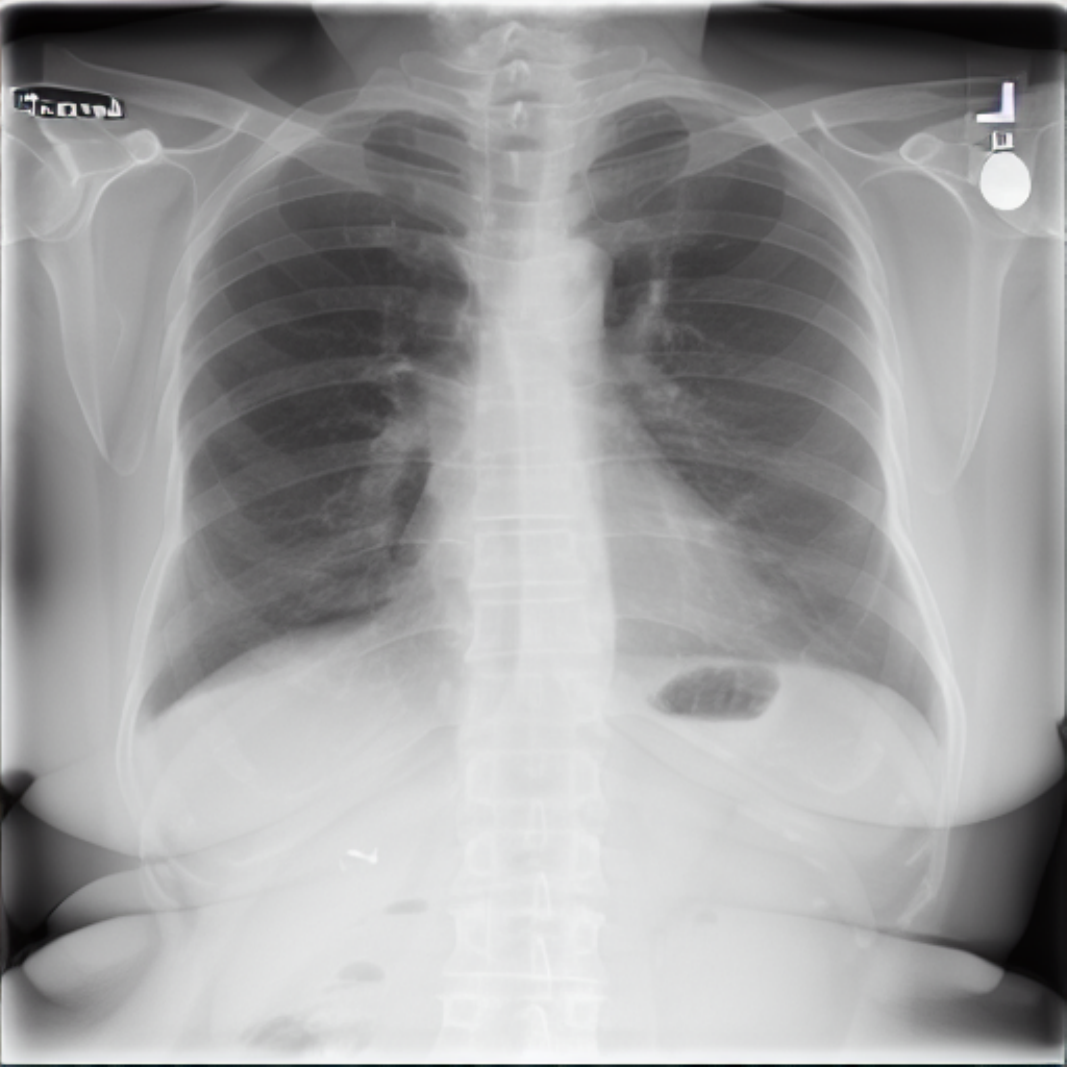}
    \end{minipage}
    \begin{minipage}[t]{0.16\linewidth}
        \centering
        \includegraphics[width=\linewidth]{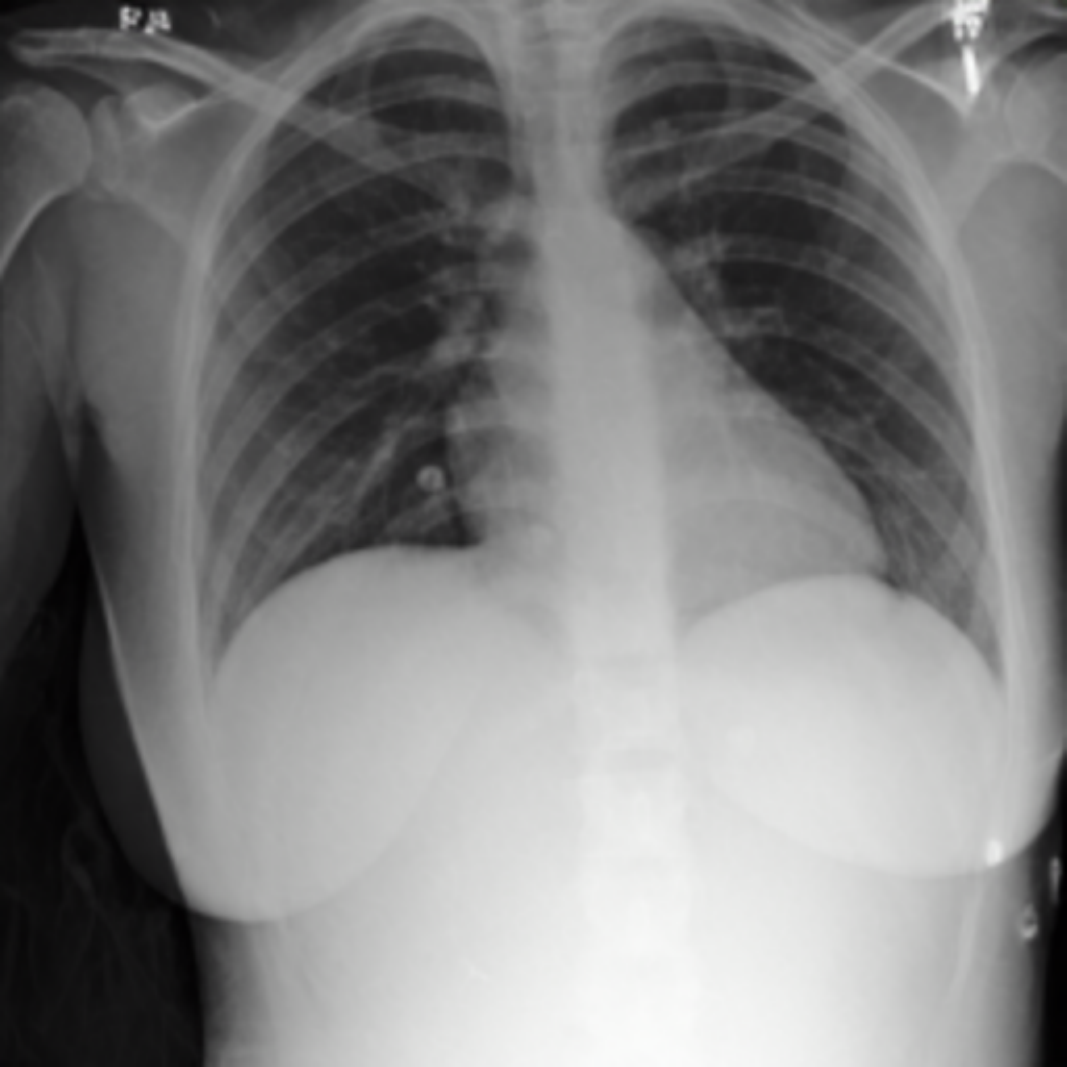}
    \end{minipage}
    \begin{minipage}[t]{0.16\linewidth}
        \centering
        \includegraphics[width=\linewidth]{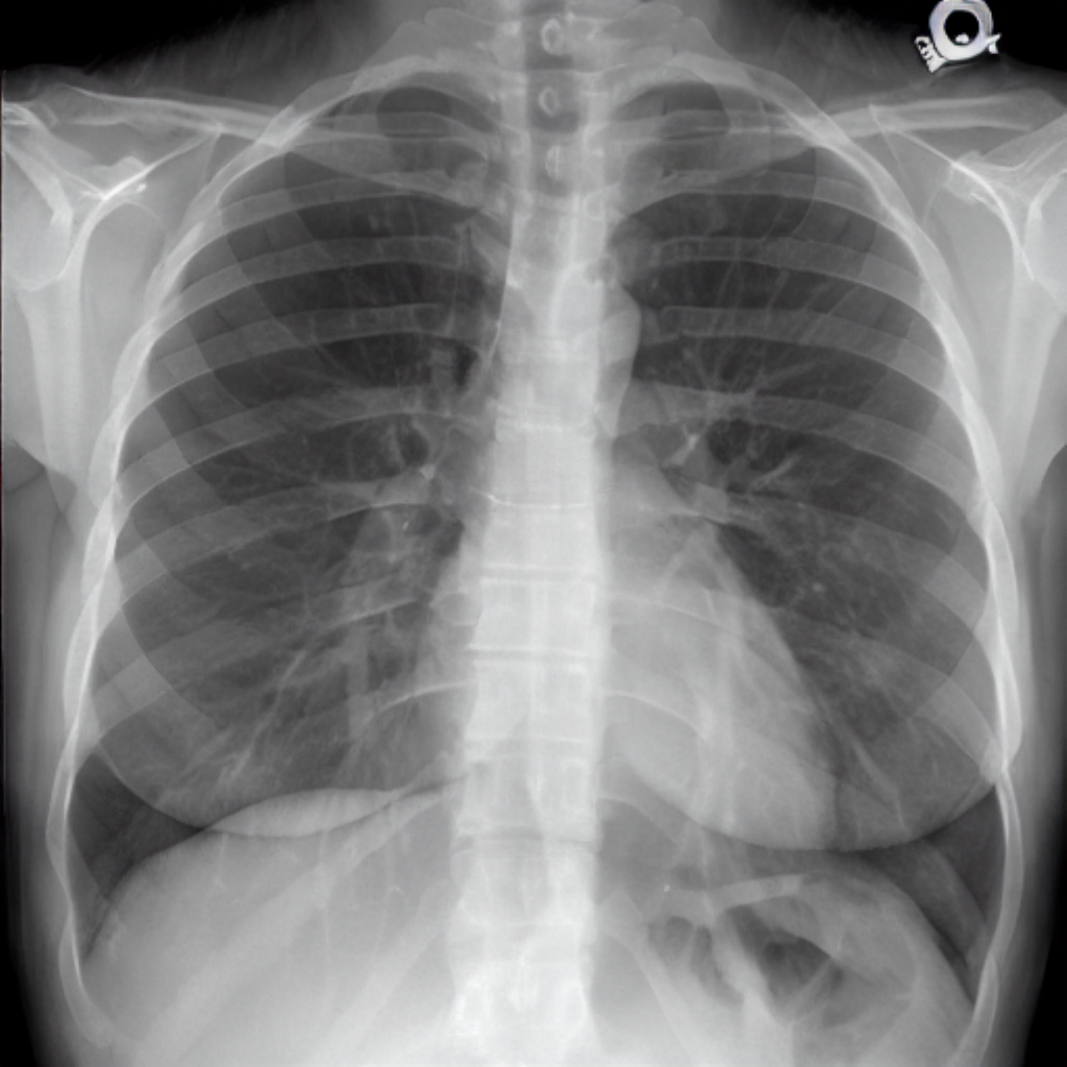}
    \end{minipage}

    \begin{minipage}[t]{0.16\linewidth}
        \centering
        \includegraphics[width=\linewidth]{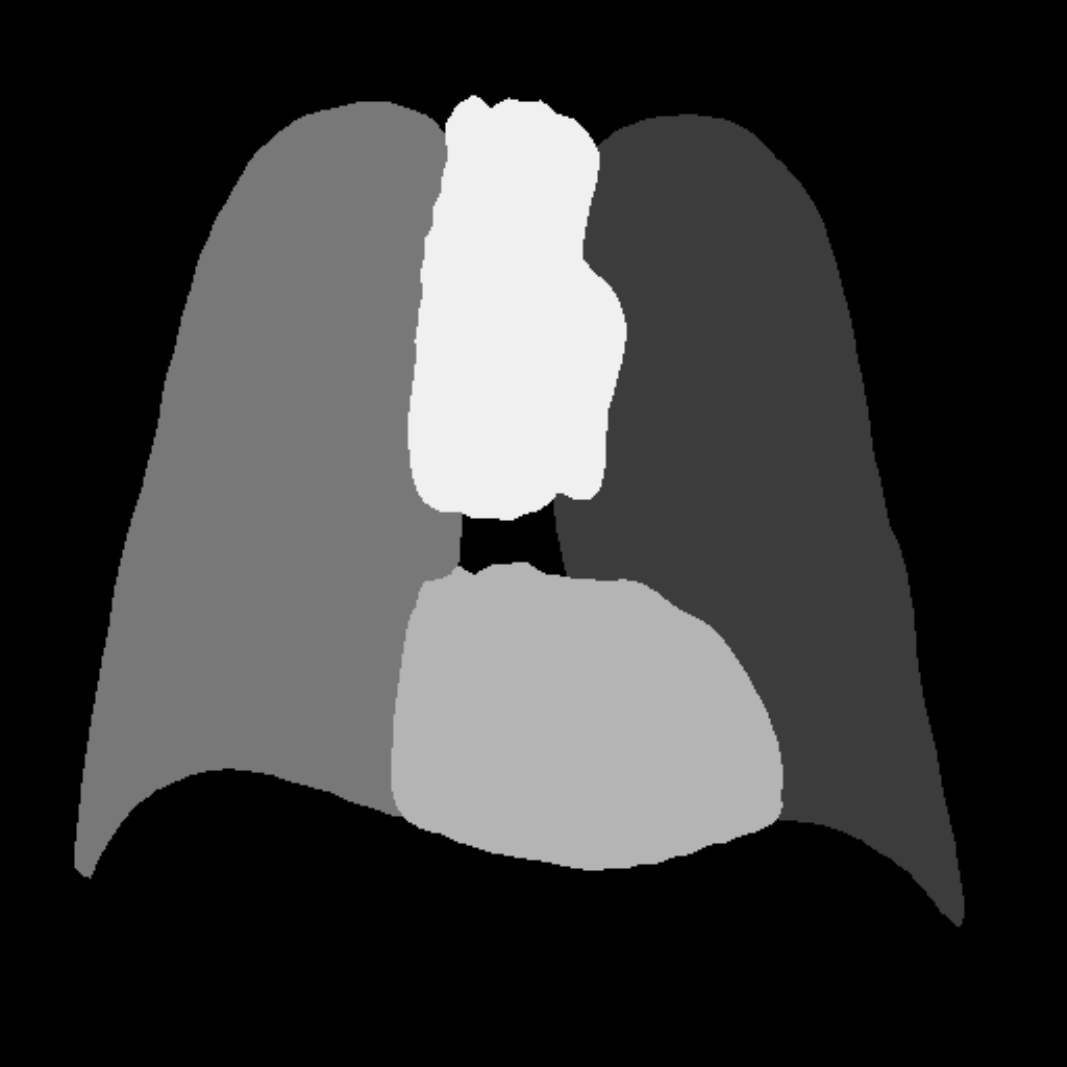}
    \end{minipage}
    \begin{minipage}[t]{0.16\linewidth}
        \centering
        \includegraphics[width=\linewidth]{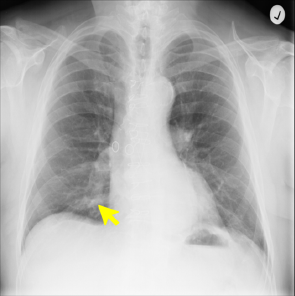}
    \end{minipage}
    \begin{minipage}[t]{0.16\linewidth}
        \centering
        \includegraphics[width=\linewidth]{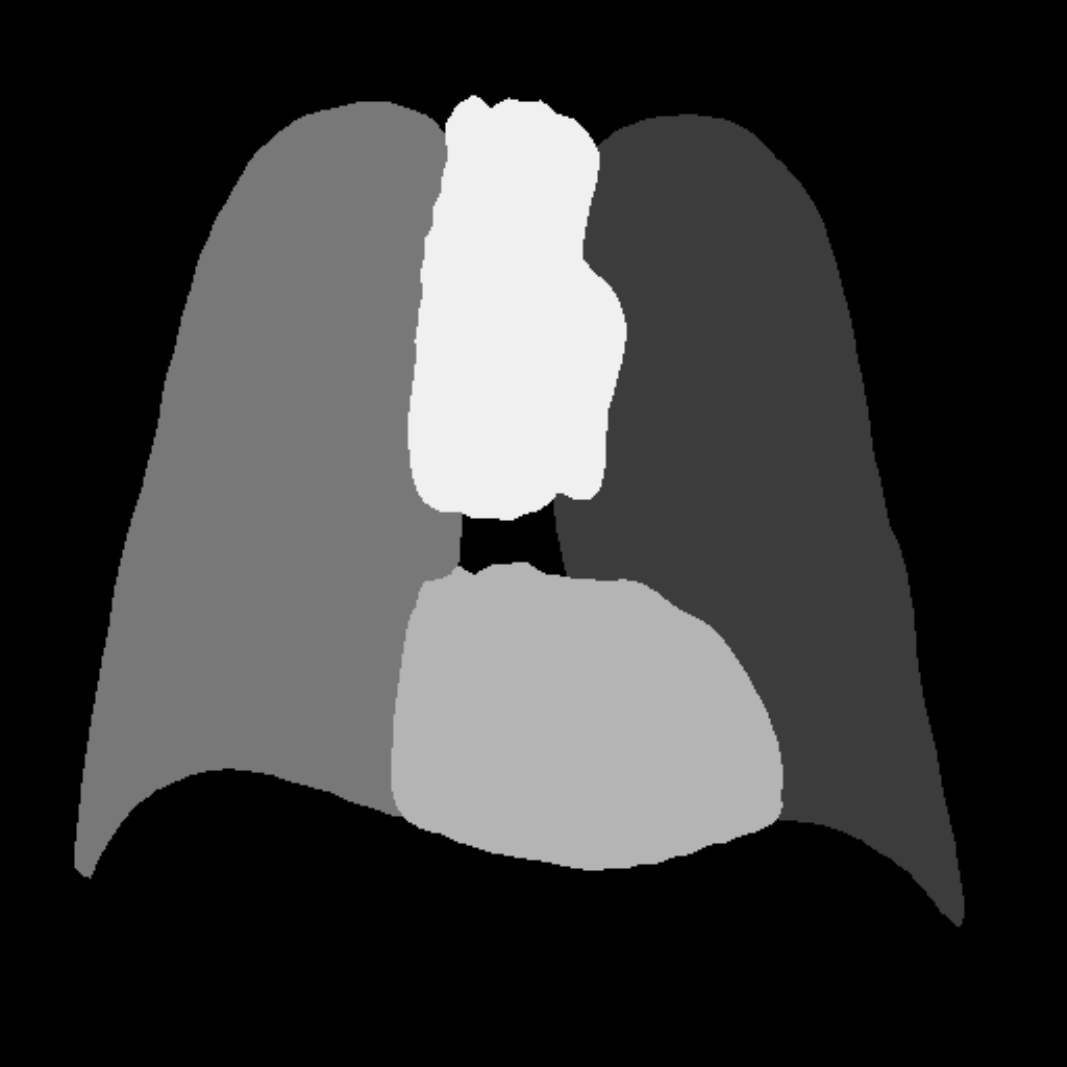}
    \end{minipage}
    \begin{minipage}[t]{0.16\linewidth}
        \centering
        \includegraphics[width=\linewidth]{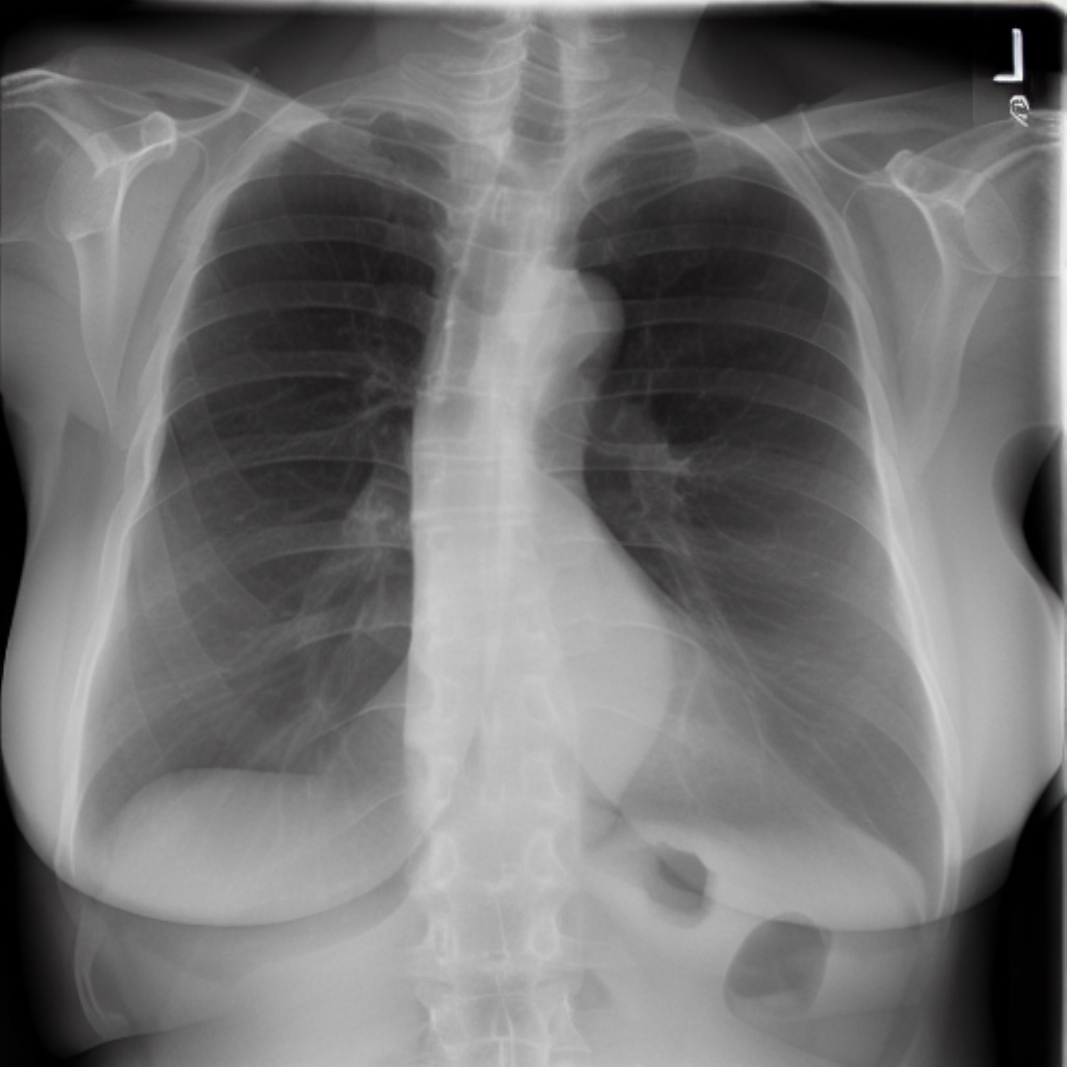}
    \end{minipage}
    \begin{minipage}[t]{0.16\linewidth}
        \centering
        \includegraphics[width=\linewidth]{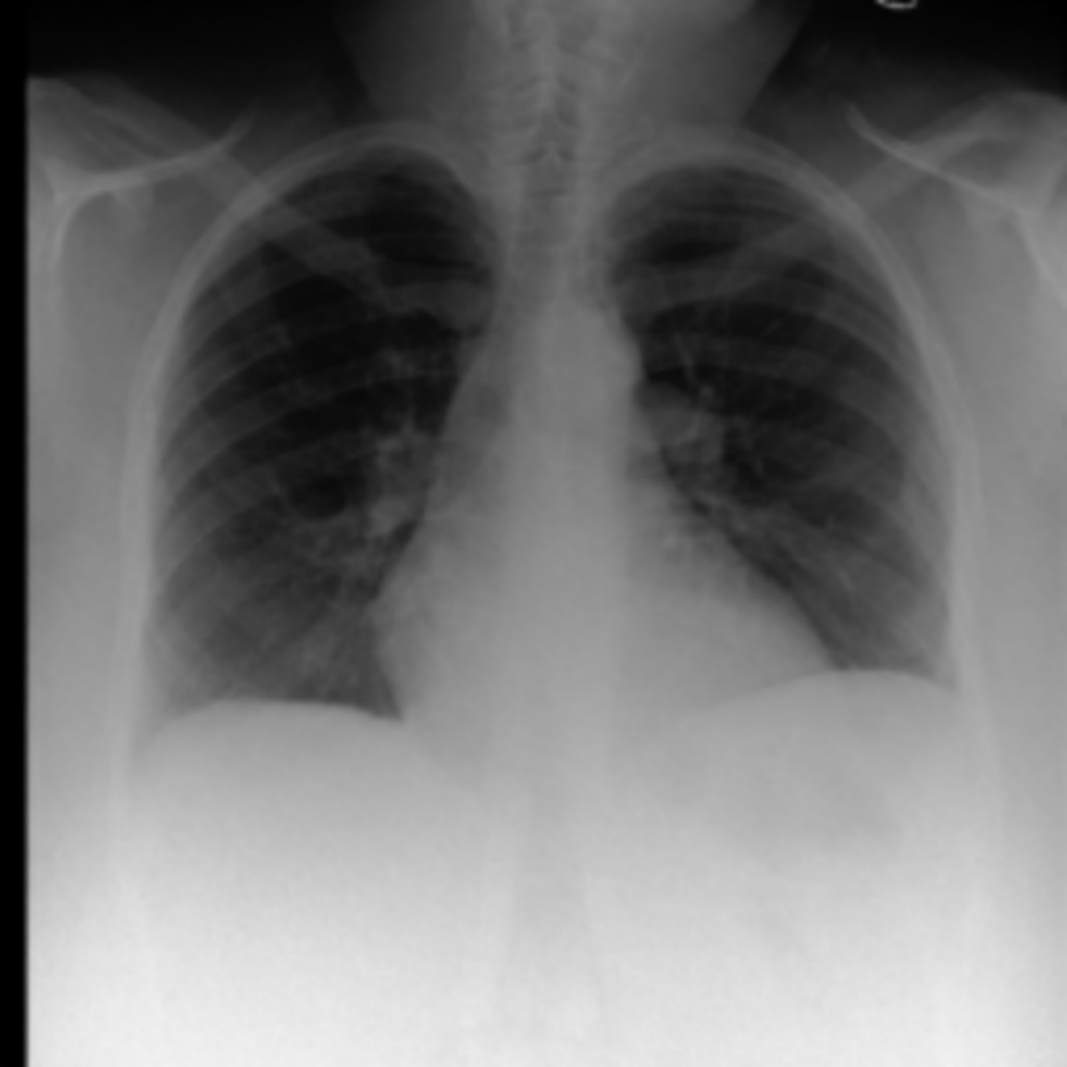}
    \end{minipage}
    \begin{minipage}[t]{0.16\linewidth}
        \centering
        \includegraphics[width=\linewidth]{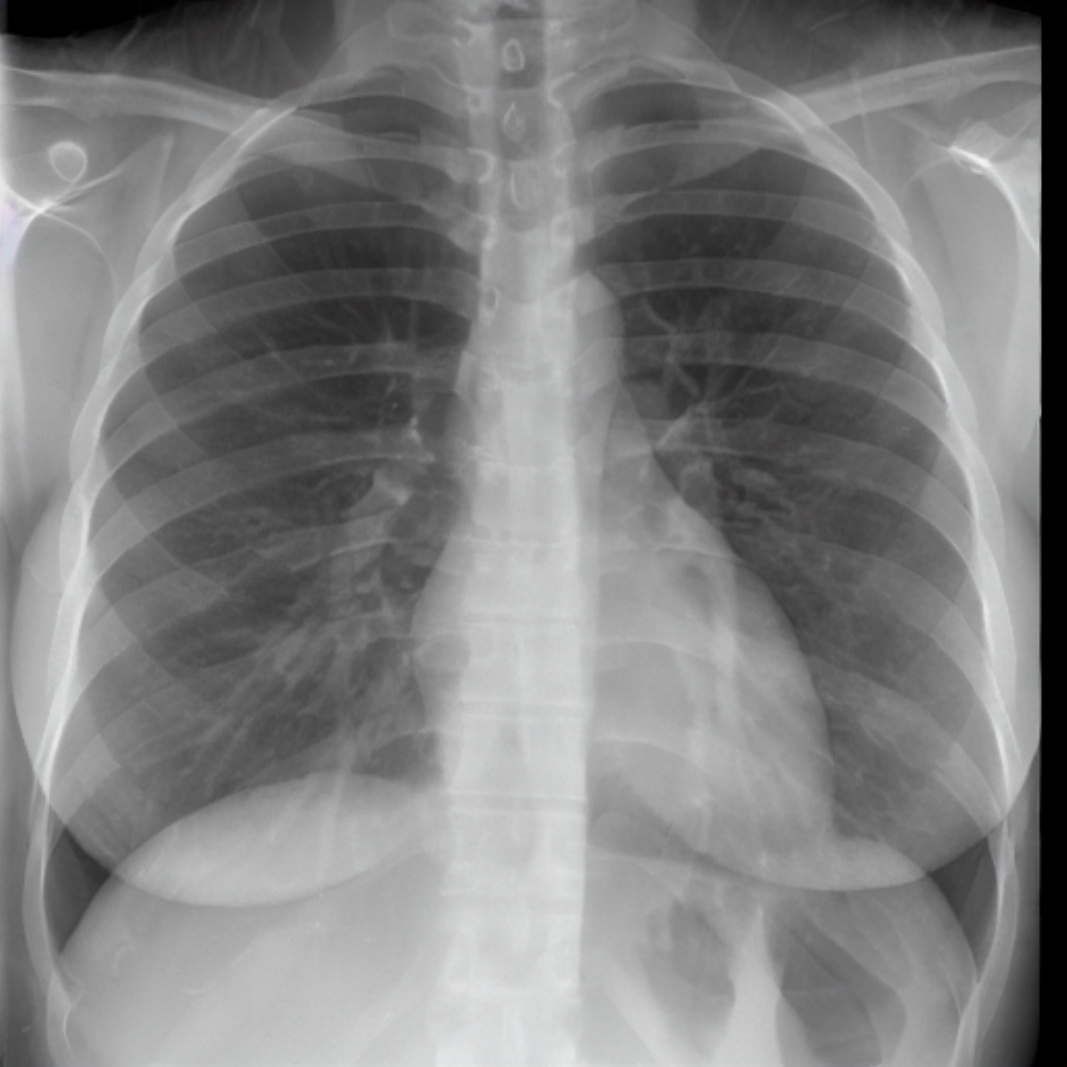}
    \end{minipage}

    \begin{minipage}[t]{0.16\linewidth}
        \centering
        \includegraphics[width=\linewidth]{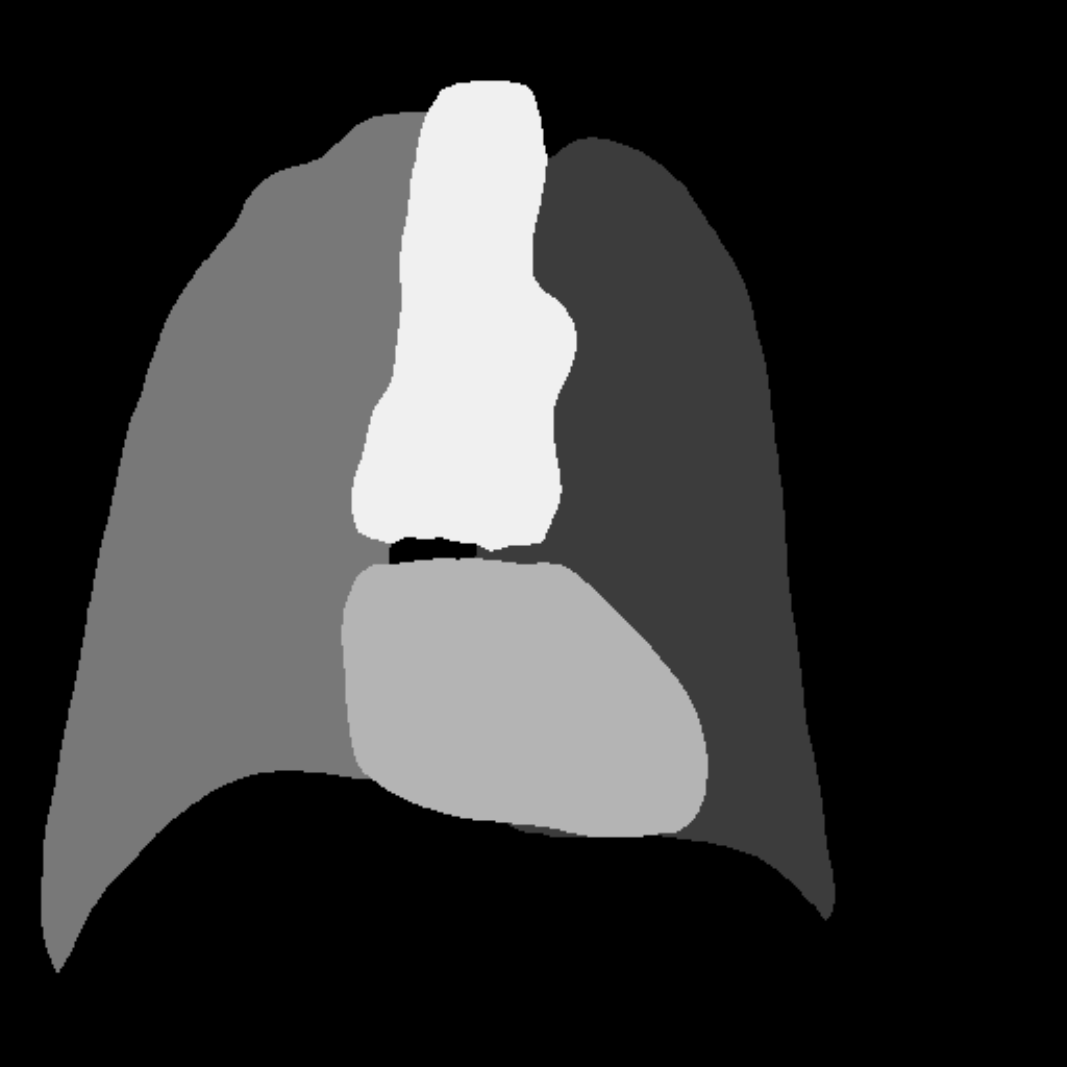}
    \end{minipage}
    \begin{minipage}[t]{0.16\linewidth}
        \centering
        \includegraphics[width=\linewidth]{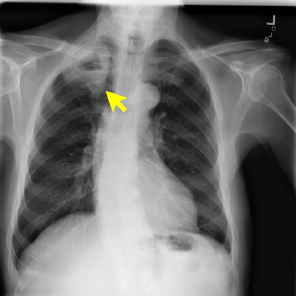}
    \end{minipage}
    \begin{minipage}[t]{0.16\linewidth}
        \centering
        \includegraphics[width=\linewidth]{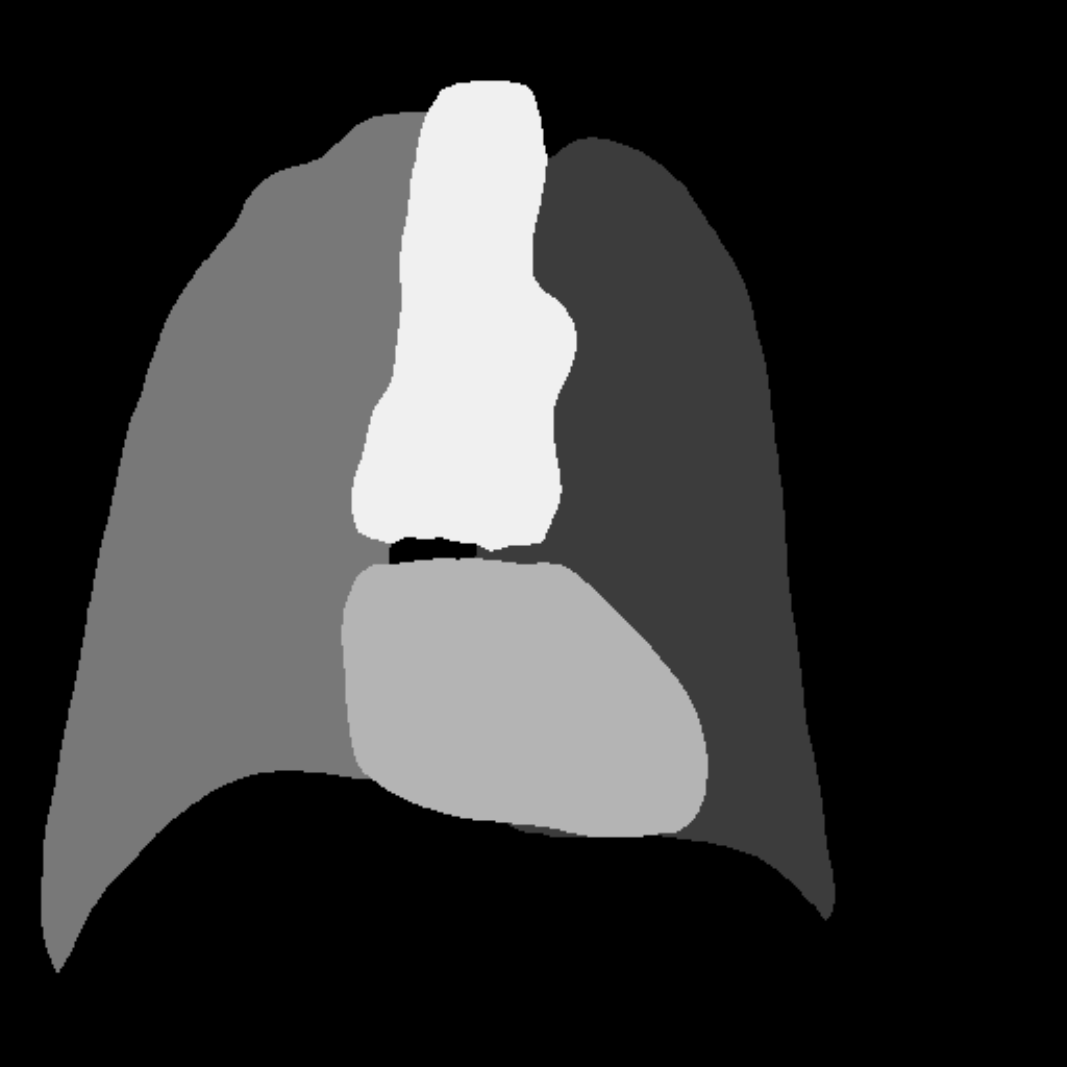}
    \end{minipage}
    \begin{minipage}[t]{0.16\linewidth}
        \centering
        \includegraphics[width=\linewidth]{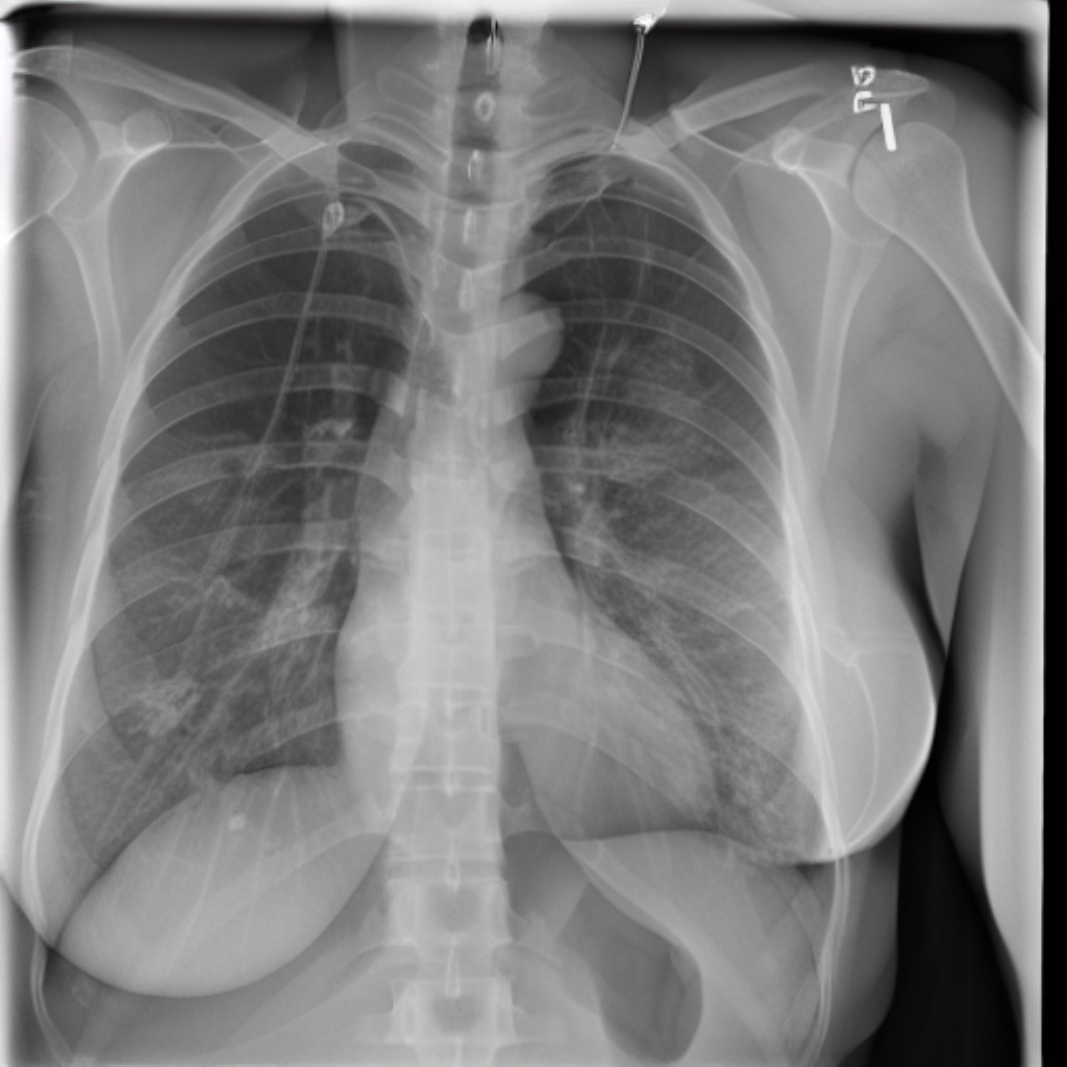}
    \end{minipage}
    \begin{minipage}[t]{0.16\linewidth}
        \centering
        \includegraphics[width=\linewidth]{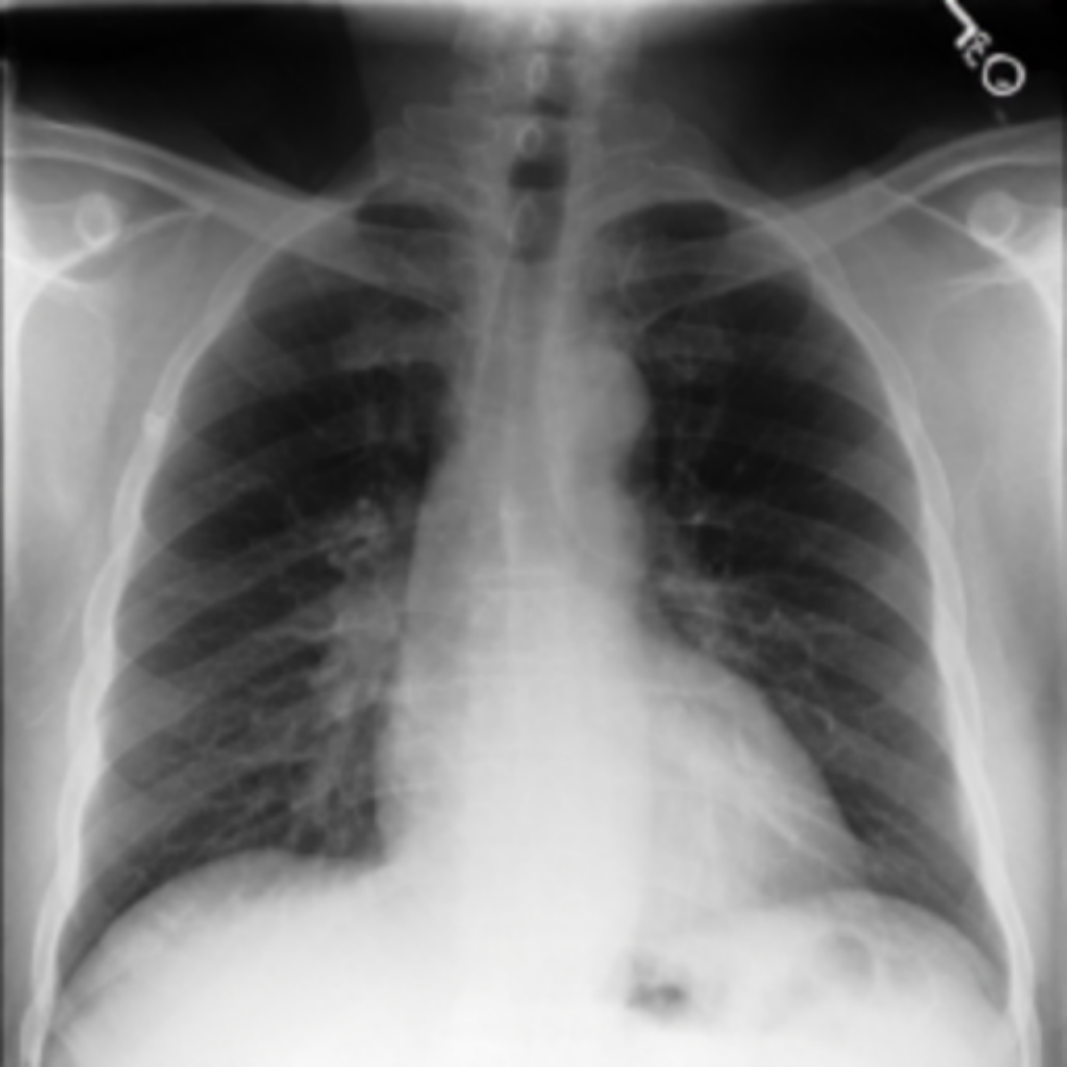}
    \end{minipage}
    \begin{minipage}[t]{0.16\linewidth}
        \centering
        \includegraphics[width=\linewidth]{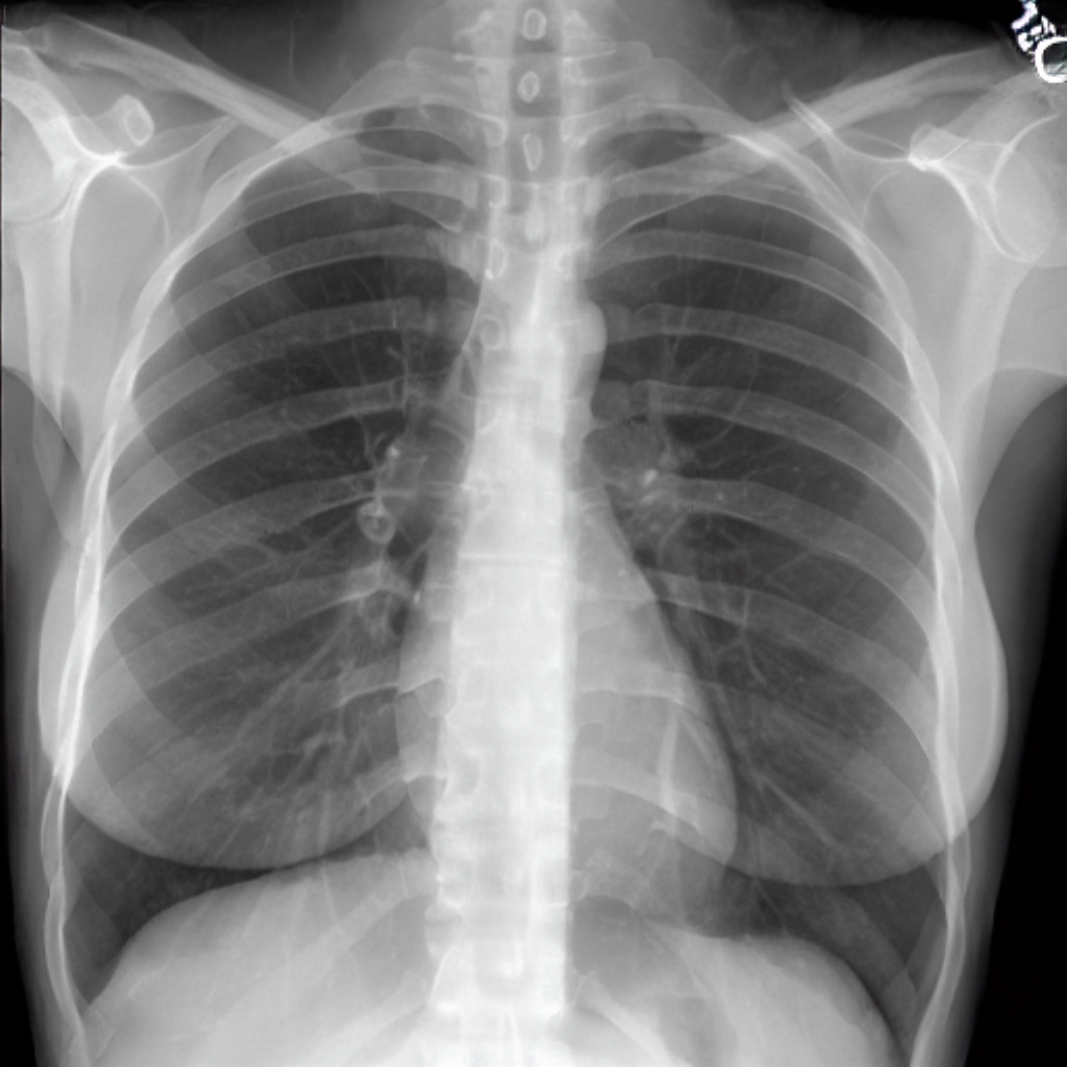}
    \end{minipage}

    \begin{minipage}[t]{0.16\linewidth}
        \centering
        \includegraphics[width=\linewidth]{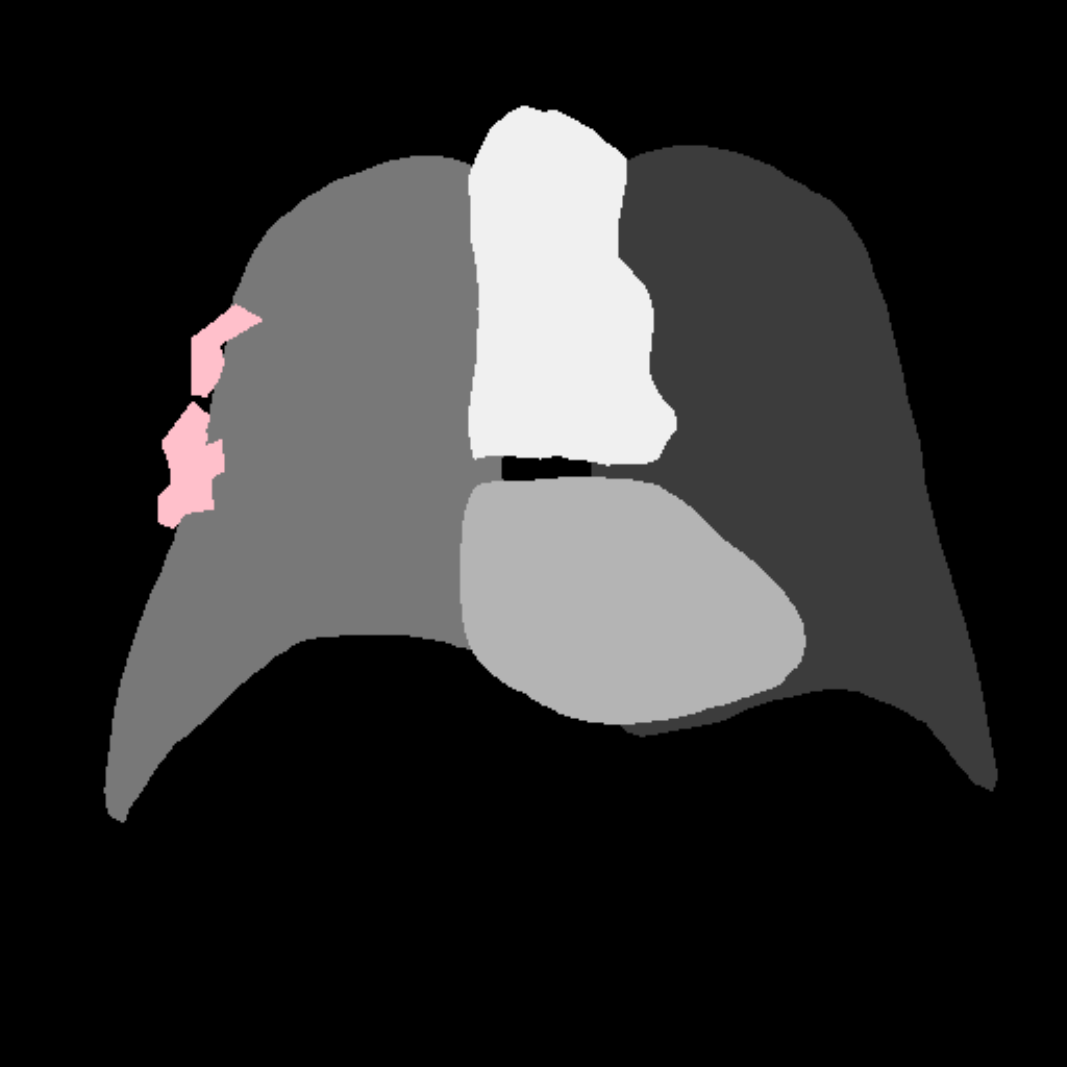}
    \end{minipage}
    \begin{minipage}[t]{0.16\linewidth}
        \centering
        \includegraphics[width=\linewidth]{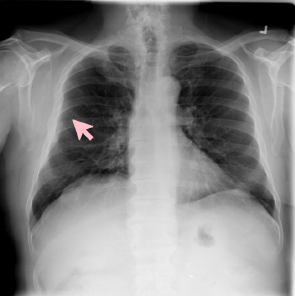}
    \end{minipage}
    \begin{minipage}[t]{0.16\linewidth}
        \centering
        \includegraphics[width=\linewidth]{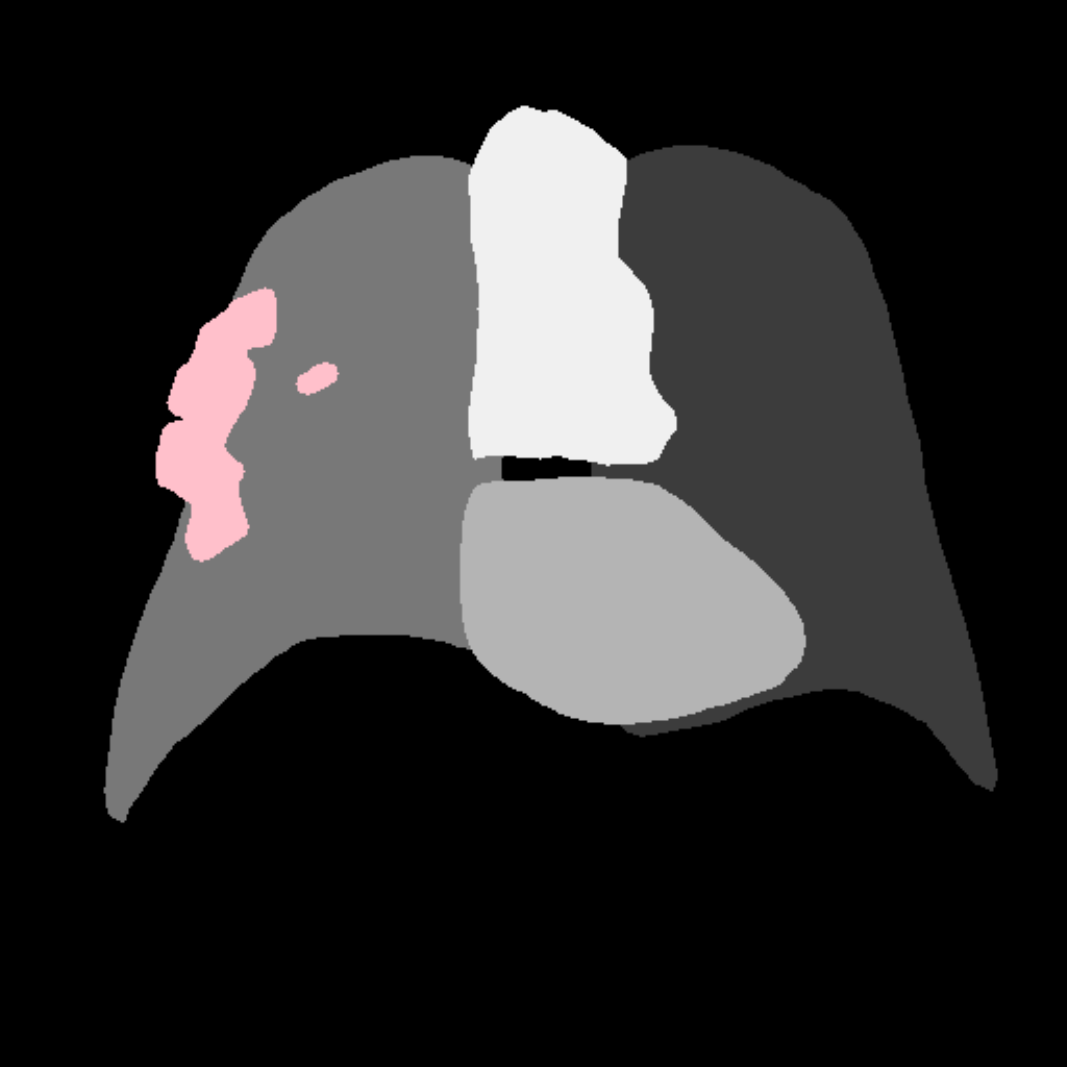}
    \end{minipage}
    \begin{minipage}[t]{0.16\linewidth}
        \centering
        \includegraphics[width=\linewidth]{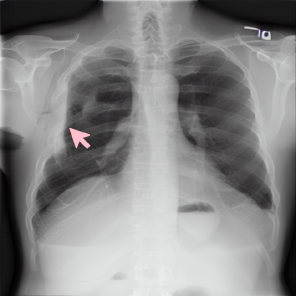}
    \end{minipage}
    \begin{minipage}[t]{0.16\linewidth}
        \centering
        \includegraphics[width=\linewidth]{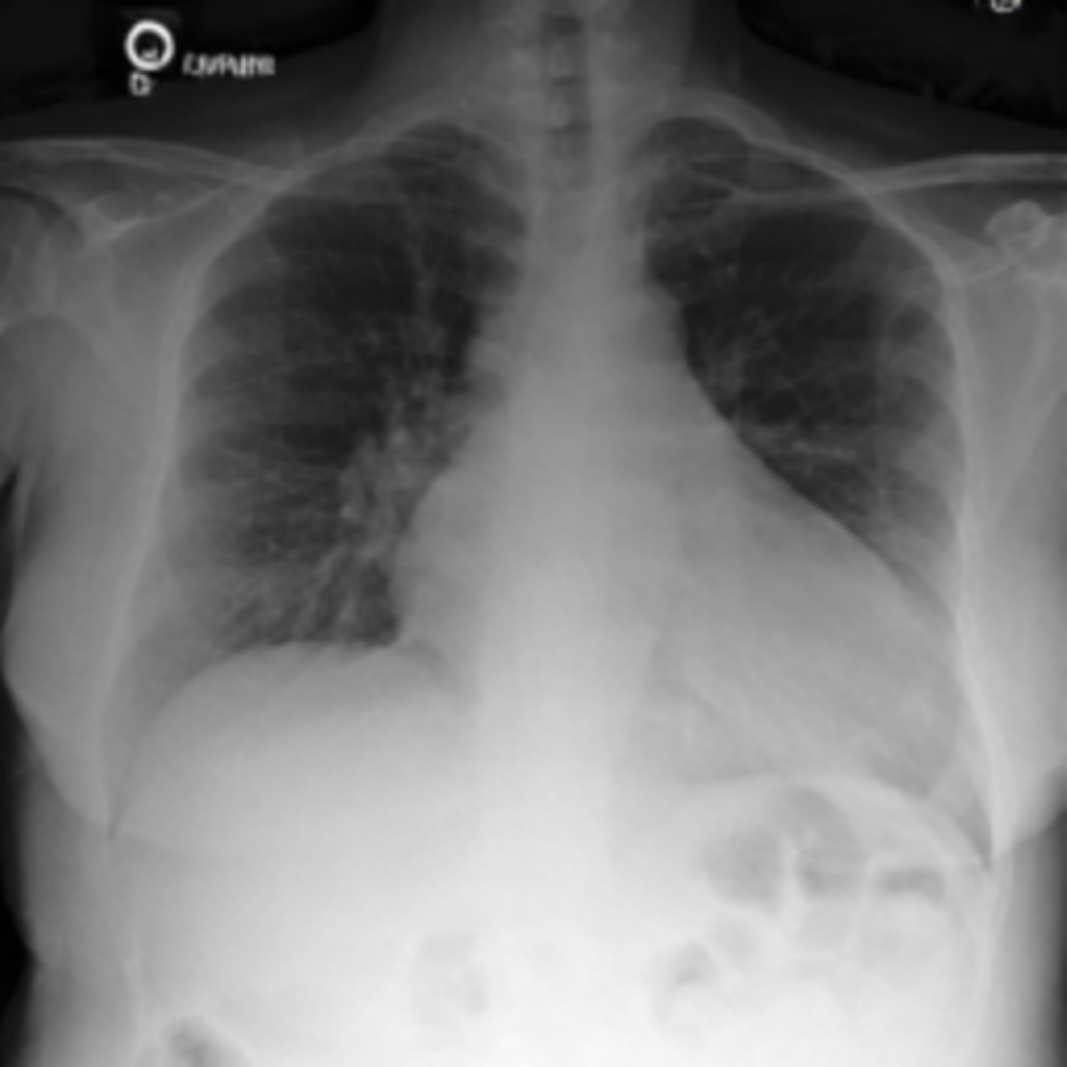}
    \end{minipage}
    \begin{minipage}[t]{0.16\linewidth}
        \centering
        \includegraphics[width=\linewidth]{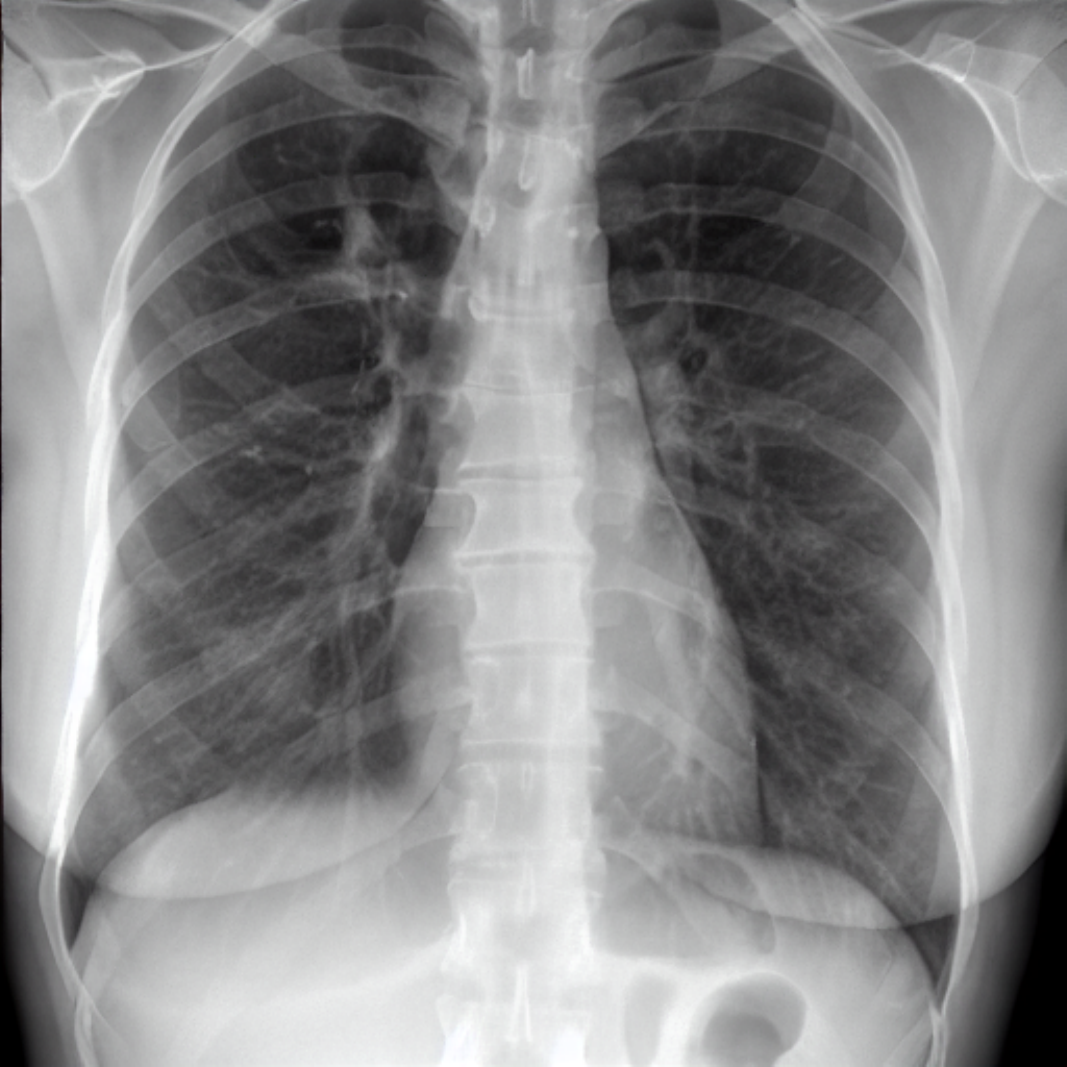}
    \end{minipage}
    
    \begin{minipage}[t]{0.16\linewidth}
        \centering
        \includegraphics[width=\linewidth]{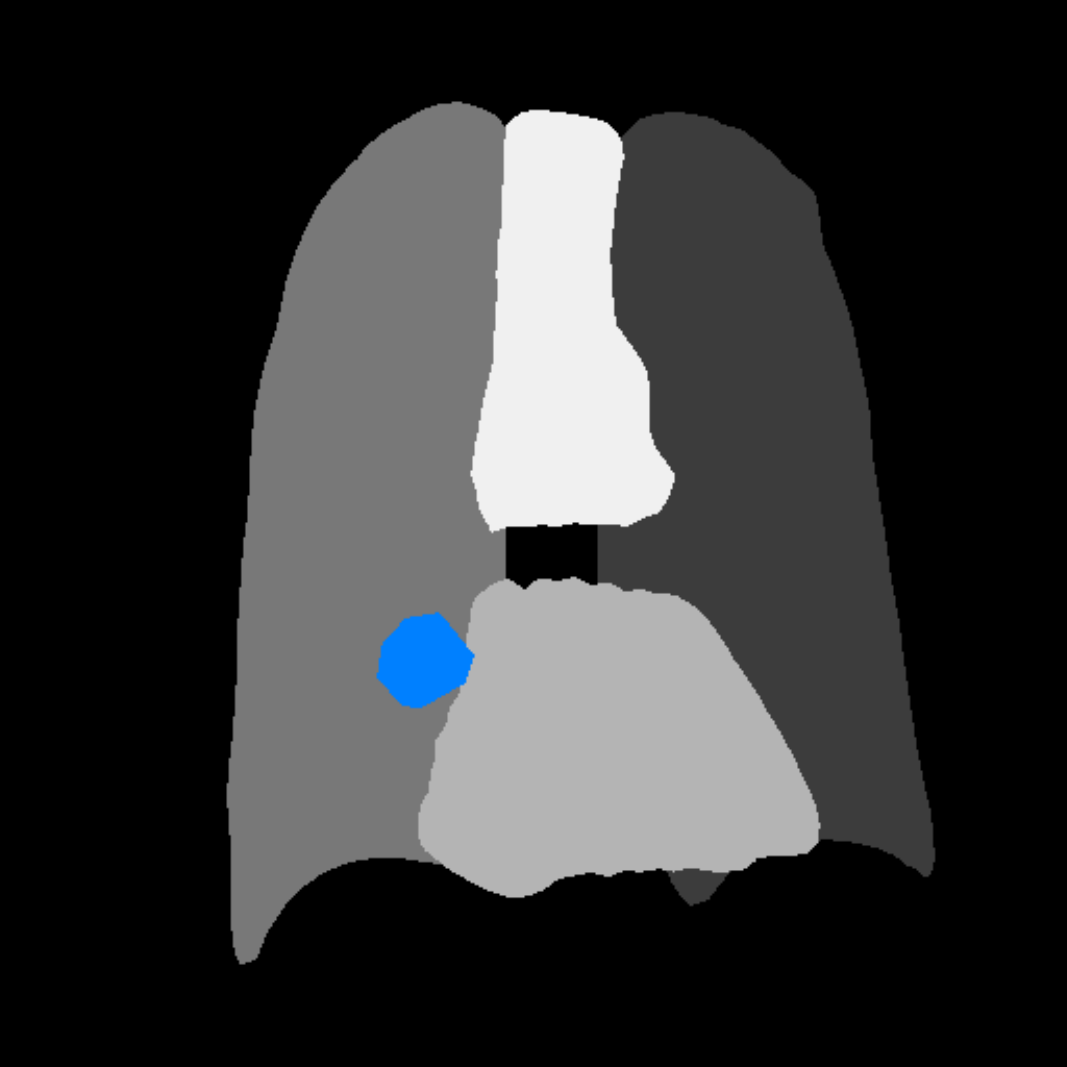}
    \end{minipage}
    \begin{minipage}[t]{0.16\linewidth}
        \centering
        \includegraphics[width=\linewidth]{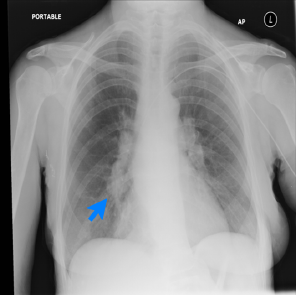}
    \end{minipage}
    \begin{minipage}[t]{0.16\linewidth}
        \centering
        \includegraphics[width=\linewidth]{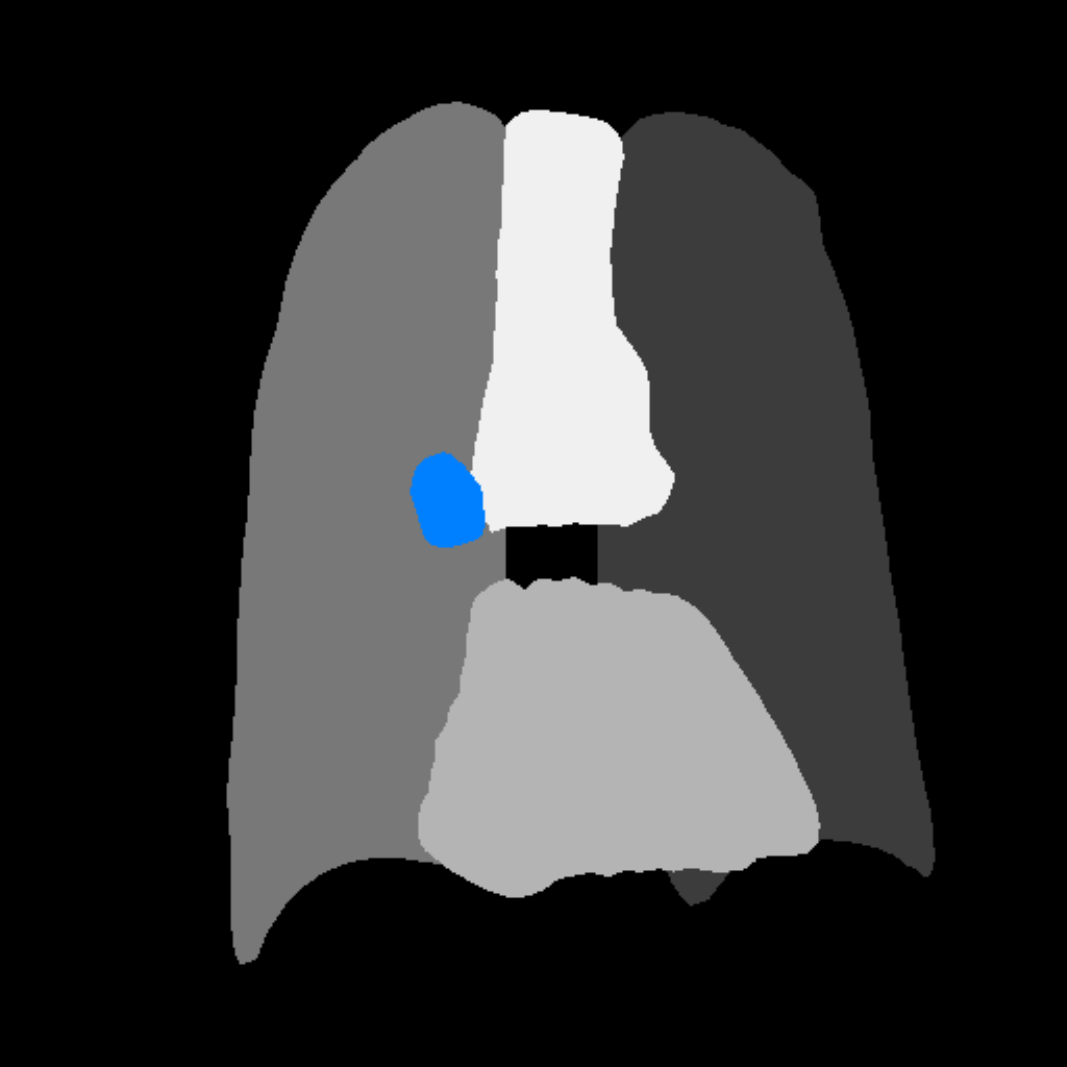}
    \end{minipage}
    \begin{minipage}[t]{0.16\linewidth}
        \centering
        \includegraphics[width=\linewidth]{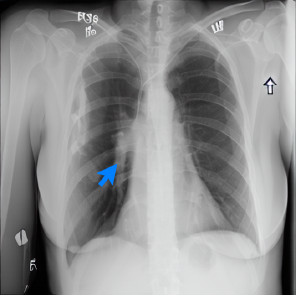}
    \end{minipage}
    \begin{minipage}[t]{0.16\linewidth}
        \centering
        \includegraphics[width=\linewidth]{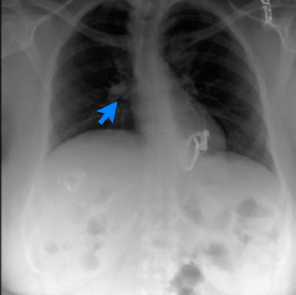}
    \end{minipage}
    \begin{minipage}[t]{0.16\linewidth}
        \centering
        \includegraphics[width=\linewidth]{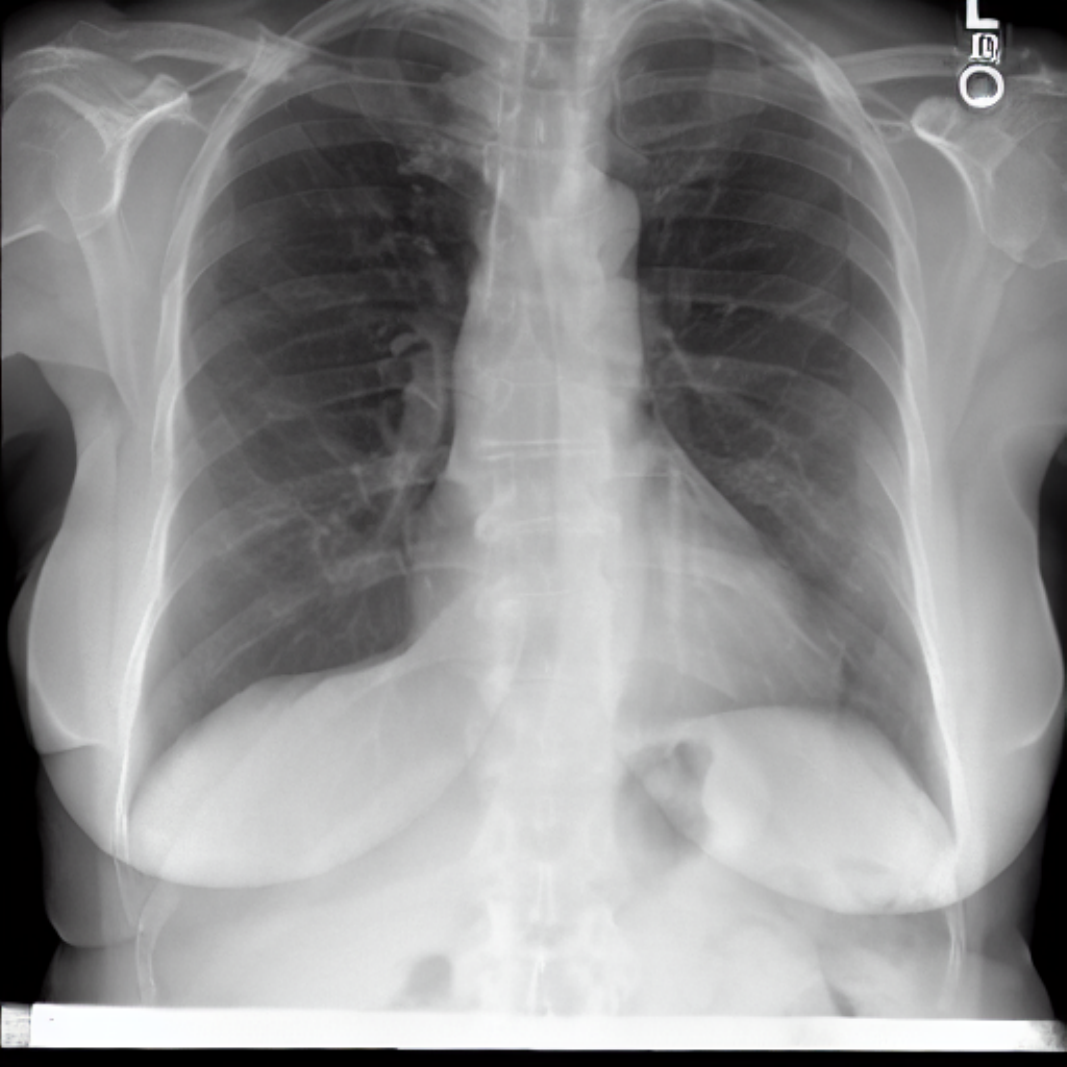}
    \end{minipage}

    \begin{minipage}[t]{0.16\linewidth}
        \centering
        \includegraphics[width=\linewidth]{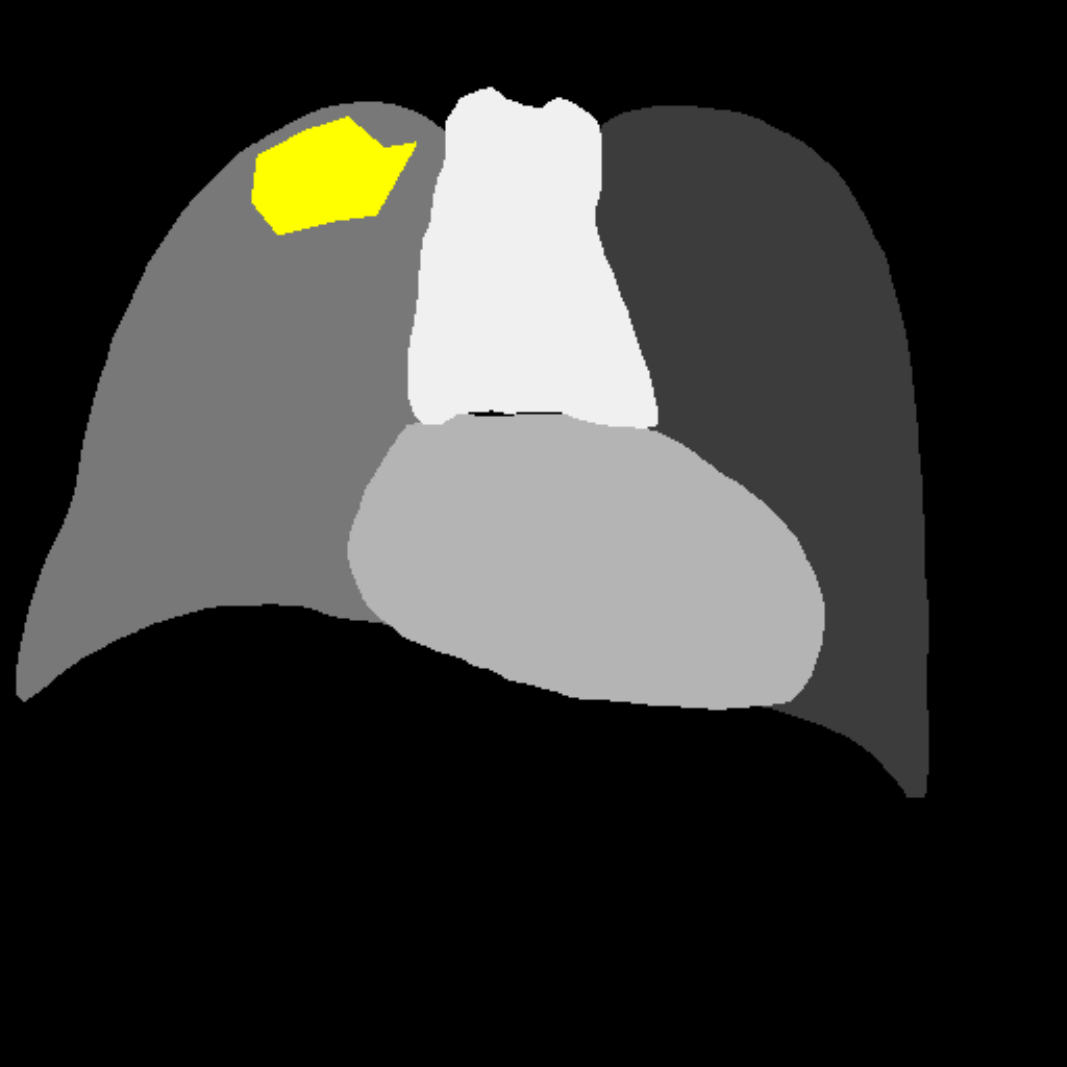}
    \end{minipage}
    \begin{minipage}[t]{0.16\linewidth}
        \centering
        \includegraphics[width=\linewidth]{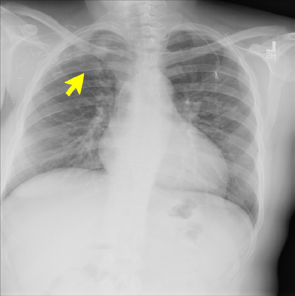}
    \end{minipage}
    \begin{minipage}[t]{0.16\linewidth}
        \centering
        \includegraphics[width=\linewidth]{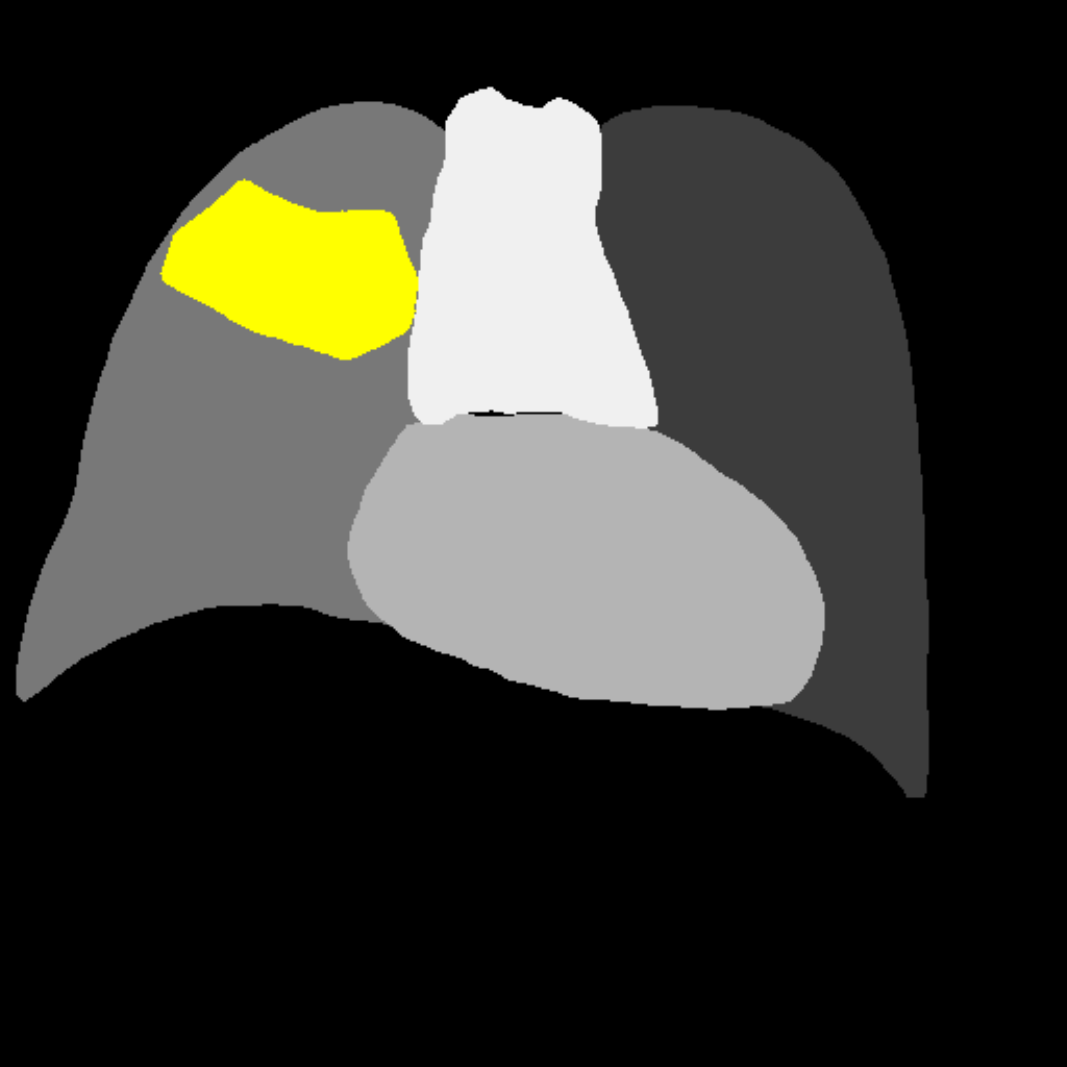}
    \end{minipage}
    \begin{minipage}[t]{0.16\linewidth}
        \centering
        \includegraphics[width=\linewidth]{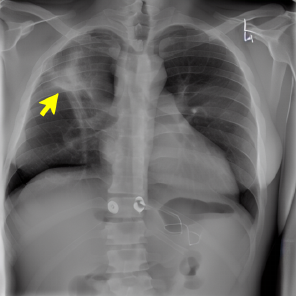}
    \end{minipage}
    \begin{minipage}[t]{0.16\linewidth}
        \centering
        \includegraphics[width=\linewidth]{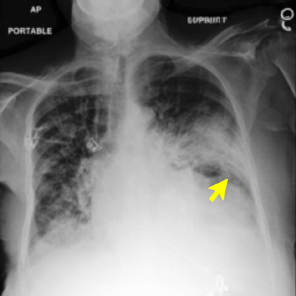}
    \end{minipage}
    \begin{minipage}[t]{0.16\linewidth}
        \centering
        \includegraphics[width=\linewidth]{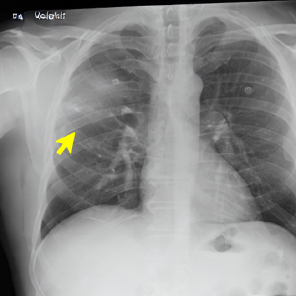}
    \end{minipage}

    \begin{minipage}[t]{0.16\linewidth}
        \centering
        \includegraphics[width=\linewidth]{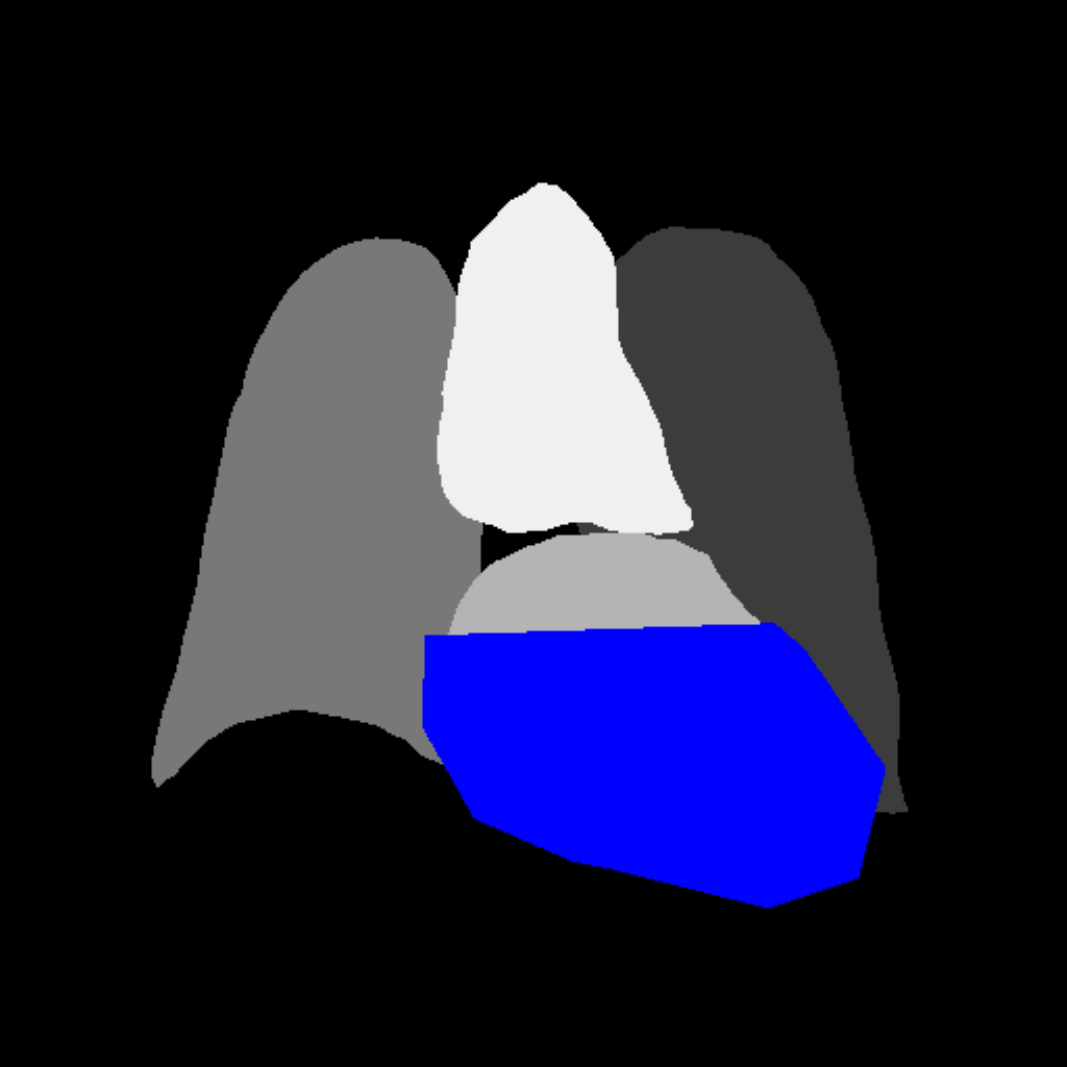}
        \centerline{\parbox{1\linewidth}{\centering(a) Real Mask}}
    \end{minipage}
    \begin{minipage}[t]{0.16\linewidth}
        \centering
        \includegraphics[width=\linewidth]{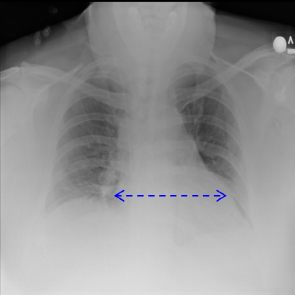}
        \centerline{\parbox{1\linewidth}{\centering(b) Real Image}}
    \end{minipage}
    \begin{minipage}[t]{0.16\linewidth}
        \centering
        \includegraphics[width=\linewidth]{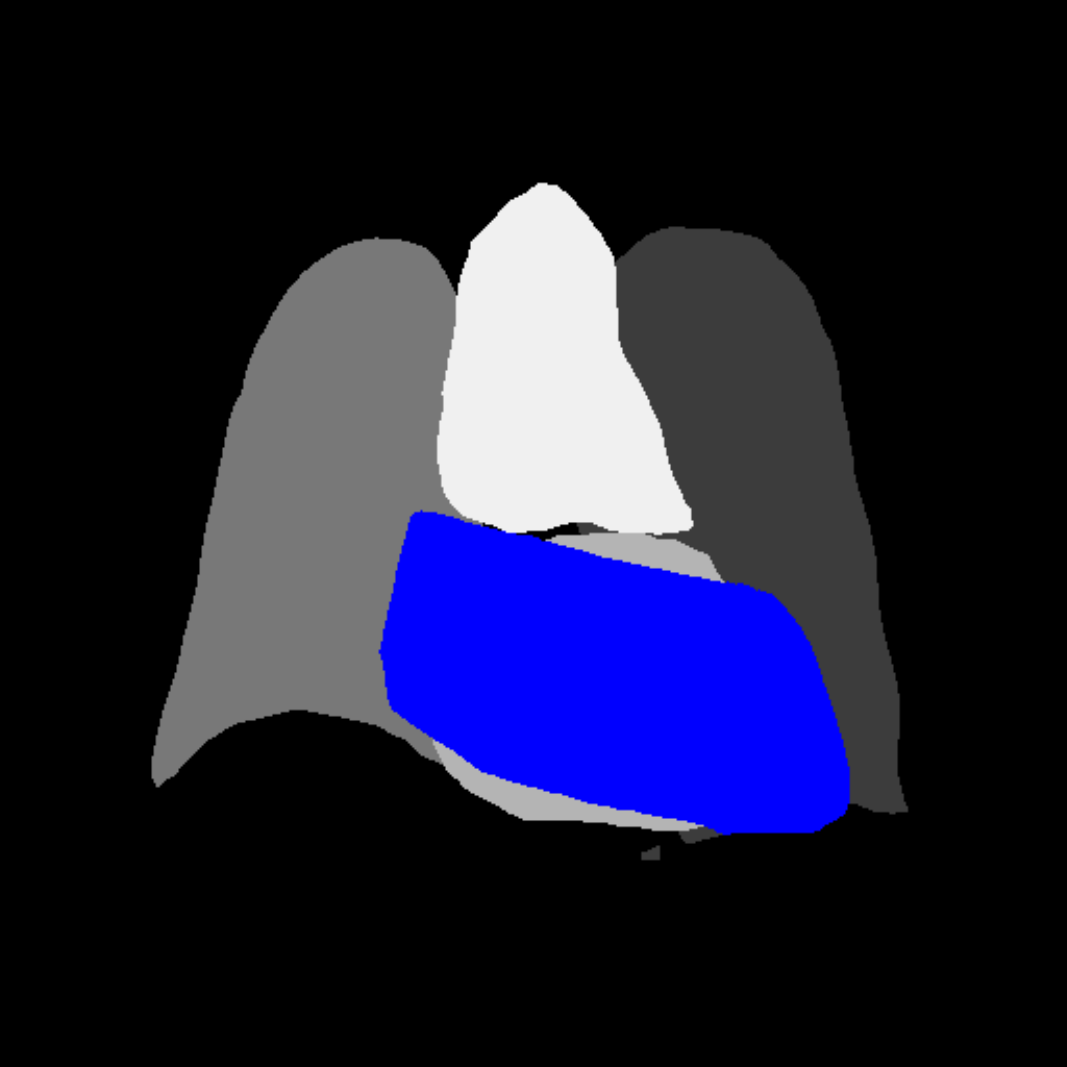}
        \centerline{\parbox{1\linewidth}{\centering(c) Syn. Mask (Ours)}}
    \end{minipage}
    \begin{minipage}[t]{0.16\linewidth}
        \centering
        \includegraphics[width=\linewidth]{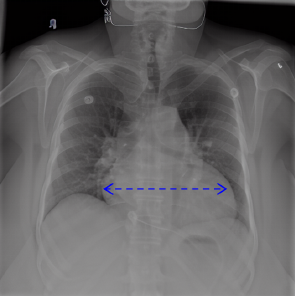}
        \centerline{\parbox{1\linewidth}{\centering(d) Syn. Image (Ours)}}
    \end{minipage}
    \begin{minipage}[t]{0.16\linewidth}
        \centering
        \includegraphics[width=\linewidth]{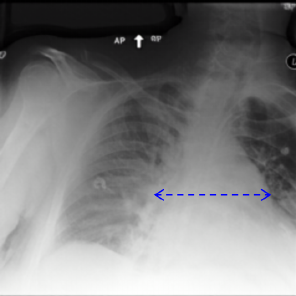}
        \centerline{\parbox{1\linewidth}{\centering(e) Cheff}}
    \end{minipage}
    \begin{minipage}[t]{0.16\linewidth}
        \centering
        \includegraphics[width=\linewidth]{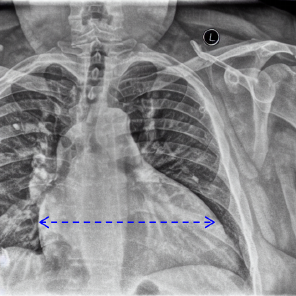}
        \centerline{\parbox{1\linewidth}{\centering(f) Roentgen}}
    \end{minipage}

\end{figure*}

\end{document}